\documentclass[]{interact}

\usepackage{epstopdf}
\usepackage{subfigure}

\usepackage{natbib}
\bibpunct[, ]{(}{)}{;}{a}{}{,}
\renewcommand\bibfont{\fontsize{10}{12}\selectfont}

\theoremstyle{plain}

\theoremstyle{definition}

\theoremstyle{remark}

\usepackage{url}
\begin{document}

\title{A Predictive Framework Integrating Multi-Scale Volatility Components and Time-Varying Quantile Spillovers: Evidence from the Cryptocurrency Market}

\author{
\name{A.~N. Author\textsuperscript{a}\thanks{CONTACT A.~N. Author. Email: latex.helpdesk@tandf.co.uk} and John Smith\textsuperscript{b}}
\affil{\textsuperscript{a}Taylor \& Francis, 4 Park Square, Milton Park, Abingdon, UK; \textsuperscript{b}Institut f\"{u}r Informatik, Albert-Ludwigs-Universit\"{a}t, Freiburg, Germany}
}

\author{
\name{Sicheng Fu\textsuperscript{a}, Fangfang Zhu\textsuperscript{a} and Xiangdong Liu\textsuperscript{a}\thanks{CONTACT Xiangdong Liu. Email: tliuxd@jnu.edu.cn}}
\affil{\textsuperscript{a}School of Economics, Jinan University, No. 601, West Huangpu Boulevard, Tianhe District, Guangzhou 510632, China}
}
\maketitle

\begin{abstract}
This paper investigates the dynamics of risk transmission in cryptocurrency markets and proposes a novel framework for volatility forecasting. The framework uncovers two key empirical facts: the asymmetric amplification of volatility spillovers in both tails, and a structural decoupling between market size and systemic importance. Building on these insights, we develop a state-adaptive volatility forecasting model by extracting time-varying quantile spillover features across different volatility components. These features are embedded into an extended Log-HAR structure, resulting in the SA-Log-HAR model. Empirical results demonstrate that the proposed model outperforms benchmark alternatives in both in-sample fitting and out-of-sample forecasting, particularly in capturing extreme volatility and tail risks with greater robustness and explanatory power.
\end{abstract}

\vspace{0.5em}

\section{Introduction}
Amid pandemic shocks, geopolitical tensions, and global monetary adjustments, cryptocurrencies have gained recognition as a distinct asset class with independent market logic and low correlation with traditional financial assets (\citet{Bor19}). Since the inception of Bitcoin, the market has grown rapidly, forming an active structure centered on Bitcoin and Ethereum (\citet{Yan23}).

Cryptocurrency markets exhibit high volatility, pronounced intraday patterns, and long memory, posing challenges for traditional GARCH-type models (\citet{Eng82,Bol86}) in capturing high-frequency dynamics. The HAR-RV model (\cite{Cor09}), incorporating multi-scale realized volatilities (daily, weekly, monthly), has become the standard for modeling long-term dependencies and shows strong predictive performance in this domain.

To enhance HAR-RV flexibility, recent studies have pursued two main extensions: (1) refining volatility decomposition using jumps (\cite{And07}), semi-variances (\cite{Bar08b}), and extreme indicators (\cite{Cle19}); and (2) incorporating macroeconomic and external information to improve explanatory power (\cite{Gup23}).

However, current models generally overlook the dynamic volatility factors driven by the internal structure of cryptocurrency markets, particularly the role of cross-asset volatility spillover effects in high-frequency environments. Given the high degree of interconnectedness and rapid information transmission among crypto assets, such spillover effects are likely to be crucial variables affecting volatility forecasting (\cite{Die09}), yet systematic incorporation into the HAR framework remains scarce.

This paper aims to integrate cross-asset volatility spillover variables into the HAR-RV model, quantifying their incremental effect on volatility forecasting in high-frequency environments, thereby providing new empirical support for revealing volatility transmission mechanisms in cryptocurrency markets.

The remainder of this paper is organized as follows. Section 2 presents the theoretical foundation of time-varying quantile spillovers and the construction methodology for state-adaptive spillover variables. Section 3 introduces all baseline models employed in this study and the methods for model combination. Section 4 conducts empirical analysis, evaluating the predictive capabilities of extended models using various statistical tests including the Model Confidence Set (MCS) test (\cite{Han03}) and out-of-sample R² test (\cite{Cam08}), and reports and discusses the results. Finally, Section 5 concludes, demonstrating that the proposed SA-Log-HAR models exhibit superior predictive performance.

\section{Methodology}\label{sec:methodology}
This paper adopts the time-varying quantile spillover framework developed by \citet{Ant20} to capture the state-dependent characteristics of cross-market risk transmission. This approach combines the Time-Varying Parameter Quantile Vector Autoregression (TVP-QVAR) model with Kalman filtering and the Generalized Forecast Error Variance Decomposition (GFEVD), producing a dynamic spillover matrix, $\tilde{\Phi}_t^g(H, \tau)$, that evolves jointly over time and quantiles. On this basis, we analyze several key connectedness measures: the Total Spillover Index (TSI), Net Spillover Index (NSI), directional spillovers (from others to asset $i$, $S_{\cdot \leftarrow i}$, and from asset $i$ to others, $S_{i \leftarrow \cdot}$), net spillovers ($S_{i,\text{net}}$), and the Net Pairwise Directional Connectedness (NPDC), which collectively offer a detailed characterization of the magnitude and directionality of inter-asset risk transmission.

\subsection{Construction of the State-Adaptive Spillover Variable}
\label{sec:construct_spillover_variable}
To integrate spillover network information into volatility forecasting, we construct a novel state-adaptive spillover variable $X_{i,t}$ , which dynamically adjusts to market conditions and captures the most influential risk transmission pathway, unlike traditional models with fixed spillover sources.
Its construction involves three main steps:

\begin{itemize}
    \item \textbf{Market State Classification:} The time series of a volatility-related characteristic variable of the target asset $i$, denoted as $V_{i,t}$ (e.g., Realized Volatility $(RV)$, Realized Semivariance $(RS)$, or Continuous Jumps $(CJ)$), is partitioned into three distinct market states: low, normal, and high. The market state variable $\mathcal{S}_{i,t}$ is defined as:
    \begin{equation} \label{eq:state_definition}
    \mathcal{S}_{i,t} =
    \begin{cases}
    \text{Low} & \text{if } V_{i,t} \leq Q_{V_i}(\tau_L) \\
    \text{High} & \text{if } V_{i,t} \geq Q_{V_i}(\tau_H) \\
    \text{Normal} & \text{otherwise}
    \end{cases}
    \end{equation}
    where $Q_{V_i}(\tau)$ denotes the $\tau$-quantile of $V_{i,t}$, with $\tau_L$ and $\tau_H$ representing the lower (e.g., 0.05) and upper (e.g., 0.95) quantile thresholds, respectively.
    
    \item \textbf{Identification of State-Dependent Spillover Sources:} For each market state, the dominant spillover contributor to asset $i$ is identified using the Net Pairwise Directional Connectedness (NPDC) index. Given quantile $\tau$, the most influential source $j^*_{(i, \tau)}$ is defined as:
    \begin{equation}
    j^*_{(i, \tau)} = \underset{j \neq i}{\operatorname{argmax}} \left( \text{NPDC}_{j \to i}(H, \tau) \right)
    \end{equation}
    This step identifies the leading spillover source under low ($j^*_{(i, \tau_L)}$), high ($j^*_{(i, \tau_H)}$), and normal ($j^*_{(i, \tau_M)}$, with $\tau_M = 0.5$) market conditions.
    
    \item \textbf{Final Construction of the Spillover Variable:} The final state-adaptive spillover variable $X_{i,t}$ is constructed by linking the identified dominant source to the contemporaneous market state. Specifically, $X_{i,t}$ is defined as the one-period lagged value of the volatility measure for the selected spillover source, conditional on the state:
    \begin{equation} \label{eq:state_adaptive_variable}
    X_{i,t} =
    \begin{cases}
    V_{j^*_{(i, \tau_L)}, t-1} & \text{if } \mathcal{S}_{i,t} = \text{Low} \\
    V_{j^*_{(i, \tau_H)}, t-1} & \text{if } \mathcal{S}_{i,t} = \text{High} \\
    V_{j^*_{(i, \tau_M)}, t-1} & \text{if } \mathcal{S}_{i,t} = \text{Normal}
    \end{cases}
    \end{equation}
\end{itemize}

The key feature of $X_{i,t}$ lies in its ability to incorporate time-varying information from the most relevant source of systemic influence, conditional on the prevailing volatility regime. This design enables the forecasting model to better reflect the heterogeneous nature of market states and the evolving structure of cross-asset risk, thereby enhancing its economic interpretability and predictive relevance.\footnote{The implementation code is available at \url{https://github.com/wangxiaobo018/State-Adaptive-Spillover.git}.}

\section{Model}\label{sec:3}

The high heterogeneity and non-normality of cryptocurrency market volatility pose stability challenges for traditional HAR-RV models under extreme values. To improve robustness, \citet{Cor09} proposed the Log-HAR-RV model, which applies a logarithmic transformation to realized volatility ($\log(RV_{i,t})$), smoothing fluctuations and better capturing long-term dependencies. Model selection can thus be adapted to volatility characteristics, with the Log-HAR-RV form particularly suited to heterogeneous markets and cross-asset linkages.

\begin{equation}
\label{eq:unified-har}
\log RV_{i,t} = \beta_0 + \sum_{k=1}^{K} \sum_{h \in \{1,5,22\}} \beta_{k,h} \log \overline{V}^{(k)}_{i,t-h} + \sum_{k=1}^{K} \gamma_k X^{\log V^{(k)}}_{j,t-1} + \epsilon_t.
\end{equation}

where: $V^{(k)}$ denotes the $k$-th volatility component; $\overline{V}^{(k)}_{i,t-h}$ represents the average value of the $k$-th component over $h$ periods (single period value when $h=1$); $K$ is the number of volatility components; and $X^{\log V^{(k)}}_{j,t-1}$ represents the quantile spillover feature of the corresponding component.

\begin{figure}[htbp]
    \centering 
    \includegraphics[width=0.4\textwidth]{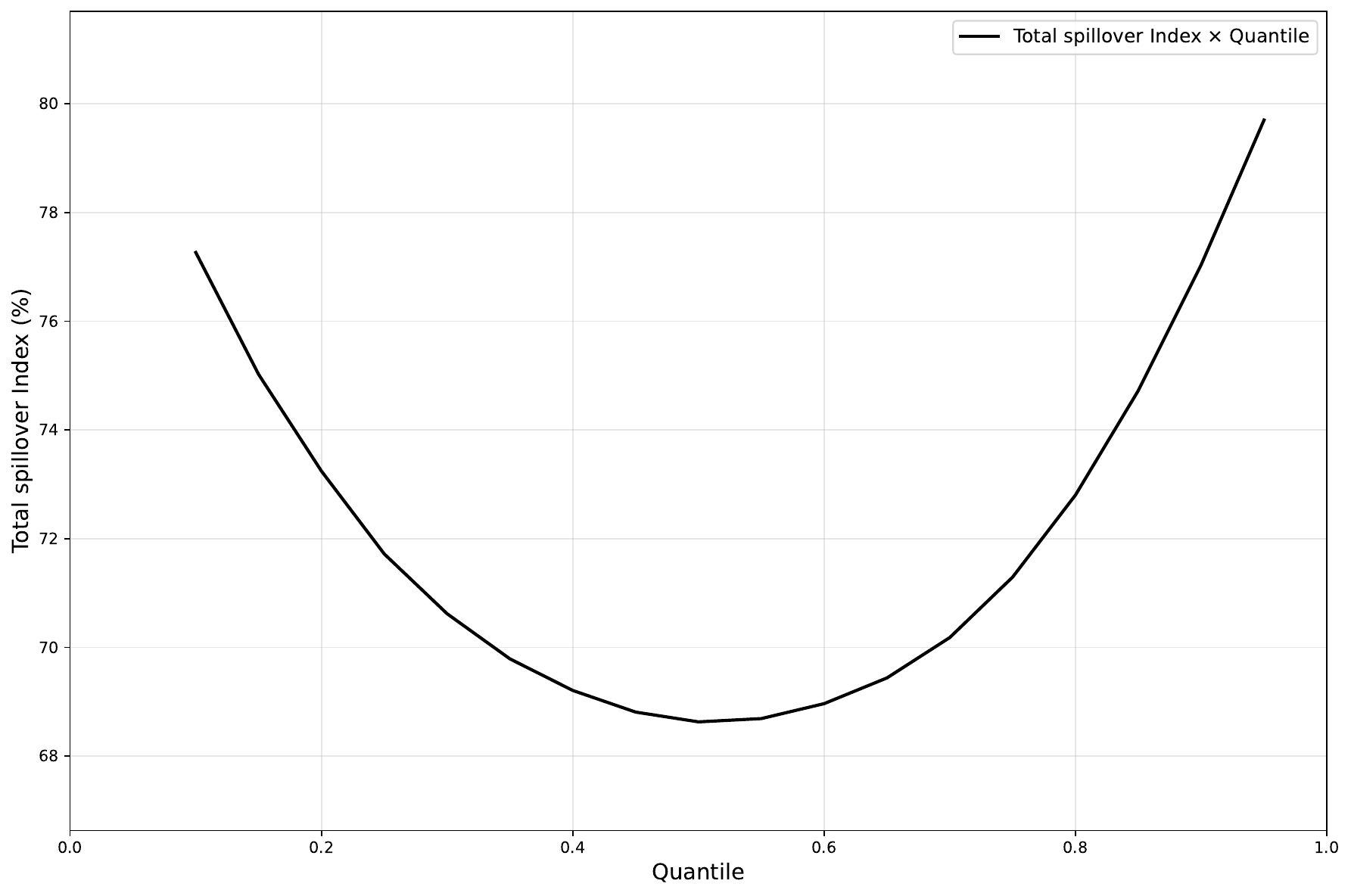} 
    \caption{Total spillovers of realized volatility at different quantiles} 
    \label{tab:u} 
\end{figure}

\section{Empirical results}\label{实证}
\subsection{In-sample parameter estimation results}\label{样本}
We use 5-minute high-frequency data for six cryptocurrencies—BTC, DASH, ETH, LTC, XLM, and XRP—sourced from Binance (\url{https://www.binance.com}), a leading global exchange. The sample spans March 28, 2019 to March 30, 2025, with the last 300 days reserved for out-of-sample forecasting and the remainder used for model estimation.

\begin{table}[!htbp]
\tbl{In-sample parameter estimation results for different volatility models}
{\resizebox{\linewidth}{!}{%
\begin{tabular}{lcccccccc}
  \toprule
  \textbf{Parameter} & {Log-HAR-RV} & {Log-HAR-CJ} & {Log-HAR-RS} &{Log-HAR-REX} & {SA-Log-HAR-RV} & {SA-Log-HAR-CJ} & {SA-Log-HAR-RS} & {SA-Log-HAR-REX} \\
  \midrule
  $\beta_0$        & 0.007$^{***}$ & 0.283$^{***}$ & 0.025 & 0.978$^{***}$ & 0.125$^{***}$ & 0.233$^{***}$ & -0.016 & 0.982$^{***}$ \\
                              & (0.002) & (0.065) & (0.051) & (0.068) & (0.051) & (0.066) & (0.054) & (0.070) \\
  \addlinespace
  $\beta_1$                   & -0.028 & 0.033 & 0.401$^{***}$ & 0.053 & 0.405$^{***}$ & 0.025 & 0.513$^{***}$ & -0.058 \\
                              & (0.027) & (0.027) & (0.057) & (0.039) & (0.039) & (0.029) & (0.075) & (0.050) \\
  \addlinespace
  $\beta_2$                   & 0.042 & -0.084 & -0.033 & 0.054 & 0.188$^{***}$ & -0.082 & -0.013 & 0.119$^{**}$ \\
                              & (0.069) & (0.062) & (0.144) & (0.096) & (0.041) & (0.062) & (0.144) & (0.050) \\
  \addlinespace
  $\beta_3$                   & 0.742$^{***}$ & 0.004 & 0.908$^{***}$ & -0.225 & 0.217$^{***}$ & 0.011 & 0.894$^{***}$ & 0.390$^{***}$ \\
                              & (0.084) & (0.094) & (0.304) & (0.152) & (0.037) & (0.094) & (0.304) & (0.079) \\
  \addlinespace
  $\beta_4$                   & & 0.506$^{***}$ & 0.242$^{***}$ & 0.103$^{**}$ & 0.107$^{***}$ & 0.418$^{***}$ & 0.051 & 0.053 \\
                              & & (0.031) & (0.055) & (0.042) & (0.034) & (0.043) & (0.078) & (0.096) \\
  \addlinespace
  $\beta_5$                   & & 0.190$^{***}$ & 0.276$^{*}$ & -0.055 & & 0.187$^{***}$ & 0.259$^{*}$ & -0.042 \\
                              & & (0.051) & (0.144) & (0.103) & & (0.051) & (0.144) & (0.102) \\
  \addlinespace
  $\beta_6$                   & & 0.227$^{***}$ & -0.580$^{*}$ & -0.051 & & 0.219$^{***}$ & -0.566$^{*}$ & 0.136 \\
                              & & (0.056) & (0.307) & (0.171) & & (0.056) & (0.306) & (0.160) \\
  \addlinespace
  $\beta_7$                   & & & & 0.394$^{***}$ & & 0.099$^{***}$ & -0.142$^{**}$ & -0.191 \\
                              & & & & (0.076) & & (0.035) & (0.063) & (0.152) \\
  \addlinespace
  $\beta_8$                   & & & & 0.142 & & 0.022 & 0.224$^{***}$ & -0.017 \\
                              & & & & (0.160) & & (0.024) & (0.067) & (0.171) \\
  \addlinespace
  $\beta_9$                   & & & & 0.440$^{**}$ & & & & 0.374$^{*}$ \\
                              & & & & (0.204) & & & & (0.206) \\
  \addlinespace
  $\beta_{10}$                & & & & & & & & -0.030 \\
                              & & & & & & & & (0.047) \\
  \addlinespace
  $\beta_{11}$                & & & & & & & & 0.144$^{***}$ \\
                              & & & & & & & & (0.043) \\
  \addlinespace
  $\beta_{12}$                & & & & & & & & -0.013 \\
                              & & & & & & & & (0.040) \\
  \midrule
  \textbf{$Adj.R^2$}     & 0.569 & 0.575 & 0.566 & 0.578 & 0.574 & 0.577 & 0.568 & 0.578 \\
  \textbf{AIC}                & 3915 & 3909  & 3947  & 3900  & 3914  & 3903  & 3939  & 3892  \\
  \textbf{BIC}                & 3949 & 3948  & 3986  & 3955  & 3941  & 3942  & 3986  & 3960  \\
  \bottomrule
\end{tabular}}}
\tabnote{{Note:} This table reports the in-sample estimation results for eight different volatility models. Standard errors are in parentheses. Asterisks $^{*}$, $^{**}$, and $^{***}$ denote significance at the 10\%, 5\%, and 1\% levels, respectively.}
\label{tab:1}
\end{table}

\begin{table}
\tbl{Summary of net spillover contributions to BTC from other cryptocurrencies under different volatility features (\%)}
{\begin{tabular*}{\textwidth}{@{}l@{\extracolsep{\fill}}lrrrrrrrr@{}} \toprule
\textbf{Cryptocurrency}  & $RV$ & $CV$ & $CJ$ & $RS^{+}$ & $RS^{-}$ & $REX^{+}$ & $REX^{-}$ & $REX^{m}$ \\ 
\midrule
DASH & $\tau=0.05$      & 12.16 & 12.38 & 14.34 & 15.55 & 15.79 & 11.83 & 11.83 & 9.41  \\
     & $\tau=0.10$      & 12.45 & 12.98 & 14.76 & 15.76 & 16.24 & 12.20 & 12.20 & 11.68 \\
     & $\tau=0.50$      & 11.72 & 12.91 & 10.33 & 14.36 & 16.14 & 11.66 & 11.66 & 11.29 \\
     & $\tau=0.90$      & 15.04 & 15.91 & 24.26 & 16.45 & 15.35 & 14.71 & 14.71 & 14.61 \\
     & $\tau=0.95$      & 16.73 & 17.09 & 19.26 & 15.90 & 16.45 & 16.52 & 16.52 & 14.39 \\
     & Cyclicality      & +4.57 & +4.71 & +4.92 & +0.36 & +0.66 & +4.68 & +4.68 & +4.98 \\ \midrule
ETH  & $\tau=0.05$      & 17.62 & 17.70 & 13.63 & 17.20 & 17.11 & 19.09 & 19.09 & 17.75 \\
     & $\tau=0.10$      & 17.87 & 17.25 & 14.67 & 17.18 & 16.54 & 18.42 & 18.42 & 17.26 \\
     & $\tau=0.50$      & 19.25 & 19.52 & 16.20 & 19.77 & 18.09 & 19.72 & 19.72 & 19.77 \\
     & $\tau=0.90$      & 15.62 & 16.31 & 13.43 & 14.85 & 16.40 & 17.76 & 17.76 & 15.93 \\
     & $\tau=0.95$      & 13.70 & 14.87 & 14.72 & 12.37 & 13.90 & 15.38 & 15.38 & 15.42 \\
     & Cyclicality      & -3.92 & -2.83 & +1.08 & -4.83 & -3.21 & -3.72 & -3.72 & -2.33 \\ \midrule
LTC  & $\tau=0.05$      & 15.02 & 14.37 & 12.88 & 16.60 & 17.44 & 15.69 & 15.69 & 13.39 \\
     & $\tau=0.10$      & 15.09 & 14.59 & 14.40 & 16.55 & 17.59 & 15.74 & 15.74 & 14.25 \\
     & $\tau=0.50$      & 15.16 & 15.82 & 14.12 & 16.54 & 16.95 & 14.26 & 14.26 & 12.85 \\
     & $\tau=0.90$      & 14.92 & 16.14 & 20.73 & 16.90 & 16.03 & 14.52 & 14.52 & 15.28 \\
     & $\tau=0.95$      & 14.67 & 15.24 & 21.20 & 15.72 & 16.93 & 15.81 & 15.81 & 14.52 \\
     & Cyclicality      & -0.35 & +0.87 & +8.32 & -0.87 & -0.51 & +0.12 & +0.12 & +1.13 \\ \midrule
XLM  & $\tau=0.05$      & 11.90 & 12.98 & 9.08  & 14.84 & 15.00 & 12.73 & 12.73 & 12.80 \\
     & $\tau=0.10$      & 12.53 & 13.16 & 8.05  & 14.91 & 15.40 & 12.64 & 12.64 & 12.22 \\
     & $\tau=0.50$      & 10.15 & 10.24 & 7.31  & 12.40 & 12.94 & 10.26 & 10.26 & 9.52  \\
     & $\tau=0.90$      & 17.64 & 18.68 & 16.46 & 18.44 & 18.34 & 14.69 & 14.69 & 17.29 \\
     & $\tau=0.95$      & 17.80 & 15.39 & 16.41 & 21.74 & 20.10 & 18.81 & 18.81 & 21.09 \\
     & Cyclicality      & +5.90 & +2.40 & +7.33 & +6.90 & +5.10 & +6.07 & +6.07 & +8.29 \\ \midrule
XRP  & $\tau=0.05$      & 17.38 & 17.60 & 9.40  & 18.38 & 18.30 & 16.98 & 16.98 & 17.98 \\
     & $\tau=0.10$      & 16.79 & 17.05 & 8.28  & 17.70 & 17.90 & 17.05 & 17.05 & 18.19 \\
     & $\tau=0.50$      & 15.39 & 14.87 & 8.07  & 14.51 & 16.22 & 14.45 & 14.45 & 14.82 \\
     & $\tau=0.90$      & 21.57 & 19.21 & 16.44 & 20.32 & 20.06 & 22.43 & 22.43 & 22.46 \\
     & $\tau=0.95$      & 24.23 & 23.13 & 17.90 & 21.81 & 20.49 & 21.10 & 21.10 & 23.33 \\
     & Cyclicality      & +6.85 & +5.53 & +8.51 & +3.44 & +2.19 & +4.12 & +4.12 & +5.36 \\ \bottomrule
\end{tabular*}}
\tabnote{Note: A positive net spillover value indicates that the major cryptocurrency is a net transmitter of spillover, while a negative value indicates it is a net receiver.}
\label{tab:2}
\end{table}

Figure \ref{tab:u} presents the total spillover results of logarithmic realized volatility at different quantiles. The results reveal significant tail dependence and asymmetry in volatility spillovers. Under the median condition of the model, system spillovers exhibit high-level stability; however, at both tails of the model, spillover effects display amplification characteristics. Whether in the left tail or right tail, system interconnectedness increases dramatically. Total spillovers for other volatility characteristics are shown in Appendix \ref{附录} Figures \ref{fig:rv_plots}-\ref{fig:rexm_plots}.

The value of the state-adaptive spillover variable constructed based on time-varying quantile spillovers is clearly demonstrated in Table \ref{tab:1}. Compared to the benchmark models, all SA-Log-HAR-X models incorporating this new variable achieve higher adjusted $R^2$ values, along with improved AIC and BIC scores. These results indicate that the newly constructed spillover variable provides significant incremental information for volatility forecasting and contributes to the development of more efficient predictive models.

\subsection{Out-of-sample forecasting results}\label{样本外预测结果}

To identify the optimal source for state-adaptive spillover variables (Section~\ref{sec:construct_spillover_variable}), we first computed each cryptocurrency’s net spillover contribution to BTC, as shown in Table~\ref{tab:2}.

The assets with the highest contributions at quantiles $\tau$ = 0.05, 0.5, and 0.95 were then selected to construct state-dependent explanatory variables. Notably, Table~\ref{tab:2} reveals a structural asymmetry in the crypto market: systemic influence is shaped more by functional centrality than market size.

While Ethereum (ETH), despite its dominant capitalization, acts mainly as a net receiver, smaller-cap assets like Ripple (XRP) and Litecoin (LTC) emerge as strong net transmitters. This indicates that volatility spillovers are driven by an asset's role in the risk network rather than its scale. Additional spillover results by quantile and volatility type are presented in Appendix~\ref{附录}, Figures~\ref{fig:rv_net_spillover_by_coin}–\ref{fig:rexmod_net_spillover_by_coin}, with contribution rates summarized in Figure~\ref{fig:contribute}.

\begin{table}[htbp]
\tbl{MCS test results for the 300-day out-of-sample forecasts}
{\begin{tabular}{@{}llcccccccc c@{}} 
\toprule
\textbf{Horizon} & \textbf{Model} & 
\multicolumn{2}{c}{$MSE$} & 
\multicolumn{2}{c}{$MAE$} & 
\multicolumn{2}{c}{$RMSE$} & 
\multicolumn{2}{c}{$QLIKE$} &
${Ratio}$ \\
\cmidrule(lr){3-4} \cmidrule(lr){5-6} \cmidrule(lr){7-8} \cmidrule(lr){9-10}
& & $T_{max}$ & $T_{R}$ & $T_{max}$ & $T_{R}$ & $T_{max}$ & $T_{R}$ & $T_{max}$ & $T_{R}$ & \\
\midrule
1 & Log-HAR-RV               & 7 & 7 & \textbf{1} & \textbf{1} & 7 & 7 & 2 & 2 & \textbf{2/8} \\
      & Log-HAR-CJ               & 11& 11& 7 & 7 & 11& 11& 11& 11& 0/8 \\
      & Log-HAR-RS               & 9 & 9 & 8 & 8 & 9 & 9 & 10& 10& 0/8 \\
      & Log-HAR-REX              & 4 & 4 & 4 & 4 & 4 & 4 & 4 & 4 & 0/8 \\
      & SA-Log-HAR-RV            & 3 & 3 & 2 & 2 & 3 & 3 & \textbf{1} & \textbf{1} & \textbf{2/8} \\
      & SA-Log-HAR-CJ            & 10& 10& 11& 11& 10& 10& 8 & 8 & 0/8 \\
      & SA-Log-HAR-RS            & \textbf{1} & \textbf{1} & 3 & 3 & \textbf{1} & \textbf{1} & 5 & 5 & \textbf{4/8} \\
      & SA-Log-HAR-REX           & 6 & 6 & 10& 10& 6 & 6 & 9 & 9 & 0/8 \\
      & Lasso-SA-Log-HAR-CJ      & 8 & 8 & 9 & 9 & 8 & 8 & 3 & 3 & 0/8 \\
      & Lasso-SA-Log-HAR-RS      & 2 & 2 & 5 & 5 & 2 & 2 & 7 & 7 & 0/8 \\
      & Lasso-SA-Log-HAR-REX     & 5 & 5 & 6 & 6 & 5 & 5 & 6 & 6 & 0/8 \\
\midrule
5 & Log-HAR-RV               & 9 & 9 & 4 & 4 & 9 & 9 & 4 & 4 & 0/8 \\
      & Log-HAR-CJ               & 3 & 3 & 6 & 6 & 3 & 3 & 7 & 7 & 0/8 \\
      & Log-HAR-RS               & 8 & 8 & 9 & 9 & 8 & 8 & 11& 11& 0/8 \\
      & Log-HAR-REX              & 11& 11& 11& 11& 11& 11& 9 & 9 & 0/8 \\
      & SA-Log-HAR-RV            & 5 & 5 & \textbf{1} & \textbf{1} & 5 & 5 & 2 & 2 & \textbf{2/8} \\
      & SA-Log-HAR-CJ            & 10& 10& 10& 10& 10& 10& 10& 10& 0/8 \\
      & SA-Log-HAR-RS            & 2 & 2 & 2 & 2 & 2 & 2 & \textbf{1} & \textbf{1} & \textbf{2/8} \\
      & SA-Log-HAR-REX           & 4 & 4 & 7 & 7 & 4 & 4 & 5 & 5 & 0/8 \\
      & Lasso-SA-Log-HAR-CJ      & 6 & 6 & 8 & 8 & 6 & 6 & 6 & 6 & 0/8 \\
      & Lasso-SA-Log-HAR-RS      & \textbf{1} & \textbf{1} & 3 & 3 & \textbf{1} & \textbf{1} & 3 & 3 & \textbf{4/8} \\
      & Lasso-SA-Log-HAR-REX     & 7 & 7 & 5 & 5 & 7 & 7 & 8 & 8 & 0/8 \\
\midrule
22 & Log-HAR-RV              & 10& 10& \textbf{1} & \textbf{1} & 10& 10& \textbf{1} & \textbf{1} & \textbf{4/8} \\
       & Log-HAR-CJ              & 11& 11& 11& 11& 11& 11& 8 & 8 & 0/8 \\
       & Log-HAR-RS              & 9 & 9 & 8 & 8 & 9 & 9 & 11& 11& 0/8 \\
       & Log-HAR-REX             & 4 & 4 & 6 & 6 & 4 & 4 & 4 & 4 & 0/8 \\
       & SA-Log-HAR-RV           & \textbf{1} & \textbf{1} & 2 & 2 & \textbf{1} & \textbf{1} & 2 & 2 & \textbf{4/8} \\
       & SA-Log-HAR-CJ           & 8 & 8 & 9 & 9 & 8 & 8 & 10& 10& 0/8 \\
       & SA-Log-HAR-RS           & 2 & 2 & 3 & 3 & 2 & 2 & 3 & 3 & 0/8 \\
       & SA-Log-HAR-REX          & 6 & 6 & 10& 10& 6 & 6 & 7 & 7 & 0/8 \\
       & Lasso-SA-Log-HAR-CJ     & 5 & 5 & 7 & 7 & 5 & 5 & 6 & 6 & 0/8 \\
       & Lasso-SA-Log-HAR-RS     & 3 & 3 & 4 & 4 & 3 & 3 & 5 & 5 & 0/8 \\
       & Lasso-SA-Log-HAR-REX    & 7 & 7 & 5 & 5 & 7 & 7 & 9 & 9 & 0/8 \\
\bottomrule
\end{tabular}}
\tabnote{Note: $T_{max}$ and $T_R$ denote two MCS test statistics. The values reflect model rankings under different loss functions, with bold 1 indicating the best. The Ratio column shows how often each model appears in the superior set across all eight tests.}
\label{tab:300}
\end{table}

Table \ref{tab:300} presents the MCS test results at the 0.25 significance level for the Log-HAR, SA-Log-HAR, and Lasso-SA-Log-HAR models. The results show that the SA-Log-HAR-RS model performs best in one-step-ahead forecasting, while the Lasso-SA-Log models exhibit the most consistent advantage in five-step forecasts. In the 22-step forecasts, the Log-HAR-RV and SA-Log-HAR-RV models share the highest number of superior outcomes. These findings can be interpreted in light of the theory proposed by \citet{Lah18}, which suggests that the Bitcoin market exhibits chaotic and multifractal properties. As a result, decomposing realized volatility into components such as jumps or signed variations may introduce substantial estimation errors in long-term forecasts.
 Moreover, the autoregressive behavior of these components tends to be unstable across different market regimes. In contrast, the more parsimonious and robust HAR-RV model, by directly modeling aggregated volatility, effectively avoids model misspecification stemming from unstable microstructure-based decompositions, thereby demonstrating greater stability and competitiveness in long-term forecasting.

\begin{table}[htbp]
\tbl{MCS test results for the 500-day out-of-sample forecasts}
{\begin{tabular}{@{}llcccccccc c@{}} 
\toprule
\textbf{Horizon} & \textbf{Model} & 
\multicolumn{2}{c}{$MSE$} & 
\multicolumn{2}{c}{$MAE$} & 
\multicolumn{2}{c}{$RMSE$} & 
\multicolumn{2}{c}{$QLIKE$} &
${Ratio}$ \\
\cmidrule(lr){3-4} \cmidrule(lr){5-6} \cmidrule(lr){7-8} \cmidrule(lr){9-10}
& & $T_{max}$ & $T_{R}$ & $T_{max}$ & $T_{R}$ & $T_{max}$ & $T_{R}$ & $T_{max}$ & $T_{R}$ & \\
\midrule
1 & Log-HAR-RV & 7 & 7 & \textbf{1} & \textbf{1} & 7 & 7 & 2 & 2 & \textbf{2/8} \\
  & Log-HAR-CJ & 11 & 11 & 7 & 7 & 11 & 11 & 11 & 11 & 0/8 \\
  & Log-HAR-RS & 9 & 9 & 8 & 8 & 9 & 9 & 10 & 10 & 0/8 \\
  & Log-HAR-REX & 4 & 4 & 4 & 4 & 4 & 4 & 4 & 4 & 0/8 \\
  & SA-Log-HAR-RV & 3 & 3 & 2 & 2 & 3 & 3 & \textbf{1} & \textbf{1} & \textbf{2/8} \\
  & SA-Log-HAR-CJ & 6 & 6 & 10 & 10 & 6 & 6 & 9 & 9 & 0/8 \\
  & SA-Log-HAR-RS & \textbf{1} & \textbf{1} & 3 & 3 & \textbf{1} & \textbf{1} & 5 & 5 & \textbf{4/8} \\
  & SA-Log-HAR-REX & 10 & 10 & 11 & 11 & 10 & 10 & 8 & 8 & 0/8 \\
  & Lasso-SA-Log-HAR-CJ & 8 & 8 & 9 & 9 & 8 & 8 & 3 & 3 & 0/8 \\
  & Lasso-SA-Log-HAR-RS & 5 & 5 & 6 & 6 & 5 & 5 & 7 & 7 & 0/8 \\
  & Lasso-SA-Log-HAR-REX & 2 & 2 & 5 & 5 & 2 & 2 & 6 & 6 & 0/8 \\
\midrule
5 & Log-HAR-RV & 9 & 9 & 4 & 4 & 9 & 9 & 4 & 4 & 0/8 \\
  & Log-HAR-CJ & 3 & 3 & 6 & 6 & 3 & 3 & 7 & 7 & 0/8 \\
  & Log-HAR-RS & 8 & 8 & 9 & 9 & 8 & 8 & 11 & 11 & 0/8 \\
  & Log-HAR-REX & 11 & 11 & 11 & 11 & 11 & 11 & 9 & 9 & 0/8 \\
  & SA-Log-HAR-RV & 5 & 5 & \textbf{1} & \textbf{1} & 5 & 5 & 3 & 3 & \textbf{2/8} \\
  & SA-Log-HAR-CJ & 10 & 10 & 10 & 10 & 10 & 10 & 10 & 10 & 0/8 \\
  & SA-Log-HAR-RS & 2 & 2 & 3 & 3 & 2 & 2 & 2 & 2 & 0/8 \\
  & SA-Log-HAR-REX & 4 & 4 & 7 & 7 & 4 & 4 & 5 & 5 & 0/8 \\
  & Lasso-SA-Log-HAR-CJ & 6 & 6 & 8 & 8 & 6 & 6 & 6 & 6 & 0/8 \\
  & Lasso-SA-Log-HAR-RS & 7 & 7 & 5 & 5 & 7 & 7 & 8 & 8 & 0/8 \\
  & Lasso-SA-Log-HAR-REX & \textbf{1} & \textbf{1} & \textbf{1} & \textbf{1} & \textbf{1} & \textbf{1} & \textbf{1} & \textbf{1} & \textbf{8/8} \\
\midrule
22 & Log-HAR-RV & 2 & 2 & \textbf{1} & \textbf{1} & 2 & 2 & \textbf{1} & \textbf{1} & \textbf{4/8} \\
   & Log-HAR-CJ & 11 & 11 & 11 & 11 & 11 & 11 & 8 & 8 & 0/8 \\
   & Log-HAR-RS & 10 & 10 & 8 & 8 & 10 & 10 & 11 & 11 & 0/8 \\
   & Log-HAR-REX & 5 & 5 & 5 & 5 & 5 & 5 & 5 & 5 & 0/8 \\
   & SA-Log-HAR-RV & \textbf{1} & \textbf{1} & 2 & 2 & \textbf{1} & \textbf{1} & 2 & 2 & \textbf{4/8} \\
   & SA-Log-HAR-CJ & 9 & 9 & 9 & 9 & 9 & 9 & 10 & 10 & 0/8 \\
   & SA-Log-HAR-RS & 3 & 3 & 3 & 3 & 3 & 3 & 3 & 3 & 0/8 \\
   & SA-Log-HAR-REX & 7 & 7 & 10 & 10 & 7 & 7 & 7 & 7 & 0/8 \\
   & Lasso-SA-Log-HAR-CJ & 6 & 6 & 7 & 7 & 6 & 6 & 6 & 6 & 0/8 \\
   & Lasso-SA-Log-HAR-RS & 8 & 8 & 6 & 6 & 8 & 8 & 9 & 9 & 0/8 \\
   & Lasso-SA-Log-HAR-REX & 4 & 4 & 4 & 4 & 4 & 4 & 4 & 4 & 0/8 \\
\bottomrule
\end{tabular}}
\tabnote{Note: $T_{max}$ and $T_R$ denote two MCS test statistics. The values reflect model rankings under different loss functions, with bold 1 indicating the best. The “Ratio” column shows how often each model appears in the superior set across all eight tests.
}
\label{tab:500}
\end{table}

\subsection{robustness checks}
\begin{enumerate}
    \item \textbf{Varying the rolling window size.} In the main analysis (Section \ref{样本}) the final 300 days of the sample are used for rolling forecasts. To validate the model's reliability under different sample schemes, we adopt a fixed estimation window of 1000 days (from February 20, 2021, to November 16, 2023) and use the subsequent 500 days (from November 17, 2023, to March 30, 2025) as the forecast evaluation period.

    \item \textbf{Using an alternative volatility measure.} While the preceding sections primarily use realized volatility , we assess the model's robustness by substituting the realized kernel $(RK) $estimator proposed by \cite{Bar08a}. The $RK$ is known to be more robust to market microstructure noise and serves as a reliable alternative measure of volatility.

    \item \textbf{Applying an alternative evaluation metric.} To further corroborate the robustness of our forecasting performance, we employ the out-of-sample $R^2_{OOS}$ statistic proposed by \cite{Cam08}. This metric provides a standardized measure of predictive power compared to a benchmark model.
\end{enumerate}

\begin{table}[htbp]
\tbl{MCS test results of each model under the realized kernel estimator}
{\begin{tabular}{@{}llcccccccc c@{}} 
\toprule
\textbf{Horizon} & \textbf{Model} & 
\multicolumn{2}{c}{$MSE$} & 
\multicolumn{2}{c}{$MAE$} & 
\multicolumn{2}{c}{$RMSE$} & 
\multicolumn{2}{c}{$QLIKE$} &
${Ratio}$ \\
\cmidrule(lr){3-4} \cmidrule(lr){5-6} \cmidrule(lr){7-8} \cmidrule(lr){9-10}
& & $T_{max}$ & $T_{R}$ & $T_{max}$ & $T_{R}$ & $T_{max}$ & $T_{R}$ & $T_{max}$ & $T_{R}$ & \\
\midrule
1 & Log-HAR-RV & 5 & 5 & 5 & 5 & 5 & 5 & 9 & 9 & 0/8 \\
  & Log-HAR-CJ & 3 & 3 & \textbf{1} & \textbf{1} & 3 & 3 & 6 & 6 & \textbf{2/8} \\
  & Log-HAR-RS & 6 & 6 & 4 & 4 & 6 & 6 & 11 & 11 & 0/8 \\
  & Log-HAR-REX & 8 & 8 & 8 & 8 & 8 & 8 & 5 & 5 & 0/8 \\
  & SA-Log-HAR-RV & 4 & 4 & 6 & 6 & 4 & 4 & 8 & 8 & 0/8 \\
  & SA-Log-HAR-CJ & 11 & 11 & 11 & 11 & 11 & 11 & 3 & 3 & 0/8 \\
  & SA-Log-HAR-RS & 7 & 7 & 7 & 7 & 7 & 7 & 4 & 4 & 0/8 \\
  & SA-Log-HAR-REX & 2 & 2 & 3 & 3 & 2 & 2 & 7 & 7 & 0/8 \\
  & Lasso-SA-Log-HAR-CJ & \textbf{1} & \textbf{1} & 2 & 2 & \textbf{1} & \textbf{1} & 10 & 10 & \textbf{4/8} \\
  & Lasso-SA-Log-HAR-RS & 10 & 10 & 10 & 10 & 10 & 10 & 2 & 2 & 0/8 \\
  & Lasso-SA-Log-HAR-REX & 9 & 9 & 9 & 9 & 9 & 9 & \textbf{1} & \textbf{1} & \textbf{2/8} \\
\midrule
5 & Log-HAR-RV & 10 & 10 & 10 & 10 & 10 & 10 & 10 & 10 & 0/8 \\
  & Log-HAR-CJ & 2 & 2 & \textbf{1} & \textbf{1} & \textbf{1} & \textbf{1} & 7 & 7 & \textbf{4/8} \\
  & Log-HAR-RS & 4 & 4 & 4 & 4 & 4 & 4 & 8 & 8 & 0/8 \\
  & Log-HAR-REX & 11 & 11 & 11 & 11 & 11 & 11 & 11 & 11 & 0/8 \\
  & SA-Log-HAR-RV & 9 & 9 & 9 & 9 & 9 & 9 & 9 & 9 & 0/8 \\
  & SA-Log-HAR-CJ & 8 & 8 & 8 & 8 & 8 & 8 & 3 & 3 & 0/8 \\
  & SA-Log-HAR-RS & 6 & 6 & 5 & 5 & 6 & 6 & 4 & 4 & 0/8 \\
  & SA-Log-HAR-REX & 3 & 3 & 2 & 2 & 3 & 3 & 6 & 6 & 0/8 \\
  & Lasso-SA-Log-HAR-CJ & \textbf{1} & \textbf{1} & 3 & 3 & 2 & 2 & 5 & 5 & \textbf{2/8} \\
  & Lasso-SA-Log-HAR-RS & 7 & 7 & 7 & 7 & 7 & 7 & 2 & 2 & 0/8 \\
  & Lasso-SA-Log-HAR-REX & 5 & 5 & 6 & 6 & 5 & 5 & \textbf{1} & \textbf{1} & \textbf{2/8} \\
\midrule
22 & Log-HAR-RV & 2 & 2 & \textbf{1} & \textbf{1} & 2 & 2 & 11 & 11 & \textbf{2/8} \\
   & Log-HAR-CJ & 6 & 6 & 7 & 7 & 6 & 6 & 7 & 7 & 0/8 \\
   & Log-HAR-RS & 7 & 7 & 6 & 6 & 7 & 7 & 9 & 9 & 0/8 \\
   & Log-HAR-REX & 11 & 11 & 11 & 11 & 11 & 11 & 8 & 8 & 0/8 \\
   & SA-Log-HAR-RV & 8 & 8 & 8 & 8 & 8 & 8 & 10 & 10 & 0/8 \\
   & SA-Log-HAR-CJ & 10 & 10 & 10 & 10 & 10 & 10 & 3 & 3 & 0/8 \\
   & SA-Log-HAR-RS & \textbf{1} & \textbf{1} & 2 & 2 & \textbf{1} & \textbf{1} & 4 & 4 & \textbf{4/8} \\
   & SA-Log-HAR-REX & 5 & 5 & 5 & 5 & 5 & 5 & 6 & 6 & 0/8 \\
   & Lasso-SA-Log-HAR-CJ & 4 & 4 & 4 & 4 & 4 & 4 & 5 & 5 & 0/8 \\
   & Lasso-SA-Log-HAR-RS & 9 & 9 & 9 & 9 & 9 & 9 & 2 & 2 & 0/8 \\
   & Lasso-SA-Log-HAR-REX & 3 & 3 & 3 & 3 & 3 & 3 & \textbf{1} & \textbf{1} & \textbf{2/8} \\
\bottomrule
\end{tabular}}
\tabnote{Note: $T_{max}$ and $T_R$ denote two MCS test statistics. The values reflect model rankings under different loss functions, with bold 1 indicating the best. The “Ratio” column shows how often each model appears in the superior set across all eight tests.}
\label{tab:rk}
\end{table}

To assess the robustness of the findings, this paper alters the forecasting horizon and the fixed rolling window, and replaces the dependent variable from Realized Volatility  to the more noise-resistant Realized Kernel Volatility (RK). As shown in Tables \ref{tab:500} and \ref{tab:rk}, the ranking of model performance remains largely unchanged. The Lasso-SA-Log-HAR and SA-Log-HAR families continue to demonstrate superior performance in the short- and medium-term forecasts, with Lasso-Log-HAR-REX standing out most prominently in the medium term. For long-term forecasting, the results again confirm the robustness of the simpler Log-HAR-RV and SA-Log-HAR-RV models, highlighting the advantage of model parsimony in dealing with long-horizon uncertainty.

Table~\ref{tab:r2} reports out-of-sample $R^2$ ($R^2_{\mathrm{oos}}$) test results using GARCH as the benchmark. HAR-based models significantly outperform GARCH, with $R^2_{\mathrm{oos}}$ values around 0.77, while GARCH often yields negative values—highlighting the superiority of the HAR framework. Within the HAR family, models enhanced with state-adaptive spillover variables (SA) and Lasso regularization further boost predictive accuracy. The Lasso-SA-Log-HAR-RS model achieves the highest  $R^2_{\mathrm{oos}}$ under both 300- and 500-day windows (0.776 and 0.751). All HAR-type models show CW test p-values of 0.000, confirming statistically significant improvements. Despite a slight decline in performance over longer horizons, the Lasso-SA-Log-HAR-RS model consistently leads, demonstrating strong robustness.

\begin{table}[!htb]
\tbl{Out-of-sample $R^2$ evaluation for 300-day and 500-day forecasts}
{\resizebox{\textwidth}{!}{
\begin{tabular}{l *{4}{c} @{\hspace{0.5cm}} *{4}{c}}
\toprule
 & \multicolumn{4}{c}{300 days} & \multicolumn{4}{c}{500 days} \\
\cmidrule(lr){2-5} \cmidrule(lr){6-9}
Models & $R^2_{\mathrm{oos}}$ & MSPE-adjust & CW-Stat & $p$-value & $R^2_{\mathrm{oos}}$ & MSPE-adjust & CW-Stat & $p$-value \\
\midrule
\multicolumn{9}{l}{\textbf{GARCH type models}} \\
APARCH      & -0.225 & 0.005 & 0.019 & 0.492 & -0.126 & 0.223 & 1.813 & 0.035 \\
EGARCH      & -0.014 & 0.833 & 3.421 & 0.000 & -0.776 & 0.226 & 0.415 & 0.339 \\
FIGARCH     &  0.243 & 0.985 & 5.026 & 0.000 &  0.118 & 0.618 & 5.666 & 0.000 \\
GJR-GARCH   &  0.153 & 0.732 & 4.307 & 0.000 & -0.022 & 0.426 & 2.665 & 0.004 \\
\midrule
\multicolumn{9}{l}{\textbf{HAR type models}} \\
Log-HAR-RV           &  0.773 & 4.307 & 5.220 & 0.000 &  0.744 & 3.522 & 7.347 & 0.000 \\
Log-HAR-CJ           &  0.770 & 4.284 & 5.294 & 0.000 &  0.741 & 3.522 & 7.439 & 0.000 \\
Log-HAR-RS           &  0.775 & 4.247 & 5.130 & 0.000 &  0.749 & 3.472 & 7.307 & 0.000 \\
Log-HAR-REX          &  0.772 & 4.272 & 5.274 & 0.000 &  0.746 & 3.558 & 7.482 & 0.000 \\
SA-Log-HAR-RV        &  0.775 & 4.294 & 5.197 & 0.000 &  0.745 & 3.518 & 7.327 & 0.000 \\
SA-Log-HAR-CJ        &  0.771 & 4.275 & 5.268 & 0.000 &  0.743 & 3.517 & 7.433 & 0.000 \\
SA-Log-HAR-RS        &  0.776 & 4.246 & 5.140 & 0.000 &  0.751 & 3.485 & 7.303 & 0.000 \\
SA-Log-HAR-REX       &  0.773 & 4.275 & 5.266 & 0.000 &  0.747 & 3.560 & 7.481 & 0.000 \\
Lasso-SA-Log-HAR-CJ  &  0.772 & 4.271 & 5.265 & 0.000 &  0.744 & 3.516 & 7.423 & 0.000 \\
Lasso-SA-Log-HAR-RS  & \textbf{0.776} & 4.228 & 5.107 & 0.000 & \textbf{0.751} & 3.474 & 7.241 & 0.000 \\
Lasso-SA-Log-HAR-REX &  0.774 & 4.280 & 5.229 & 0.000 &  0.747 & 3.550 & 7.433 & 0.000 \\
\bottomrule
\end{tabular}}}
\tabnote{Note: $R^2_{\mathrm{oos}}$ denotes the out-of-sample coefficient of determination, measuring the improvement in predictive ability relative to the benchmark model. MSPE-adjust is the adjusted mean squared prediction error. CW-Stat is the t-statistic for the Clark-West test, testing whether the model is significantly superior to the nested benchmark model, with its $p$-value reflecting statistical significance. All values are rounded to three decimal places, with bold font indicating the optimal $R^2_{\mathrm{oos}}$ values.}
\label{tab:r2}
\end{table}

\section{Conclusion}\label{Conclusion}

This paper proposes a novel modeling framework that incorporates time-varying quantile spillover effects into realized volatility modeling. By identifying key features that significantly influence Bitcoin's realized volatility at different quantile levels, we construct state-adaptive spillover variables. Empirical results demonstrate that the extended Log-HAR models—namely, SA-Log-HAR and Lasso-SA-Log-HAR—consistently outperform benchmark models in both in-sample fitting and out-of-sample forecasting, exhibiting superior accuracy and stability. Robustness checks further confirm the reliability of the proposed approach.

In addition, the framework uncovers two critical empirical facts about the cryptocurrency market: (1) the presence of asymmetric two-tail amplification effects in volatility spillovers, indicating significant differences in the transmission mechanisms of downside and upside risks; and (2) a structural decoupling between market size and systemic importance, suggesting that cryptocurrencies with large market capitalizations are not necessarily central in the risk transmission network. These findings enrich our understanding of the structural characteristics of risk in cryptocurrency markets.

In long-term forecasting tasks, both the Log-HAR-RV and SA-Log-HAR-RV models demonstrate outstanding predictive performance, potentially attributable to the chaotic and multifractal nature of cryptocurrency markets—thereby providing strong empirical support for the theoretical insights of \citet{Lah18}.

In summary, this study not only extends the applicability of the Log-HAR class of models in volatility forecasting but also emphasizes the critical role of internal market spillover mechanisms in enhancing predictive performance. The findings offer more targeted and forward-looking insights for investors and policymakers.

\section*{Disclosure statement}
No potential conflict of interest was reported by the authors.

\section*{Funding}

This work is not supported by funding

\section*{Data available}
The 5-min high-frequency data supporting this study, spanning the period from 28 March  2019 to 30 March 2025, are available from Binance \url{www.binance.com} and from the corresponding author upon request. The code for implementing the time-varying quantile spillover model is publicly available {at:} 
 \url{https://github.com/wangxiaobo018/State-Adaptive-Spillover}  . The full text code uses Python version 3.12.4.







\section{Appendices}\label{附录}

\appendix
\section{Total spillovers of volatility features across different quantiles}

\begin{figure}[htbp]
\centering
\includegraphics[width=0.32\textwidth]{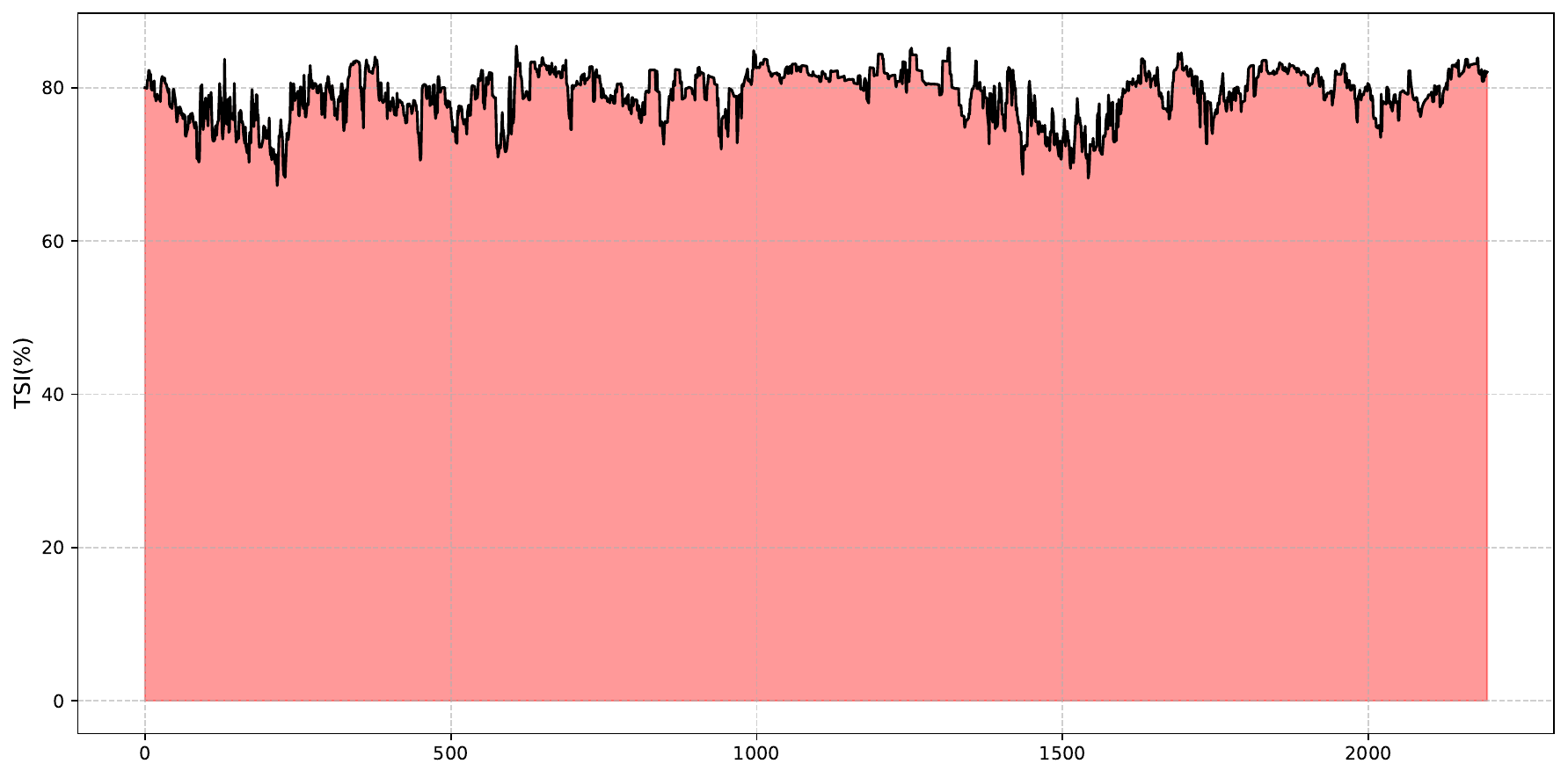}
\includegraphics[width=0.32\textwidth]{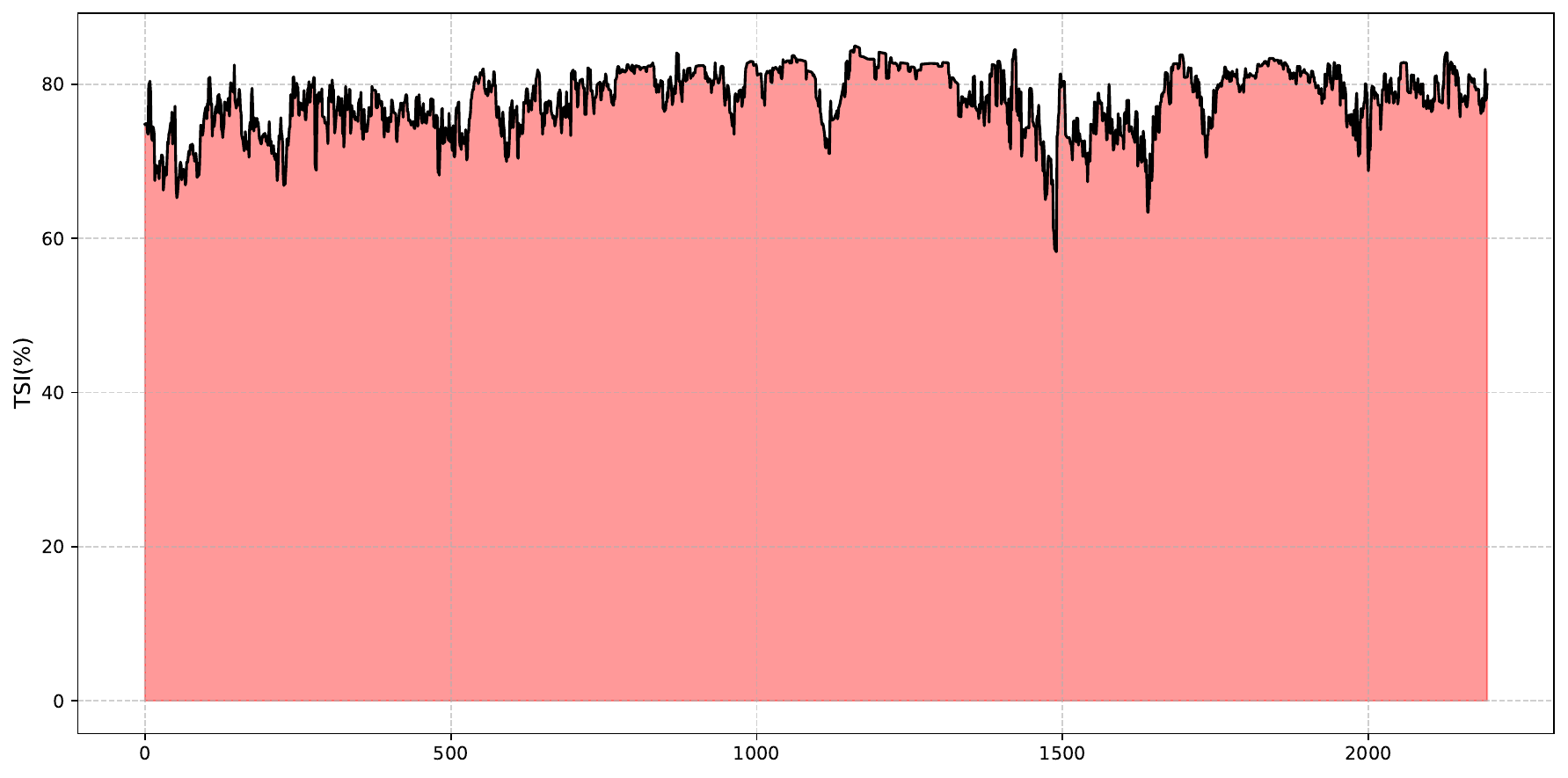}
\includegraphics[width=0.32\textwidth]{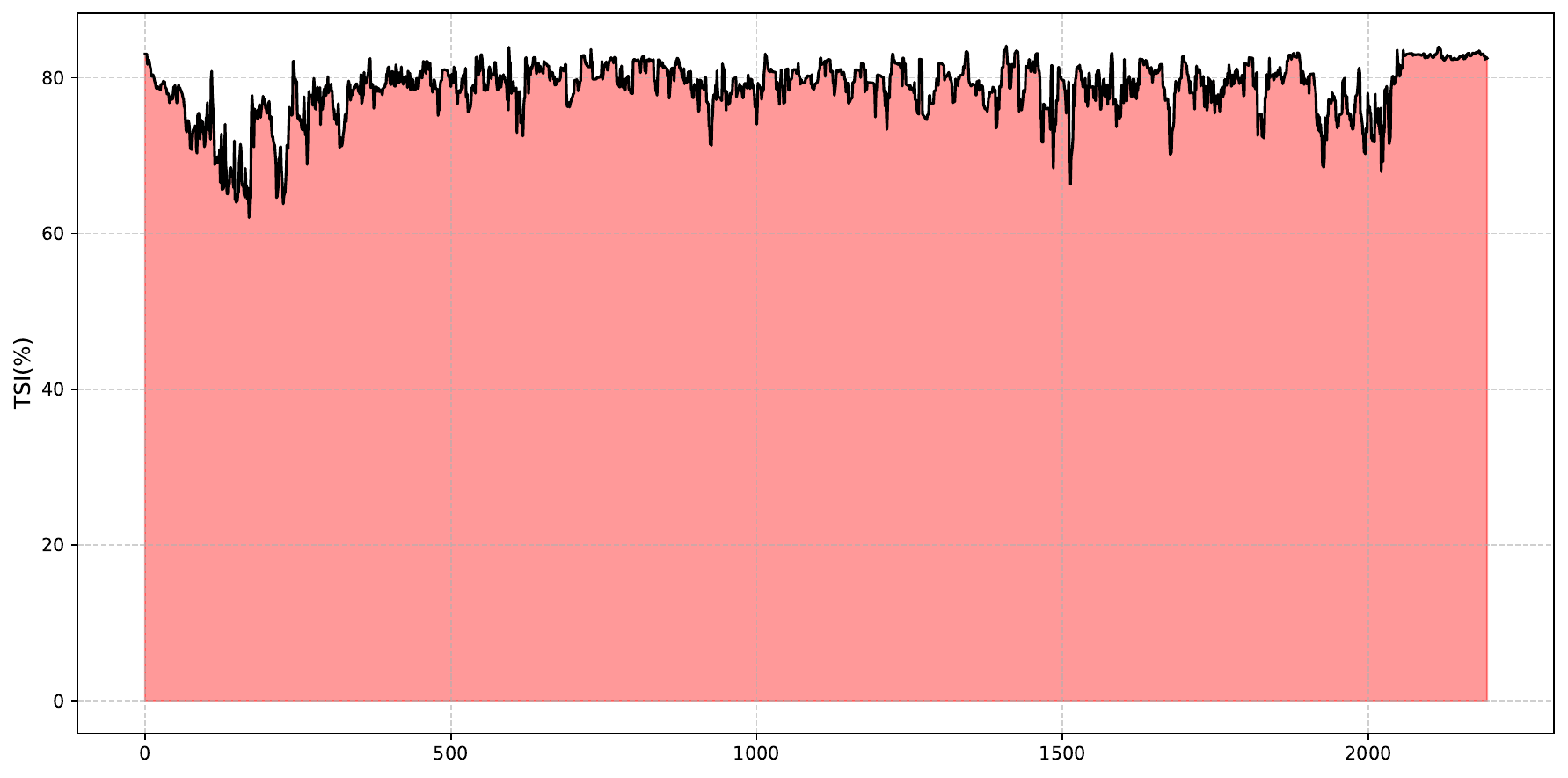}\\
\hspace{0.32\textwidth} \hspace{0.32\textwidth} \hspace{0.32\textwidth}\\
(a) 0.05 quantile \hspace{0.2\textwidth} (b) 0.5 quantile \hspace{0.2\textwidth} (c) 0.95 quantile
\caption{Total spillover of $RV$ at different quantiles}
\label{fig:rv_plots}
\end{figure}

\begin{figure}[htbp]
\centering
\includegraphics[width=0.32\textwidth]{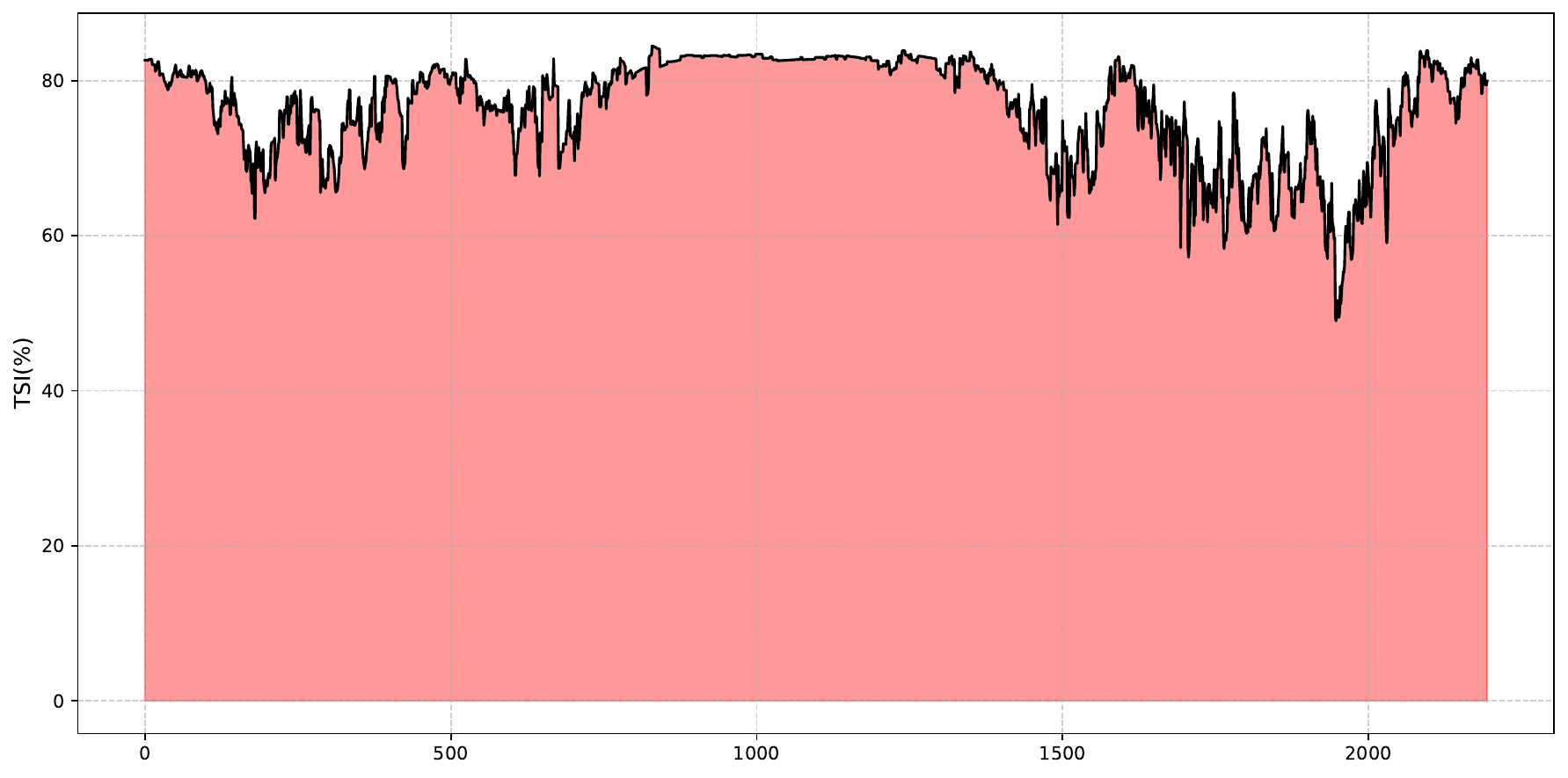}
\includegraphics[width=0.32\textwidth]{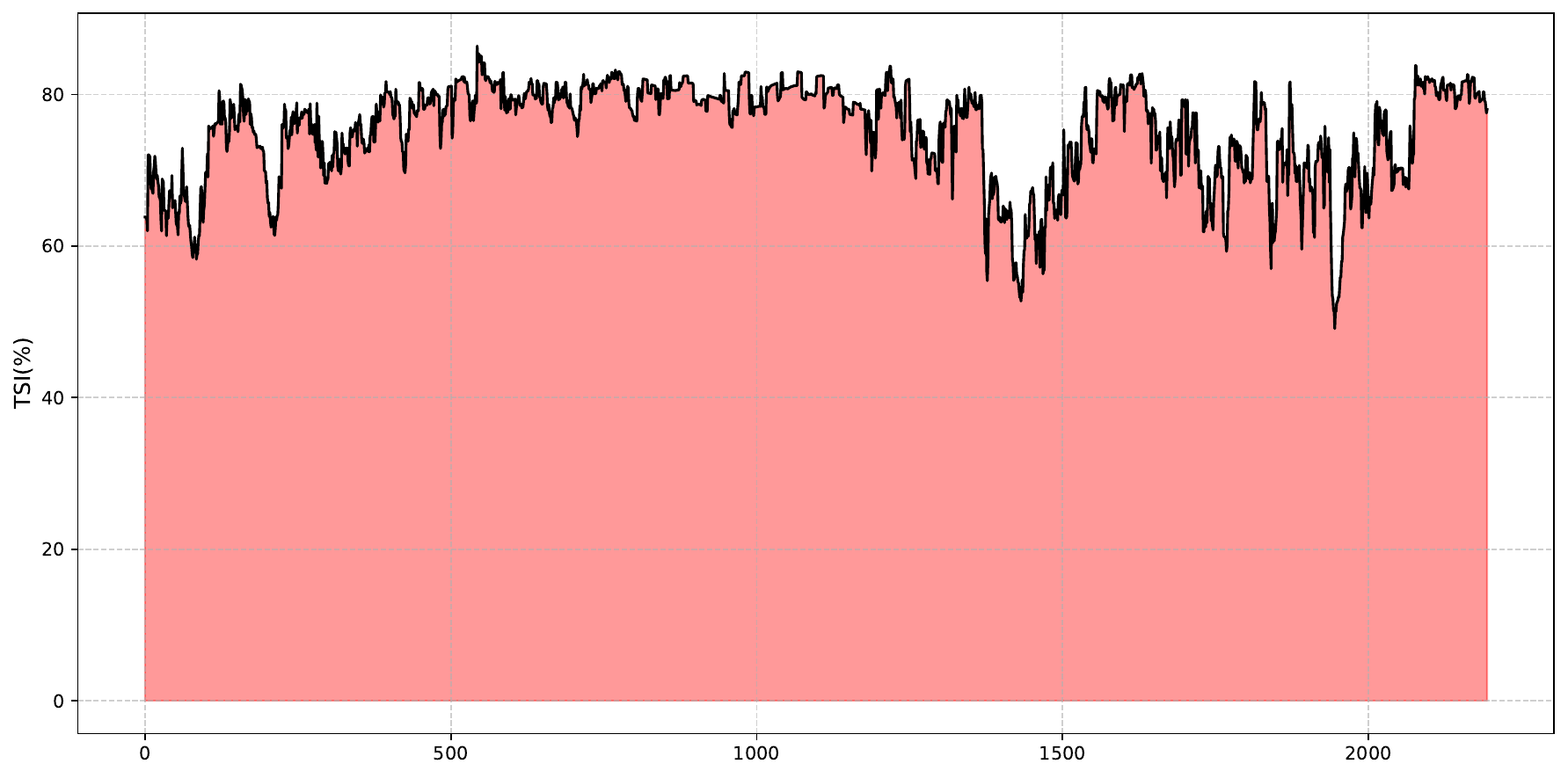}
\includegraphics[width=0.32\textwidth]{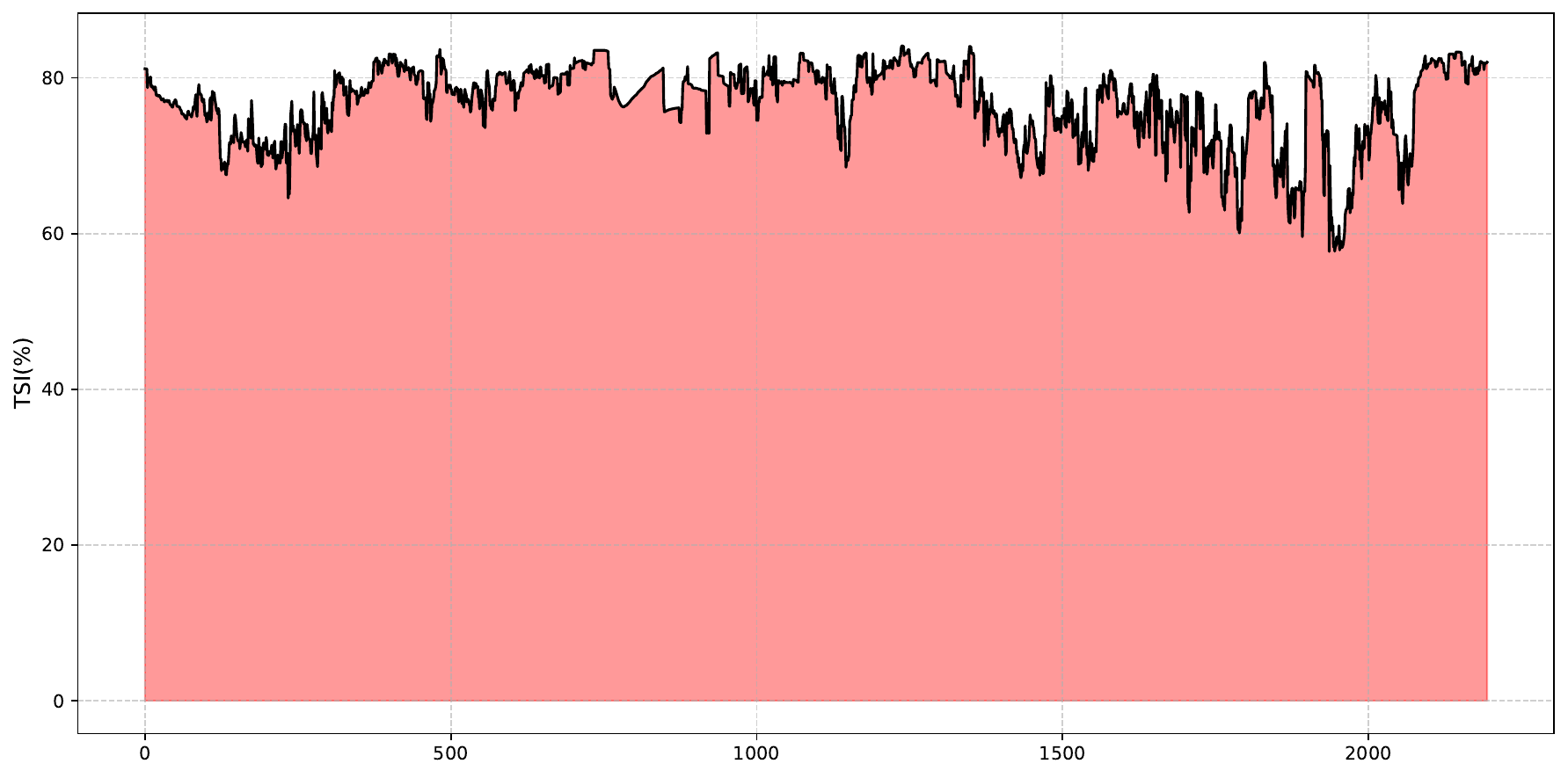}\\
\hspace{0.32\textwidth} \hspace{0.32\textwidth} \hspace{0.32\textwidth}\\
(a) 0.05 quantile \hspace{0.2\textwidth} (b) 0.5 quantile \hspace{0.2\textwidth} (c) 0.95 quantile
\caption{Total spillover of $CV$ at different quantiles}
\label{fig:ct}
\end{figure}

\begin{figure}[htbp]
\centering
\includegraphics[width=0.32\textwidth]{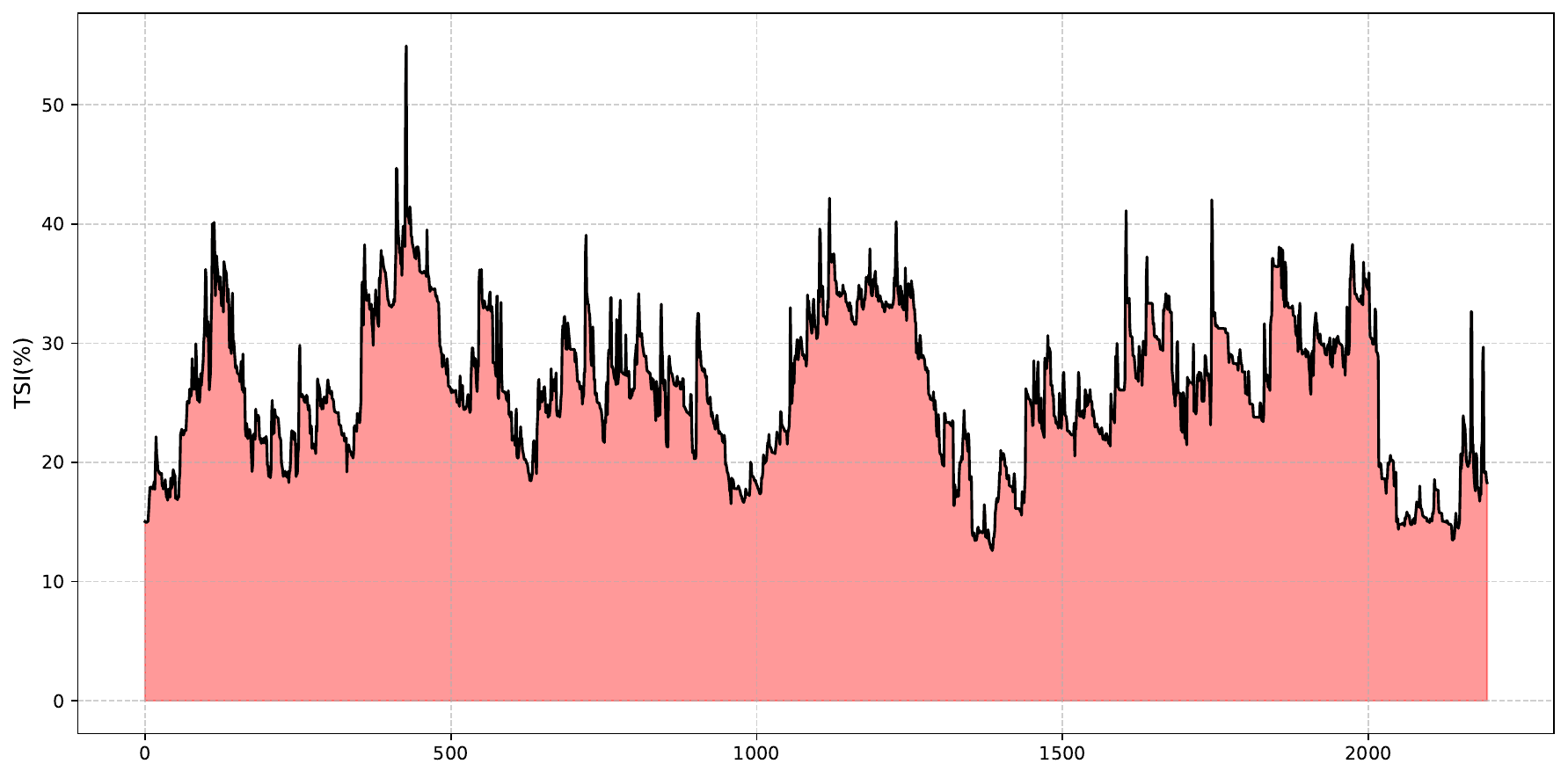}
\includegraphics[width=0.32\textwidth]{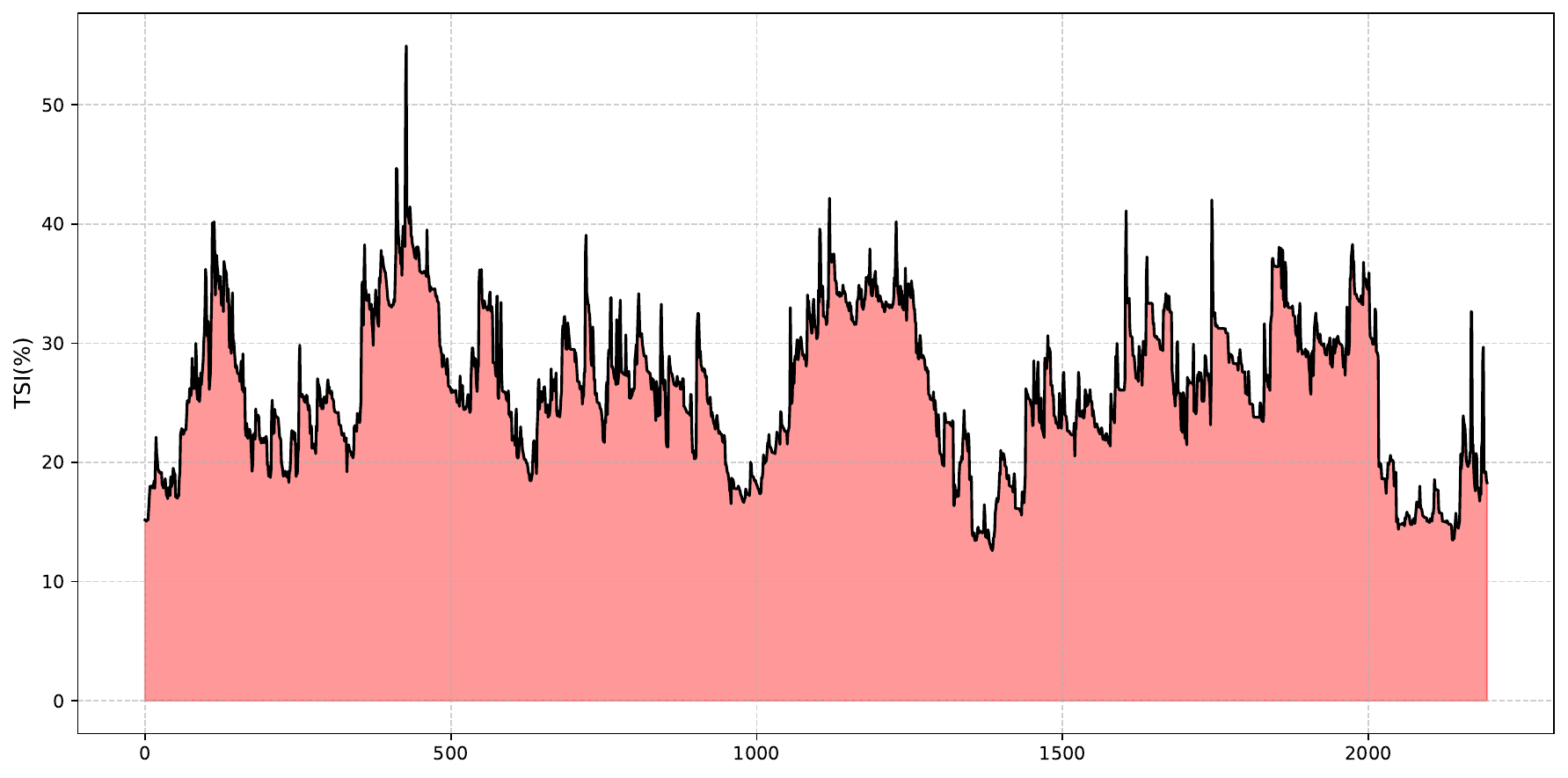}
\includegraphics[width=0.32\textwidth]{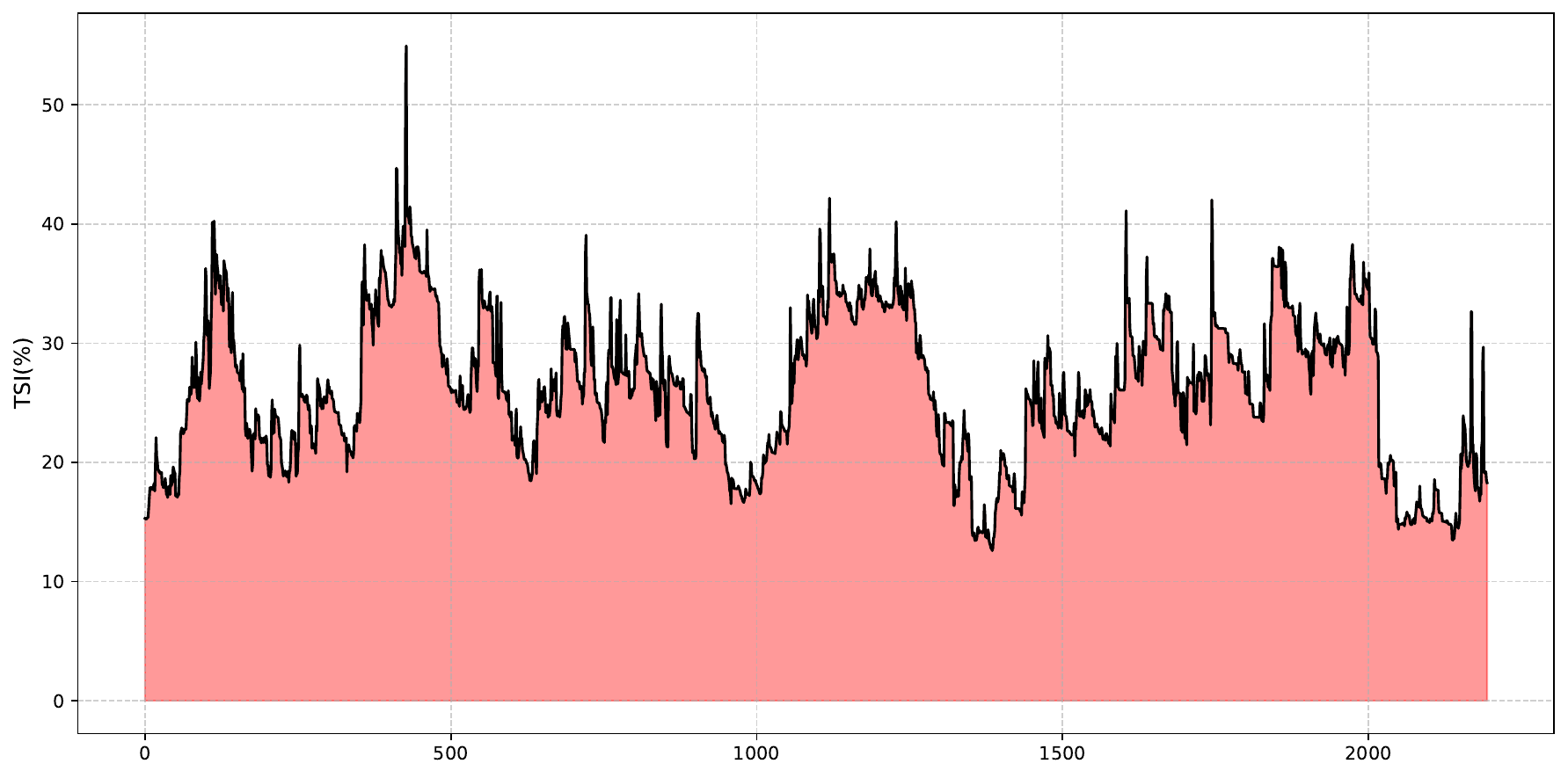}\\
\hspace{0.32\textwidth} \hspace{0.32\textwidth} \hspace{0.32\textwidth}\\
(a) 0.05 quantile \hspace{0.2\textwidth} (b) 0.5 quantile \hspace{0.2\textwidth} (c) 0.95 quantile
\caption{Total spillover of $CJ$ at different quantiles}
\label{fig:jv_plots}
\end{figure}

\begin{figure}[htbp]
\centering
\includegraphics[width=0.32\textwidth]{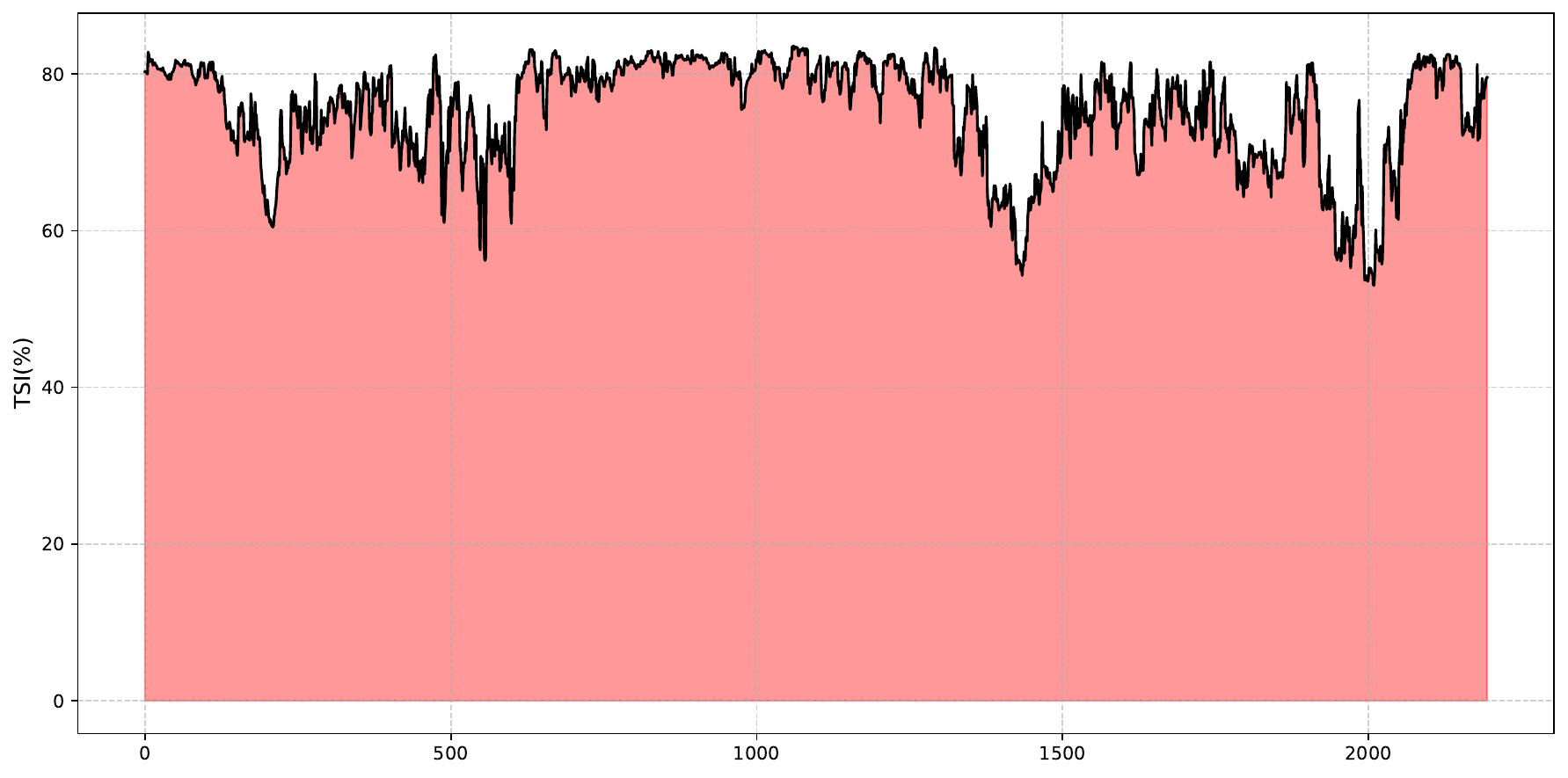}
\includegraphics[width=0.32\textwidth]{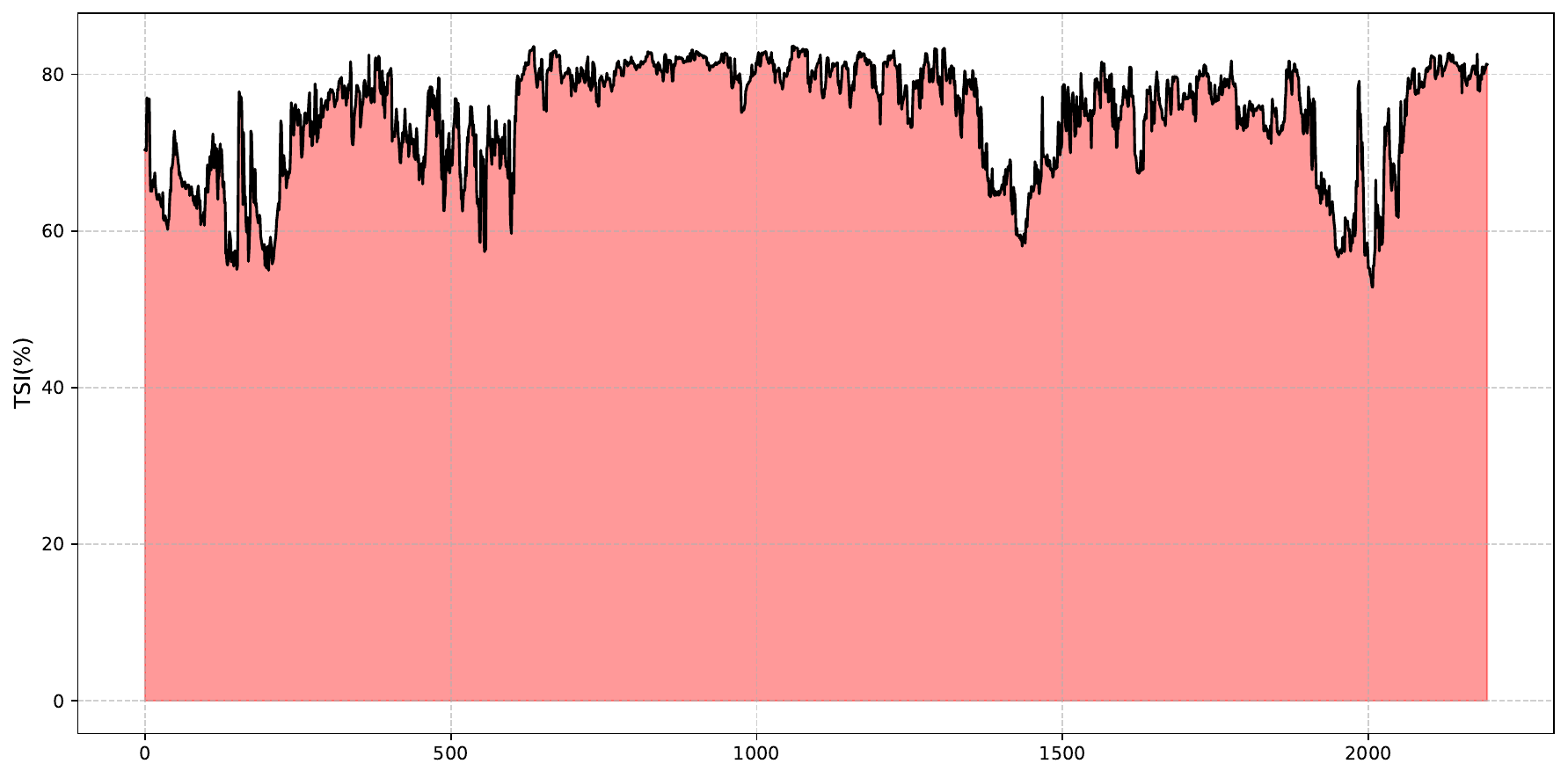}
\includegraphics[width=0.32\textwidth]{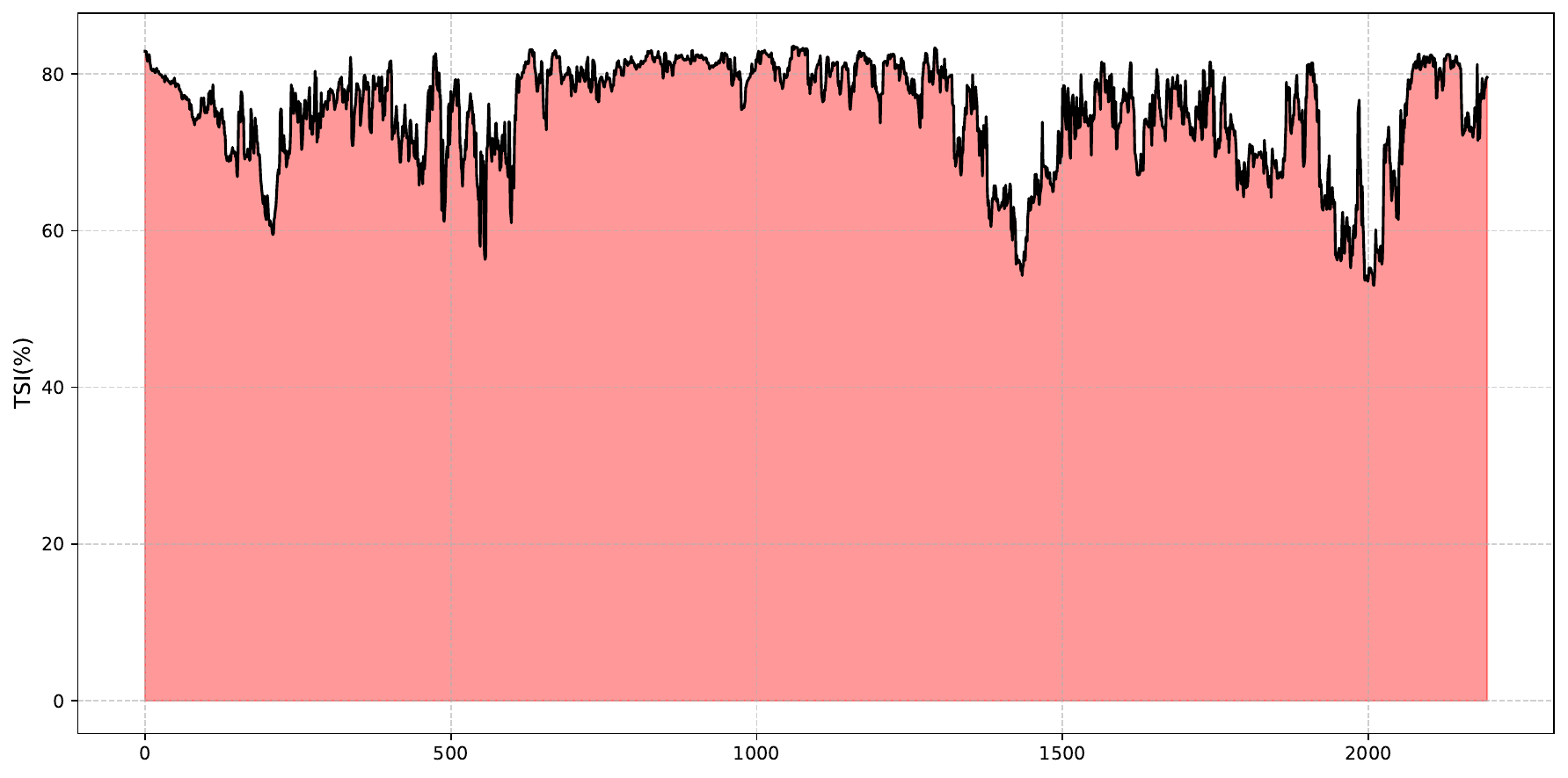}\\
\hspace{0.32\textwidth} \hspace{0.32\textwidth} \hspace{0.32\textwidth}\\
(a) 0.05 quantile \hspace{0.2\textwidth} (b) 0.5 quantile \hspace{0.2\textwidth} (c) 0.95 quantile
\caption{Total spillover of $RS^+$ at different quantiles}
\label{fig:rs+_plots}
\end{figure}

\begin{figure}[htbp]
\centering
\includegraphics[width=0.32\textwidth]{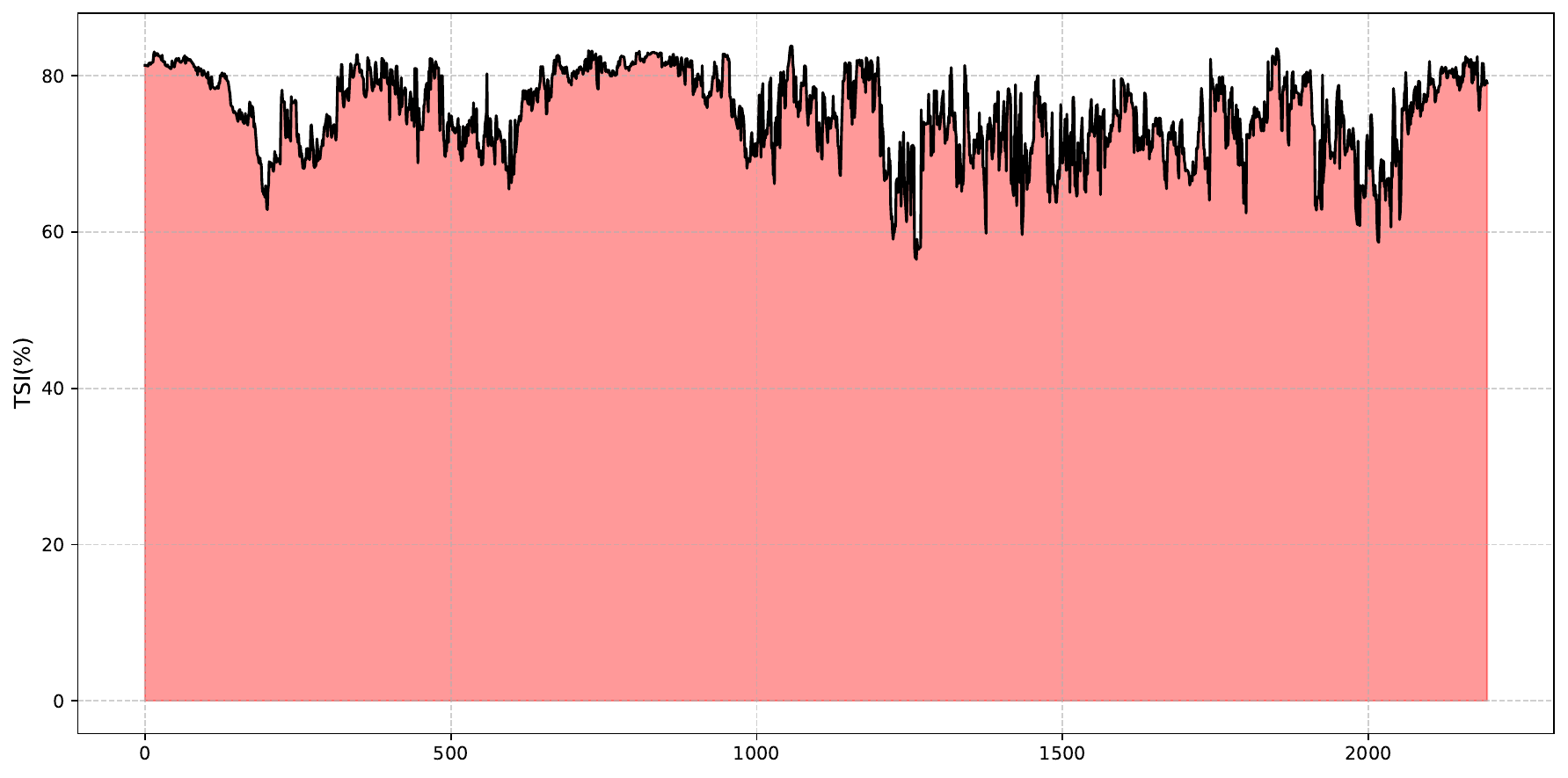}
\includegraphics[width=0.32\textwidth]{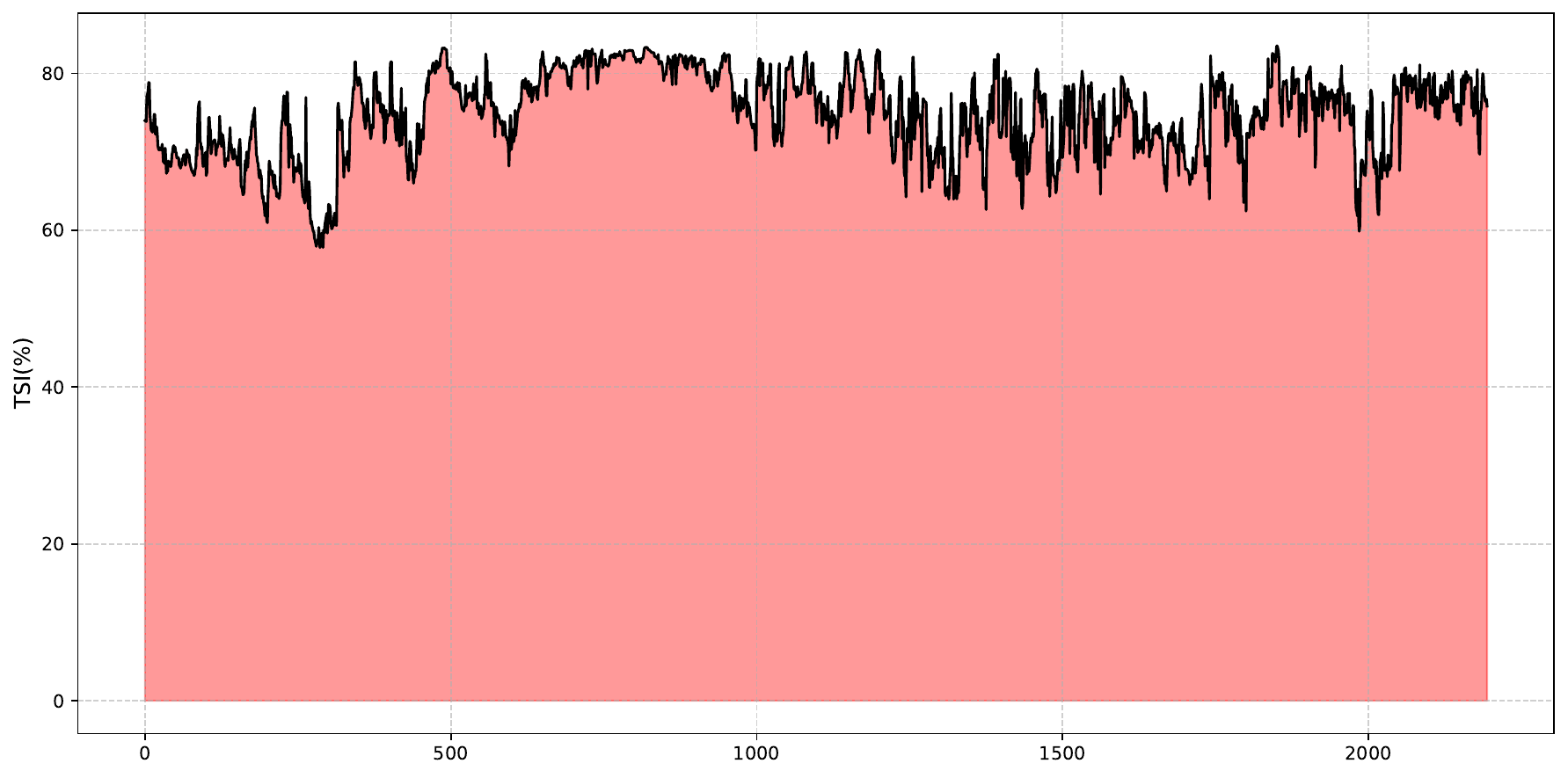}
\includegraphics[width=0.32\textwidth]{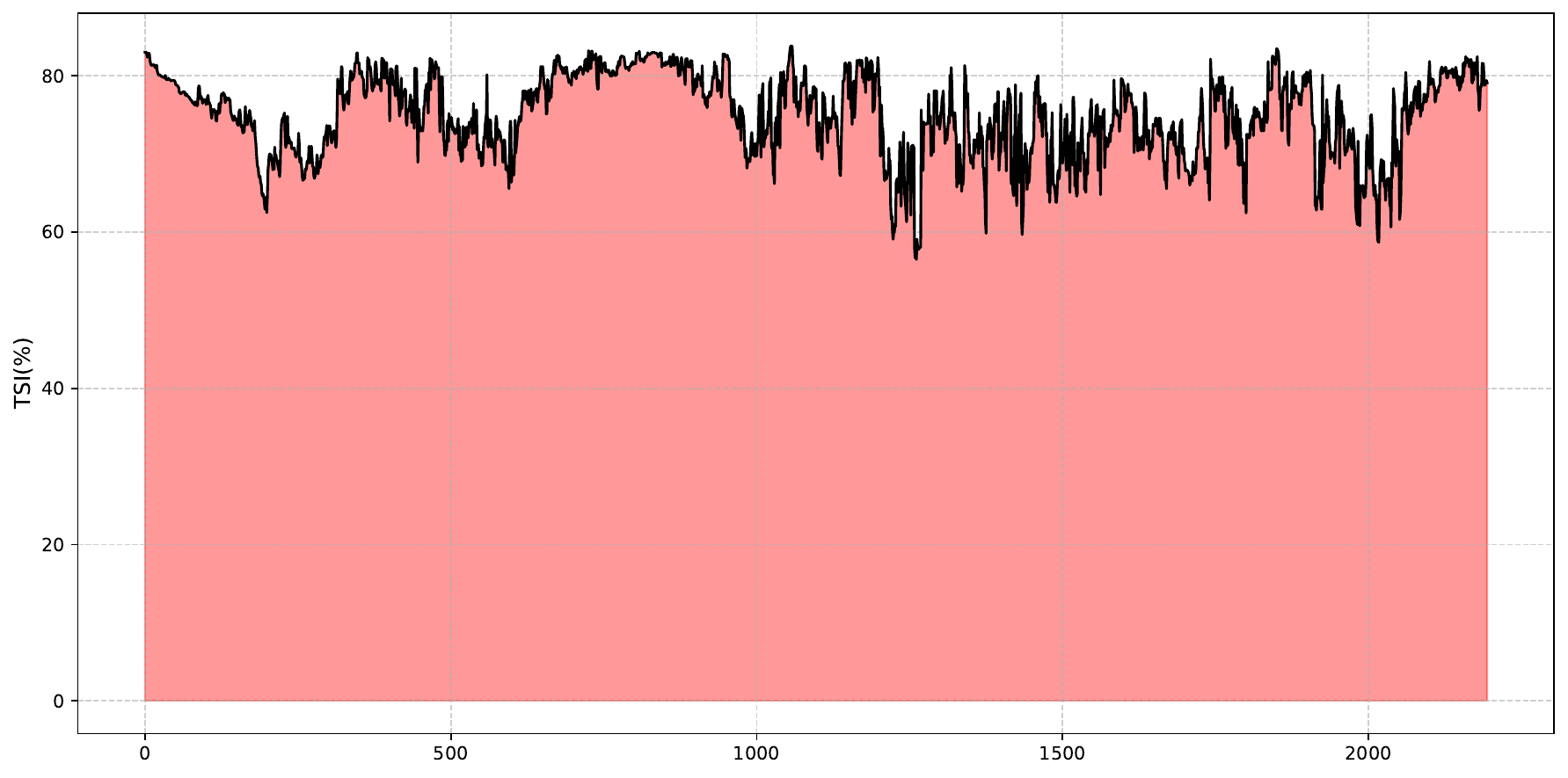}\\
\hspace{0.32\textwidth} \hspace{0.32\textwidth} \hspace{0.32\textwidth}\\
(a) 0.05 quantile \hspace{0.2\textwidth} (b) 0.5 quantile \hspace{0.2\textwidth} (c) 0.95 quantile
\caption{Total spillover of $RS^-$ at different quantiles}
\label{fig:rs-_plots}
\end{figure}

\begin{figure}[htbp]
\centering
\includegraphics[width=0.32\textwidth]{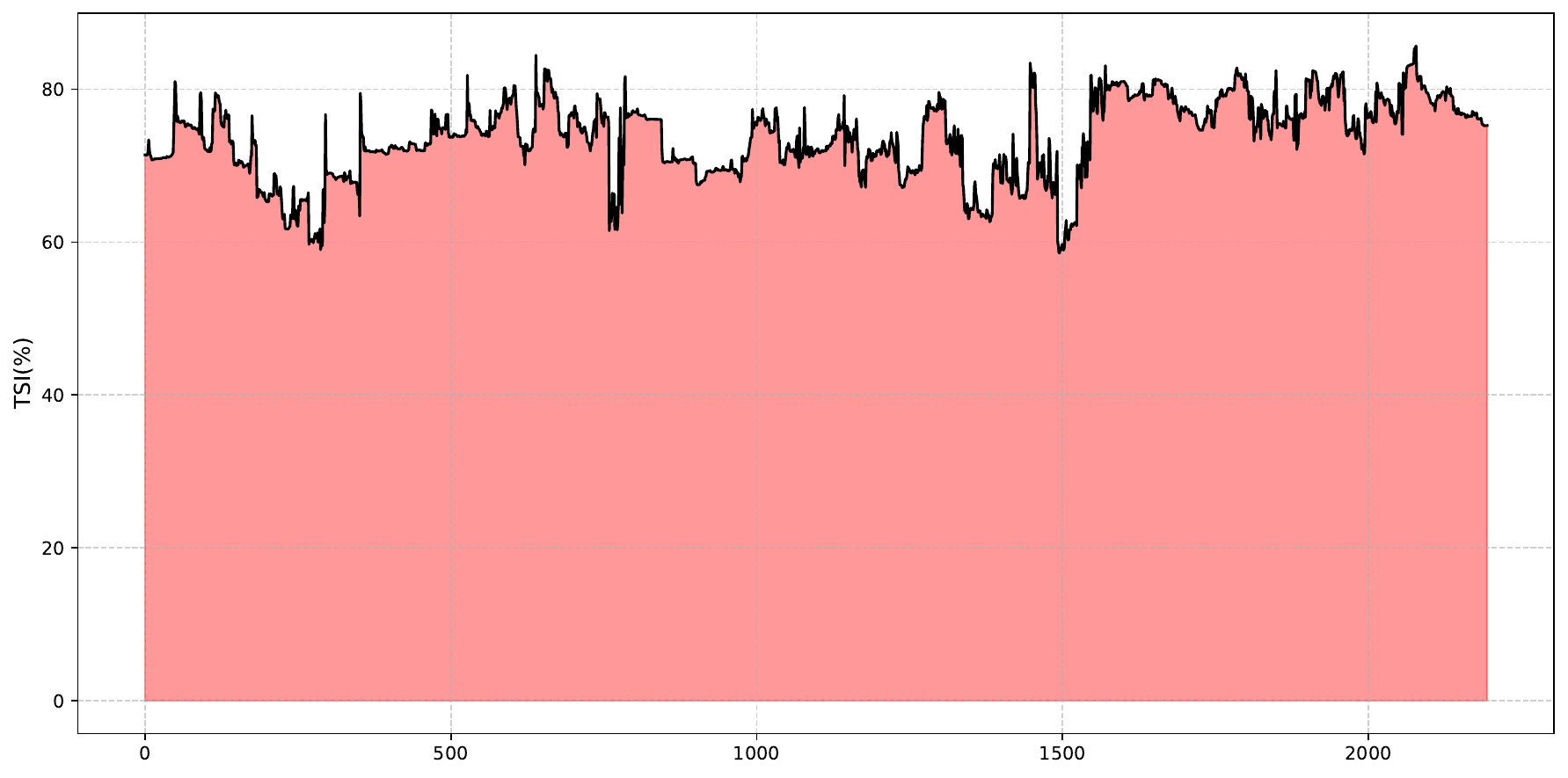}
\includegraphics[width=0.32\textwidth]{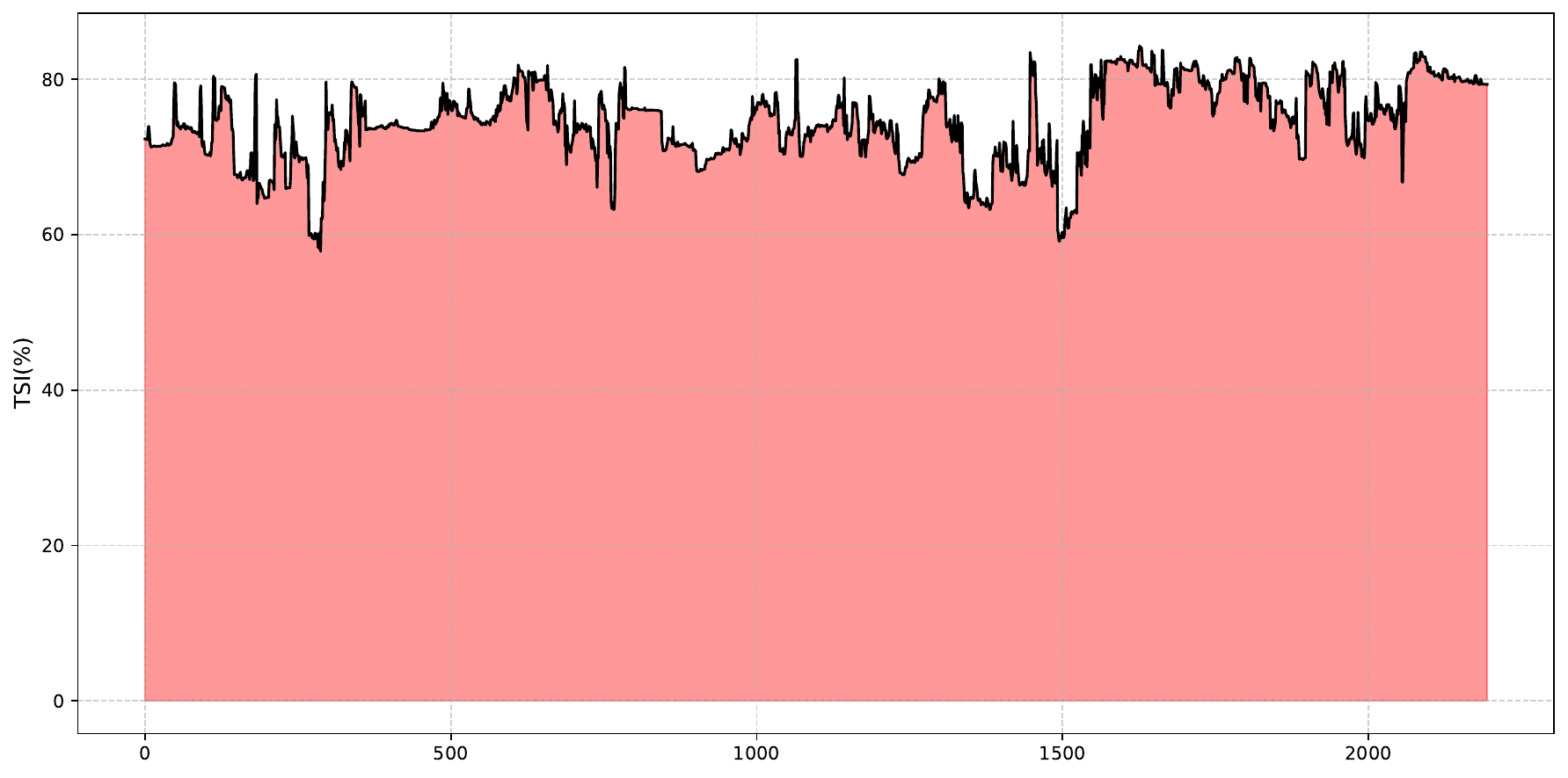}
\includegraphics[width=0.32\textwidth]{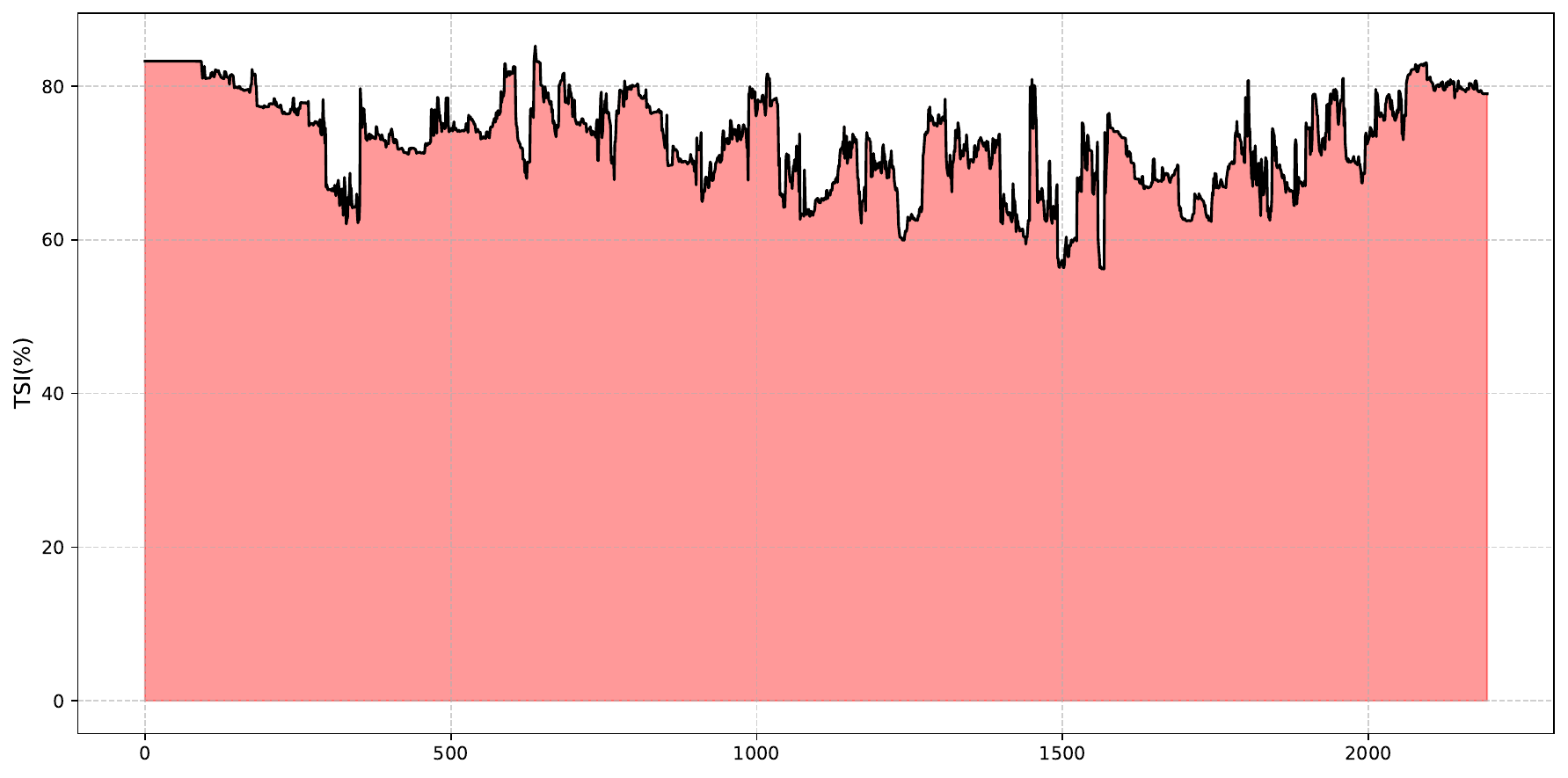}\\
\hspace{0.32\textwidth} \hspace{0.32\textwidth} \hspace{0.32\textwidth}\\
(a) 0.05 quantile \hspace{0.2\textwidth} (b) 0.5 quantile \hspace{0.2\textwidth} (c) 0.95 quantile
\caption{Total spillover of $REX^+$ at different quantiles}
\label{fig:rexp_plots}
\end{figure}

\begin{figure}[htbp]
\centering
\includegraphics[width=0.32\textwidth]{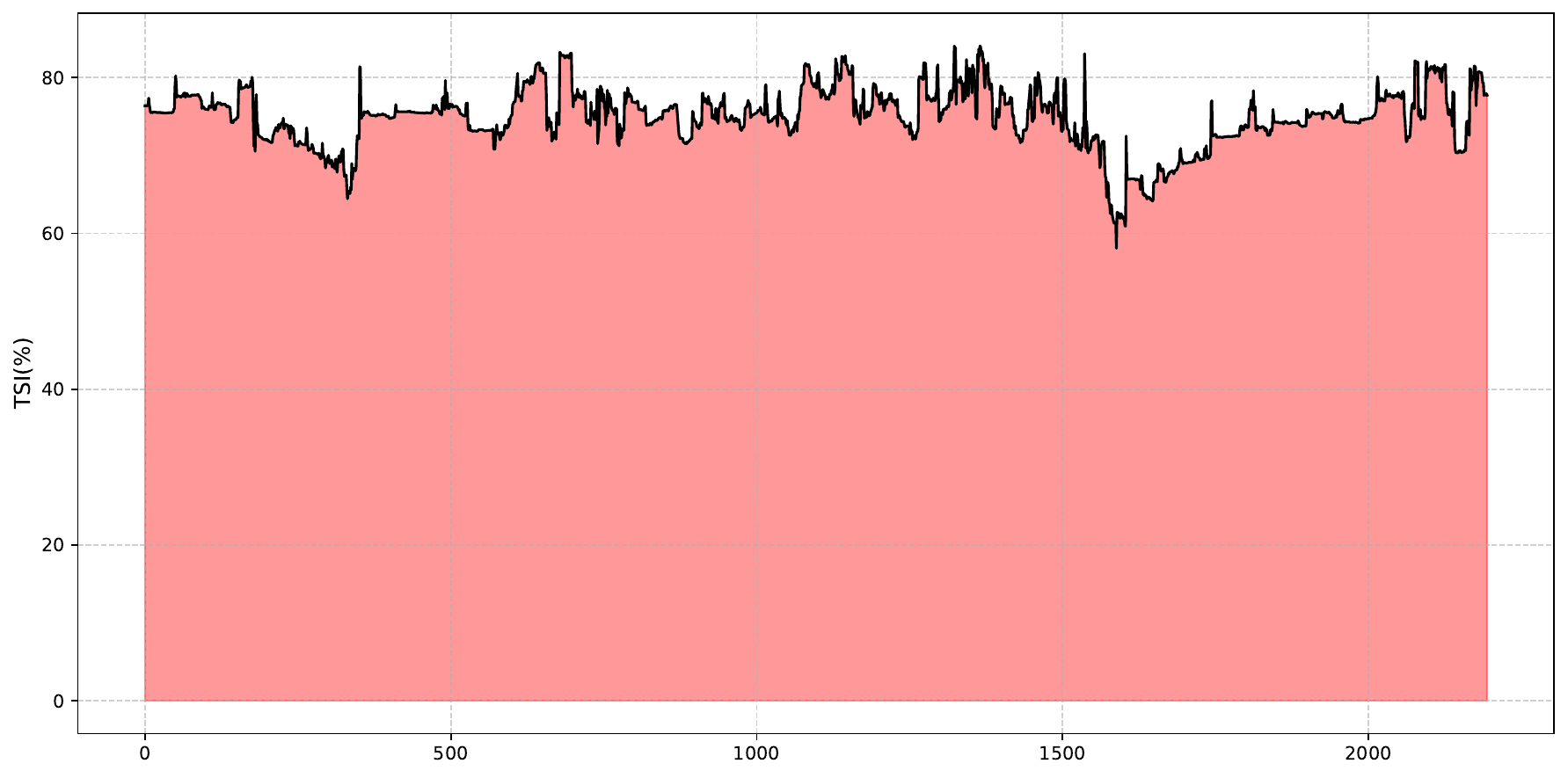}
\includegraphics[width=0.32\textwidth]{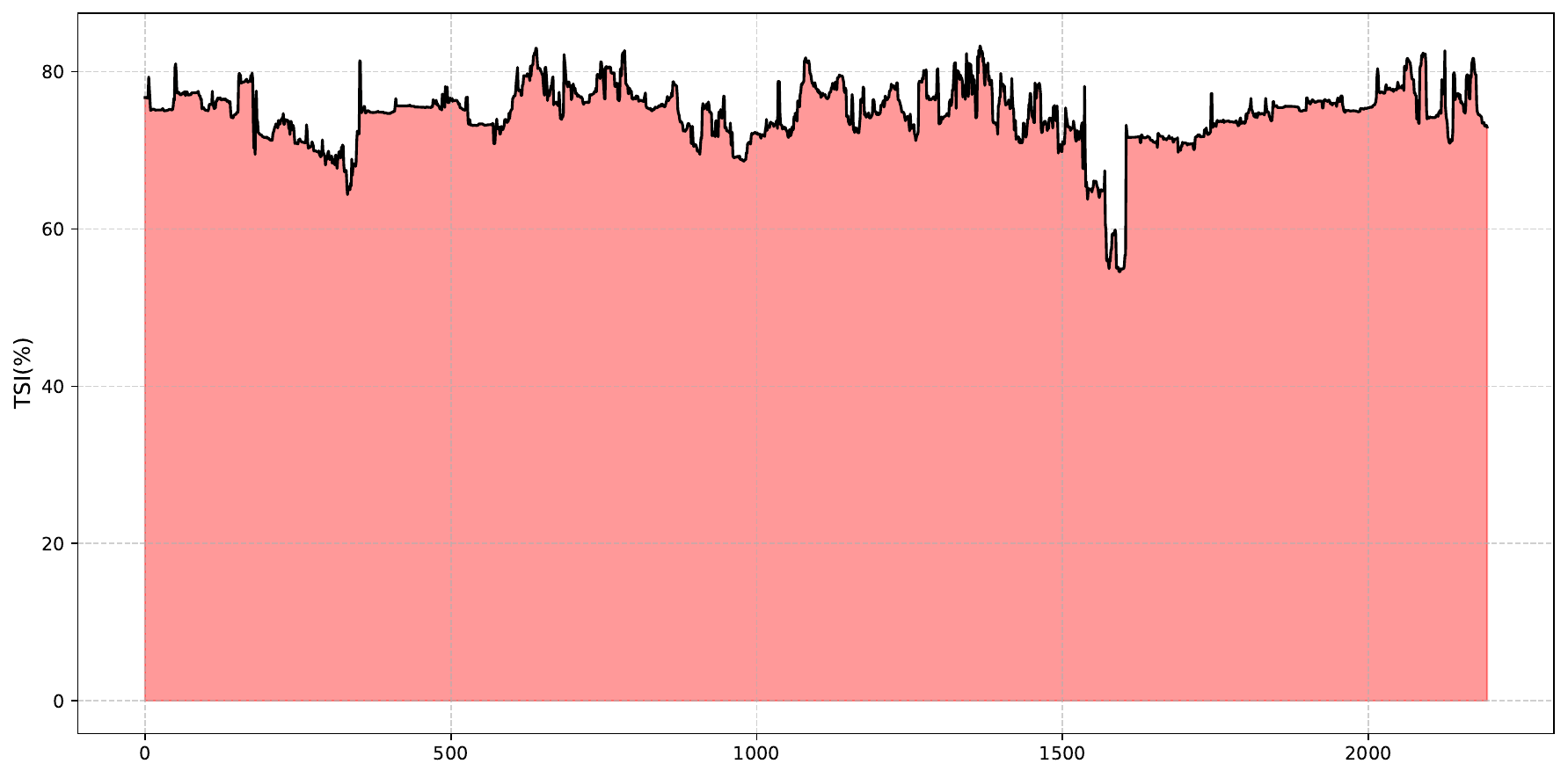}
\includegraphics[width=0.32\textwidth]{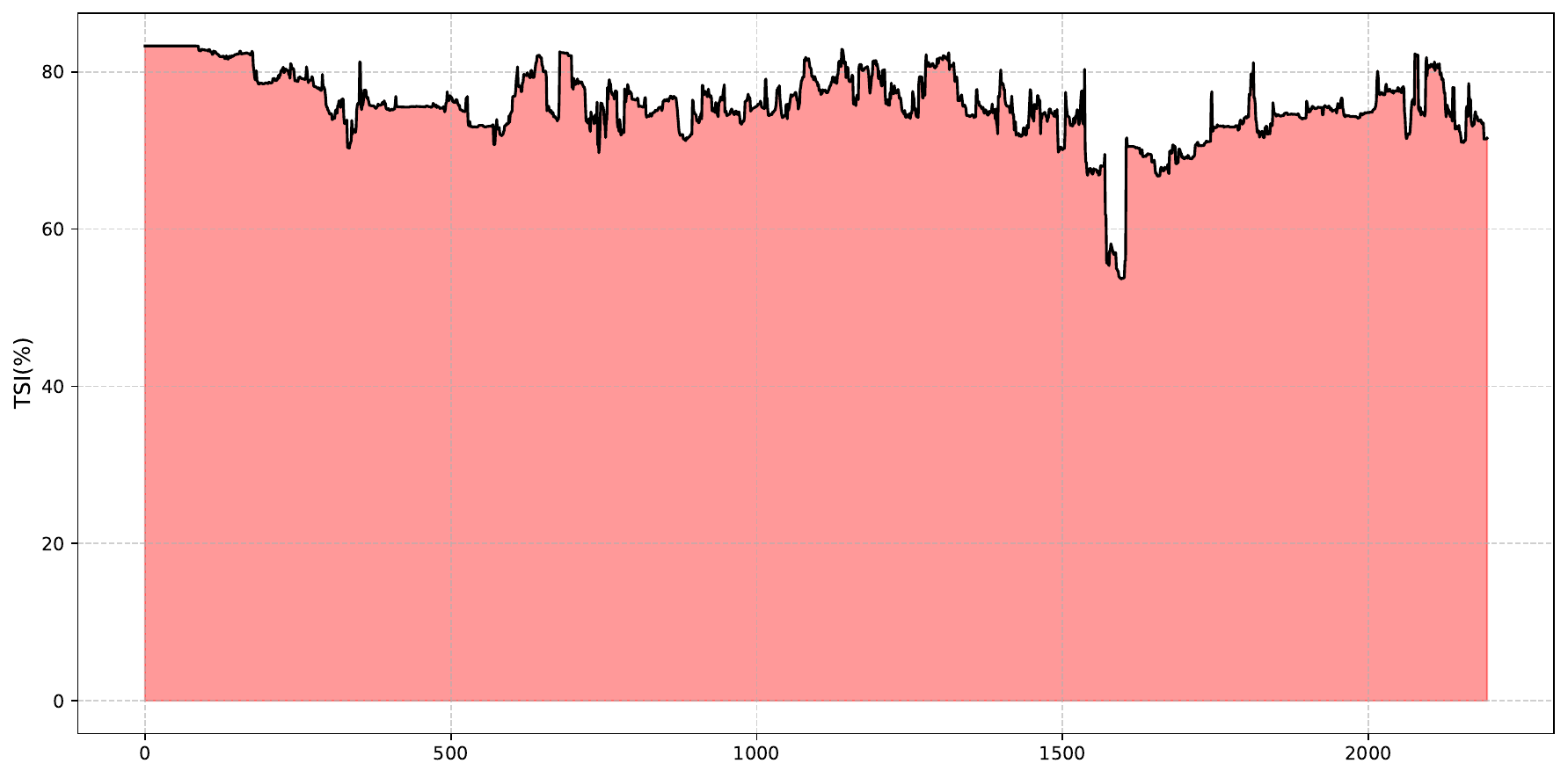}\\
\hspace{0.32\textwidth} \hspace{0.32\textwidth} \hspace{0.32\textwidth}\\
(a) 0.05 quantile \hspace{0.2\textwidth} (b) 0.5 quantile \hspace{0.2\textwidth} (c) 0.95 quantile
\caption{Total spillover of $REX^-$ at different quantiles}
\label{fig:rexd_plots}
\end{figure}

\begin{figure}[htbp]
\centering
\includegraphics[width=0.32\textwidth]{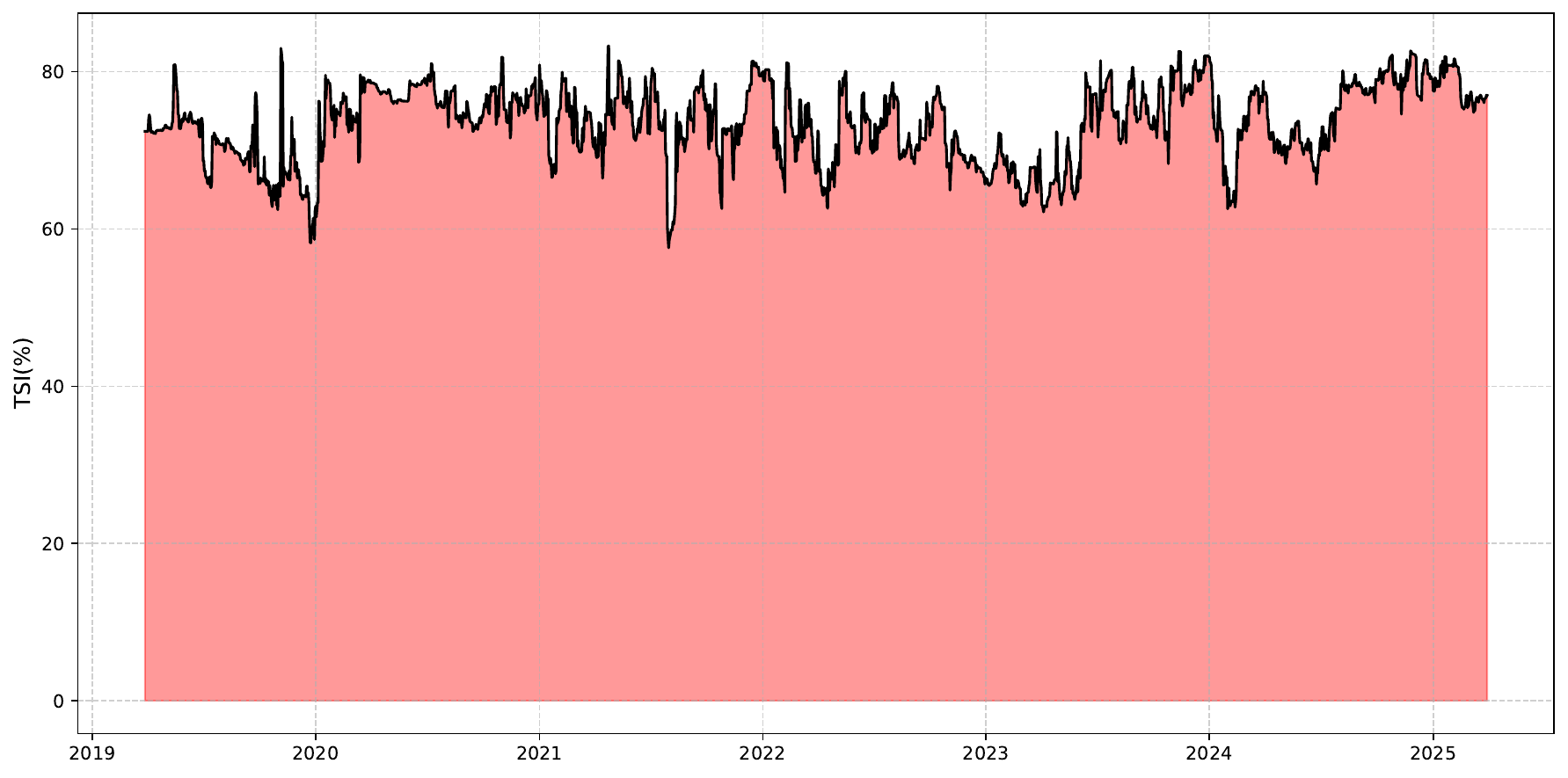}
\includegraphics[width=0.32\textwidth]{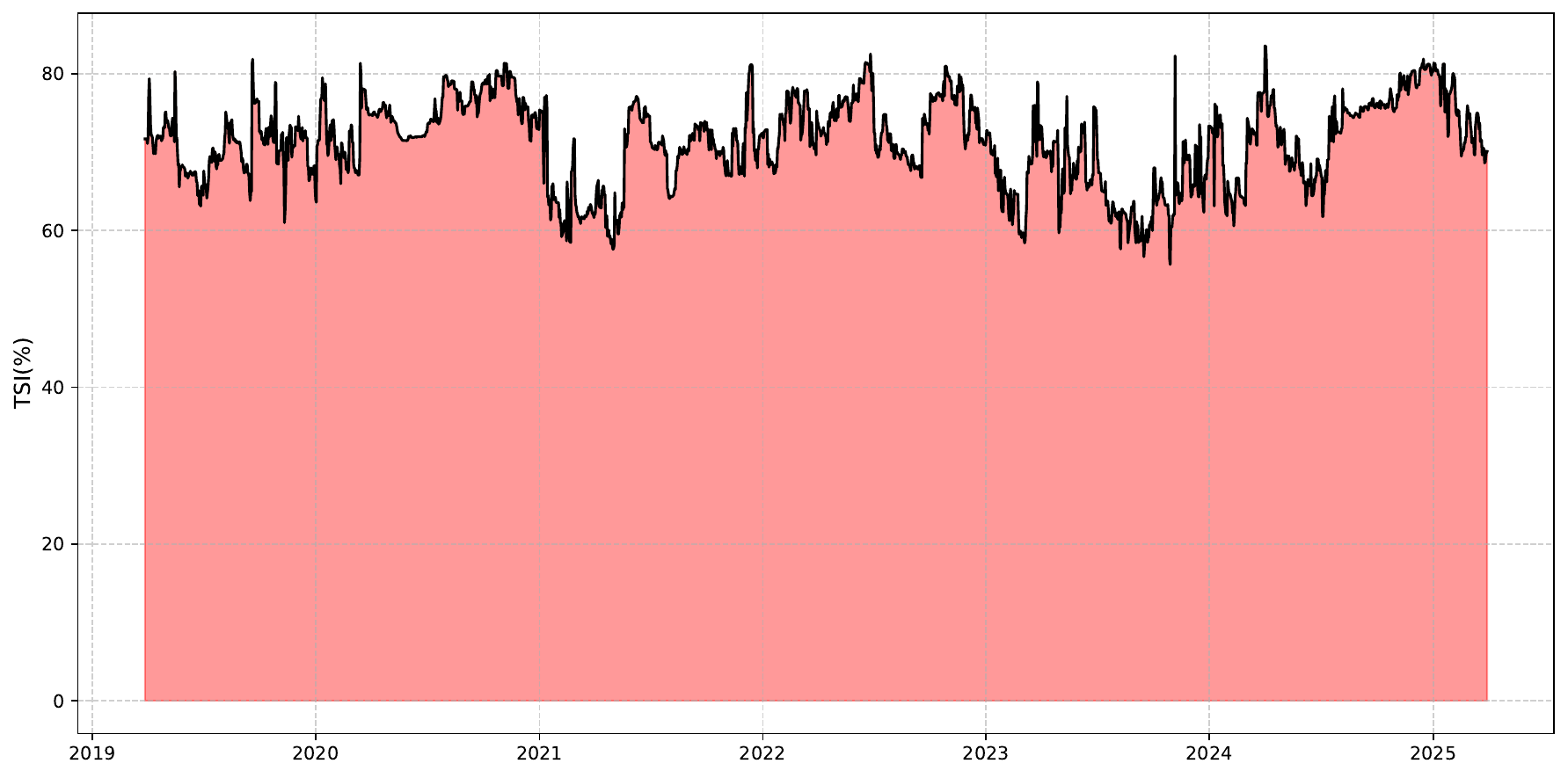}
\includegraphics[width=0.32\textwidth]{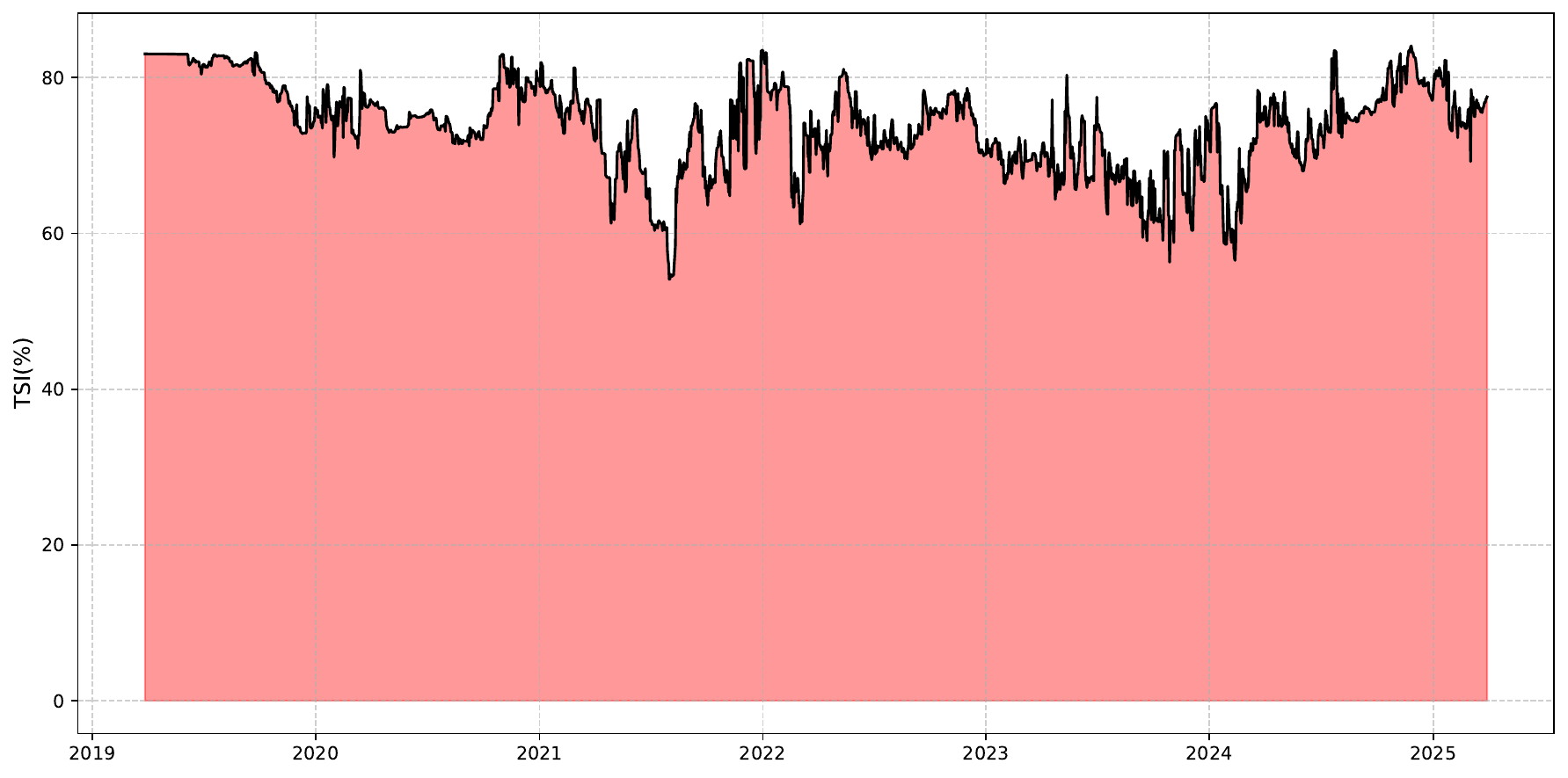}\\
\hspace{0.32\textwidth} \hspace{0.32\textwidth} \hspace{0.32\textwidth}\\
(a) 0.05 quantile \hspace{0.2\textwidth} (b) 0.5 quantile \hspace{0.2\textwidth} (c) 0.95 quantile
\caption{Total spillover of $REX^m$ at different quantiles}
\label{fig:rexm_plots}
\end{figure}

\section{Net quantile spillovers of volatility features}
\begin{figure}[p]
    \centering
    \caption{Quantile net spillovers for major cryptocurrencies using RV as the feature variable. Each row corresponds to a specific cryptocurrency, and columns represent quantiles $\tau=0.05$ (left), $\tau=0.50$ (middle), and $\tau=0.95$ (right).}
    \label{fig:rv_net_spillover_by_coin}

    \subfigure[BTC, $\tau=0.05$]{\includegraphics[width=0.32\linewidth]{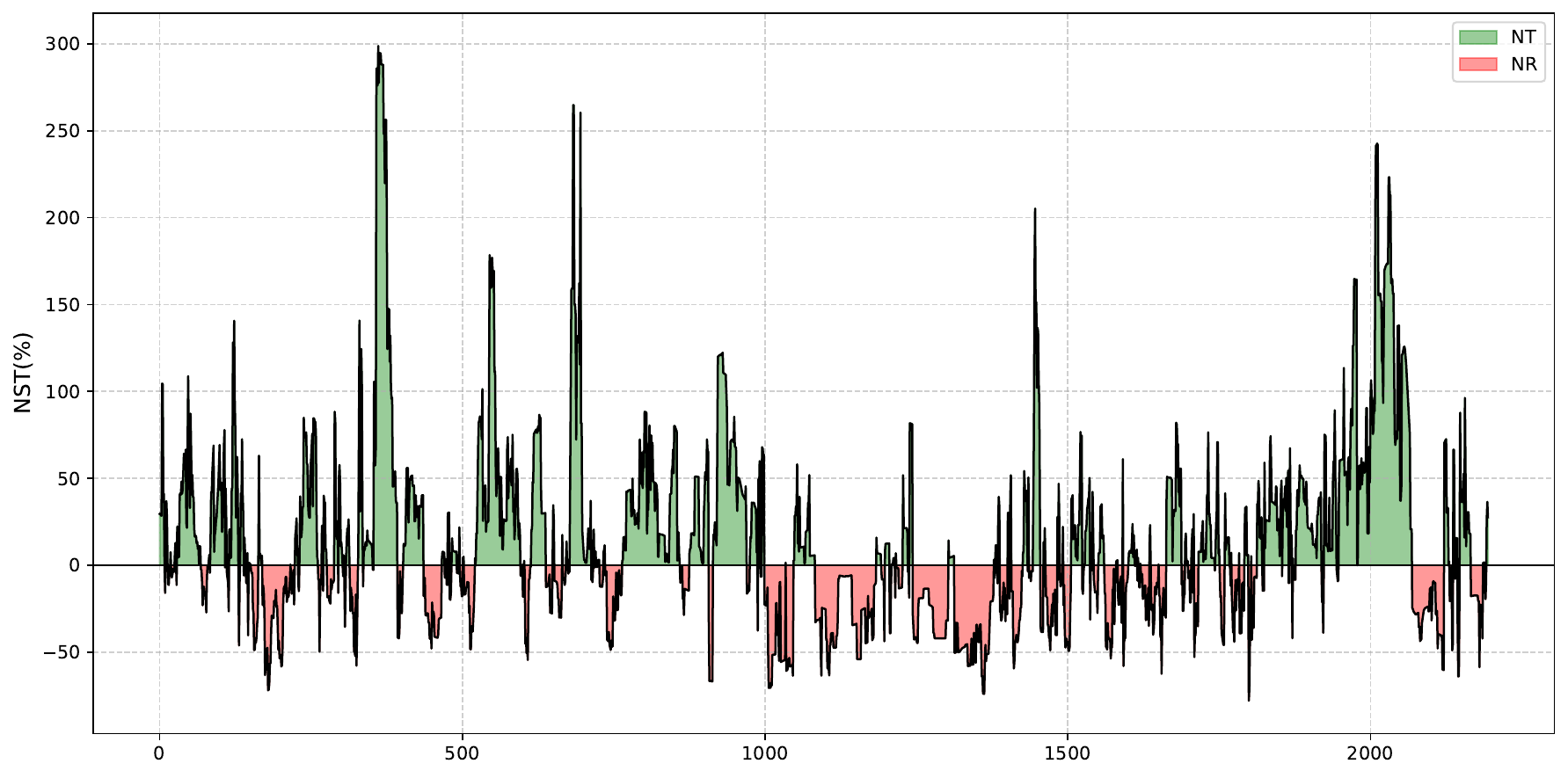}\label{fig:rv_btc_low}}\hfill
    \subfigure[BTC, $\tau=0.50$]{\includegraphics[width=0.32\linewidth]{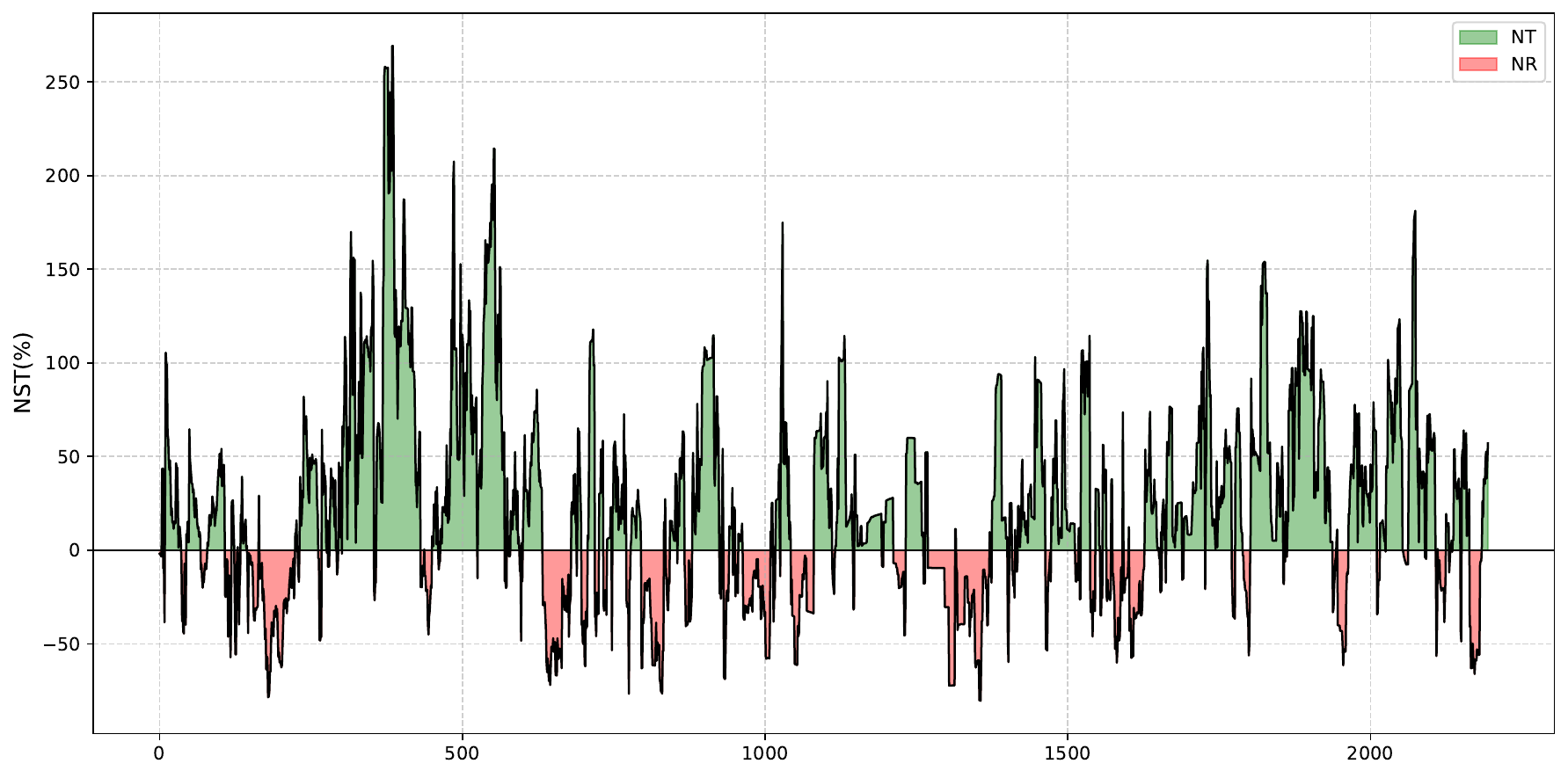}\label{fig:rv_btc_mid}}\hfill
    \subfigure[BTC, $\tau=0.95$]{\includegraphics[width=0.32\linewidth]{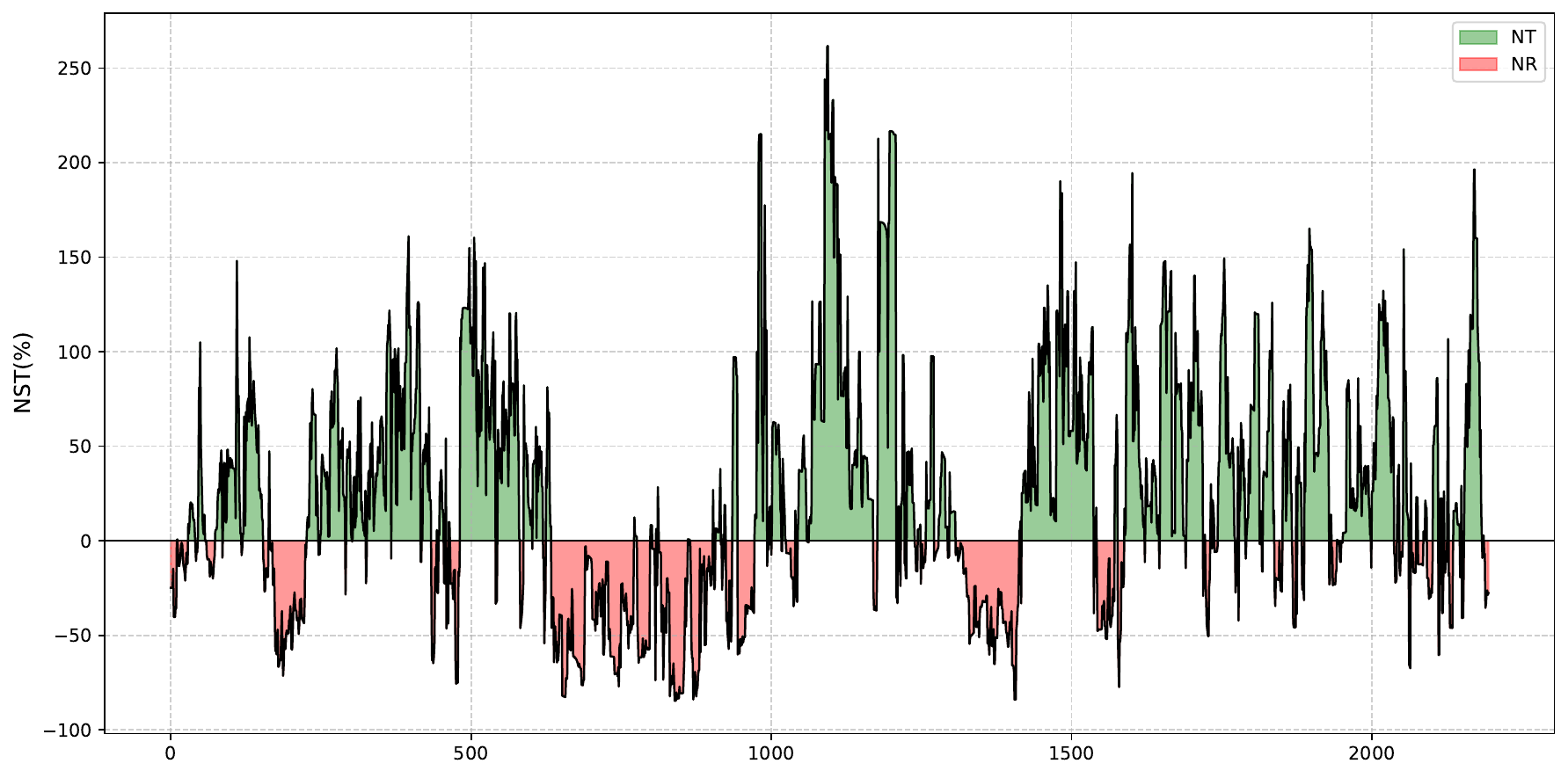}\label{fig:rv_btc_high}}
    \vspace{0.3cm}
    \subfigure[DASH, $\tau=0.05$]{\includegraphics[width=0.32\linewidth]{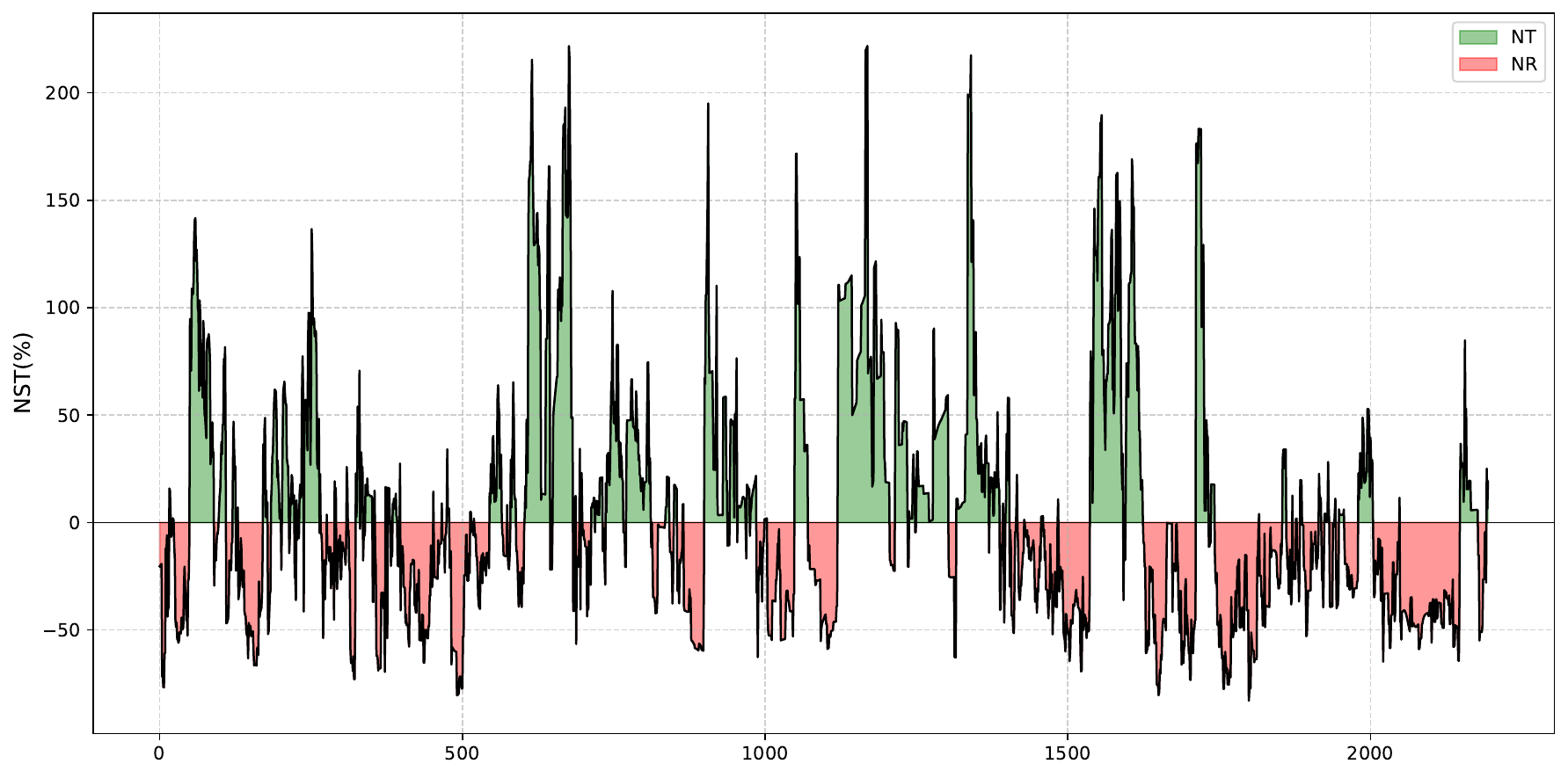}\label{fig:rv_dash_low}}\hfill
    \subfigure[DASH, $\tau=0.50$]{\includegraphics[width=0.32\linewidth]{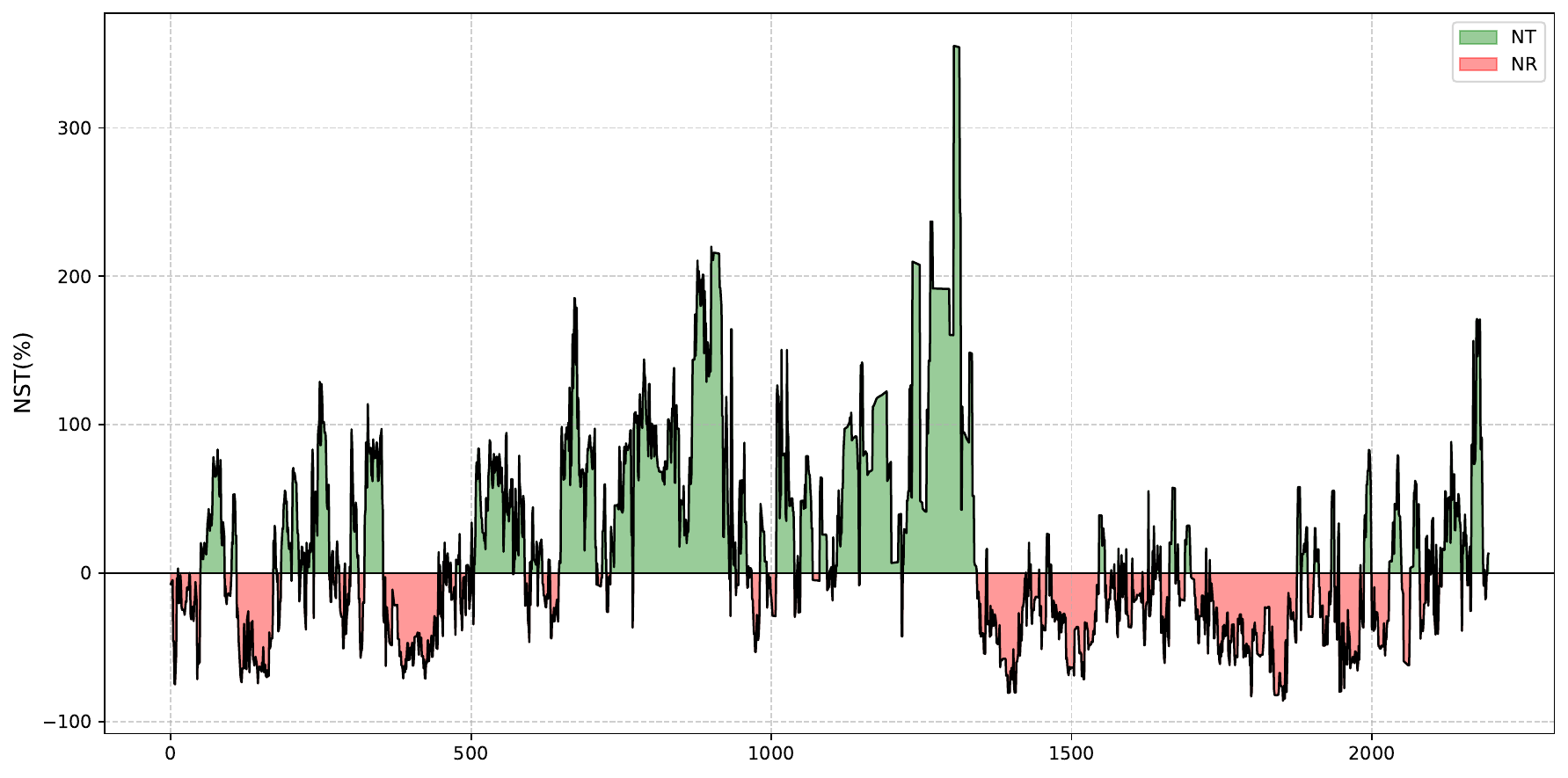}\label{fig:rv_dash_mid}}\hfill
    \subfigure[DASH, $\tau=0.95$]{\includegraphics[width=0.32\linewidth]{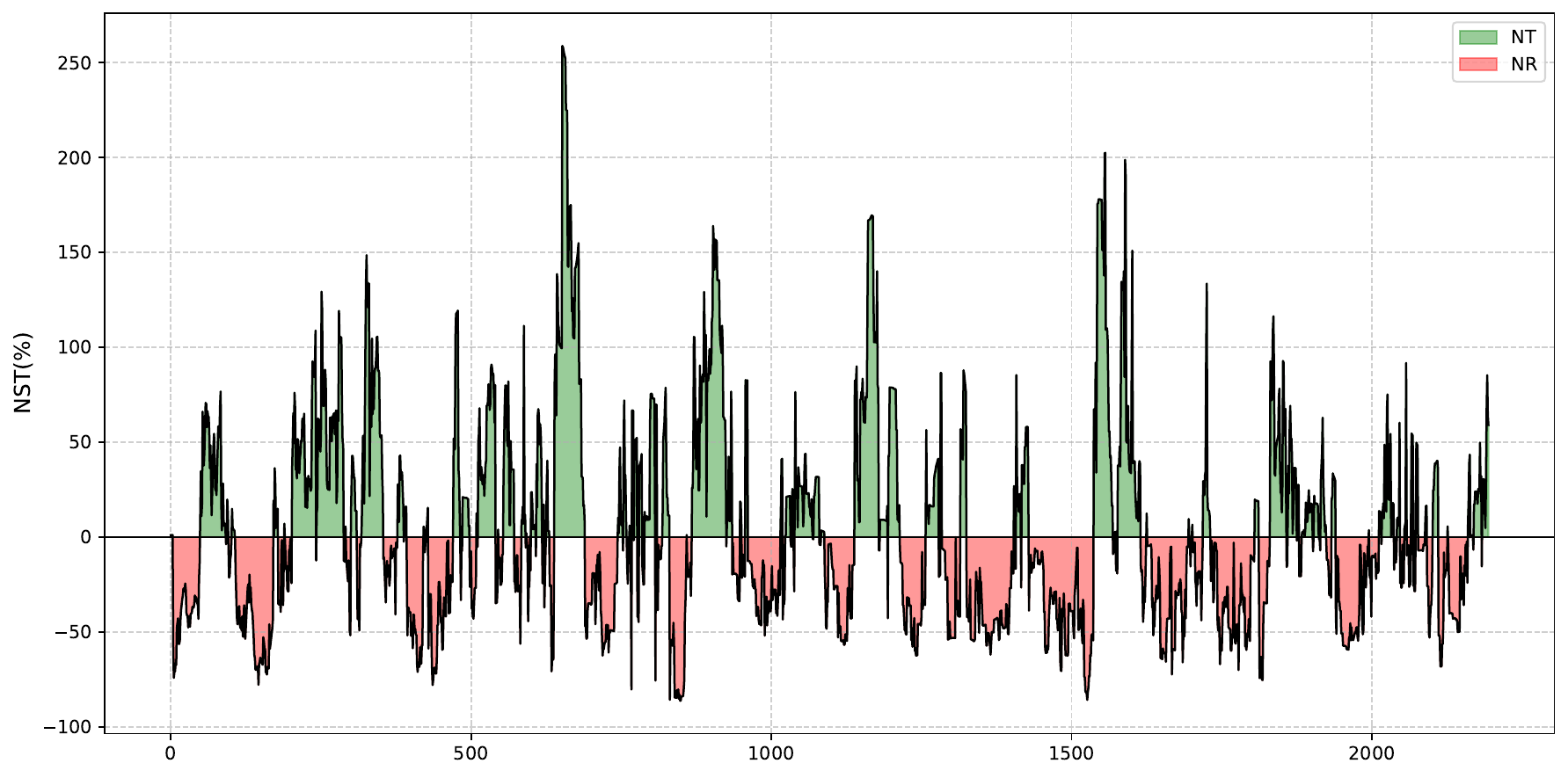}\label{fig:rv_dash_high}}
    \vspace{0.3cm}
    \subfigure[ETH, $\tau=0.05$]{\includegraphics[width=0.32\linewidth]{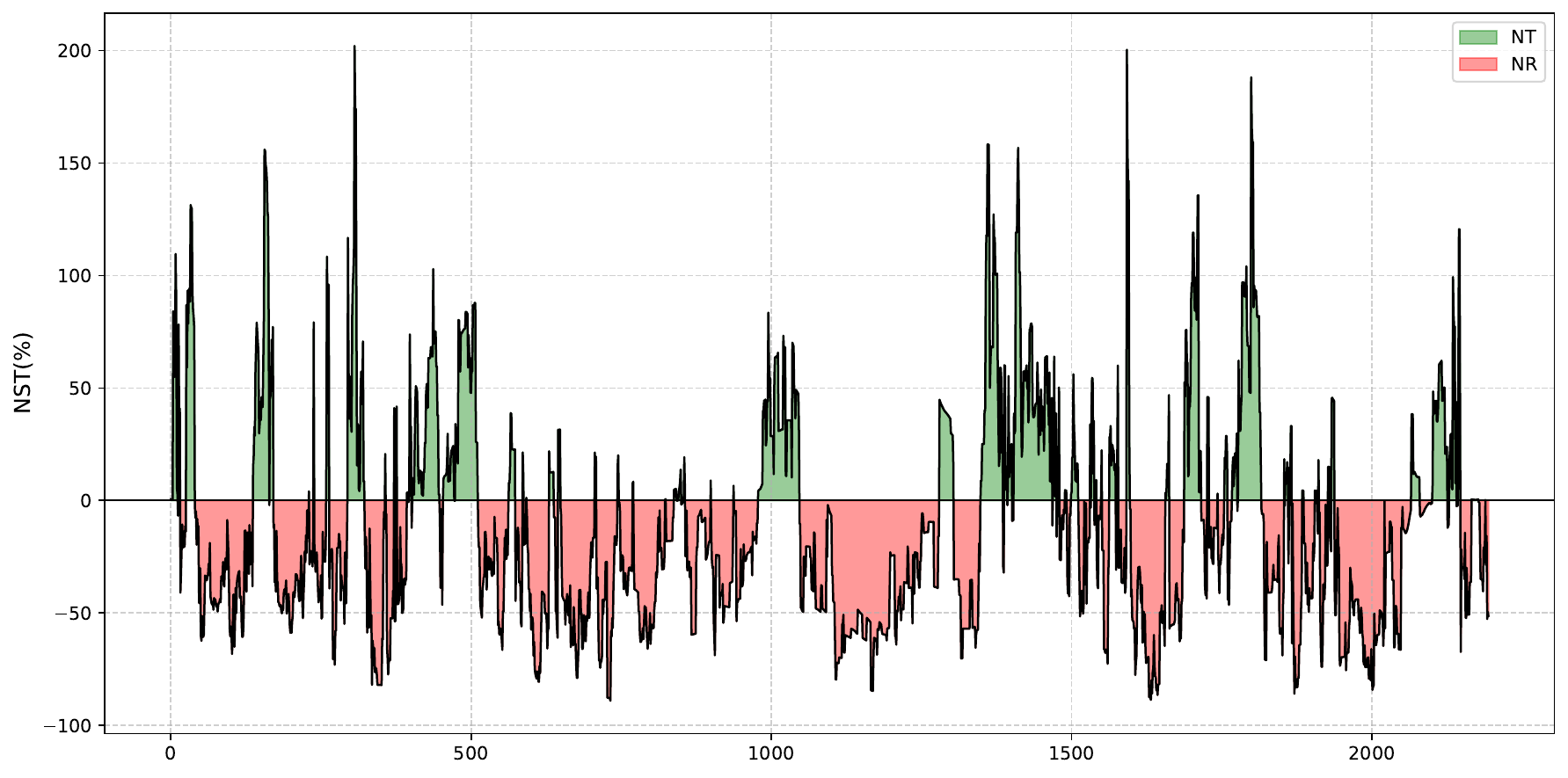}\label{fig:rv_eth_low}}\hfill
    \subfigure[ETH, $\tau=0.50$]{\includegraphics[width=0.32\linewidth]{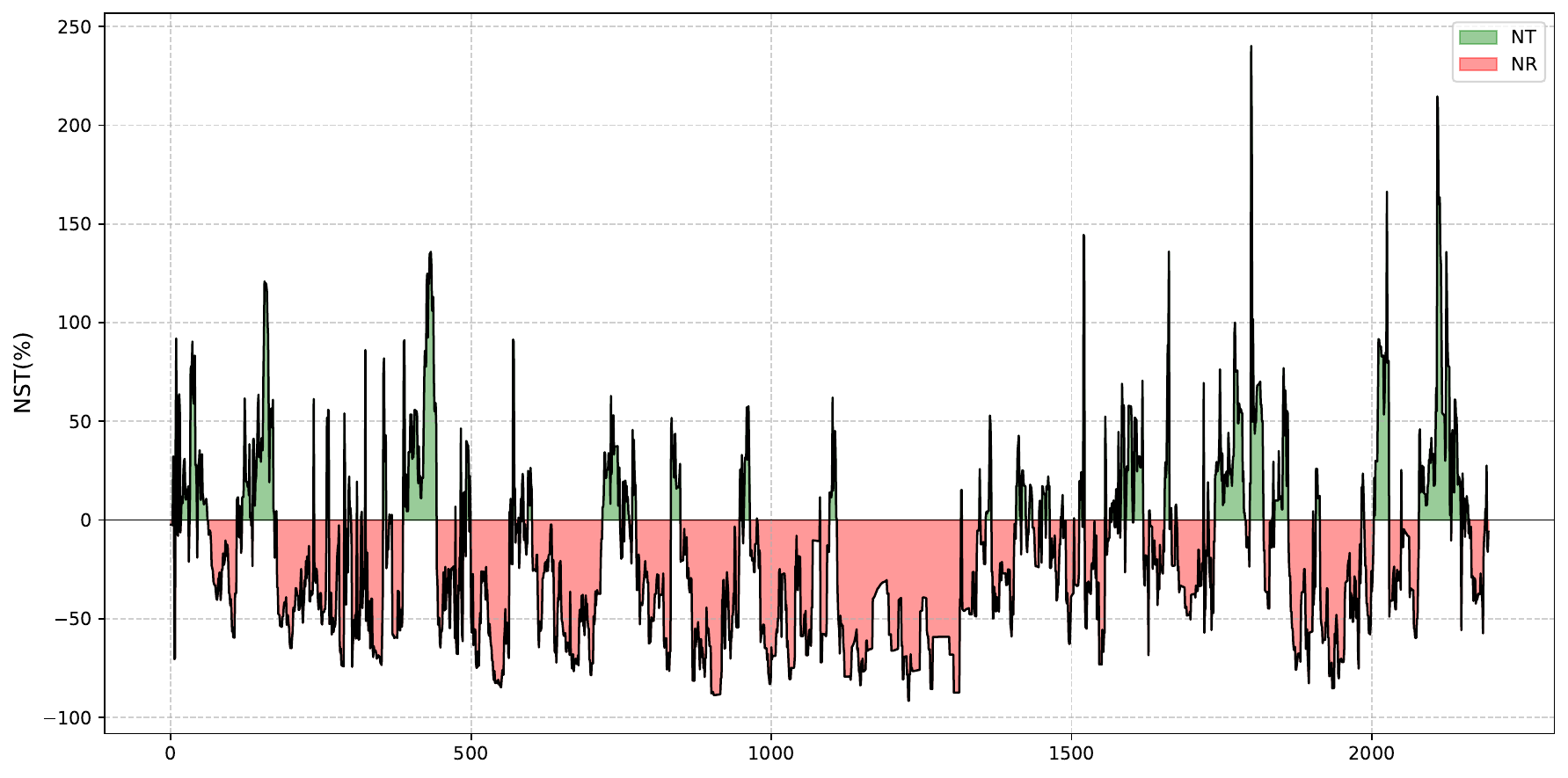}\label{fig:rv_eth_mid}}\hfill
    \subfigure[ETH, $\tau=0.95$]{\includegraphics[width=0.32\linewidth]{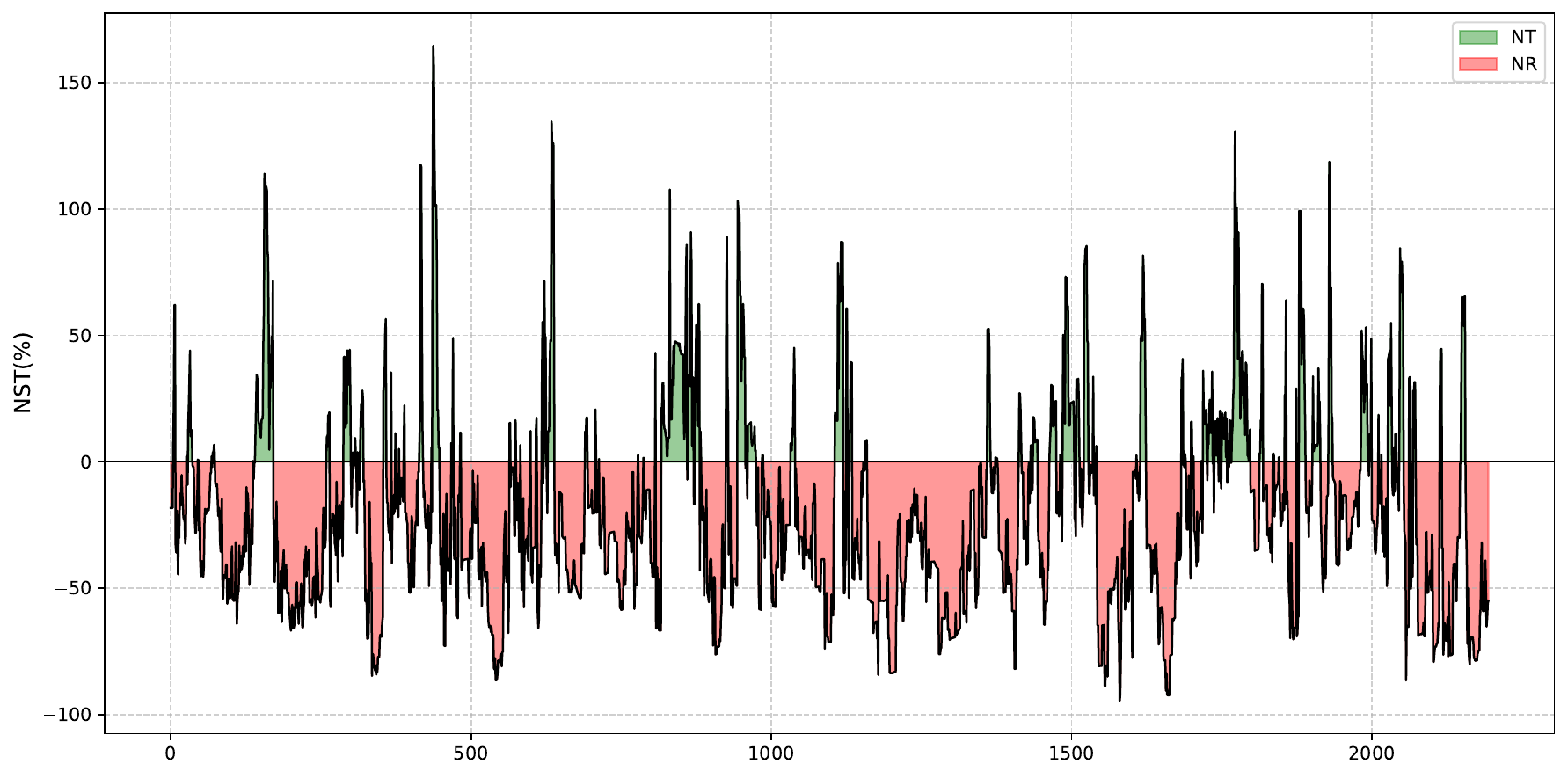}\label{fig:rv_eth_high}}
    \vspace{0.3cm}
    \subfigure[LTC, $\tau=0.05$]{\includegraphics[width=0.32\linewidth]{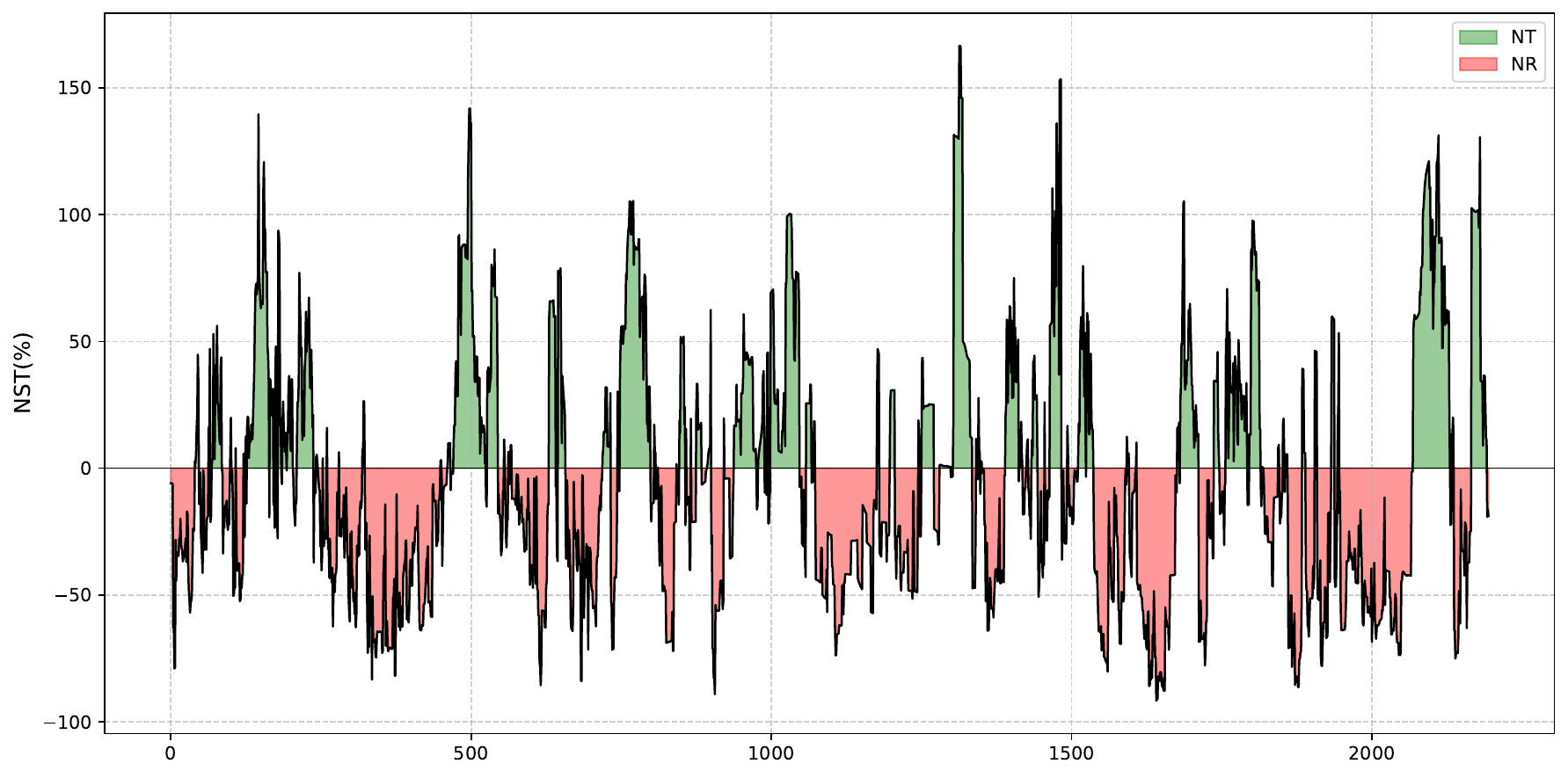}\label{fig:rv_ltc_low}}\hfill
    \subfigure[LTC, $\tau=0.50$]{\includegraphics[width=0.32\linewidth]{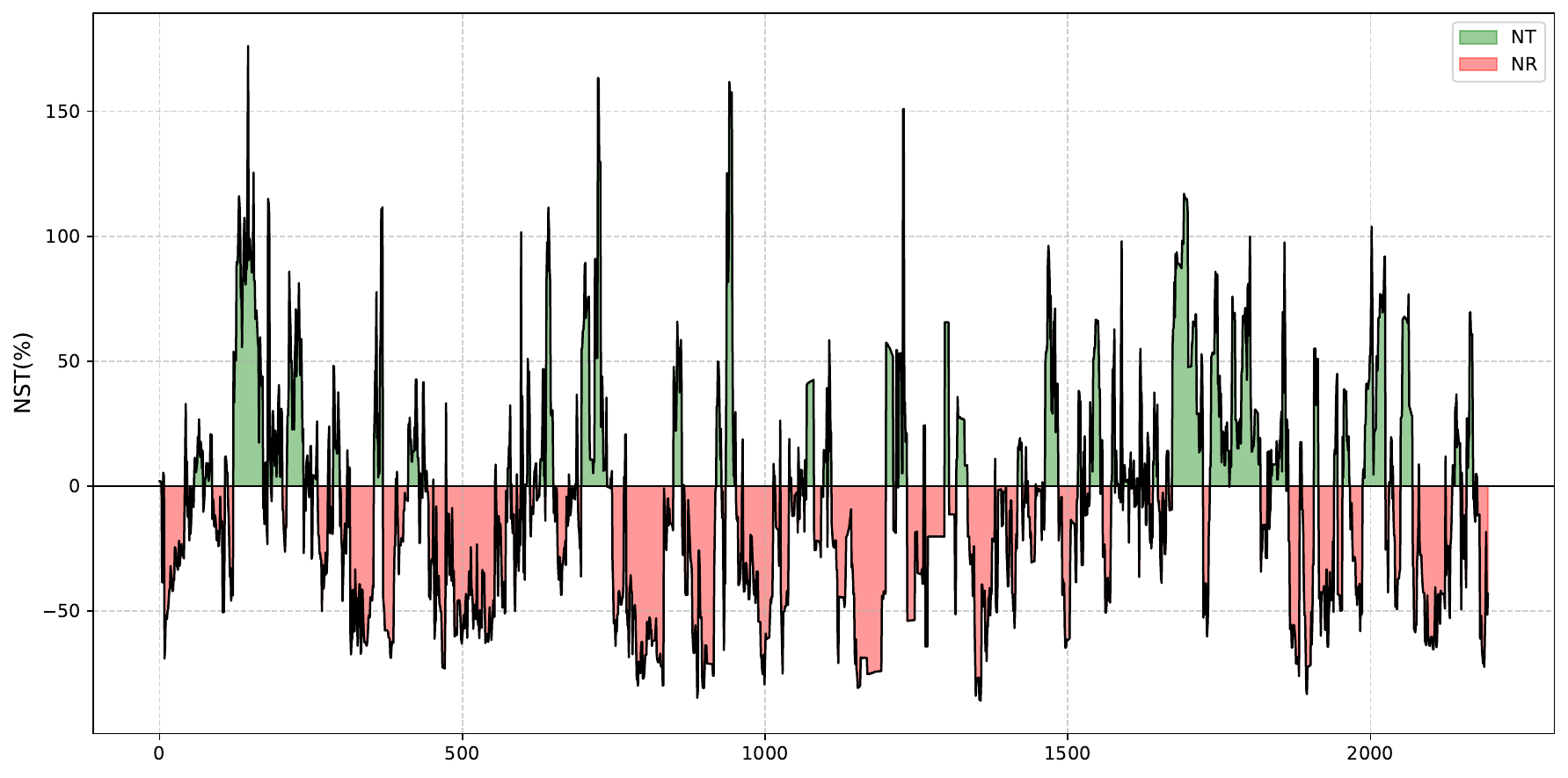}\label{fig:rv_ltc_mid}}\hfill
    \subfigure[LTC, $\tau=0.95$]{\includegraphics[width=0.32\linewidth]{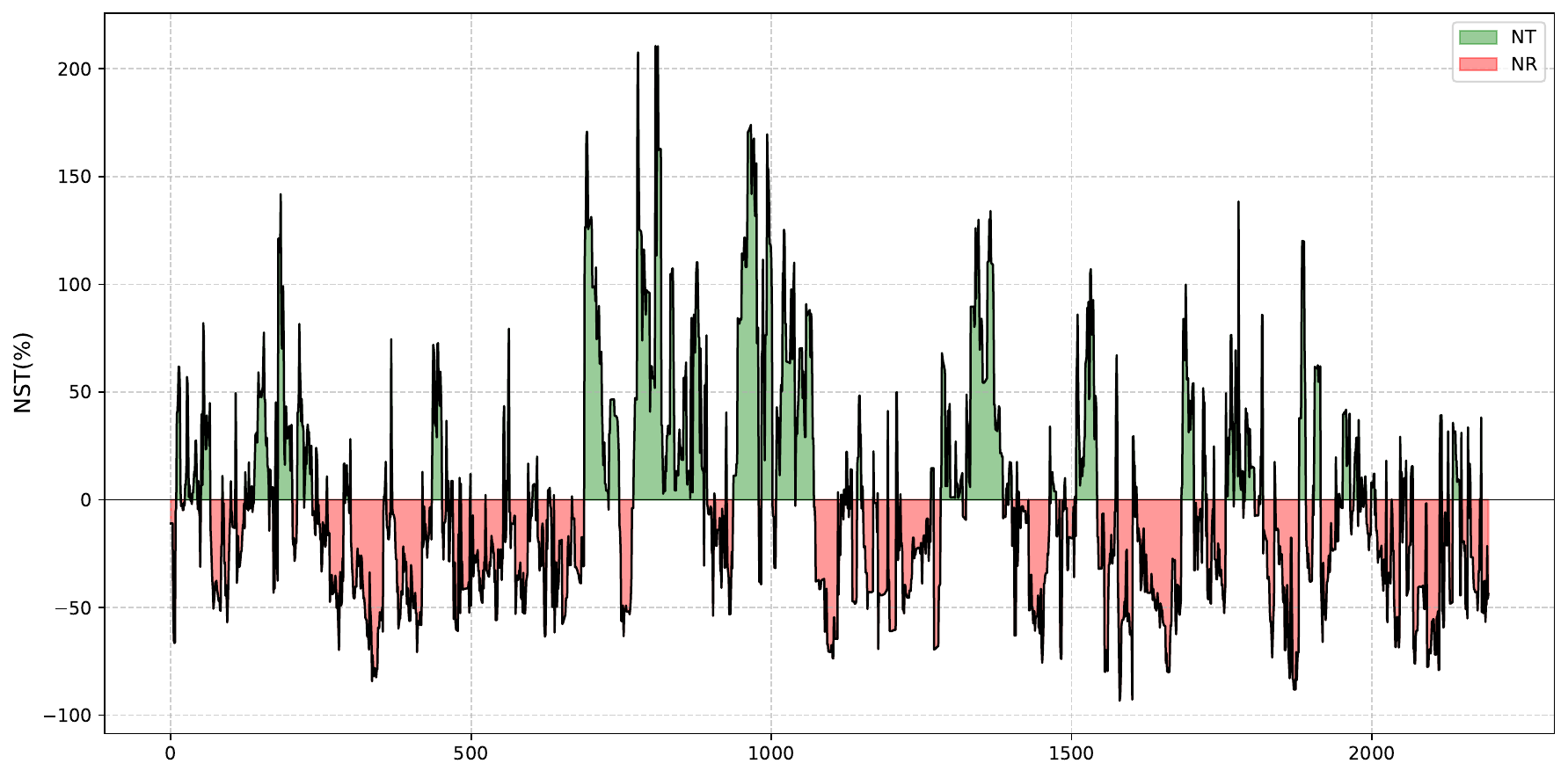}\label{fig:rv_ltc_high}}
    \vspace{0.3cm}
    \subfigure[XLM, $\tau=0.05$]{\includegraphics[width=0.32\linewidth]{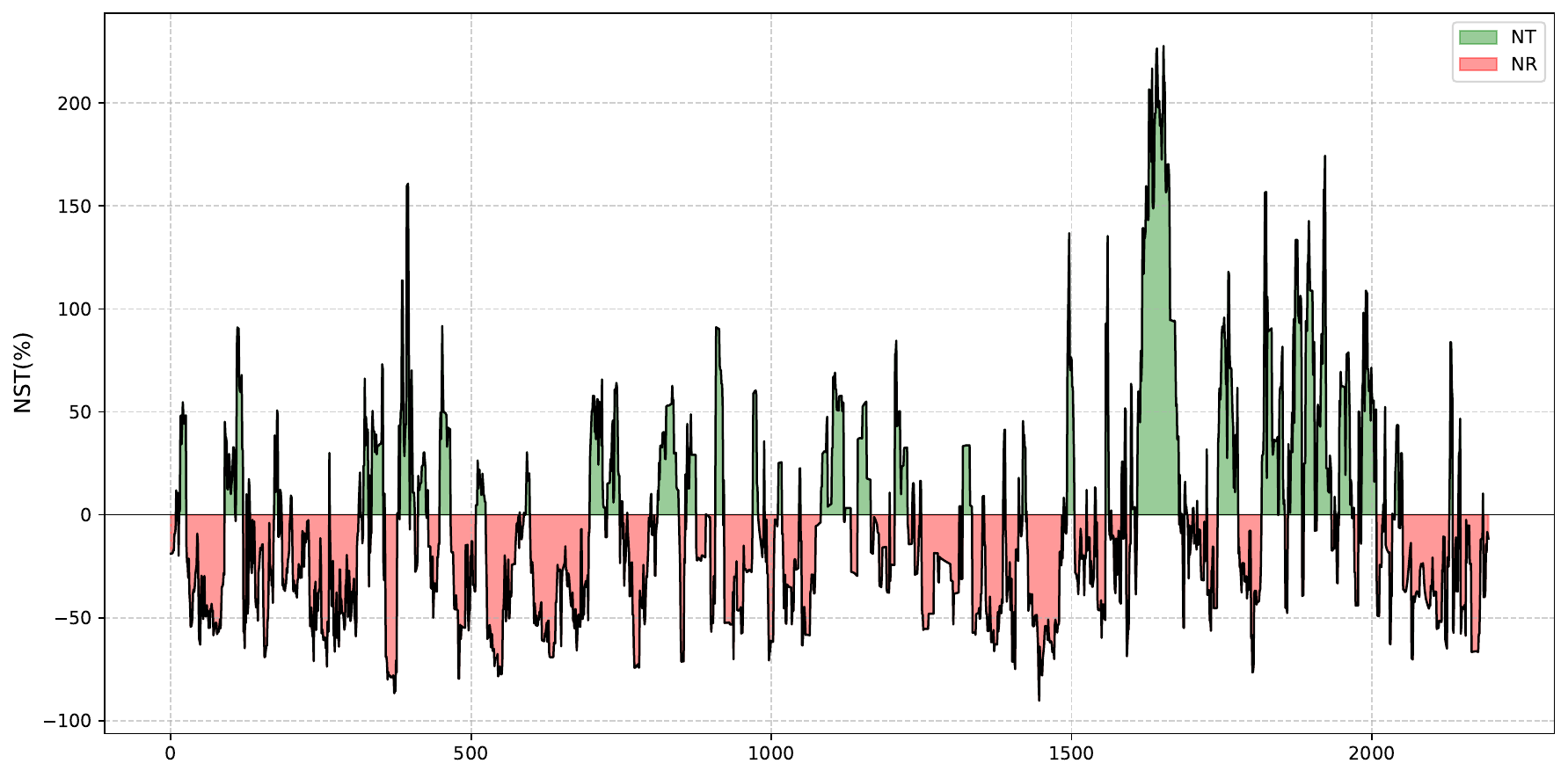}\label{fig:rv_xlm_low}}\hfill
    \subfigure[XLM, $\tau=0.50$]{\includegraphics[width=0.32\linewidth]{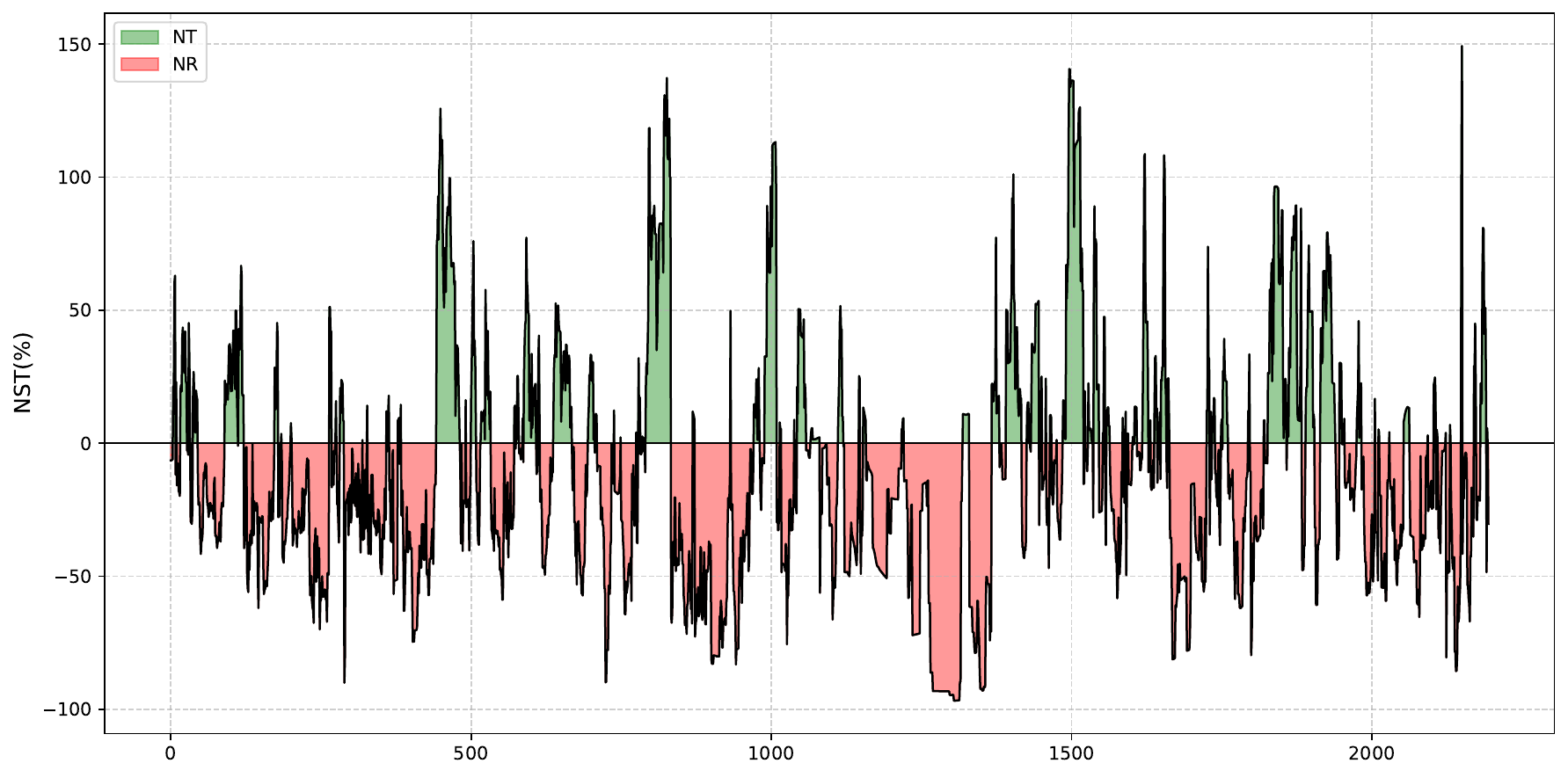}\label{fig:rv_xlm_mid}}\hfill
    \subfigure[XLM, $\tau=0.95$]{\includegraphics[width=0.32\linewidth]{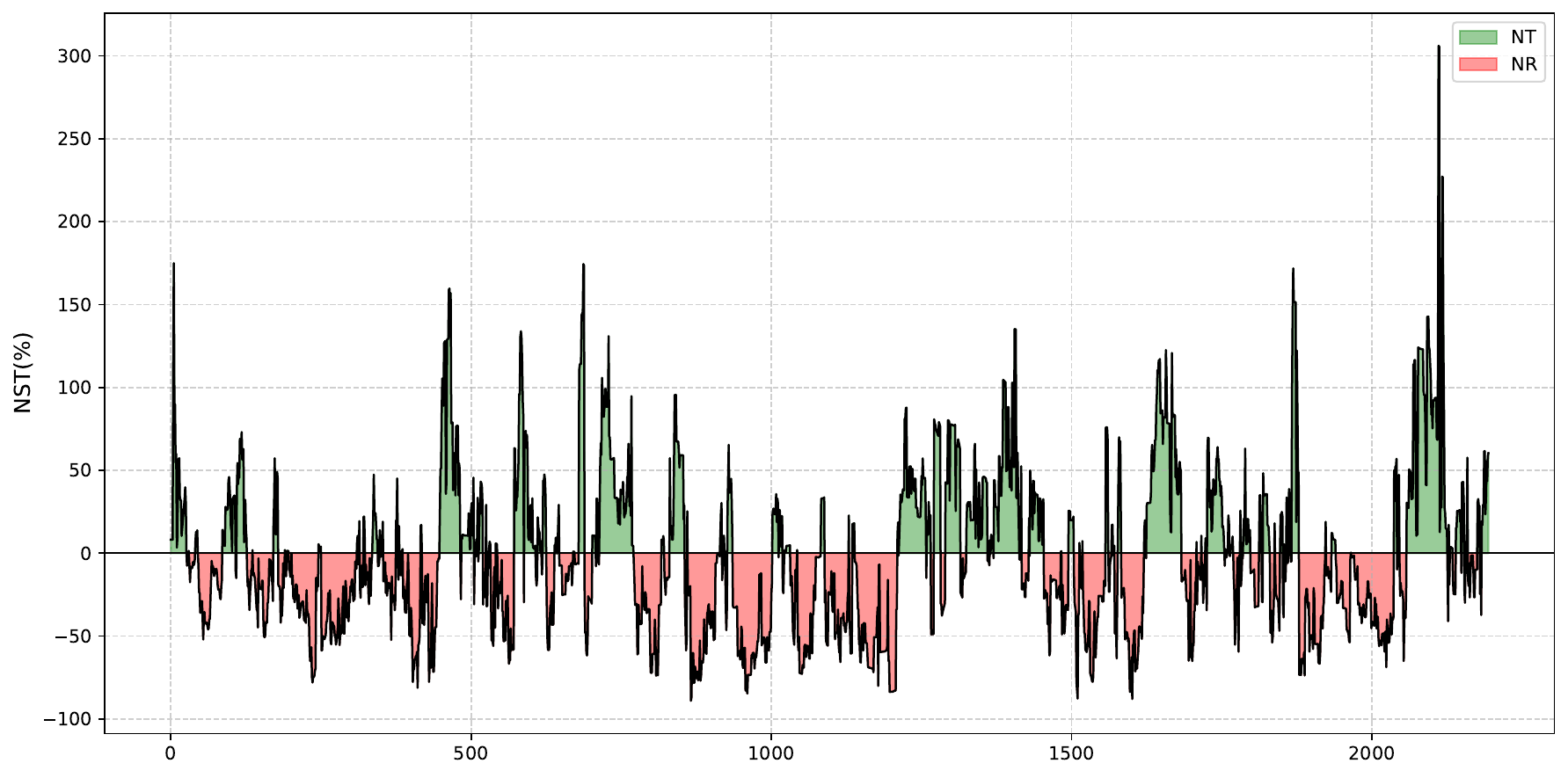}\label{fig:rv_xlm_high}}
    \vspace{0.3cm}
    \subfigure[XRP, $\tau=0.05$]{\includegraphics[width=0.32\linewidth]{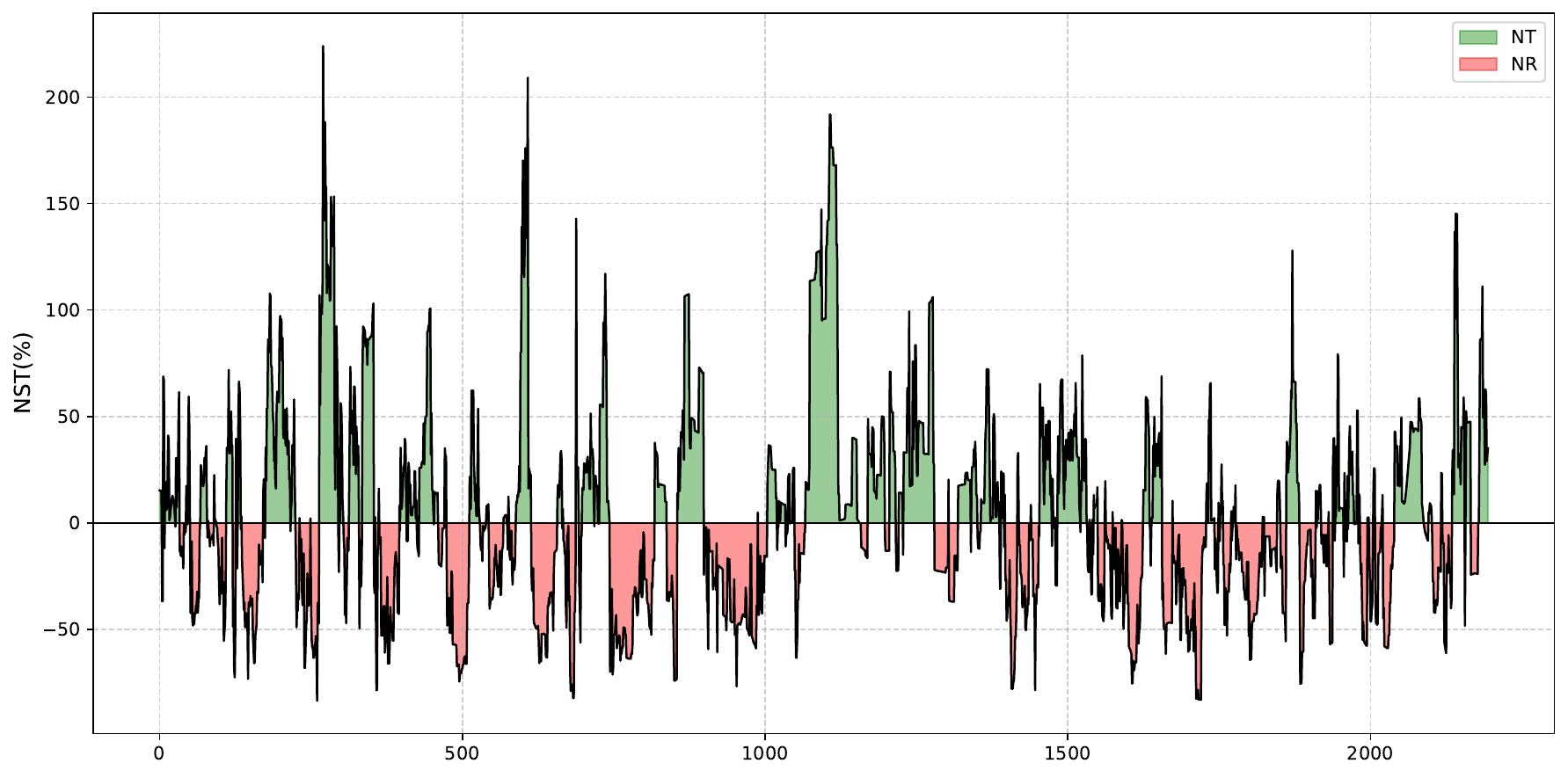}\label{fig:rv_xrp_low}}\hfill
    \subfigure[XRP, $\tau=0.50$]{\includegraphics[width=0.32\linewidth]{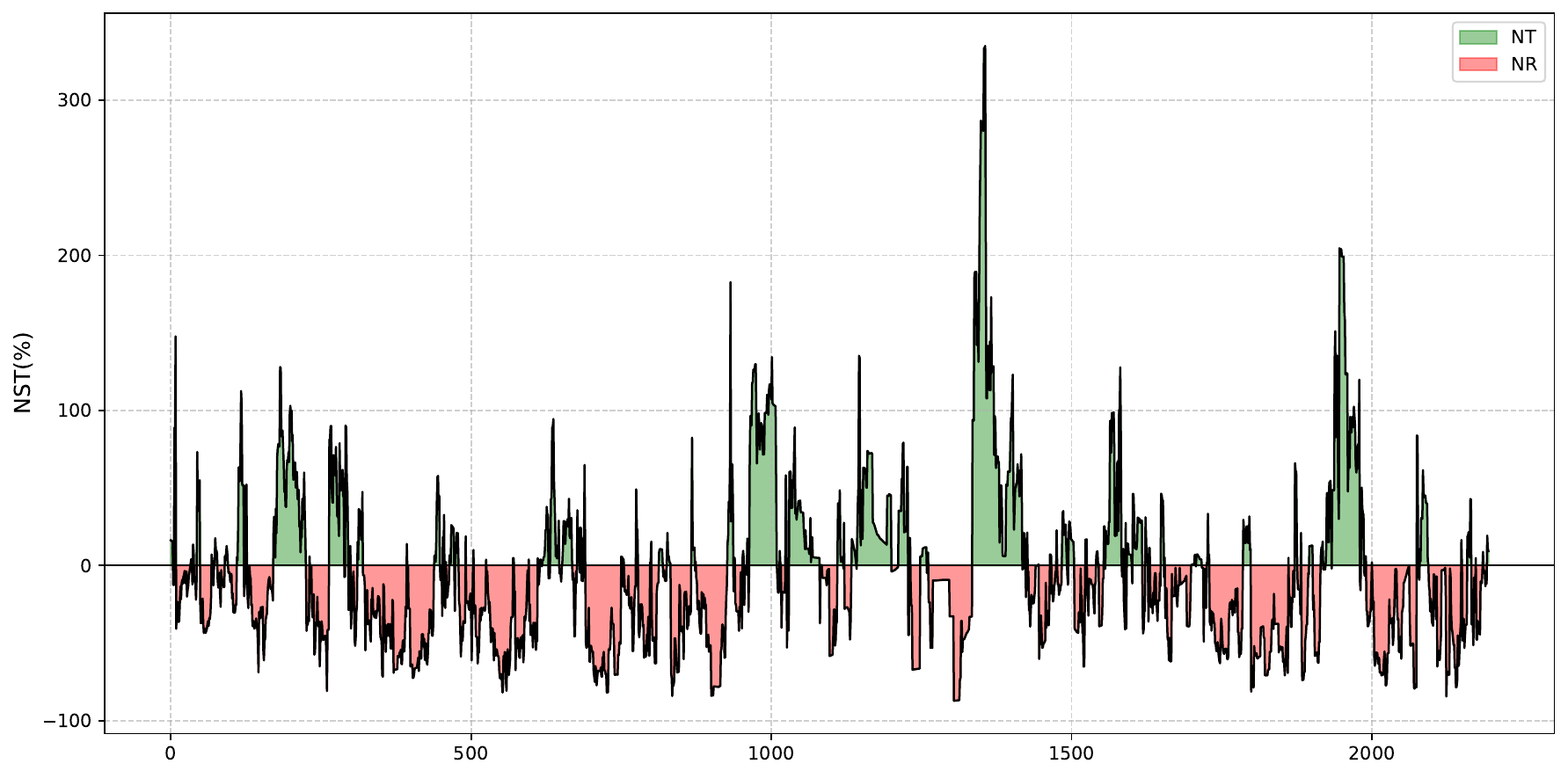}\label{fig:rv_xrp_mid}}\hfill
    \subfigure[XRP, $\tau=0.95$]{\includegraphics[width=0.32\linewidth]{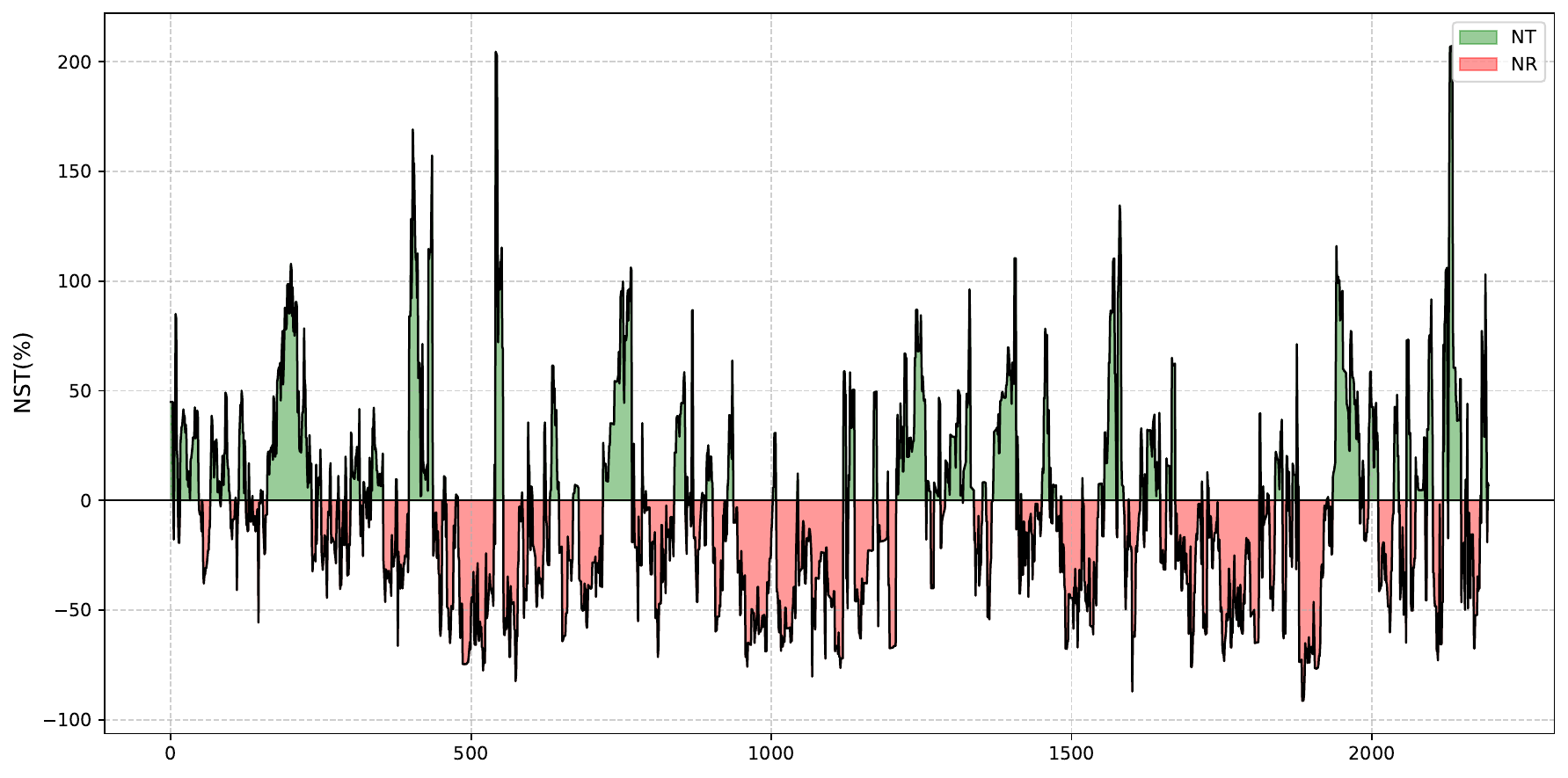}\label{fig:rv_xrp_high}}
\end{figure}

\begin{figure}[p]
    \centering
    \caption{Quantile net spillovers for major cryptocurrencies using CV as the feature variable.}
    \label{fig:cv_net_spillover_by_coin}

    \subfigure[BTC, $\tau=0.05$]{\includegraphics[width=0.32\linewidth]{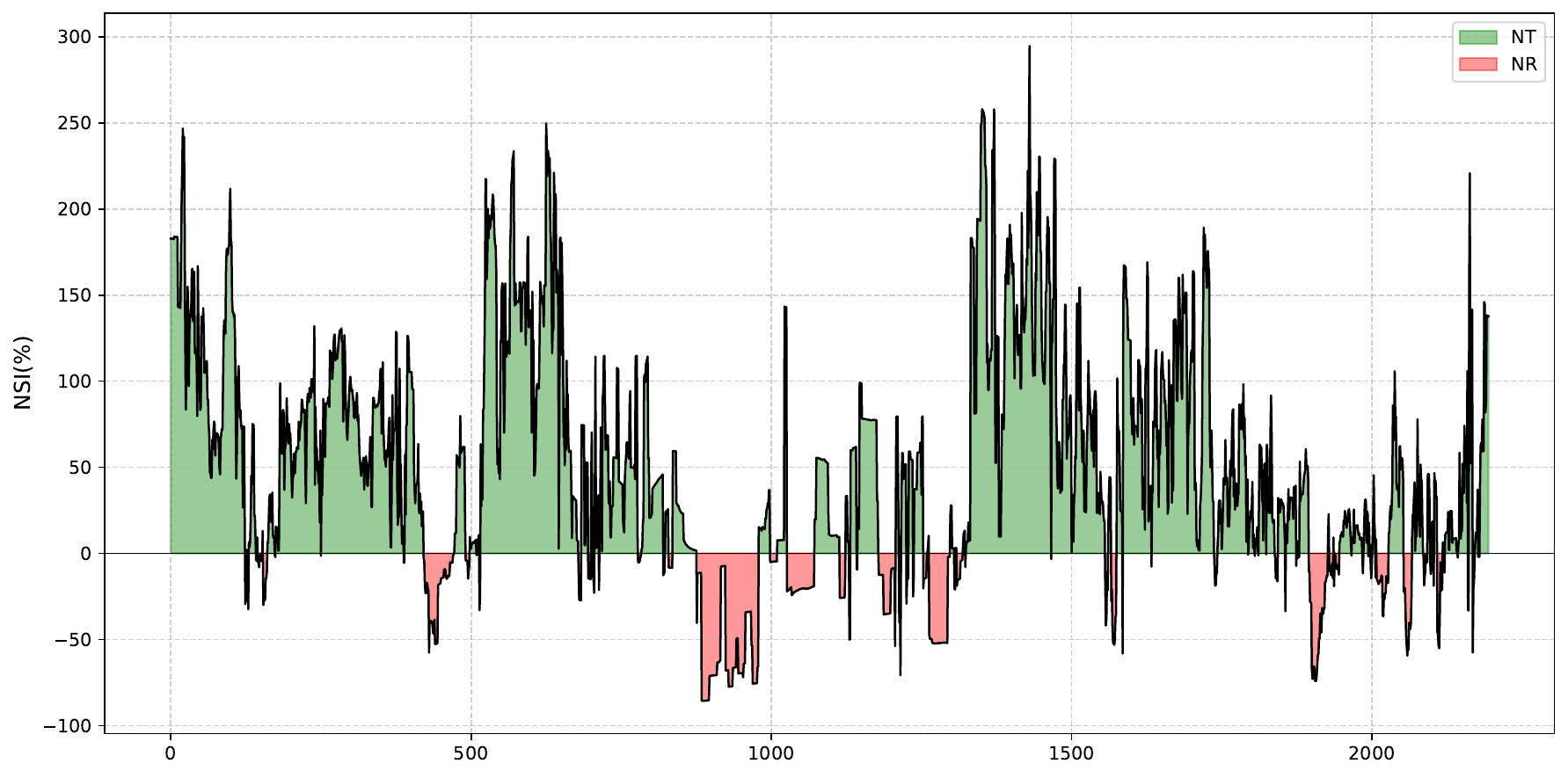}\label{fig:cv_btc_low}}\hfill
    \subfigure[BTC, $\tau=0.50$]{\includegraphics[width=0.32\linewidth]{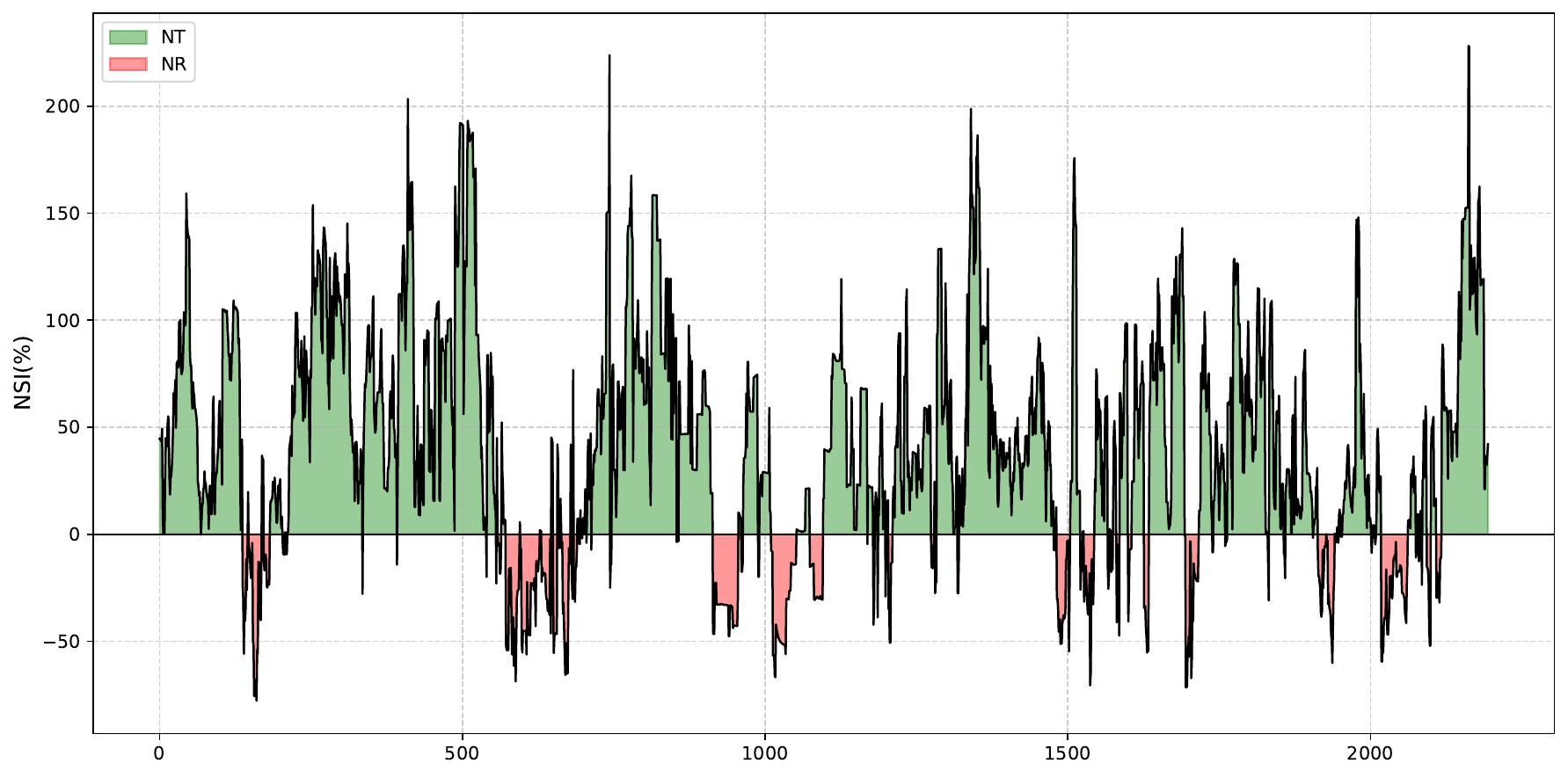}\label{fig:cv_btc_mid}}\hfill
    \subfigure[BTC, $\tau=0.95$]{\includegraphics[width=0.32\linewidth]{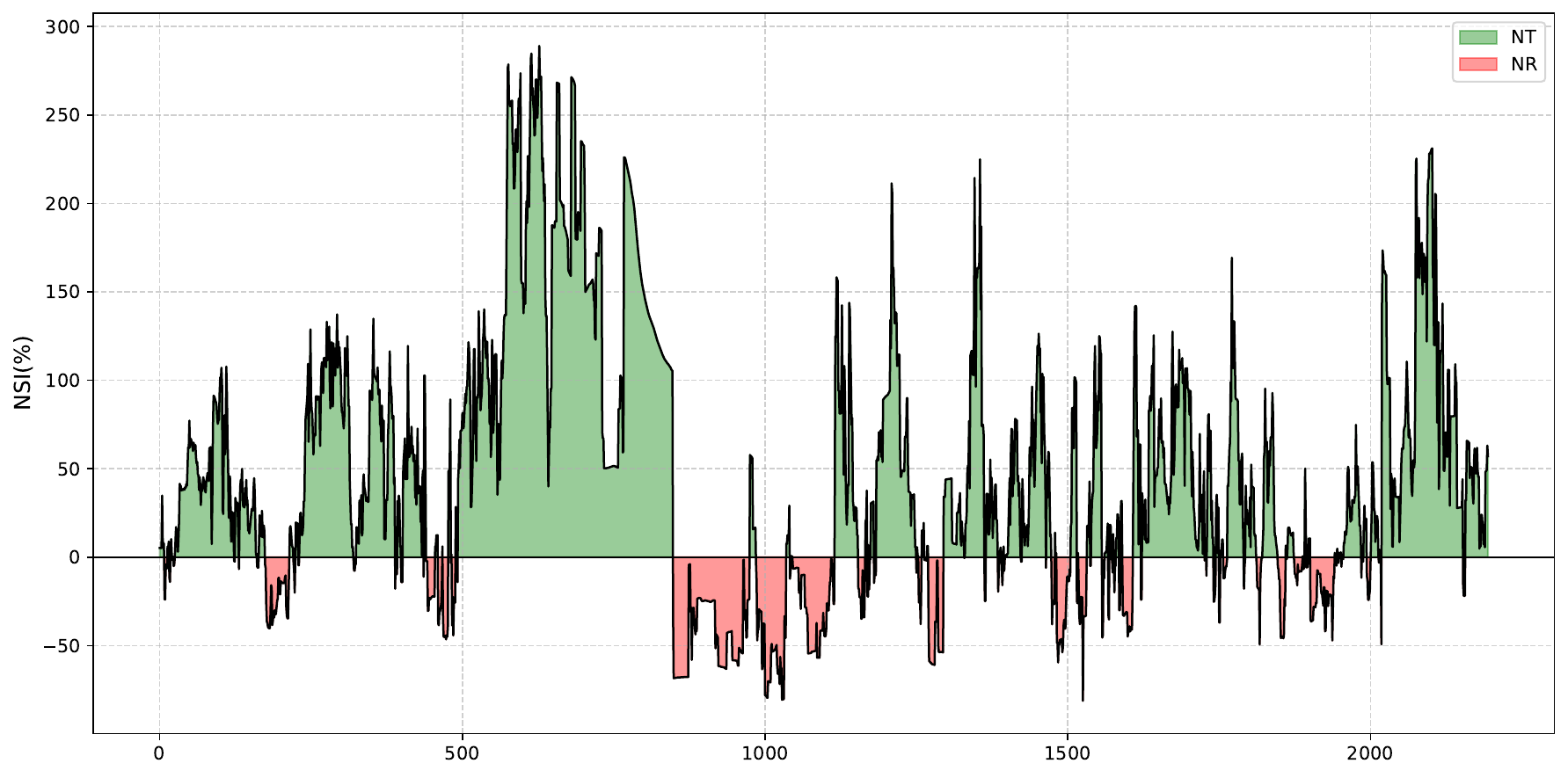}\label{fig:cv_btc_high}}
    \vspace{0.3cm}
    \subfigure[DASH, $\tau=0.05$]{\includegraphics[width=0.32\linewidth]{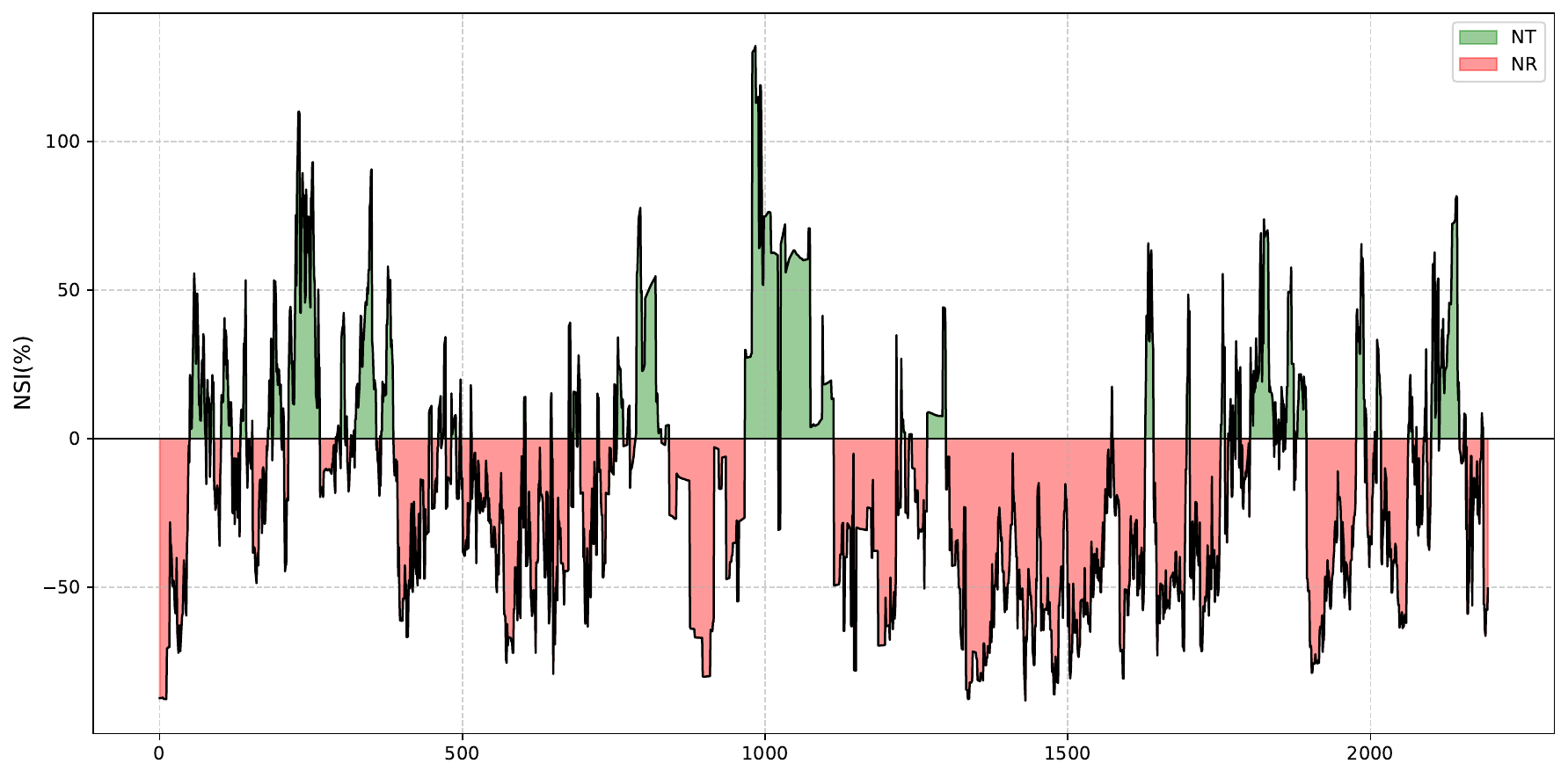}\label{fig:cv_dash_low}}\hfill
    \subfigure[DASH, $\tau=0.50$]{\includegraphics[width=0.32\linewidth]{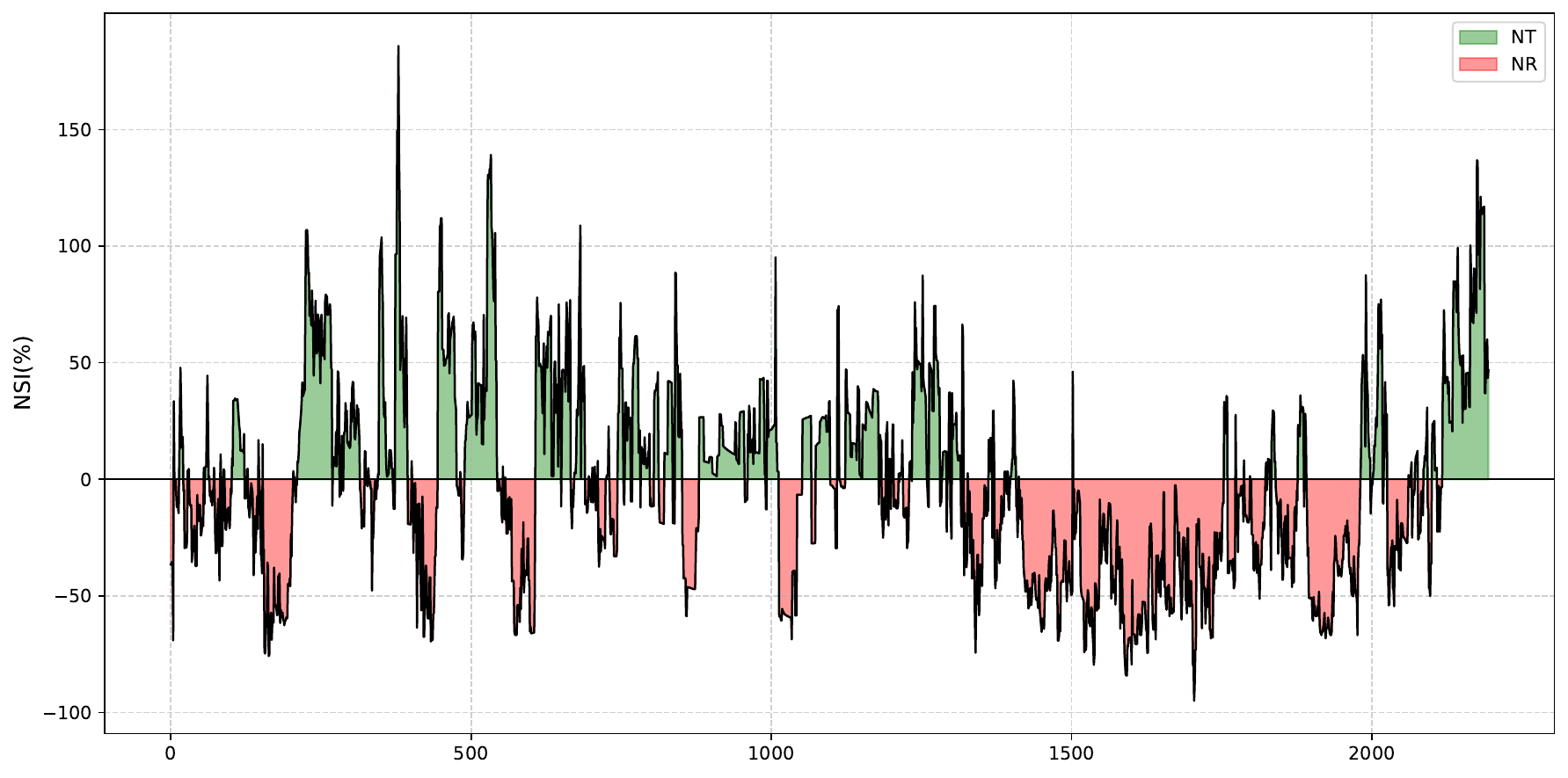}\label{fig:cv_dash_mid}}\hfill
    \subfigure[DASH, $\tau=0.95$]{\includegraphics[width=0.32\linewidth]{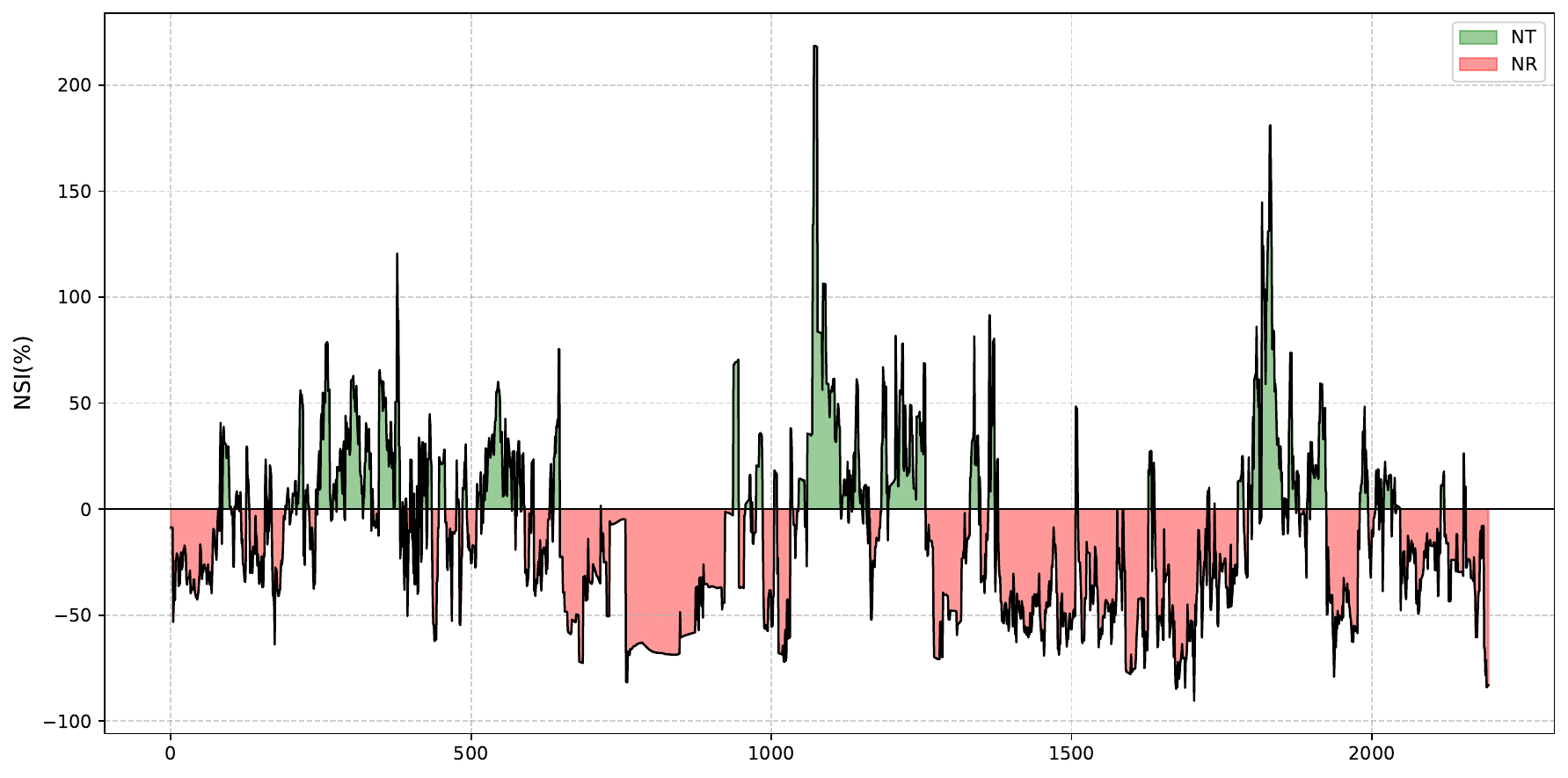}\label{fig:cv_dash_high}}
    \vspace{0.3cm}
    \subfigure[ETH, $\tau=0.05$]{\includegraphics[width=0.32\linewidth]{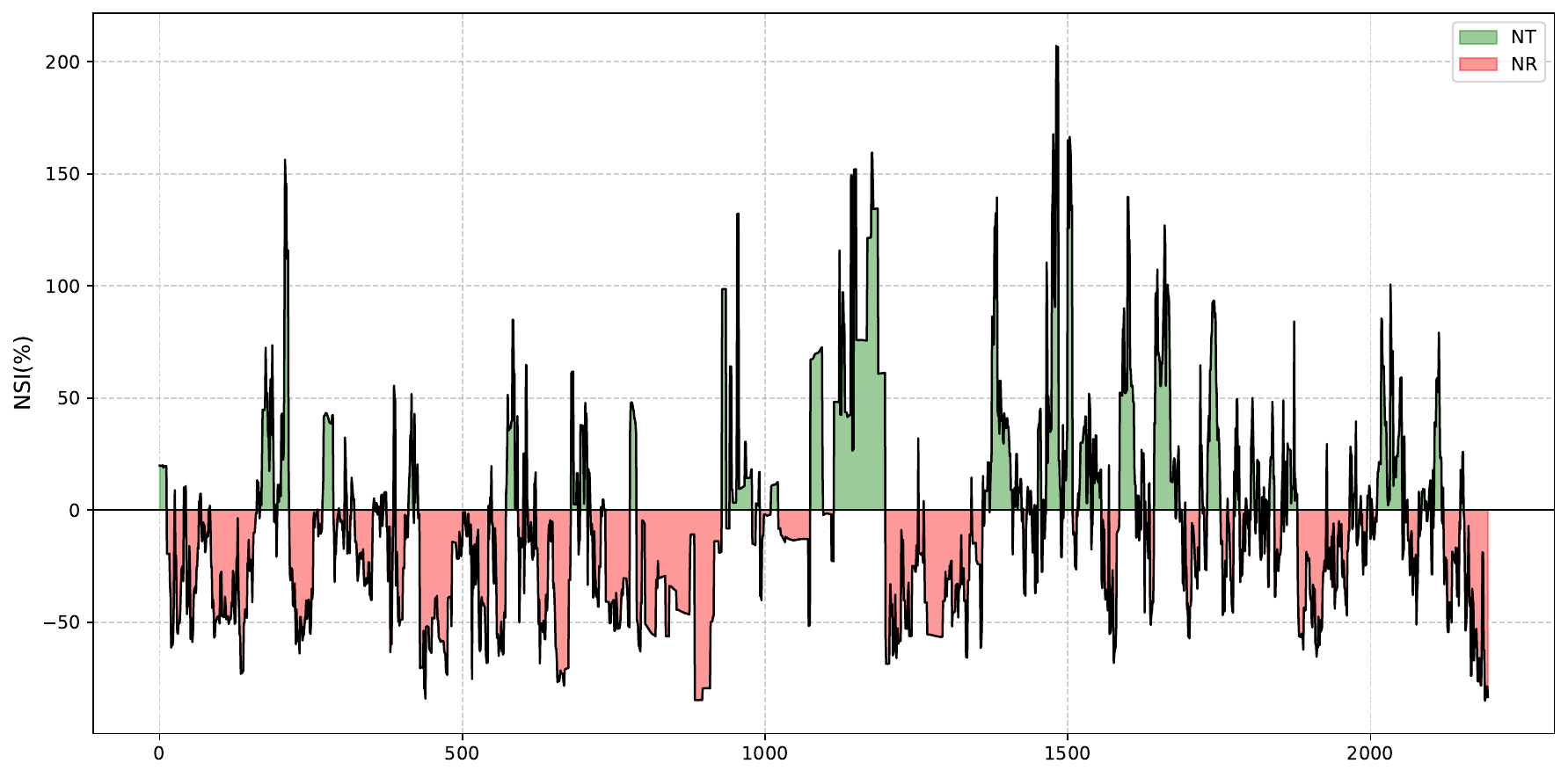}\label{fig:cv_eth_low}}\hfill
    \subfigure[ETH, $\tau=0.50$]{\includegraphics[width=0.32\linewidth]{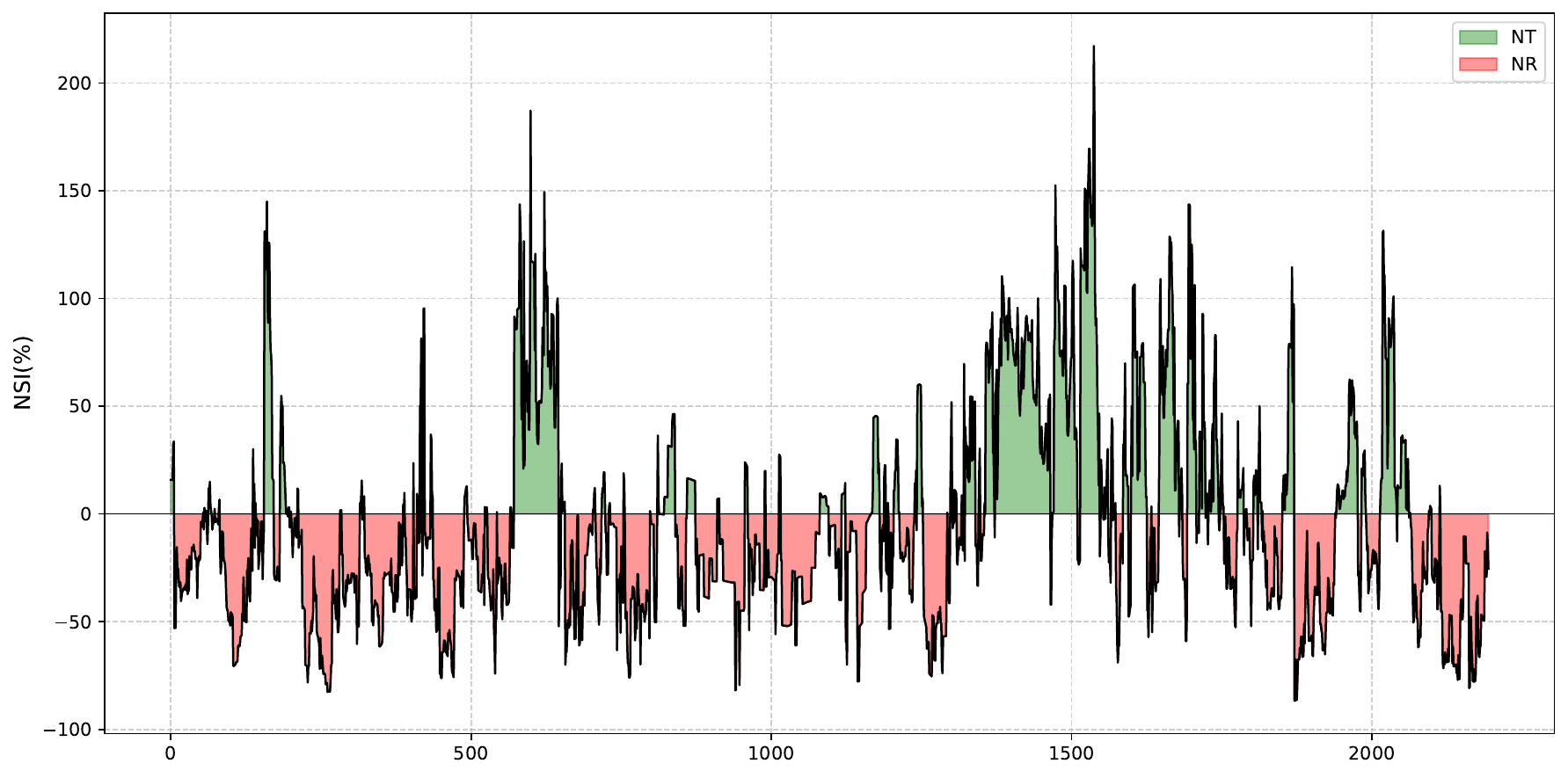}\label{fig:cv_eth_mid}}\hfill
    \subfigure[ETH, $\tau=0.95$]{\includegraphics[width=0.32\linewidth]{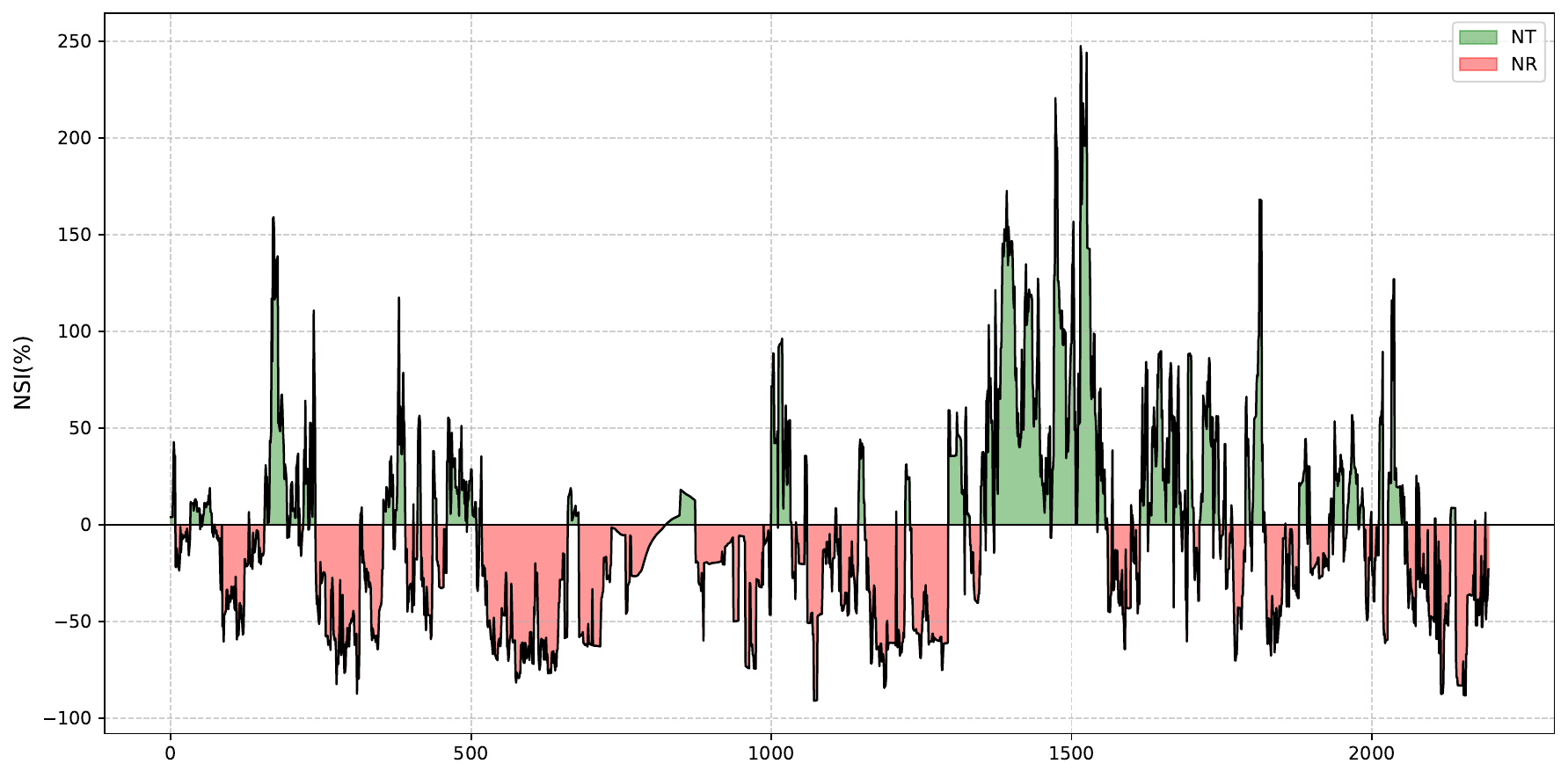}\label{fig:cv_eth_high}}
    \vspace{0.3cm}
    \subfigure[LTC, $\tau=0.05$]{\includegraphics[width=0.32\linewidth]{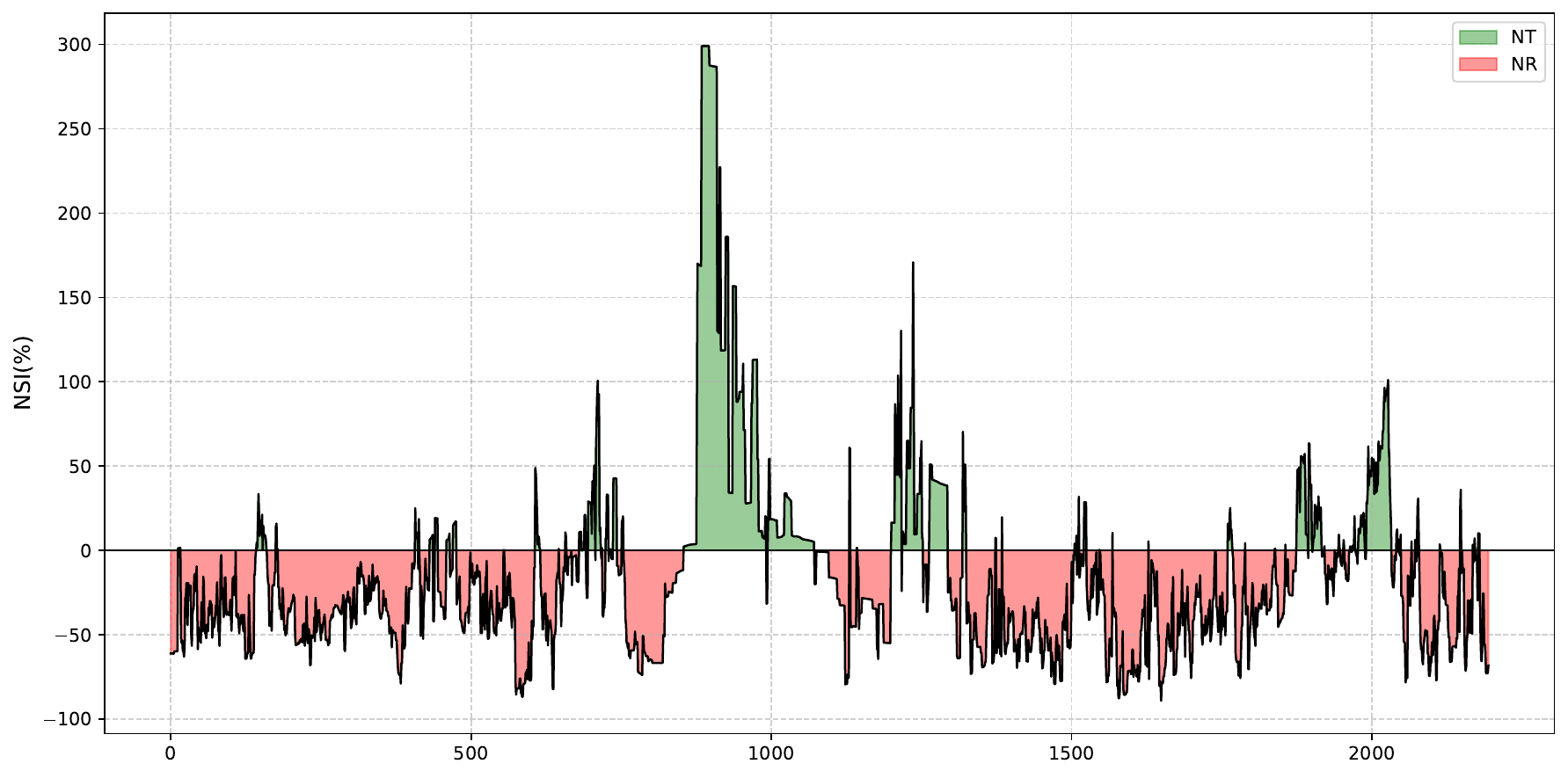}\label{fig:cv_ltc_low}}\hfill
    \subfigure[LTC, $\tau=0.50$]{\includegraphics[width=0.32\linewidth]{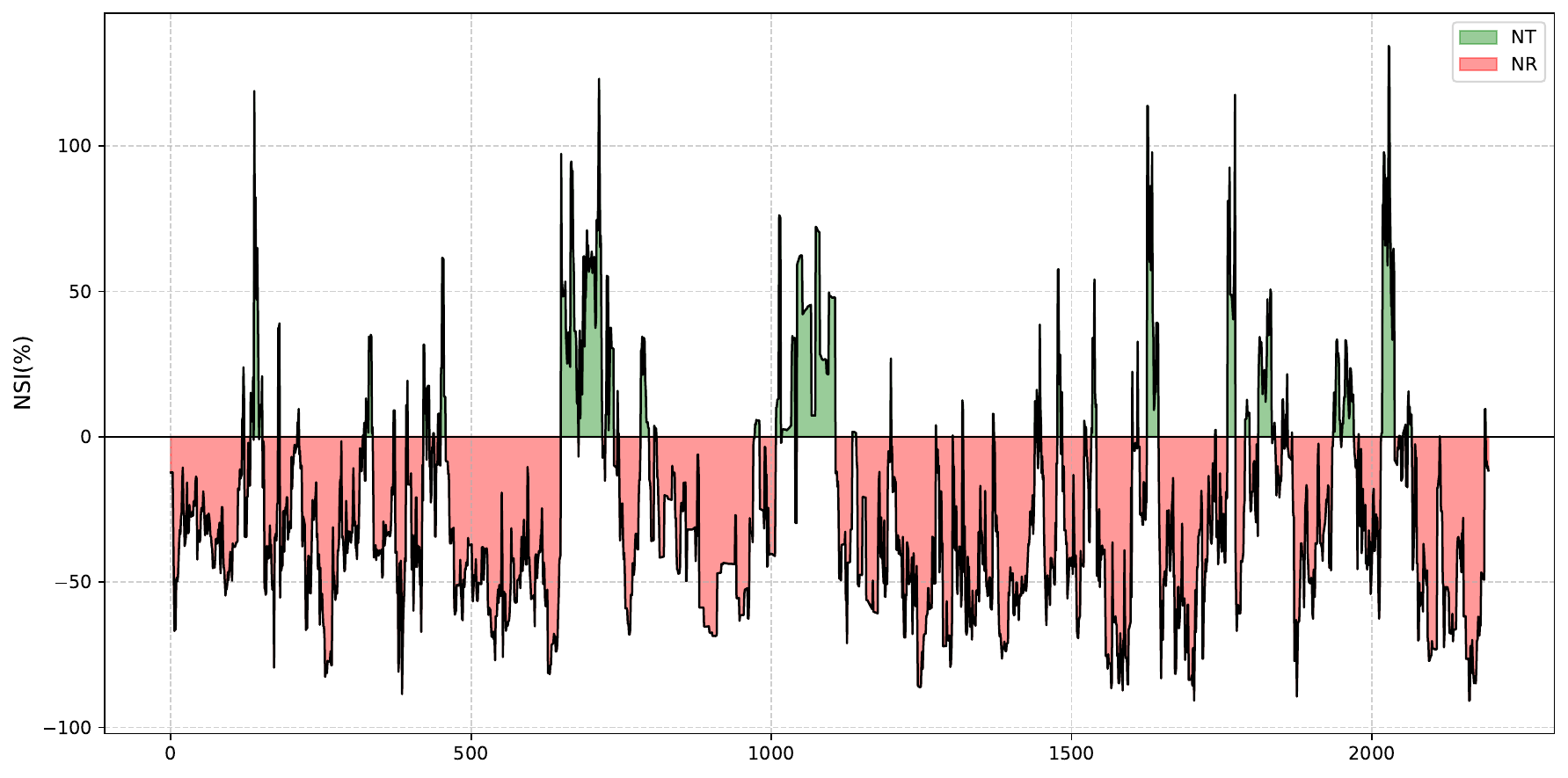}\label{fig:cv_ltc_mid}}\hfill
    \subfigure[LTC, $\tau=0.95$]{\includegraphics[width=0.32\linewidth]{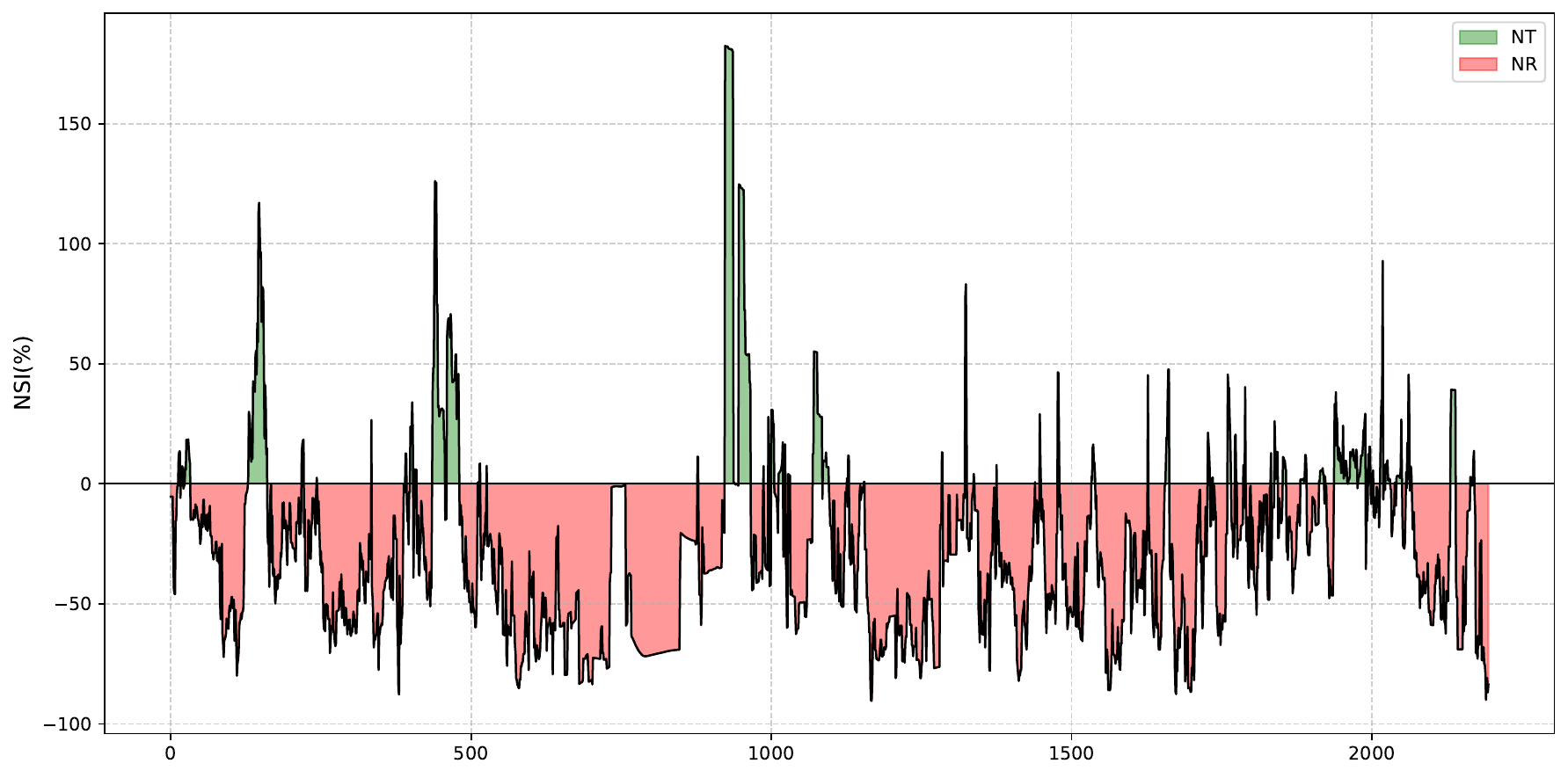}\label{fig:cv_ltc_high}}
    \vspace{0.3cm}
    \subfigure[XLM, $\tau=0.05$]{\includegraphics[width=0.32\linewidth]{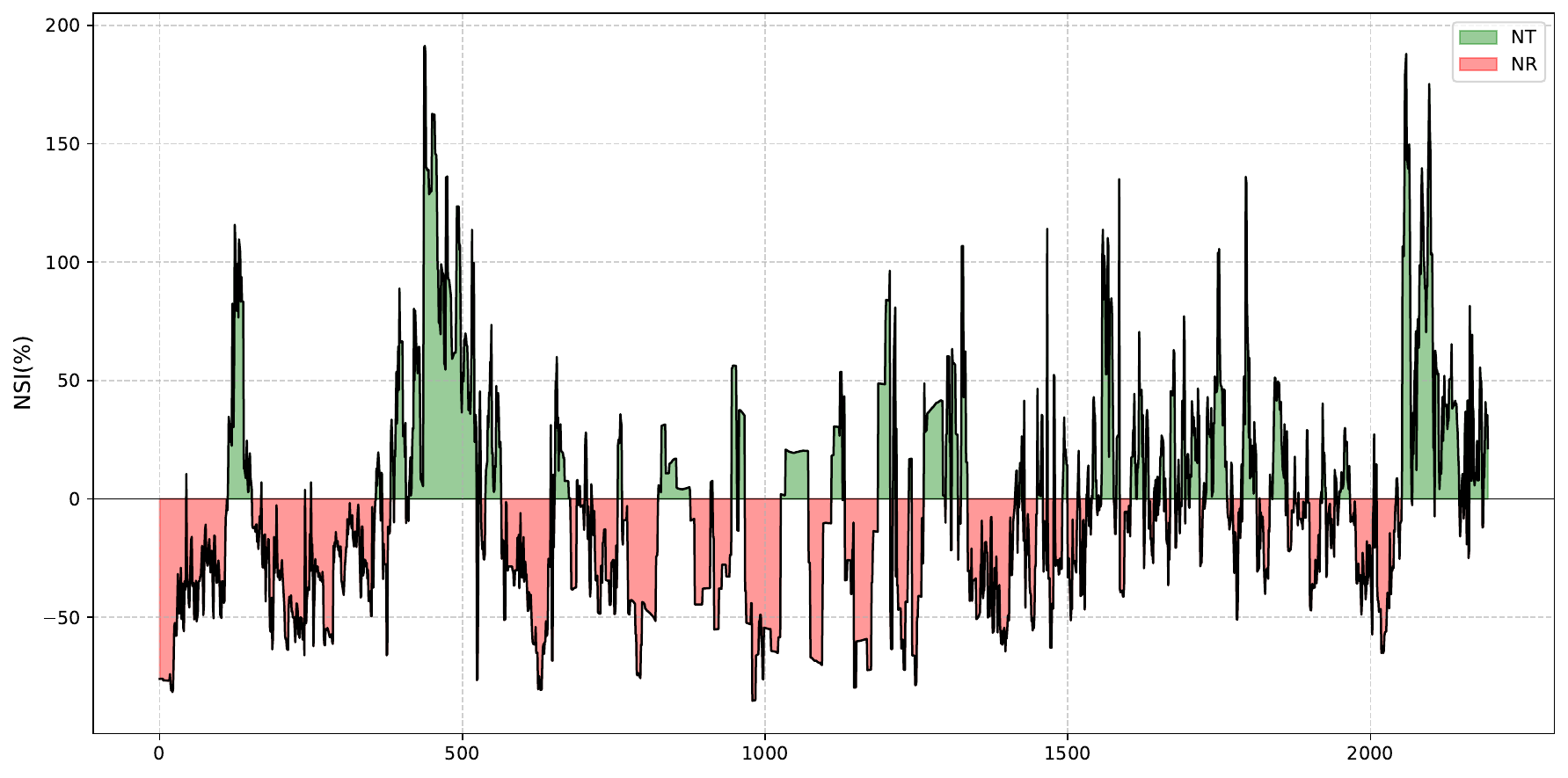}\label{fig:cv_xlm_low}}\hfill
    \subfigure[XLM, $\tau=0.50$]{\includegraphics[width=0.32\linewidth]{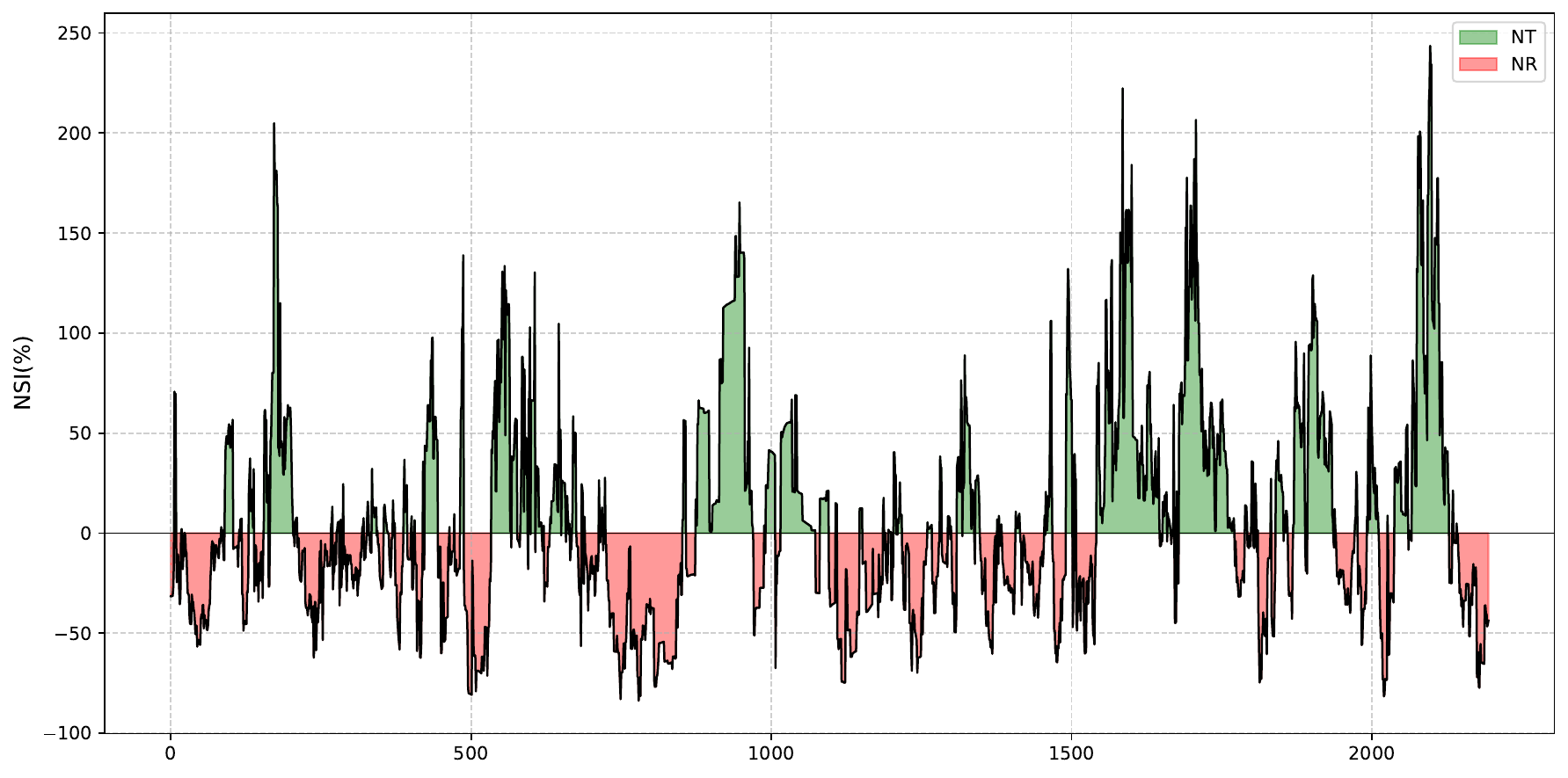}\label{fig:cv_xlm_mid}}\hfill
    \subfigure[XLM, $\tau=0.95$]{\includegraphics[width=0.32\linewidth]{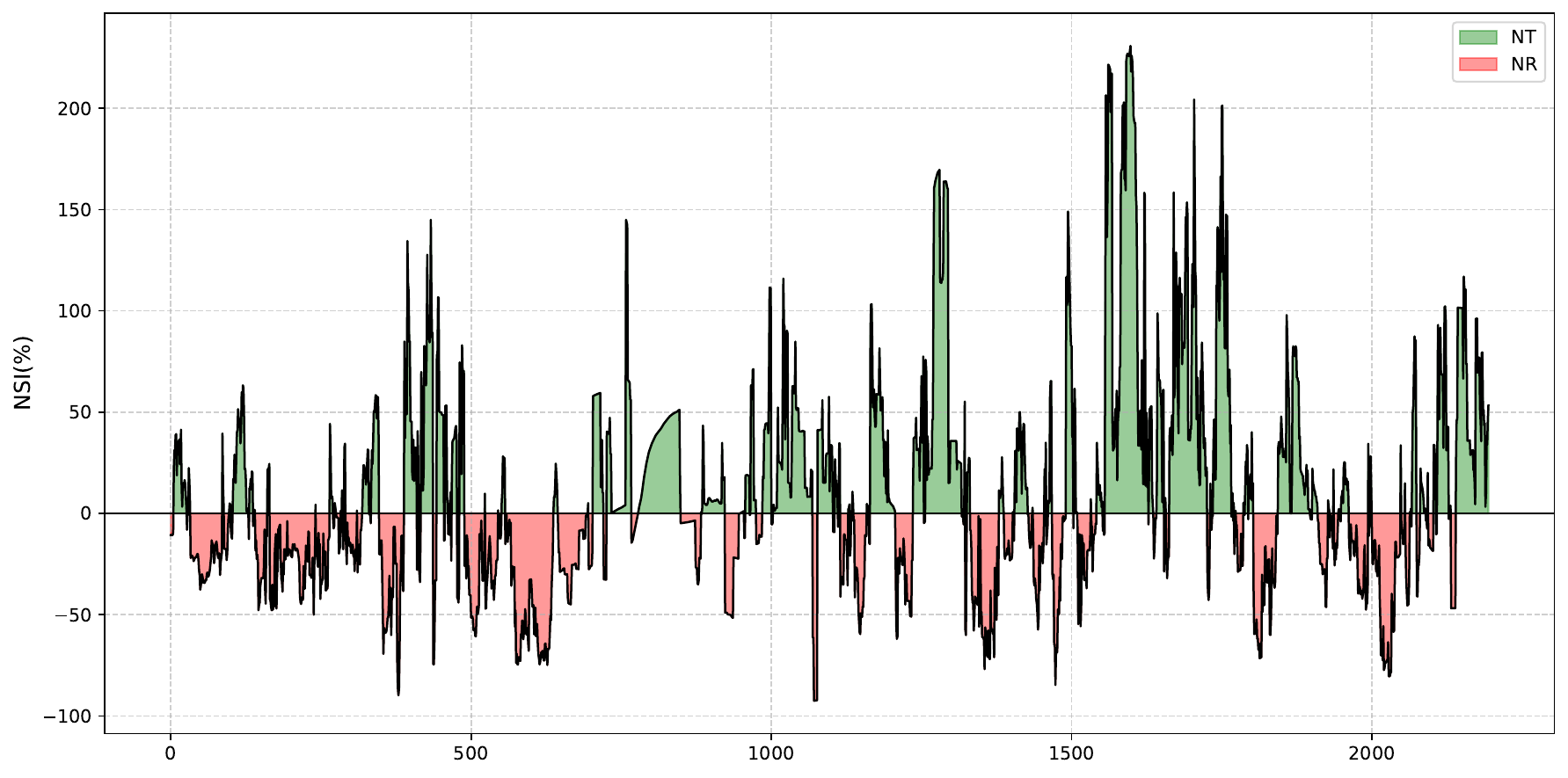}\label{fig:cv_xlm_high}}
    \vspace{0.3cm}
    \subfigure[XRP, $\tau=0.05$]{\includegraphics[width=0.32\linewidth]{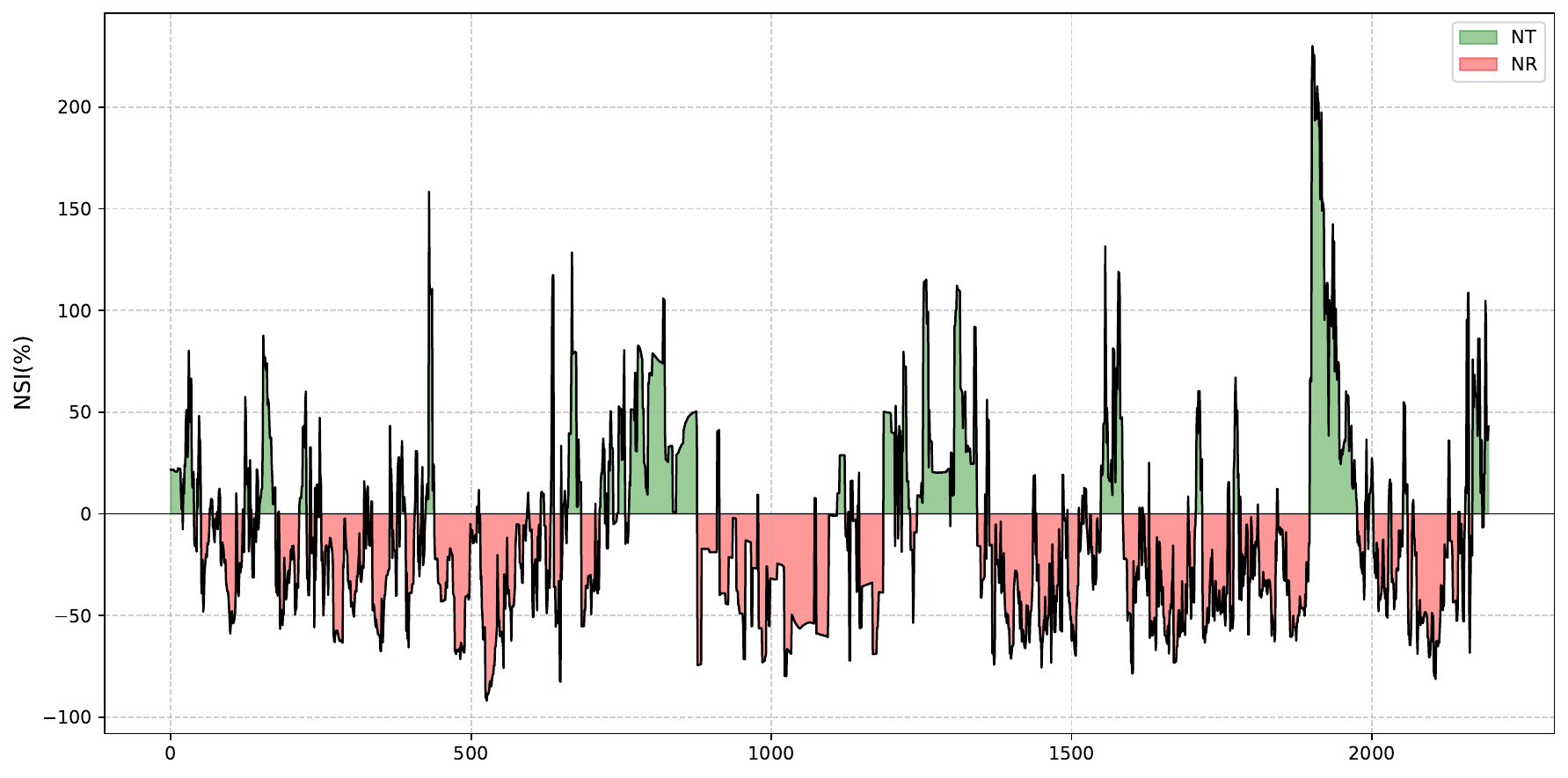}\label{fig:cv_xrp_low}}\hfill
    \subfigure[XRP, $\tau=0.50$]{\includegraphics[width=0.32\linewidth]{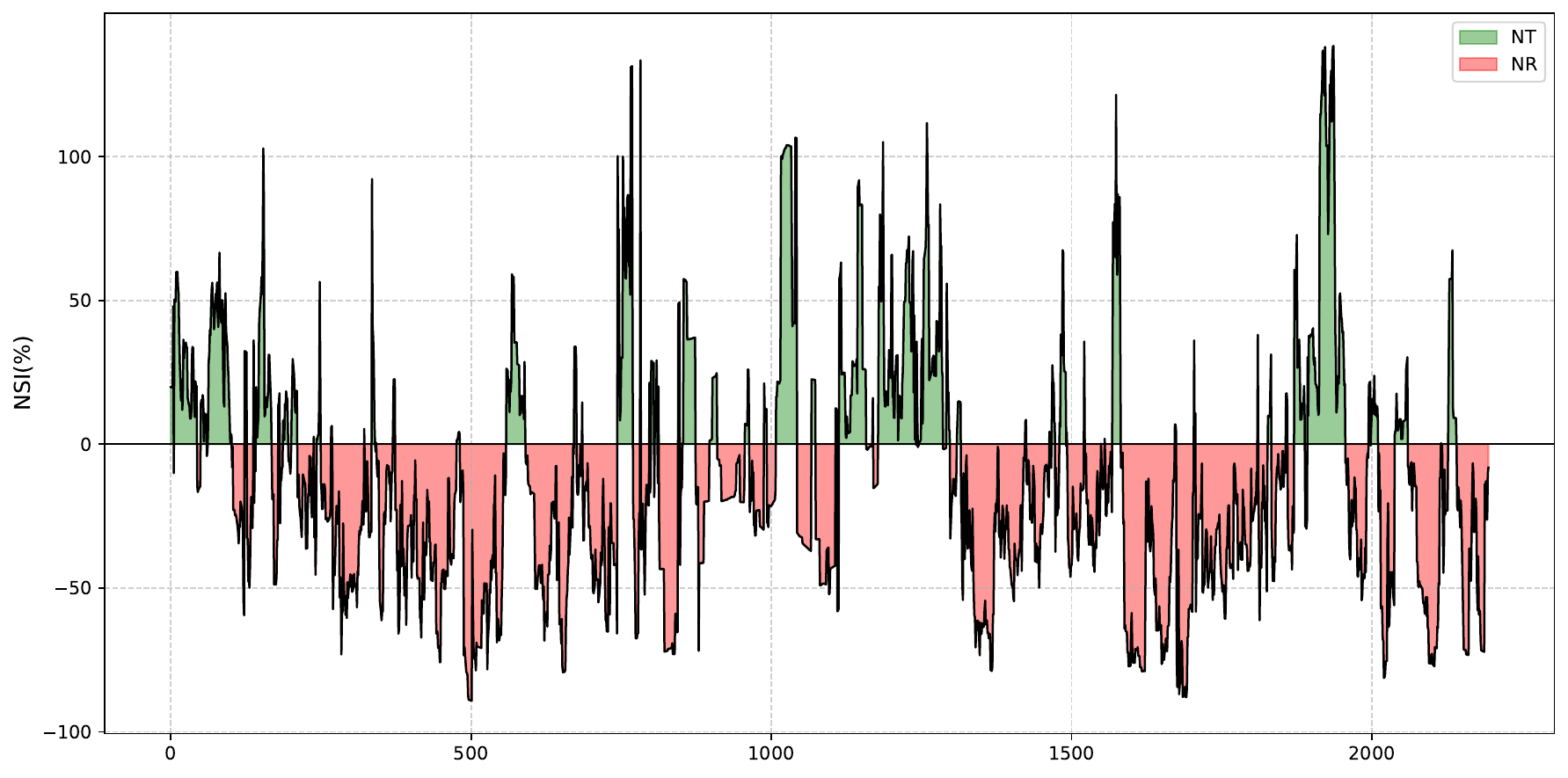}\label{fig:cv_xrp_mid}}\hfill
    \subfigure[XRP, $\tau=0.95$]{\includegraphics[width=0.32\linewidth]{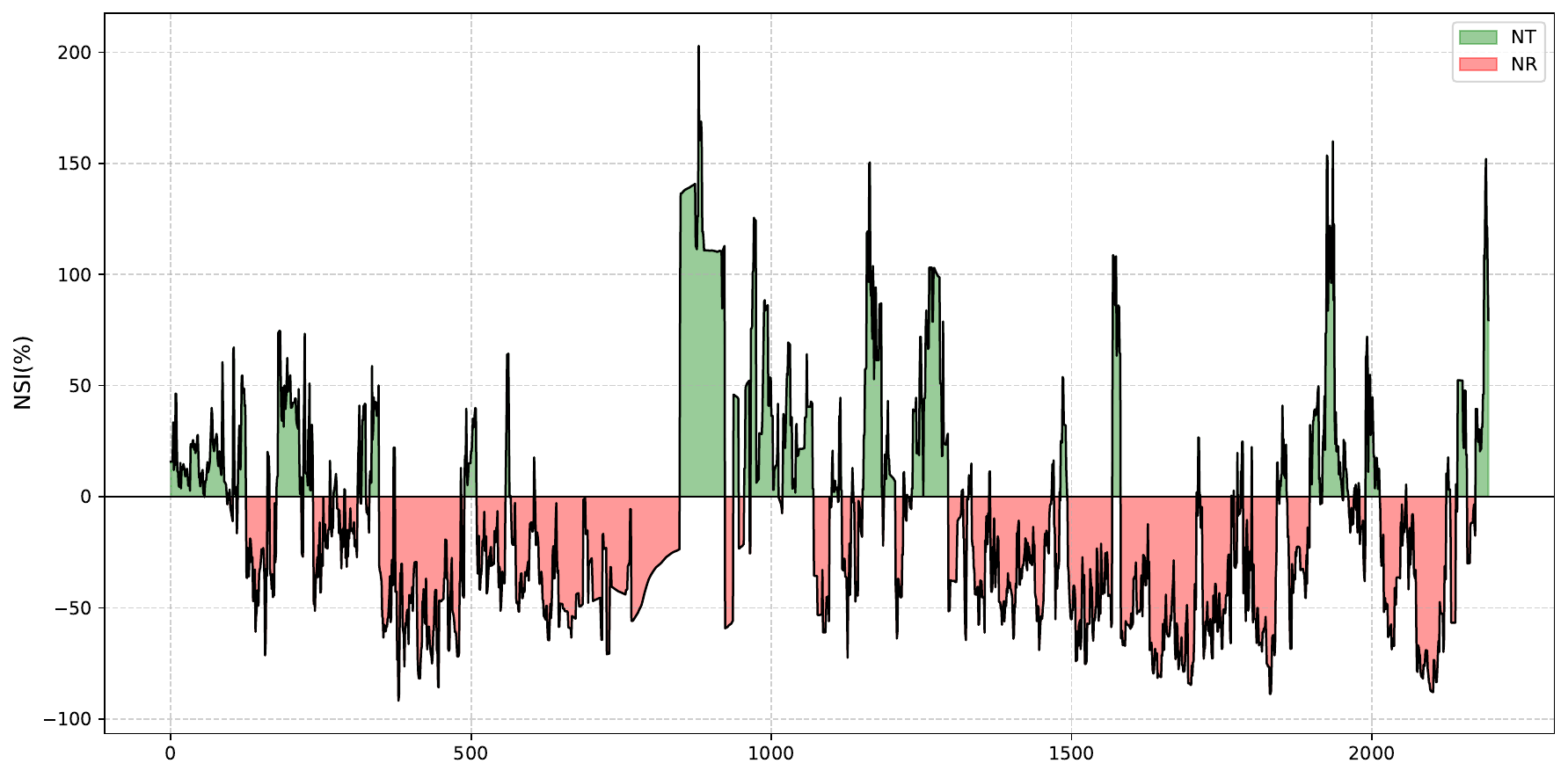}\label{fig:cv_xrp_high}}
\end{figure}

\begin{figure}[p]
    \centering
    \caption{Quantile net spillovers for major cryptocurrencies using CJ as the feature variable.}
    \label{fig:jv_net_spillover_by_coin}

    \subfigure[BTC, $\tau=0.05$]{\includegraphics[width=0.32\linewidth]{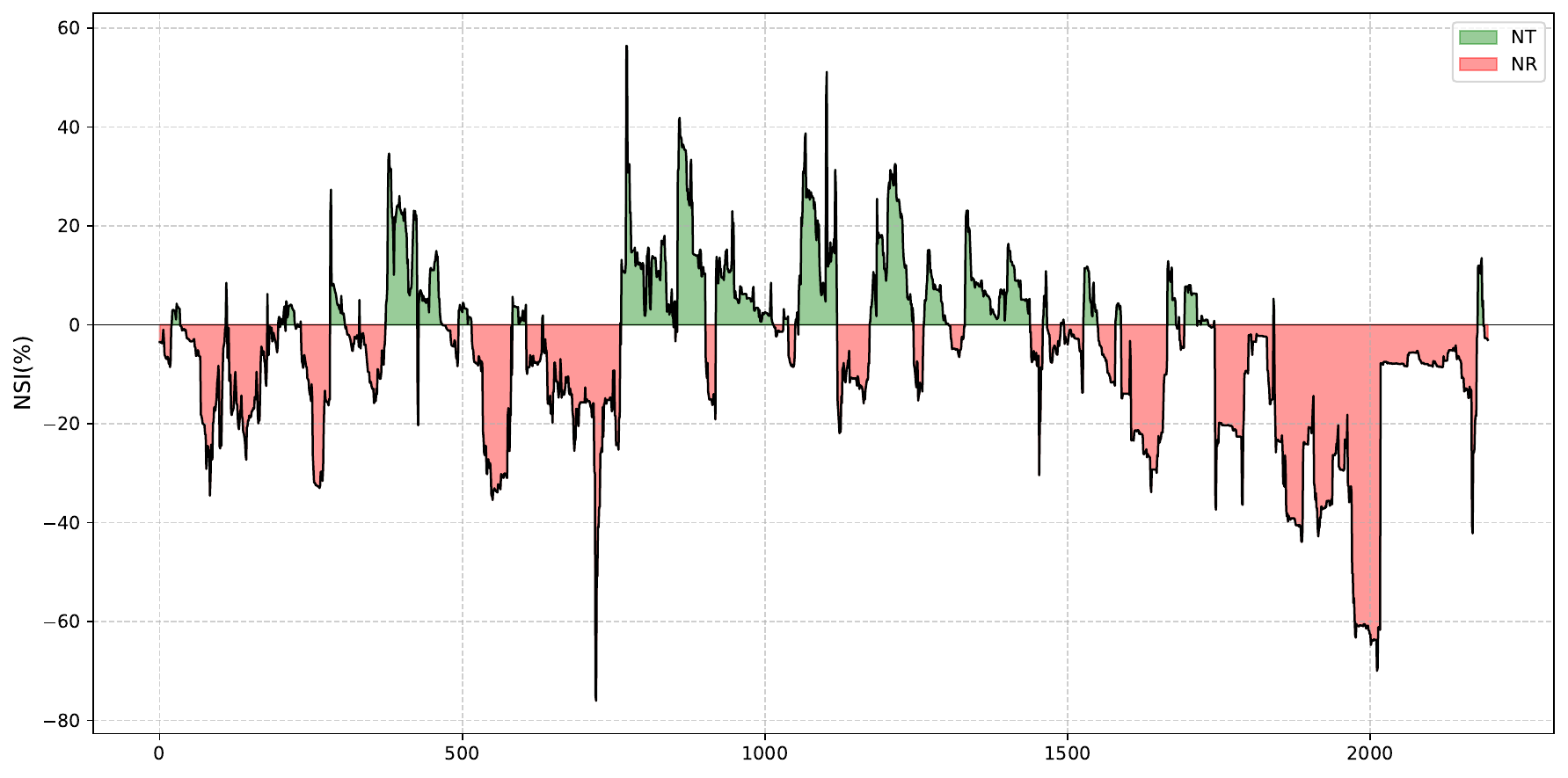}\label{fig:jv_btc_low}}\hfill
    \subfigure[BTC, $\tau=0.50$]{\includegraphics[width=0.32\linewidth]{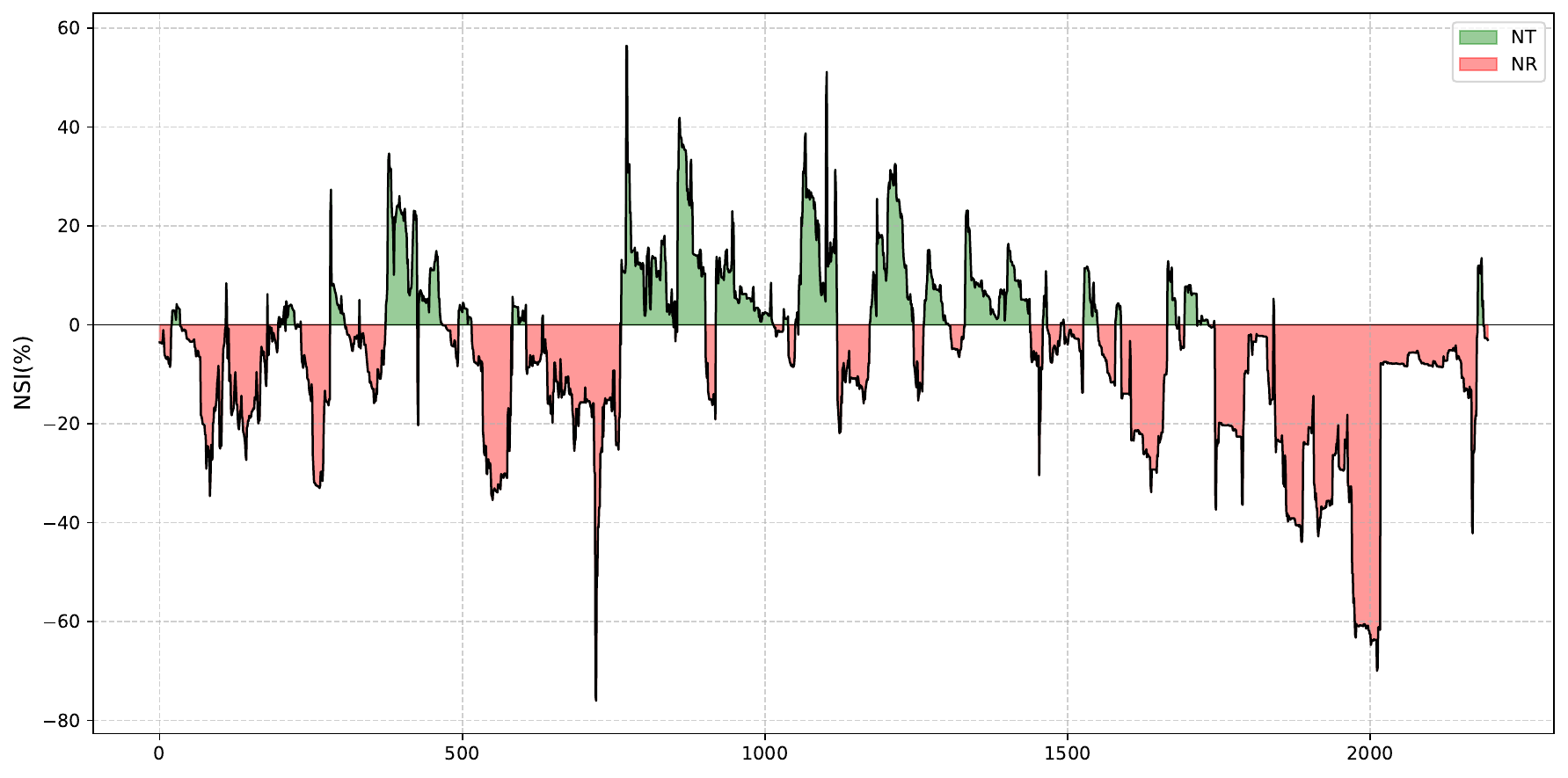}\label{fig:jv_btc_mid}}\hfill
    \subfigure[BTC, $\tau=0.95$]{\includegraphics[width=0.32\linewidth]{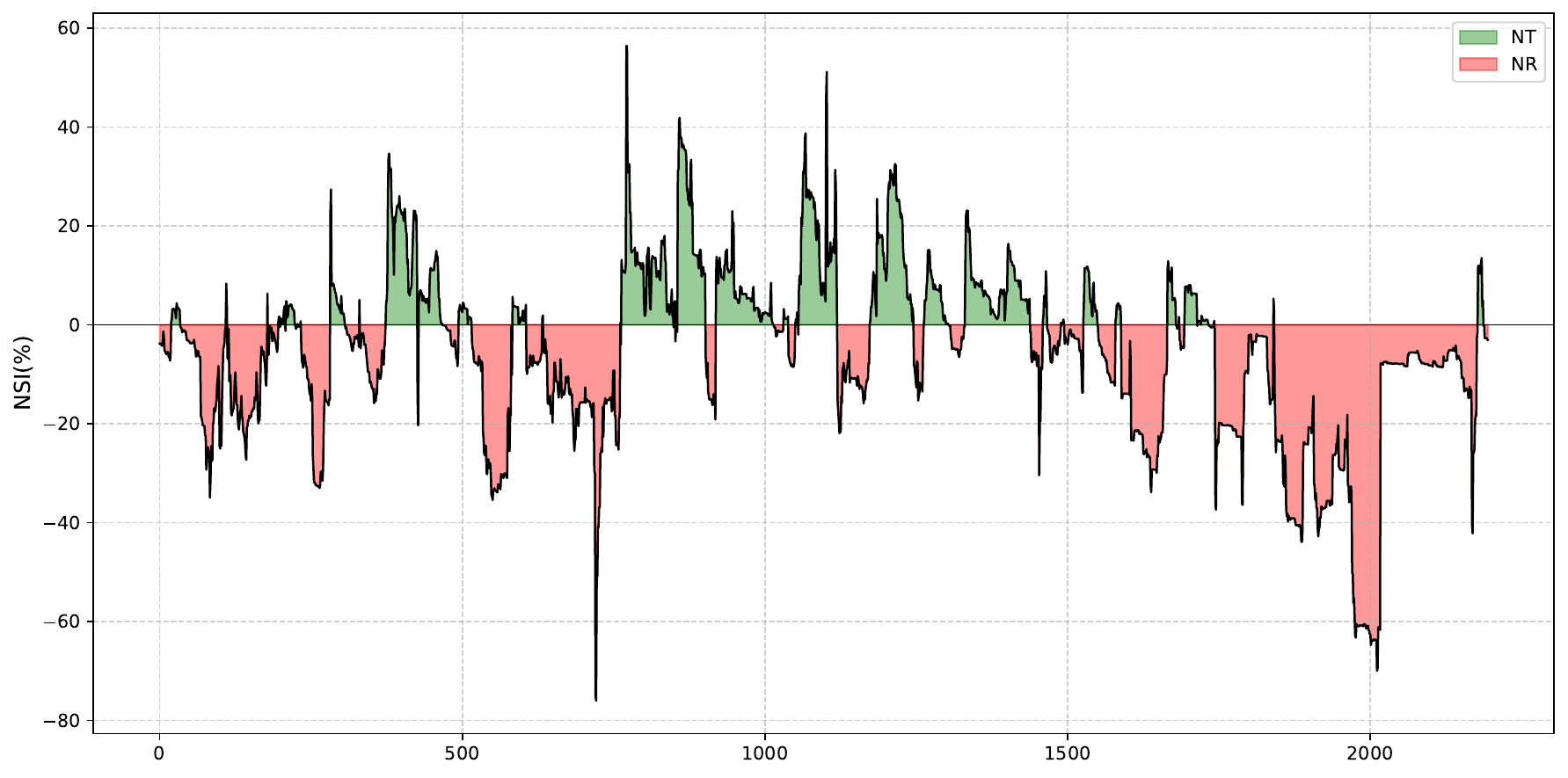}\label{fig:jv_btc_high}}
    \vspace{0.3cm}
    \subfigure[DASH, $\tau=0.05$]{\includegraphics[width=0.32\linewidth]{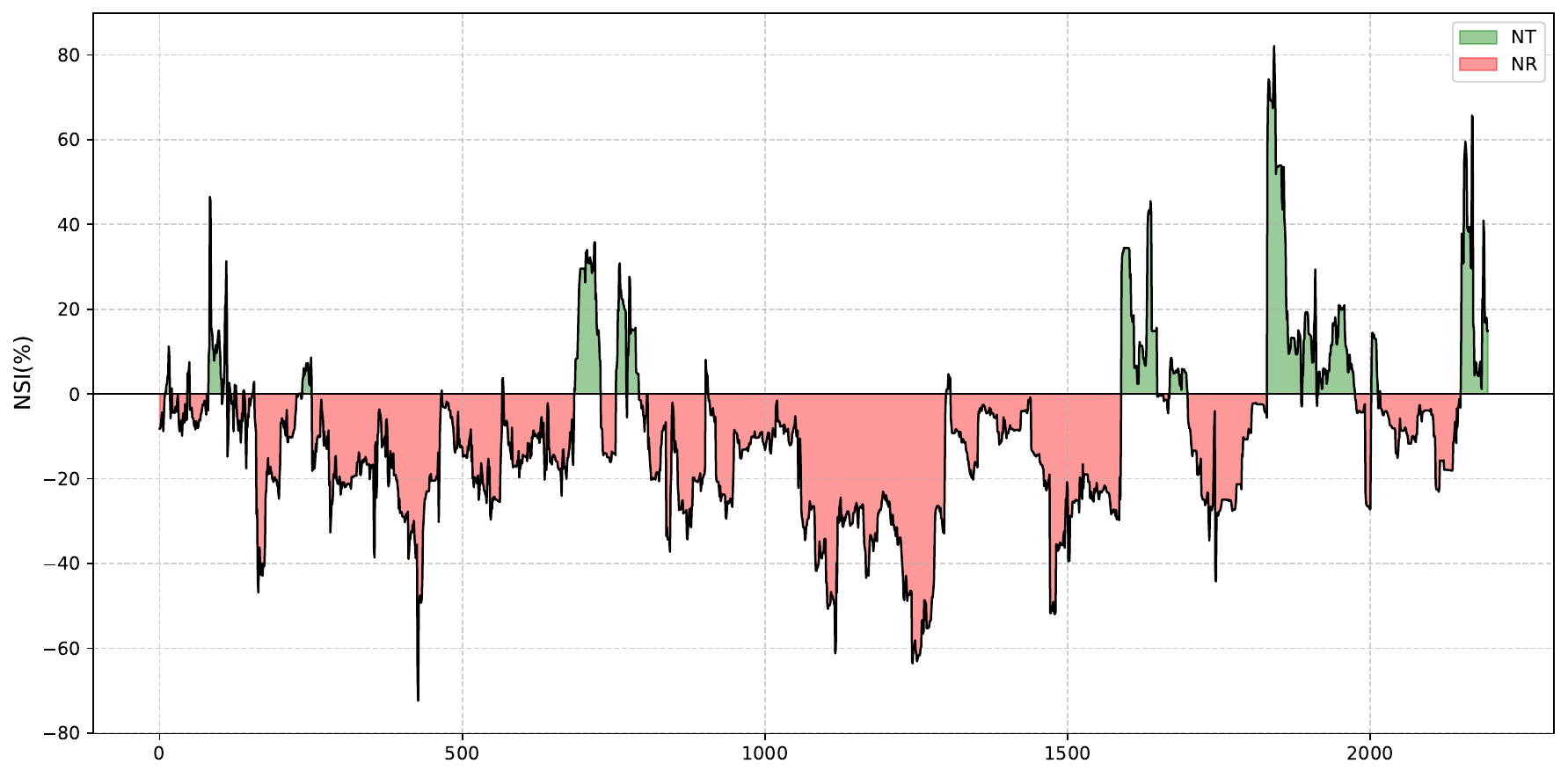}\label{fig:jv_dash_low}}\hfill
    \subfigure[DASH, $\tau=0.50$]{\includegraphics[width=0.32\linewidth]{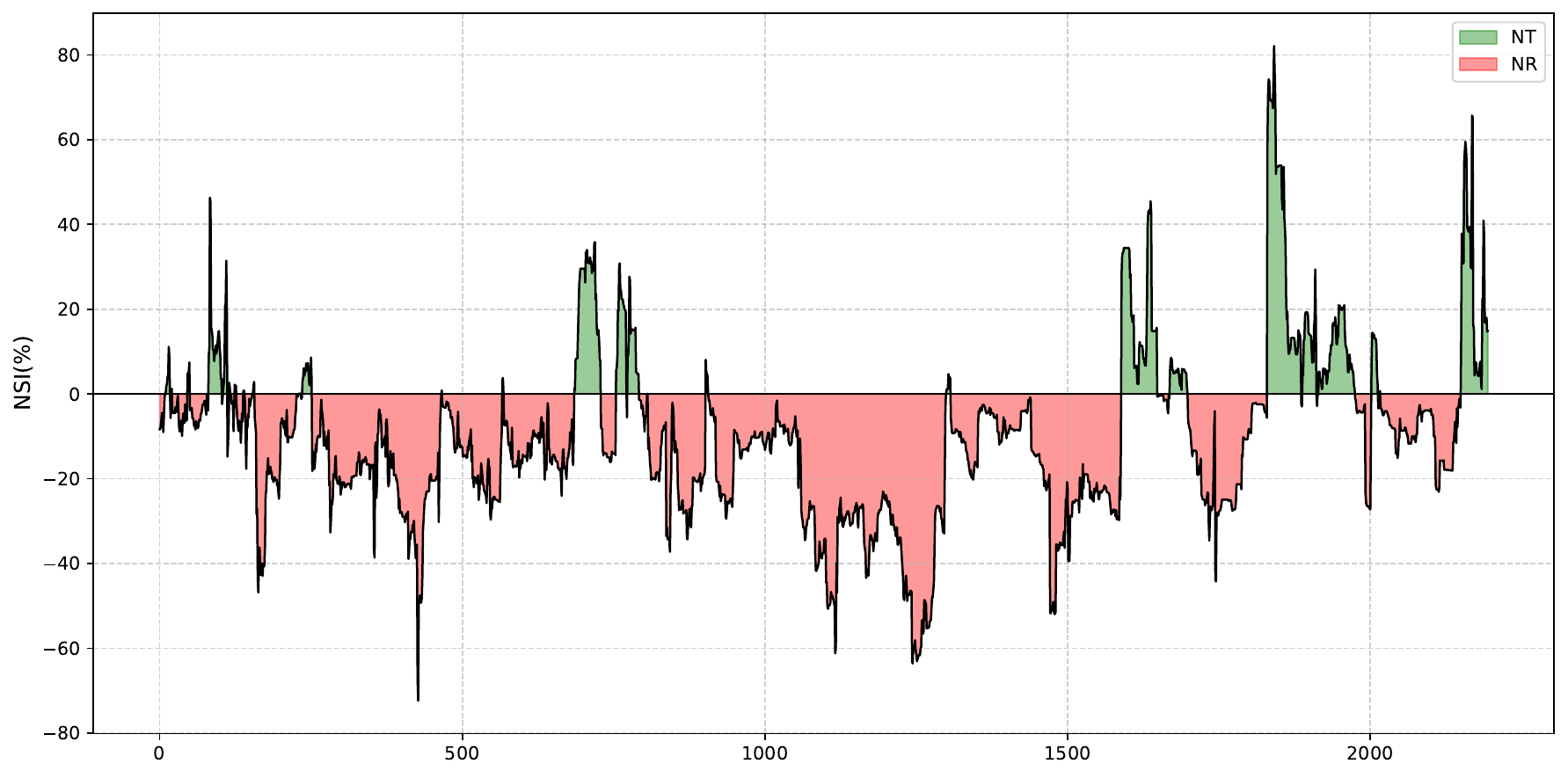}\label{fig:jv_dash_mid}}\hfill
    \subfigure[DASH, $\tau=0.95$]{\includegraphics[width=0.32\linewidth]{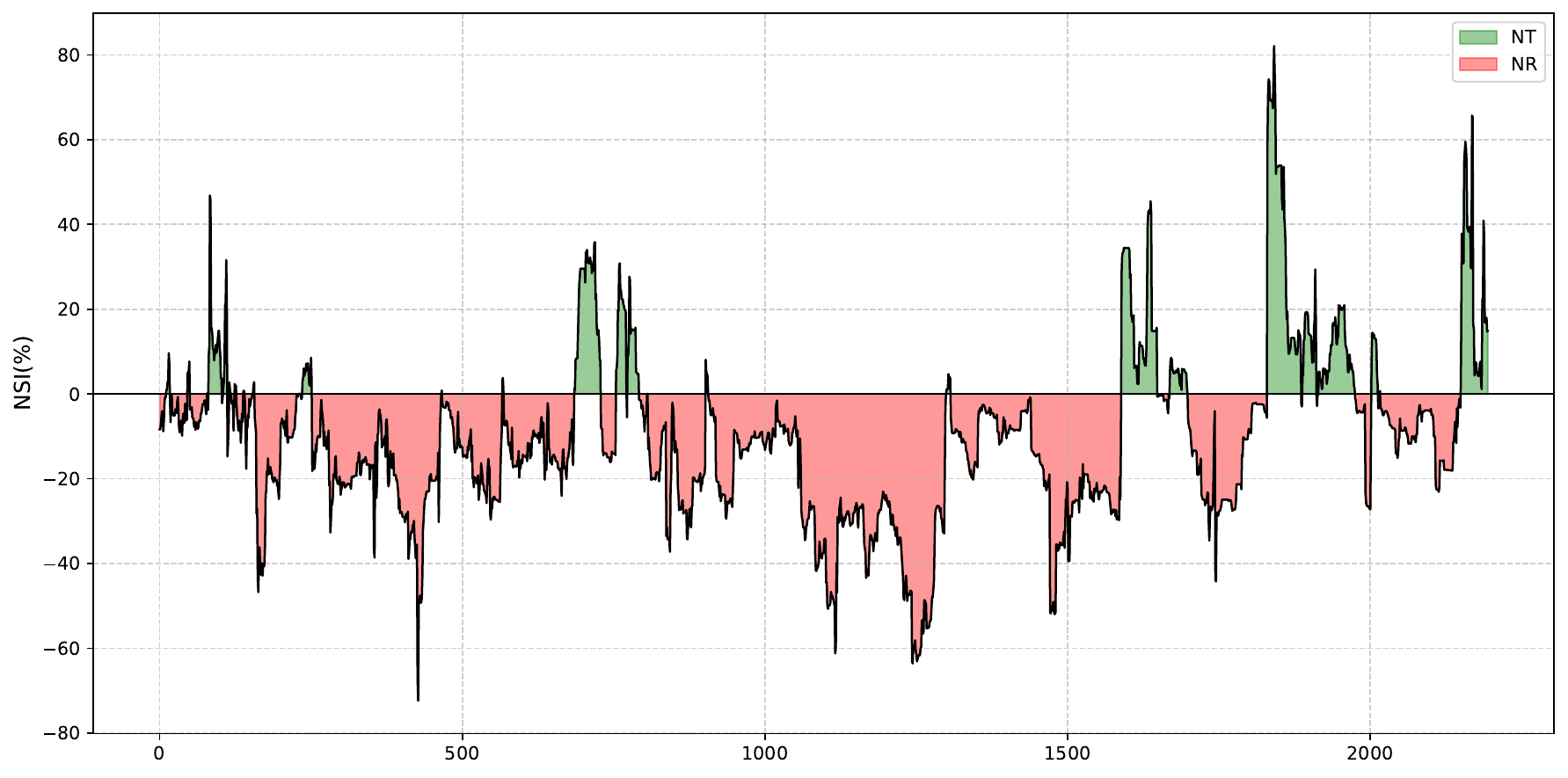}\label{fig:jv_dash_high}}
    \vspace{0.3cm}
    \subfigure[ETH, $\tau=0.05$]{\includegraphics[width=0.32\linewidth]{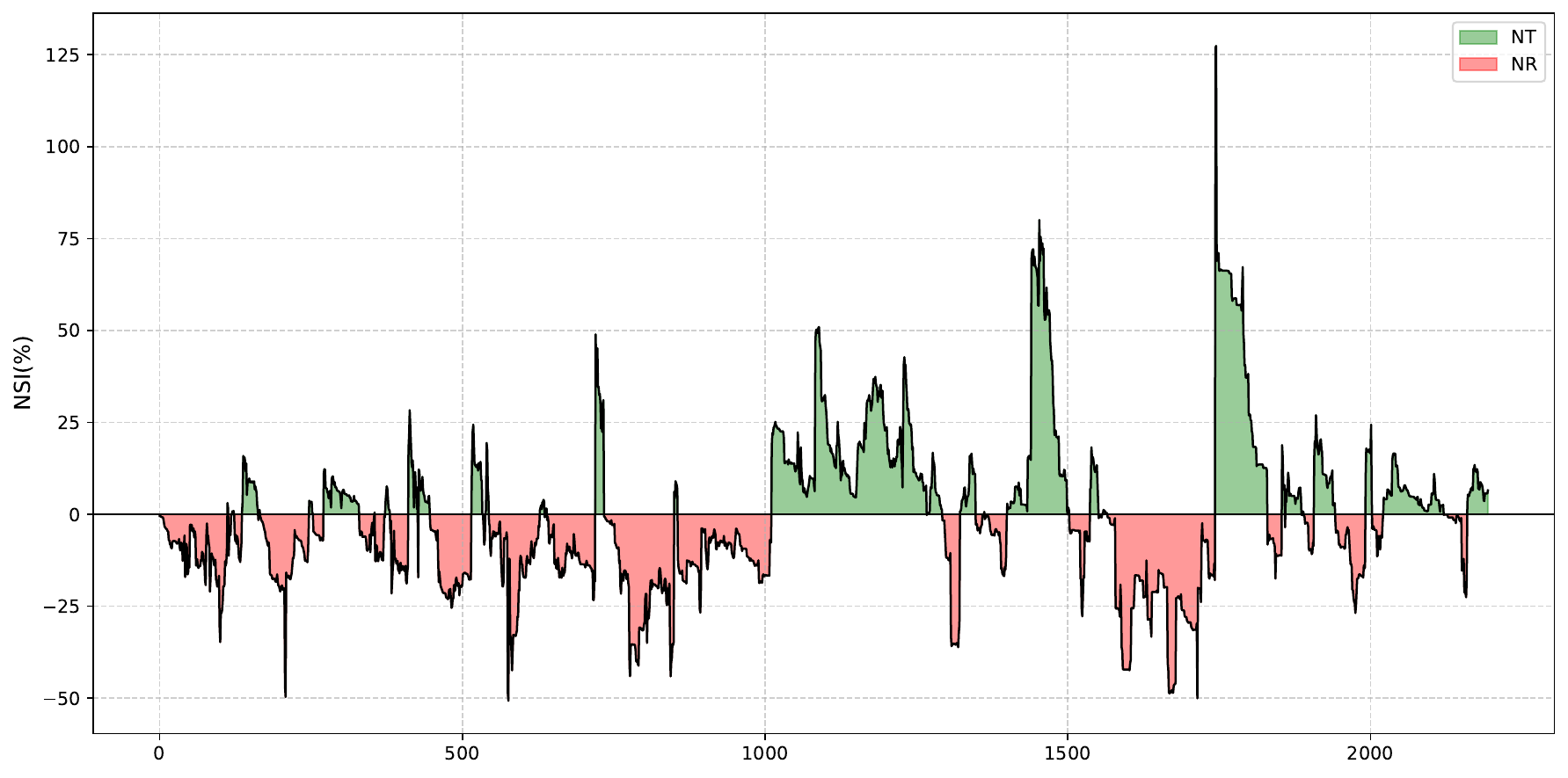}\label{fig:jv_eth_low}}\hfill
    \subfigure[ETH, $\tau=0.50$]{\includegraphics[width=0.32\linewidth]{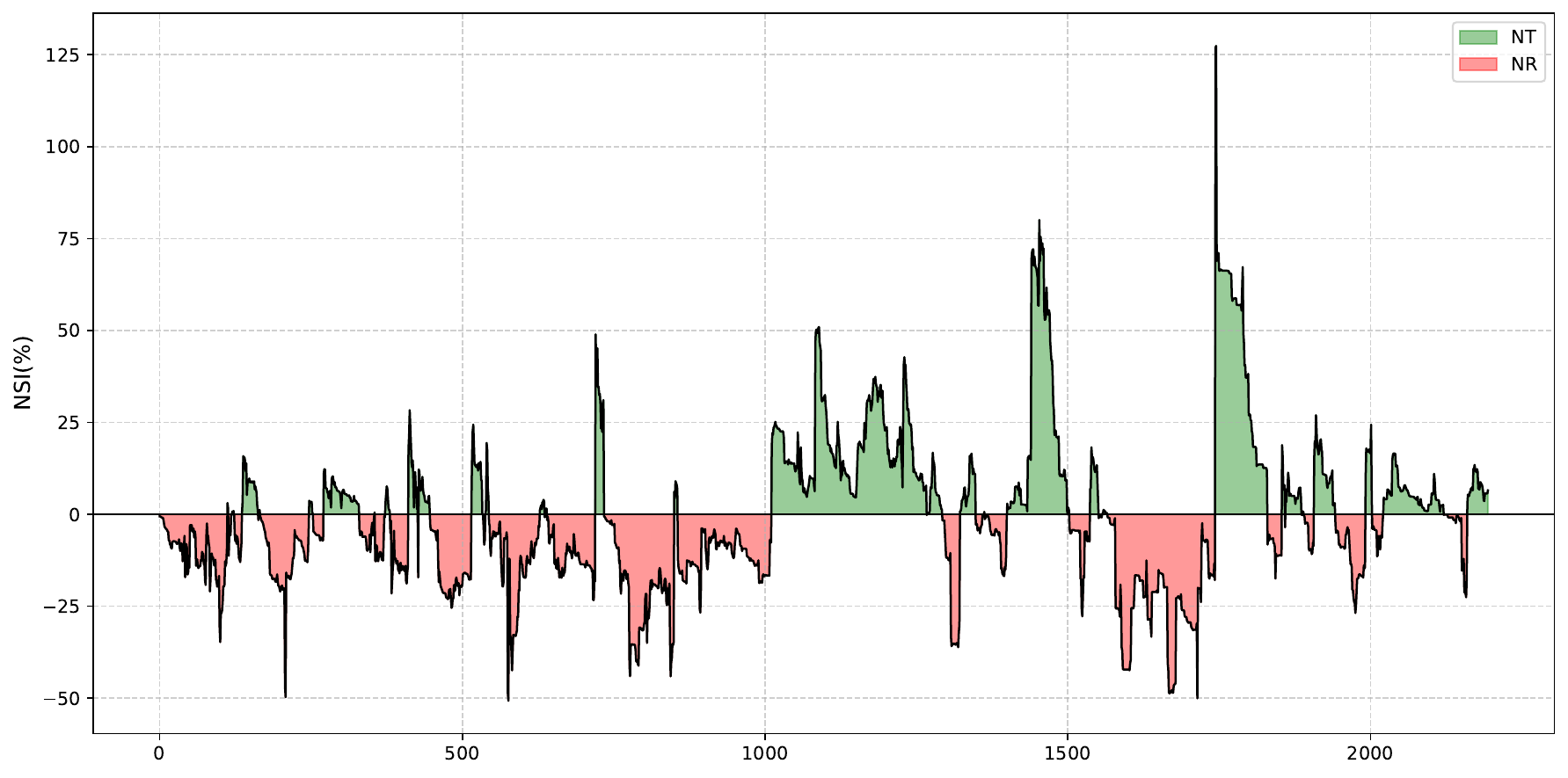}\label{fig:jv_eth_mid}}\hfill
    \subfigure[ETH, $\tau=0.95$]{\includegraphics[width=0.32\linewidth]{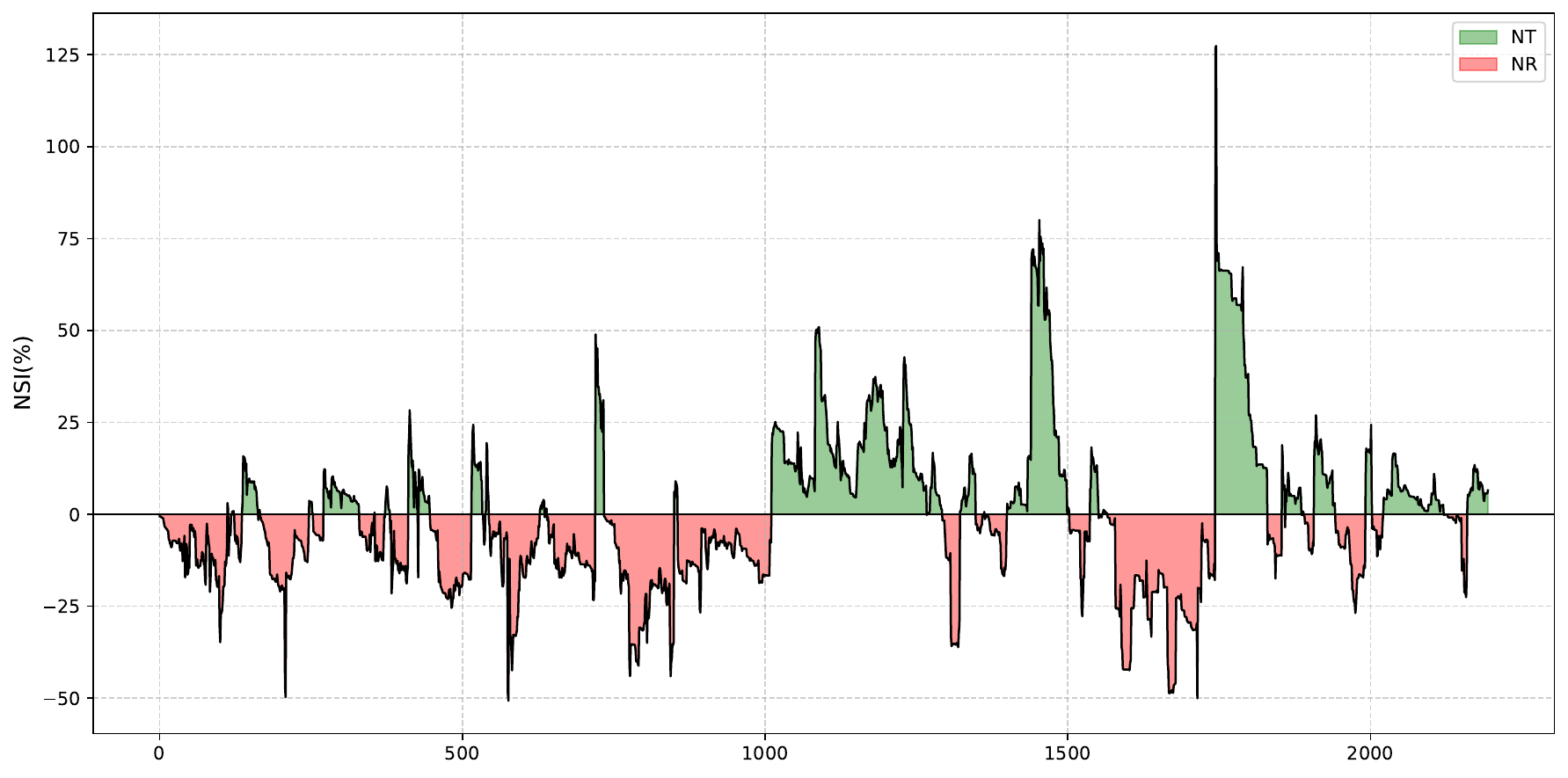}\label{fig:jv_eth_high}}
    \vspace{0.3cm}
    \subfigure[LTC, $\tau=0.05$]{\includegraphics[width=0.32\linewidth]{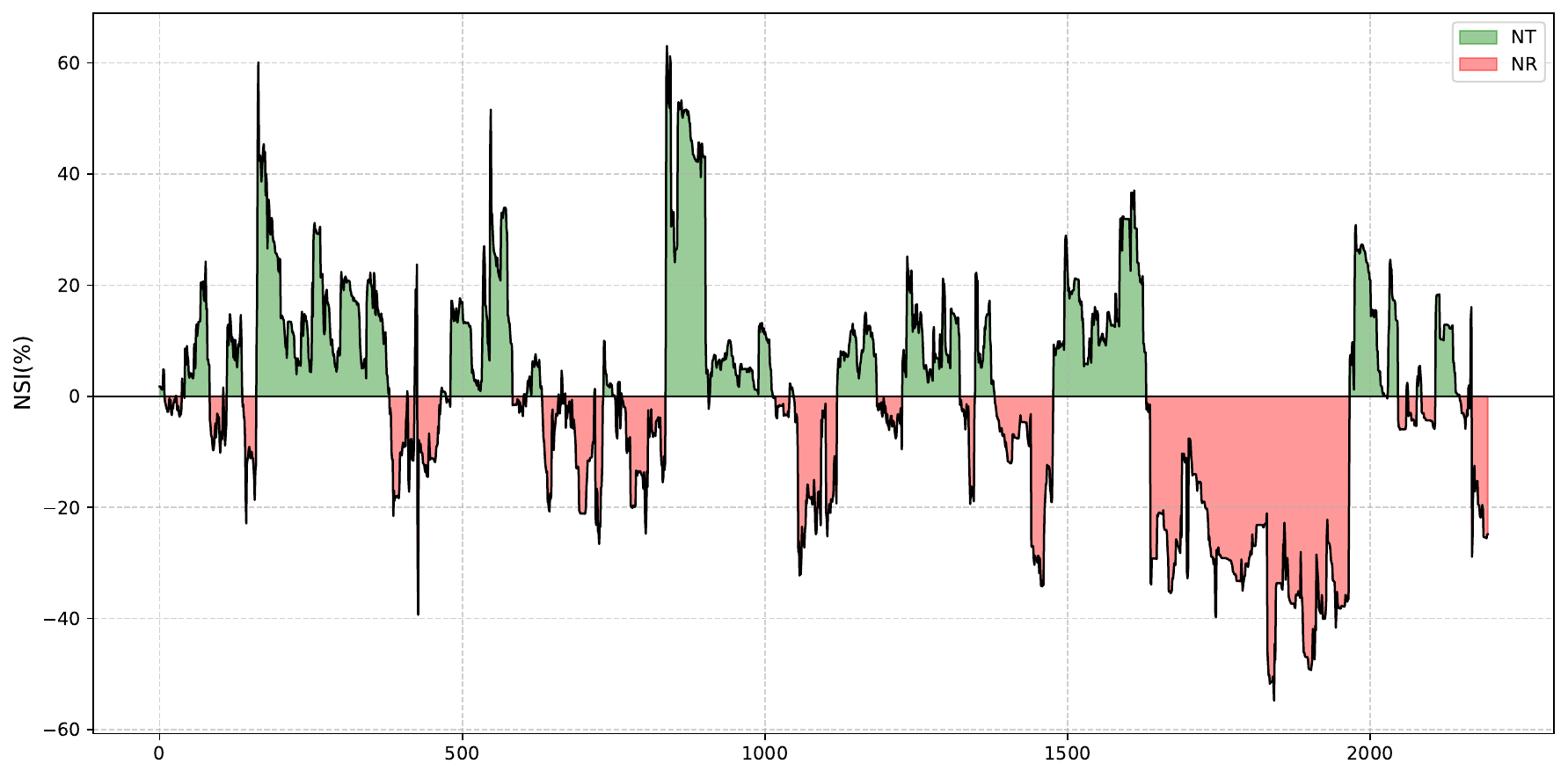}\label{fig:jv_ltc_low}}\hfill
    \subfigure[LTC, $\tau=0.50$]{\includegraphics[width=0.32\linewidth]{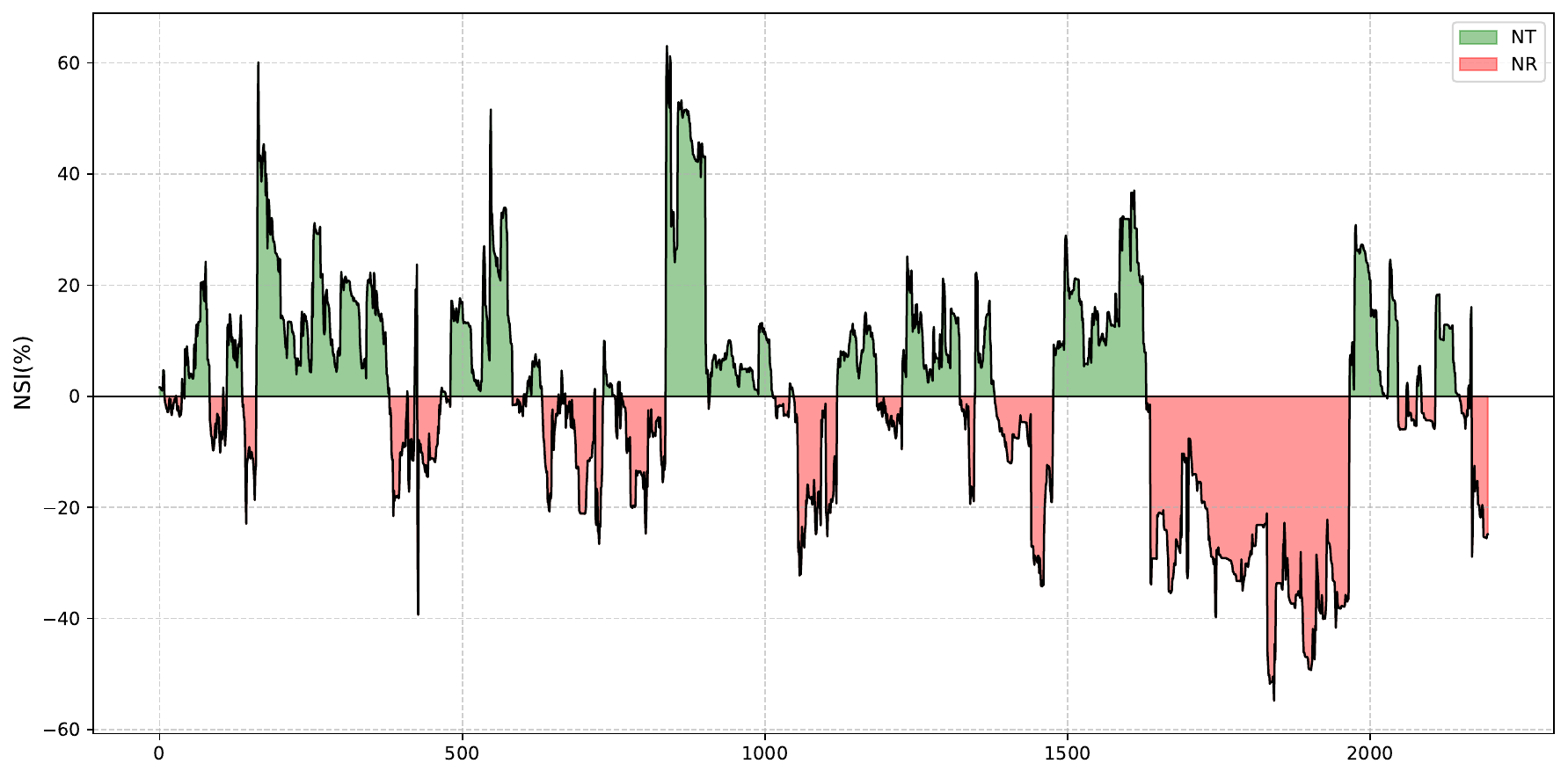}\label{fig:jv_ltc_mid}}\hfill
    \subfigure[LTC, $\tau=0.95$]{\includegraphics[width=0.32\linewidth]{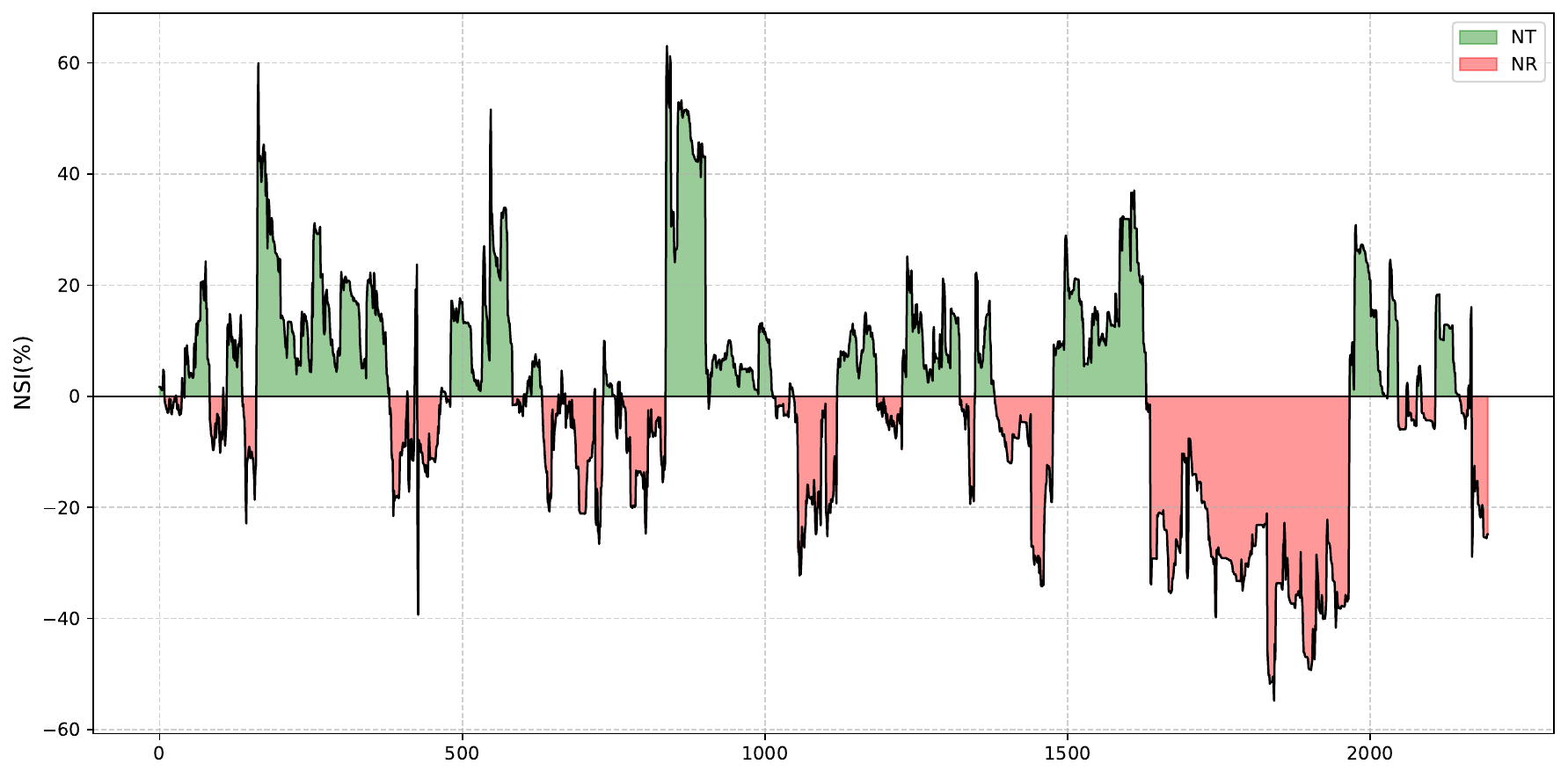}\label{fig:jv_ltc_high}}
    \vspace{0.3cm}
    \subfigure[XLM, $\tau=0.05$]{\includegraphics[width=0.32\linewidth]{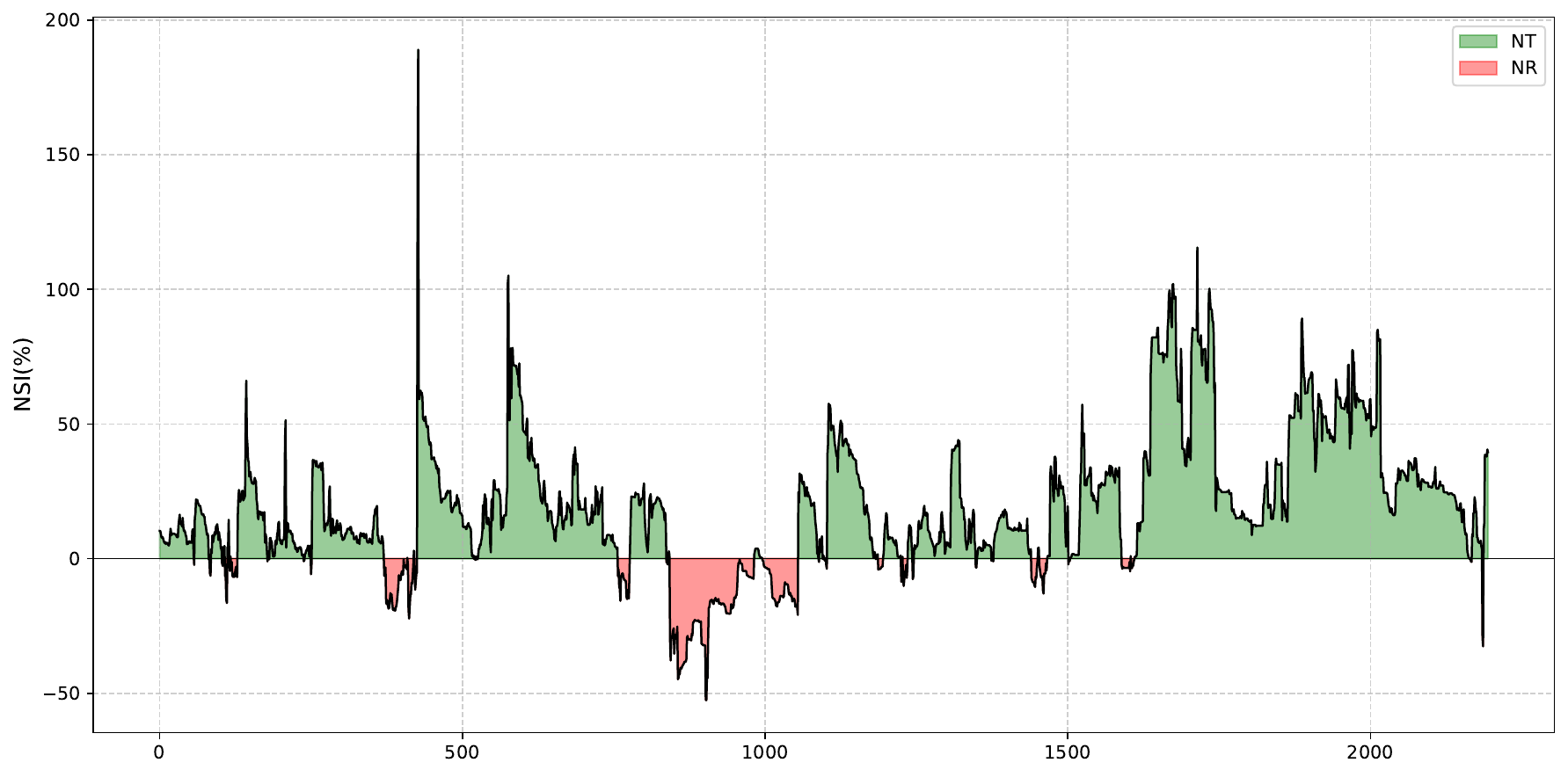}\label{fig:jv_xlm_low}}\hfill
    \subfigure[XLM, $\tau=0.50$]{\includegraphics[width=0.32\linewidth]{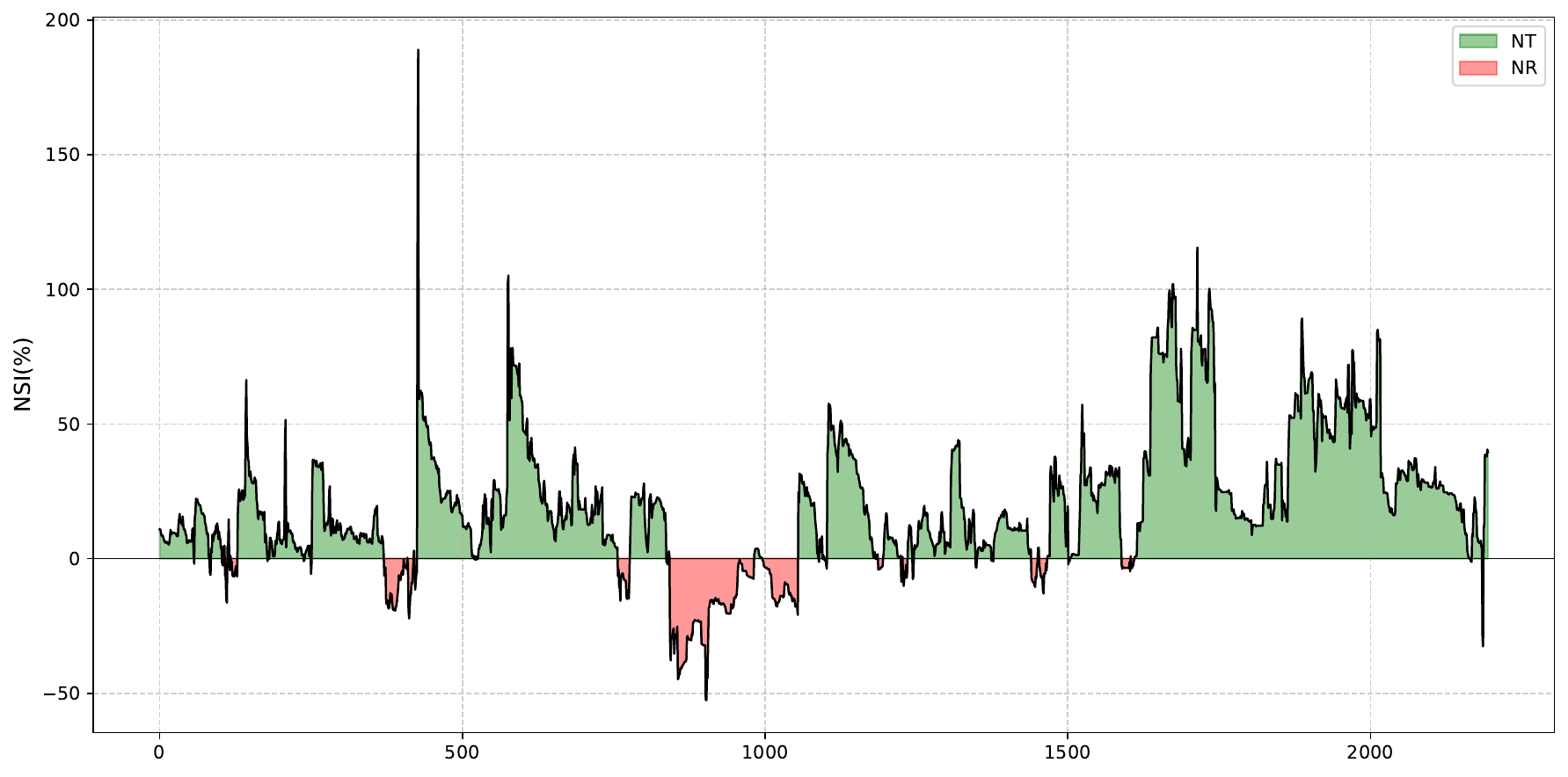}\label{fig:jv_xlm_mid}}\hfill
    \subfigure[XLM, $\tau=0.95$]{\includegraphics[width=0.32\linewidth]{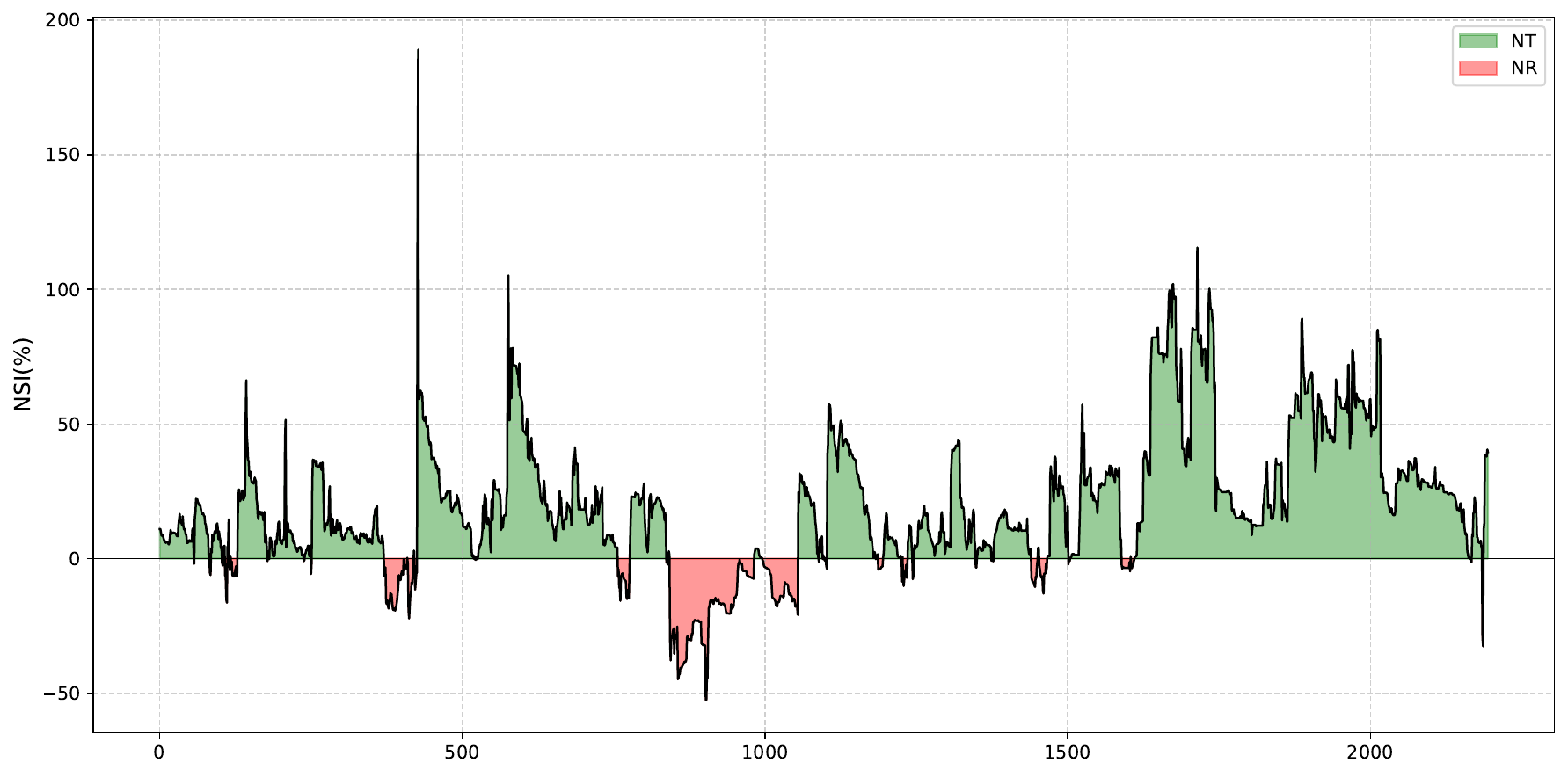}\label{fig:jv_xlm_high}}
    \vspace{0.3cm}
    \subfigure[XRP, $\tau=0.05$]{\includegraphics[width=0.32\linewidth]{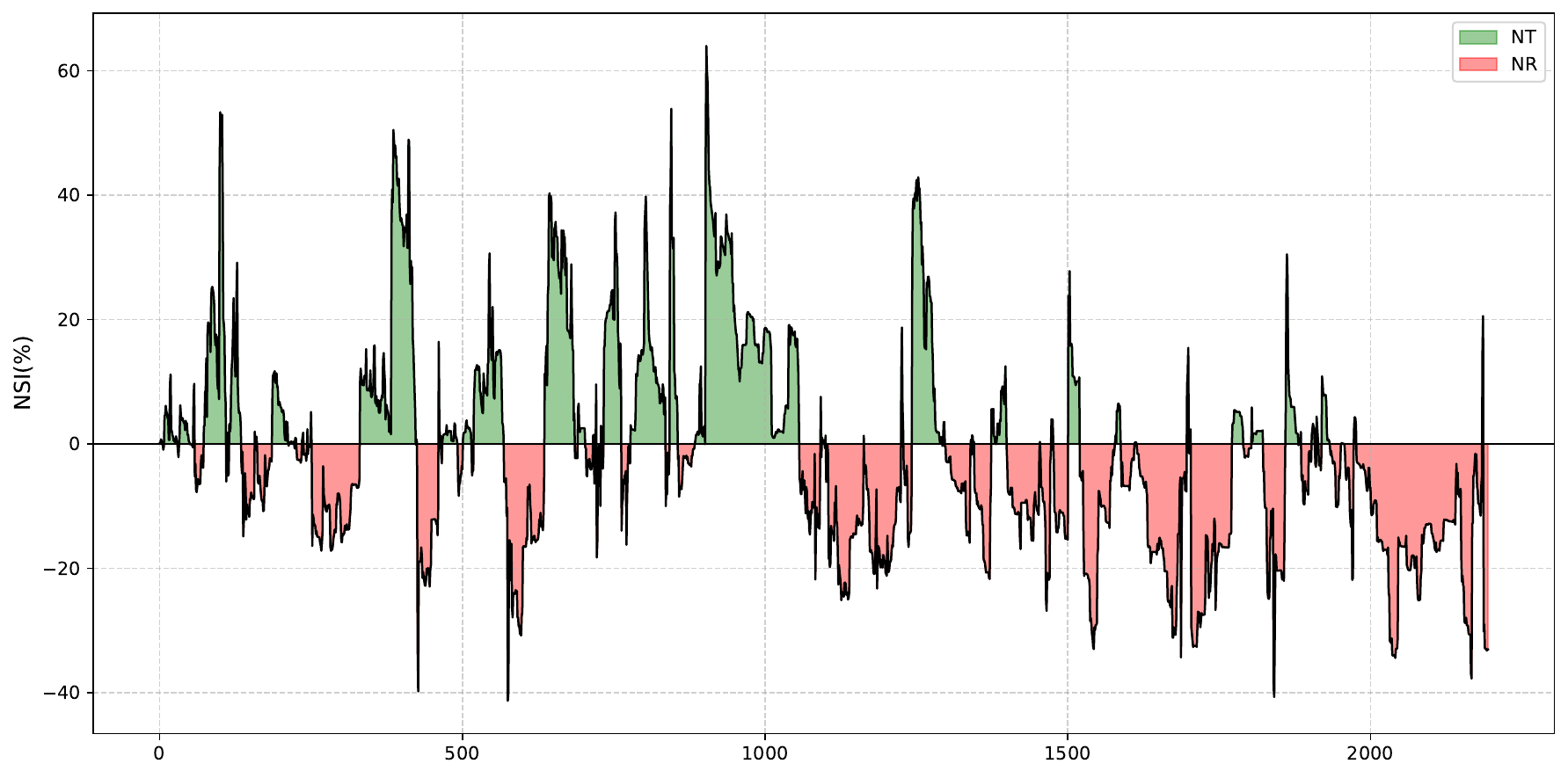}\label{fig:jv_xrp_low}}\hfill
    \subfigure[XRP, $\tau=0.50$]{\includegraphics[width=0.32\linewidth]{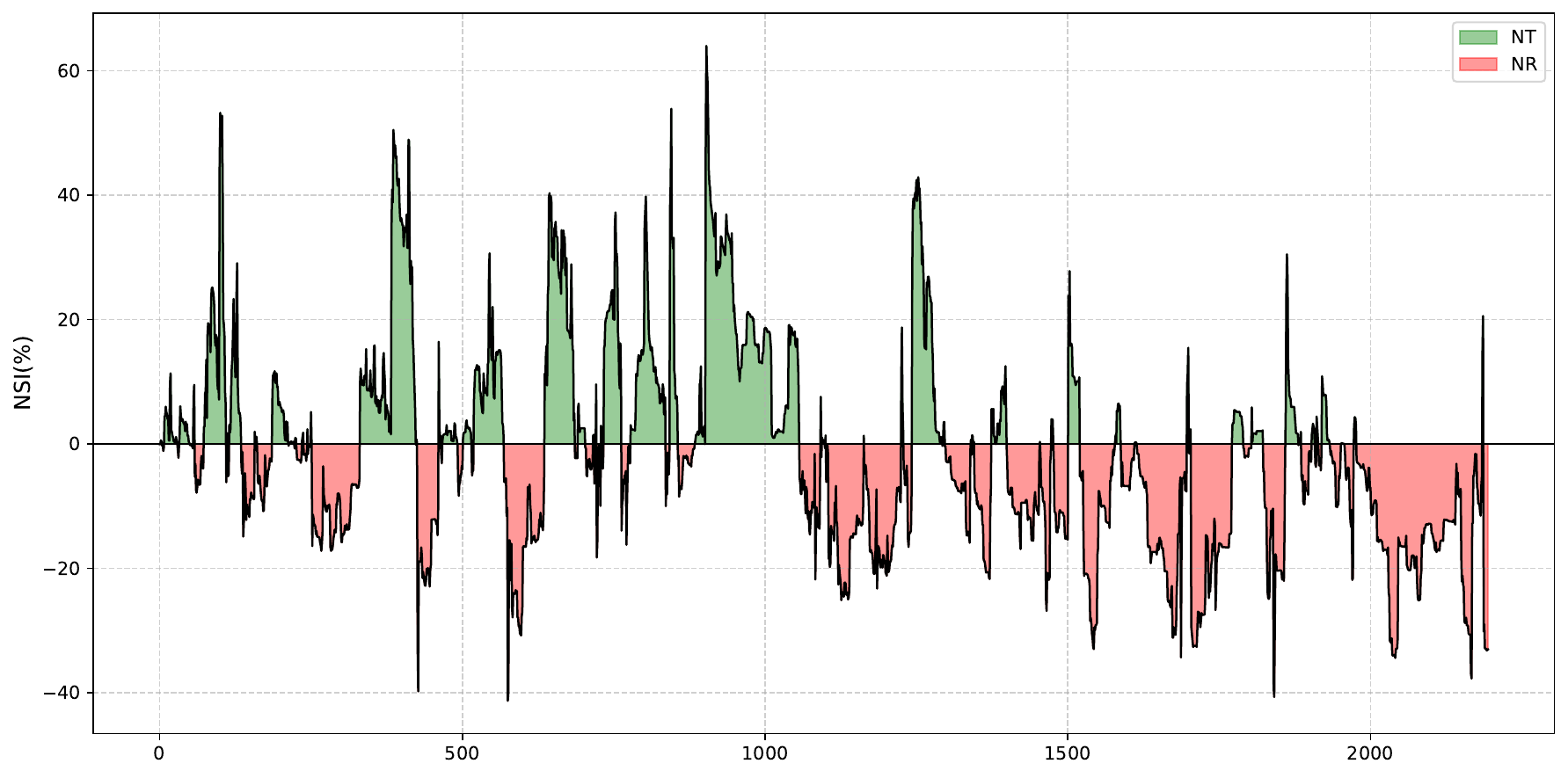}\label{fig:jv_xrp_mid}}\hfill
    \subfigure[XRP, $\tau=0.95$]{\includegraphics[width=0.32\linewidth]{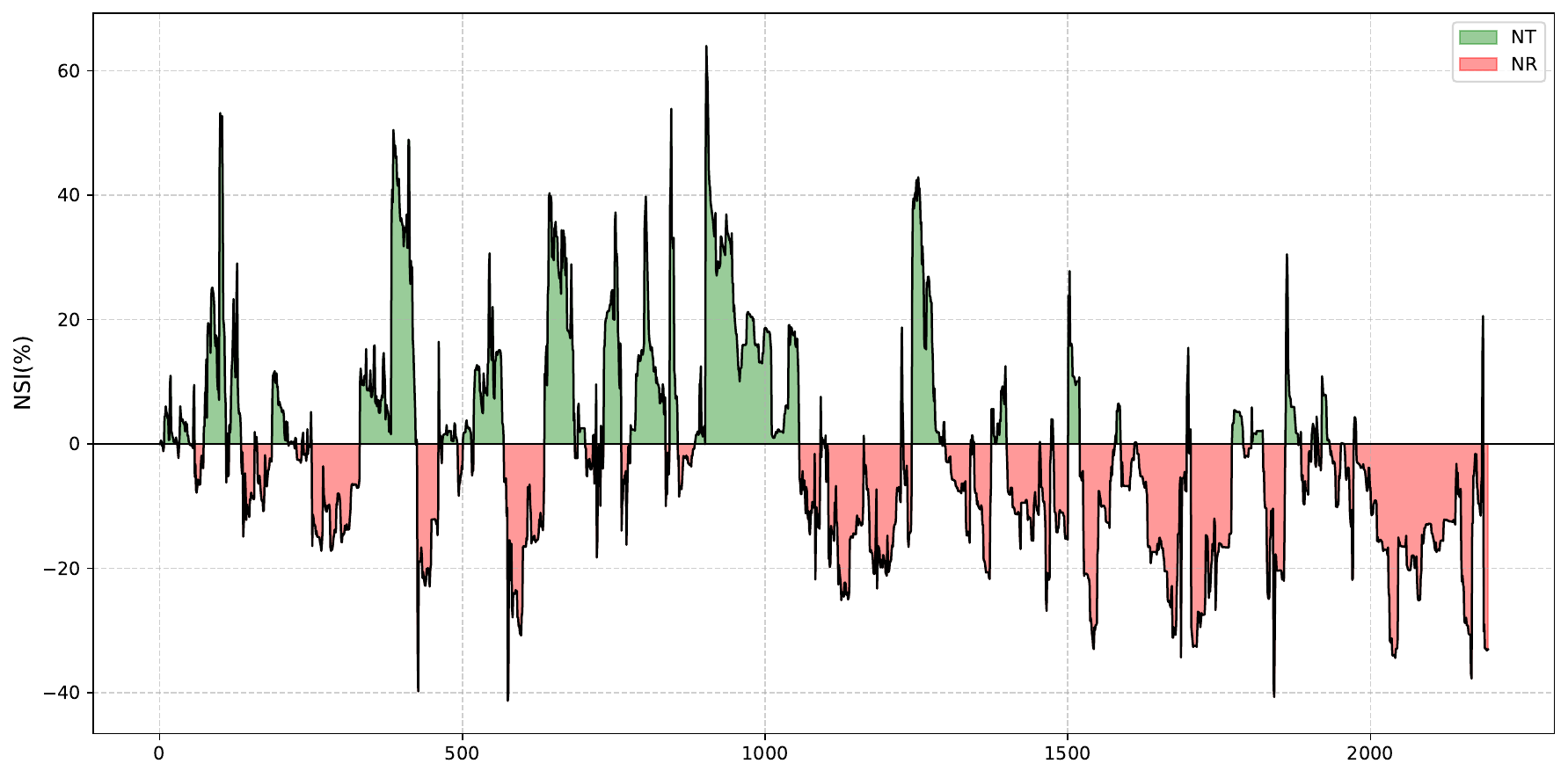}\label{fig:jv_xrp_high}}
\end{figure}

\begin{figure}[p]
    \centering
    \caption{Quantile net spillovers for major cryptocurrencies using $RS^+$ as the feature variable.}
    \label{fig:rsp_net_spillover_by_coin}

    \subfigure[BTC, $\tau=0.05$]{\includegraphics[width=0.32\linewidth]{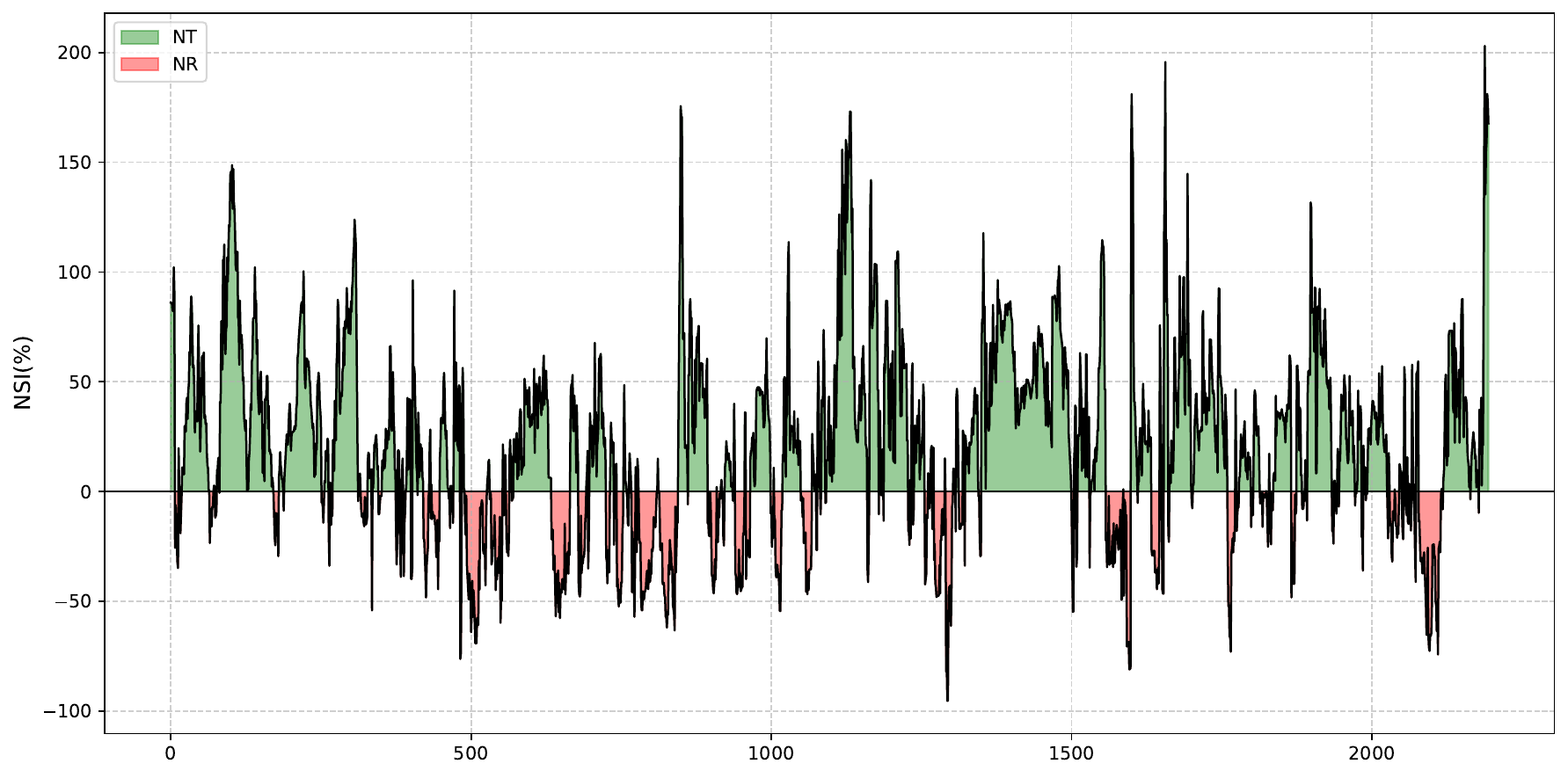}\label{fig:rsp_btc_low}}\hfill
    \subfigure[BTC, $\tau=0.50$]{\includegraphics[width=0.32\linewidth]{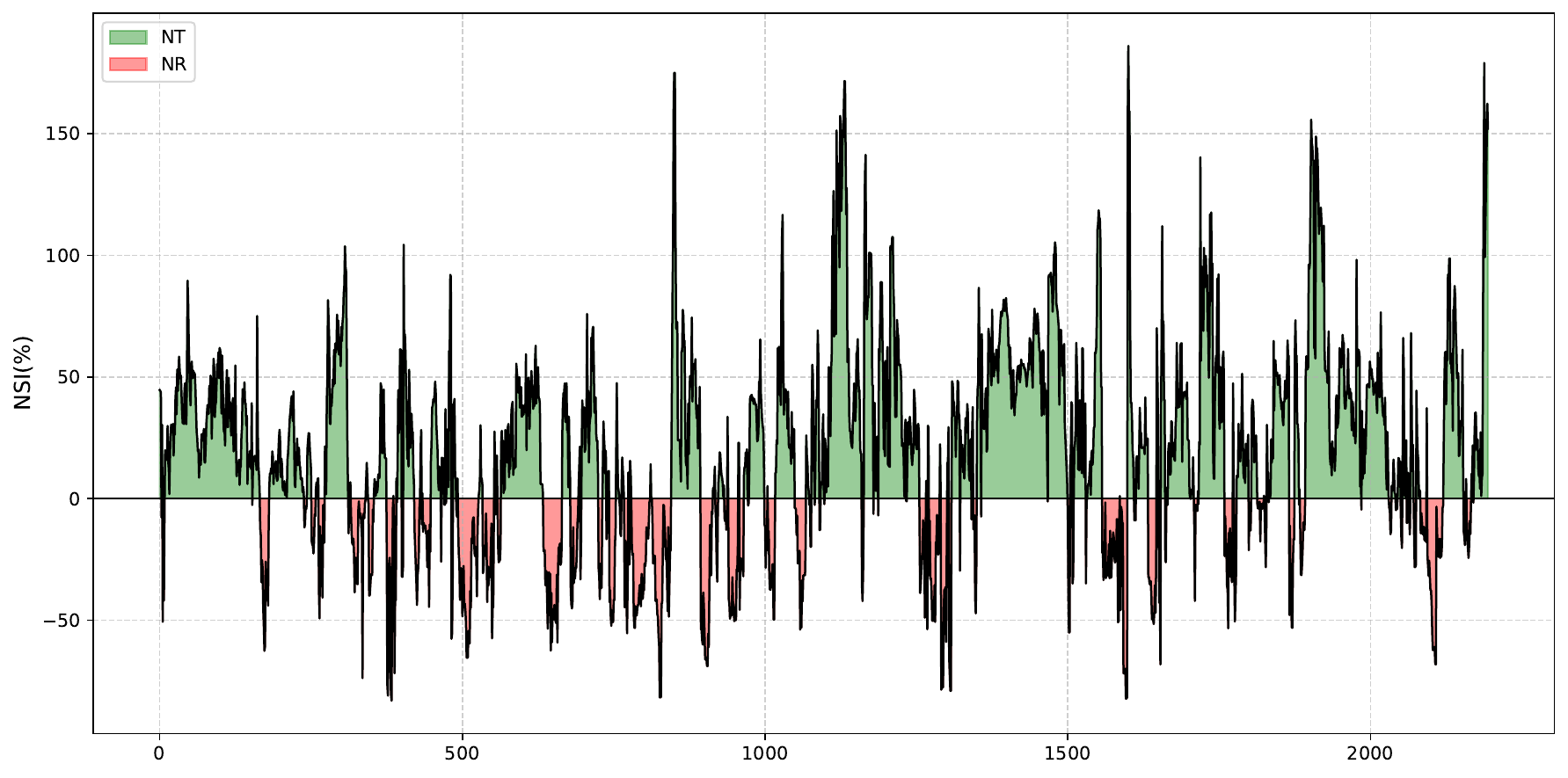}\label{fig:rsp_btc_mid}}\hfill
    \subfigure[BTC, $\tau=0.95$]{\includegraphics[width=0.32\linewidth]{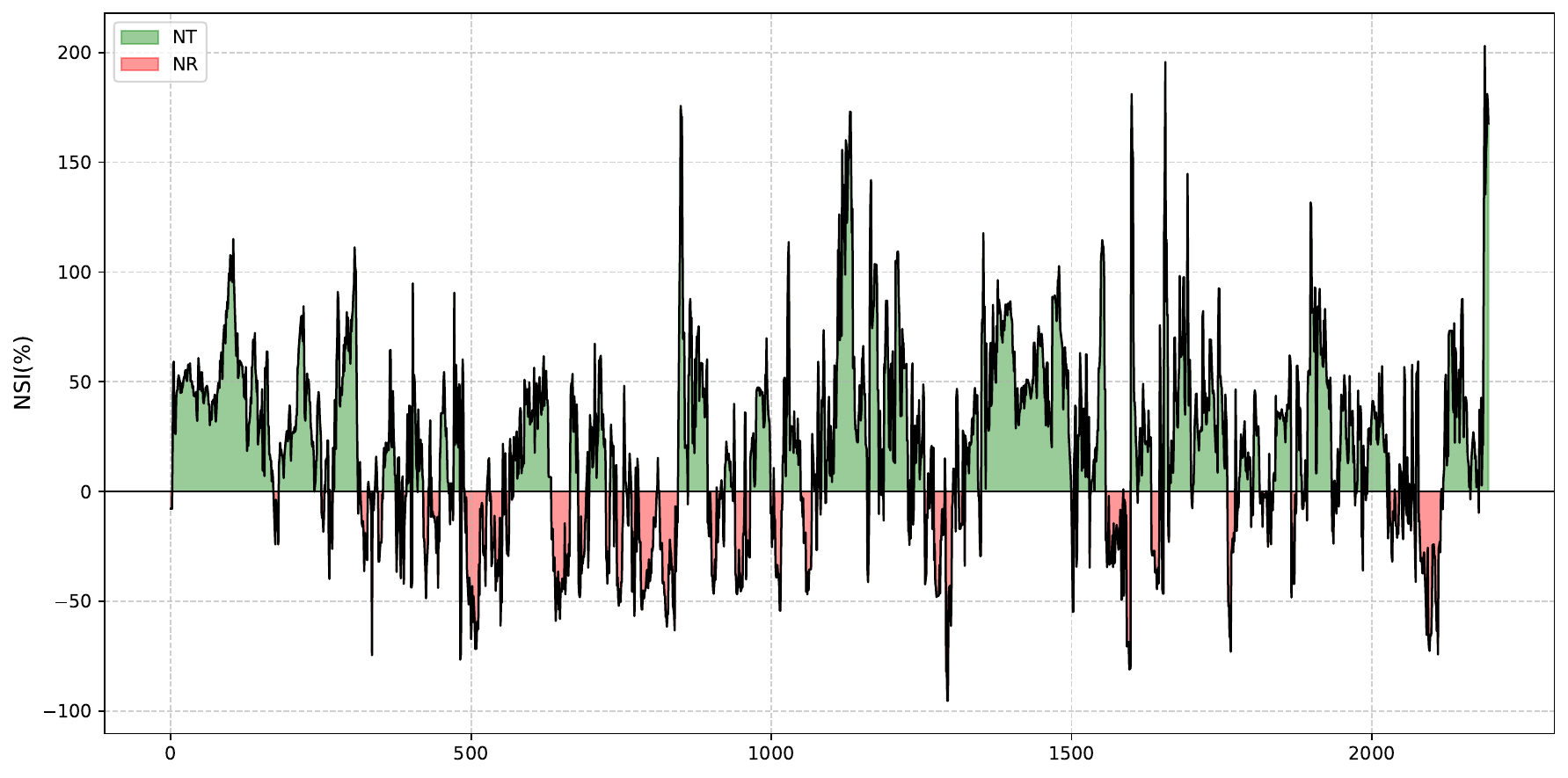}\label{fig:rsp_btc_high}}
    \vspace{0.3cm}
    \subfigure[DASH, $\tau=0.05$]{\includegraphics[width=0.32\linewidth]{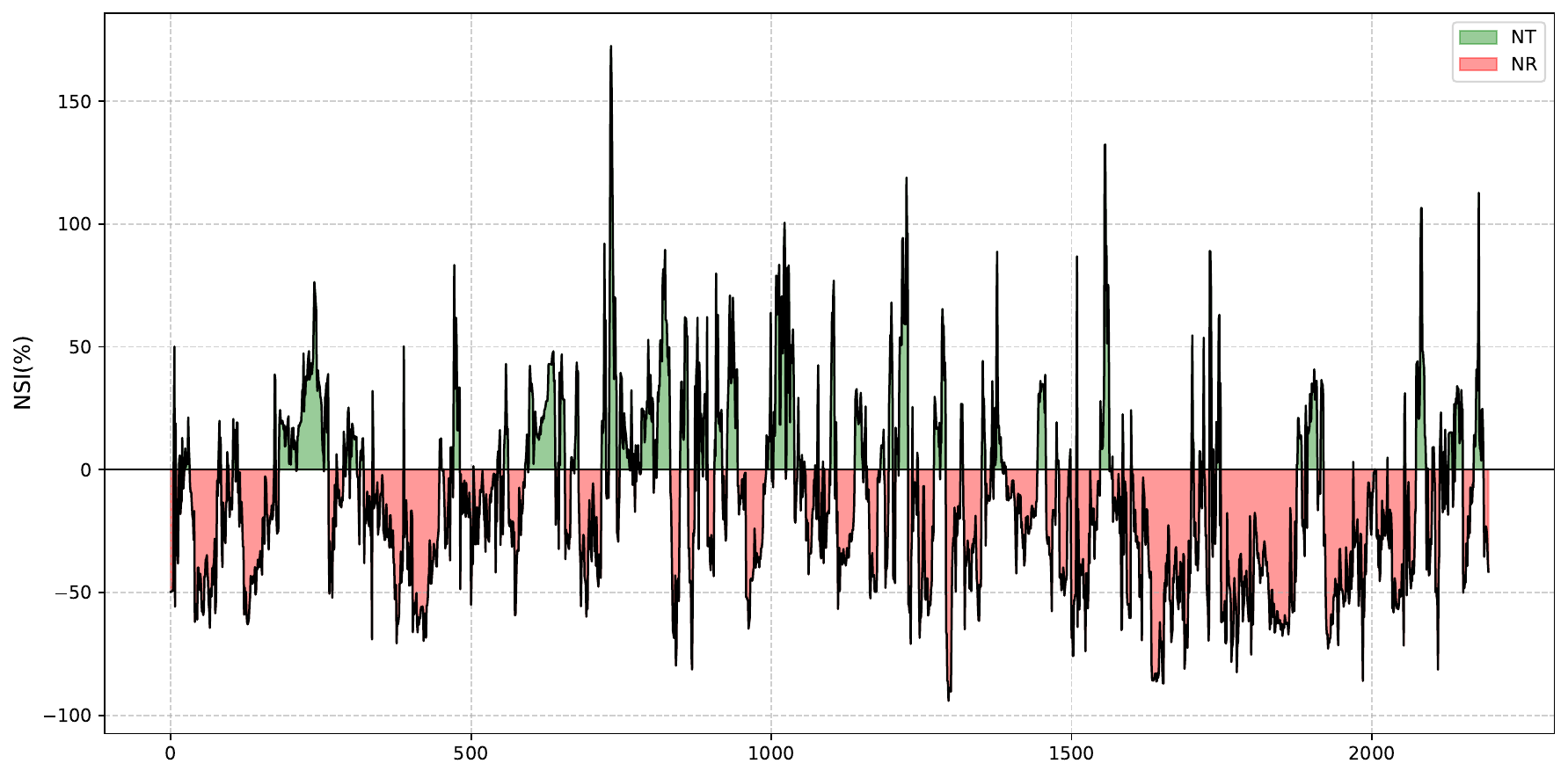}\label{fig:rsp_dash_low}}\hfill
    \subfigure[DASH, $\tau=0.50$]{\includegraphics[width=0.32\linewidth]{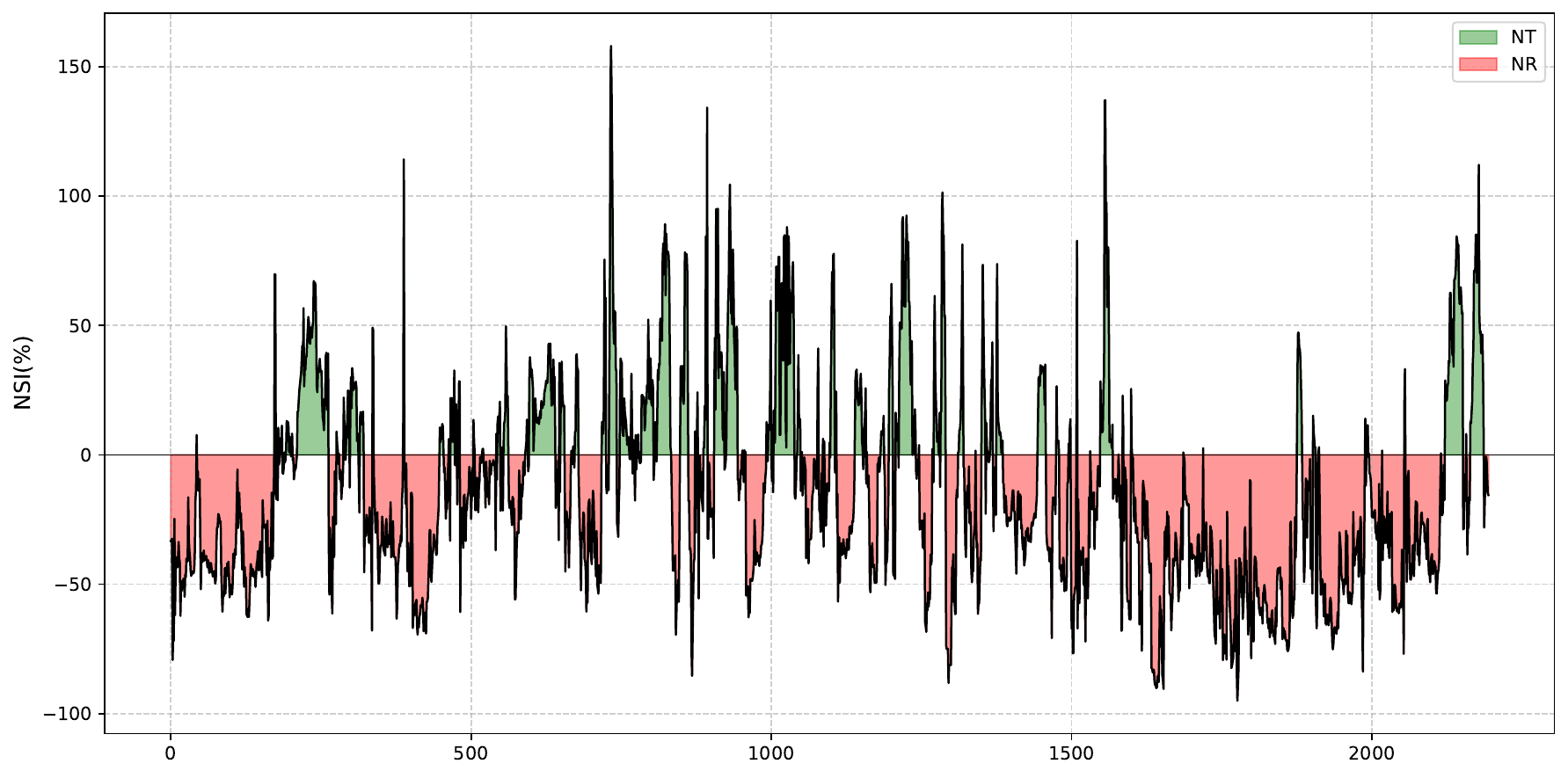}\label{fig:rsp_dash_mid}}\hfill
    \subfigure[DASH, $\tau=0.95$]{\includegraphics[width=0.32\linewidth]{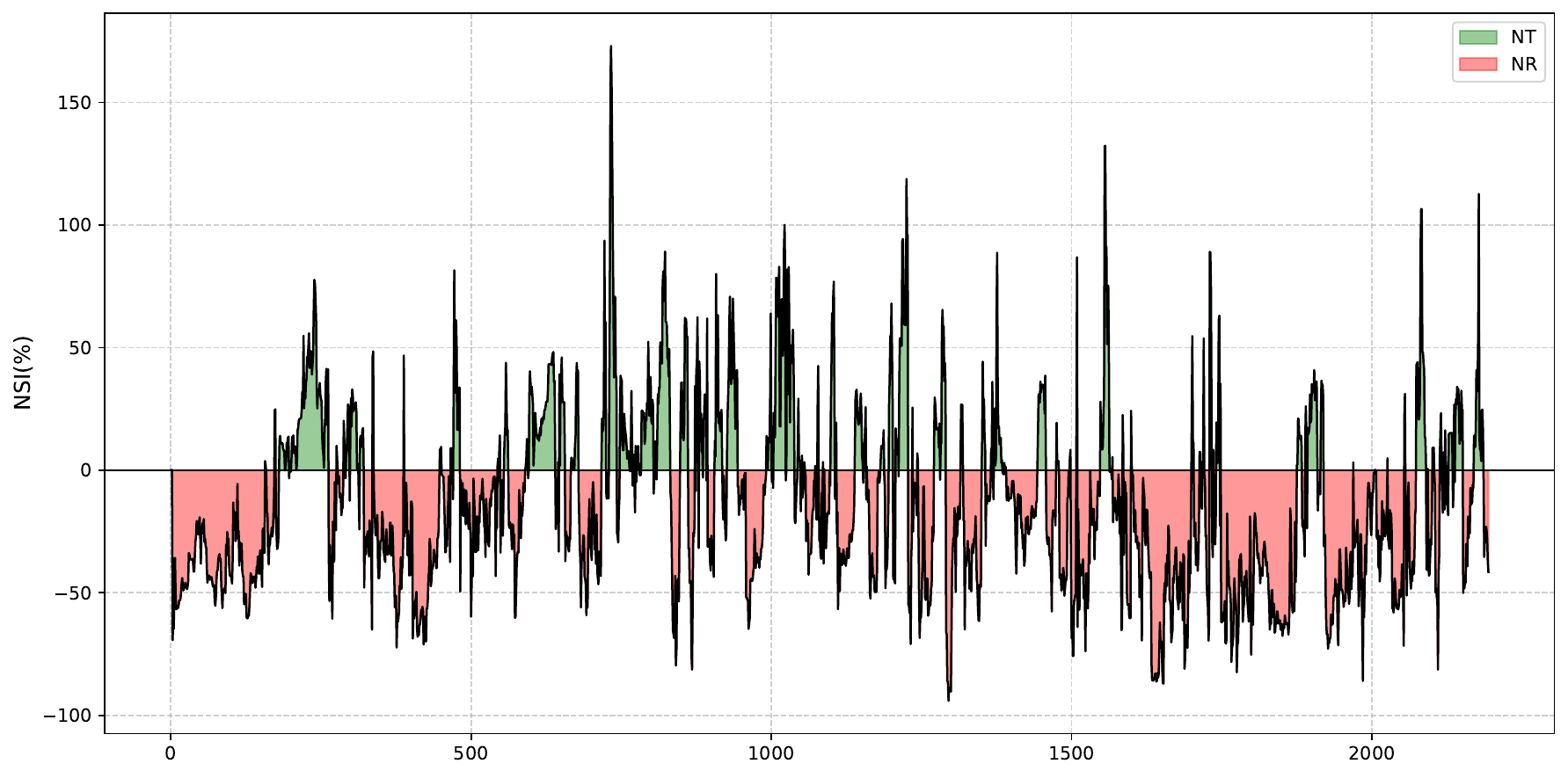}\label{fig:rsp_dash_high}}
    \vspace{0.3cm}
    \subfigure[ETH, $\tau=0.05$]{\includegraphics[width=0.32\linewidth]{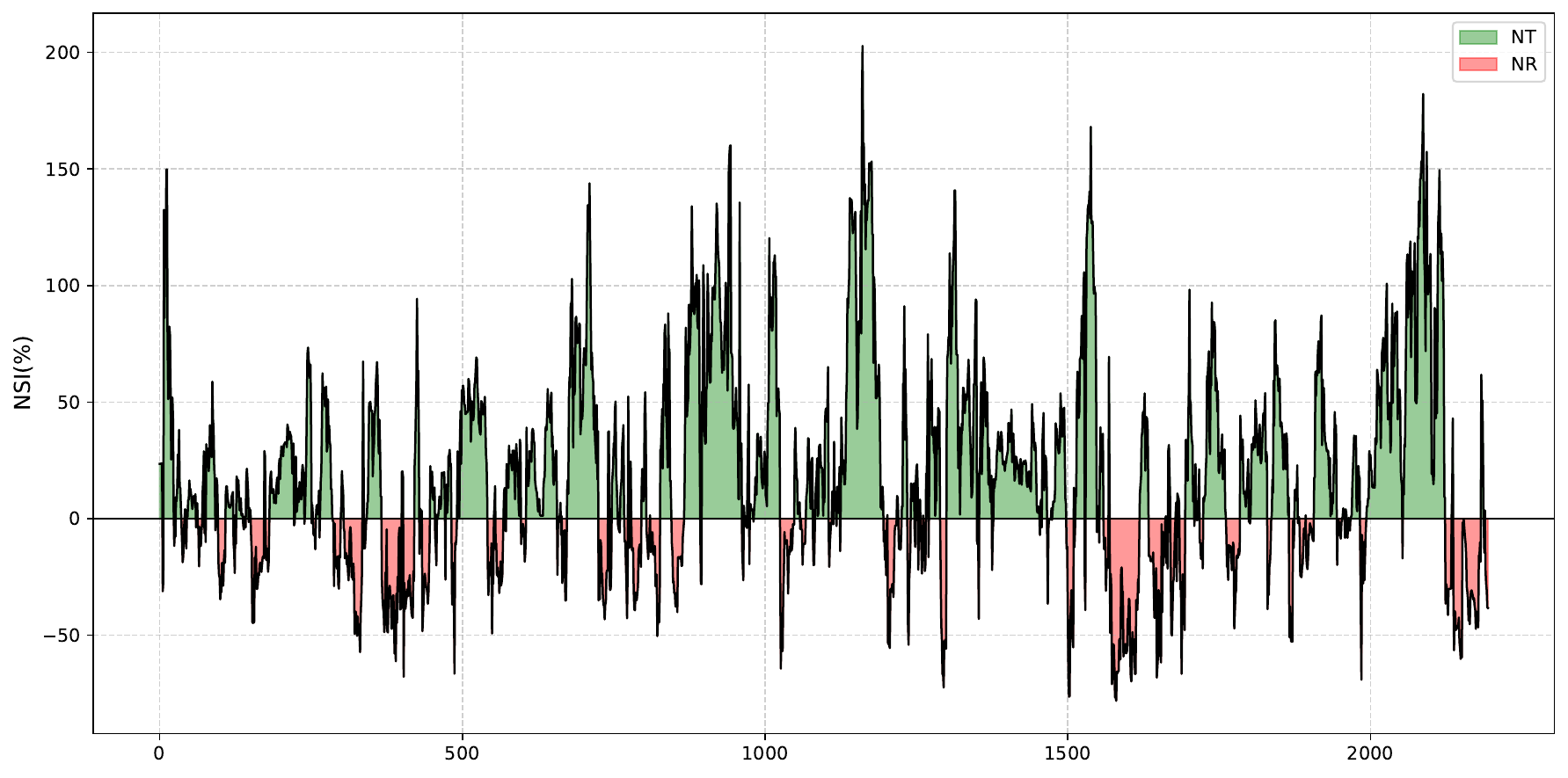}\label{fig:rsp_eth_low}}\hfill
    \subfigure[ETH, $\tau=0.50$]{\includegraphics[width=0.32\linewidth]{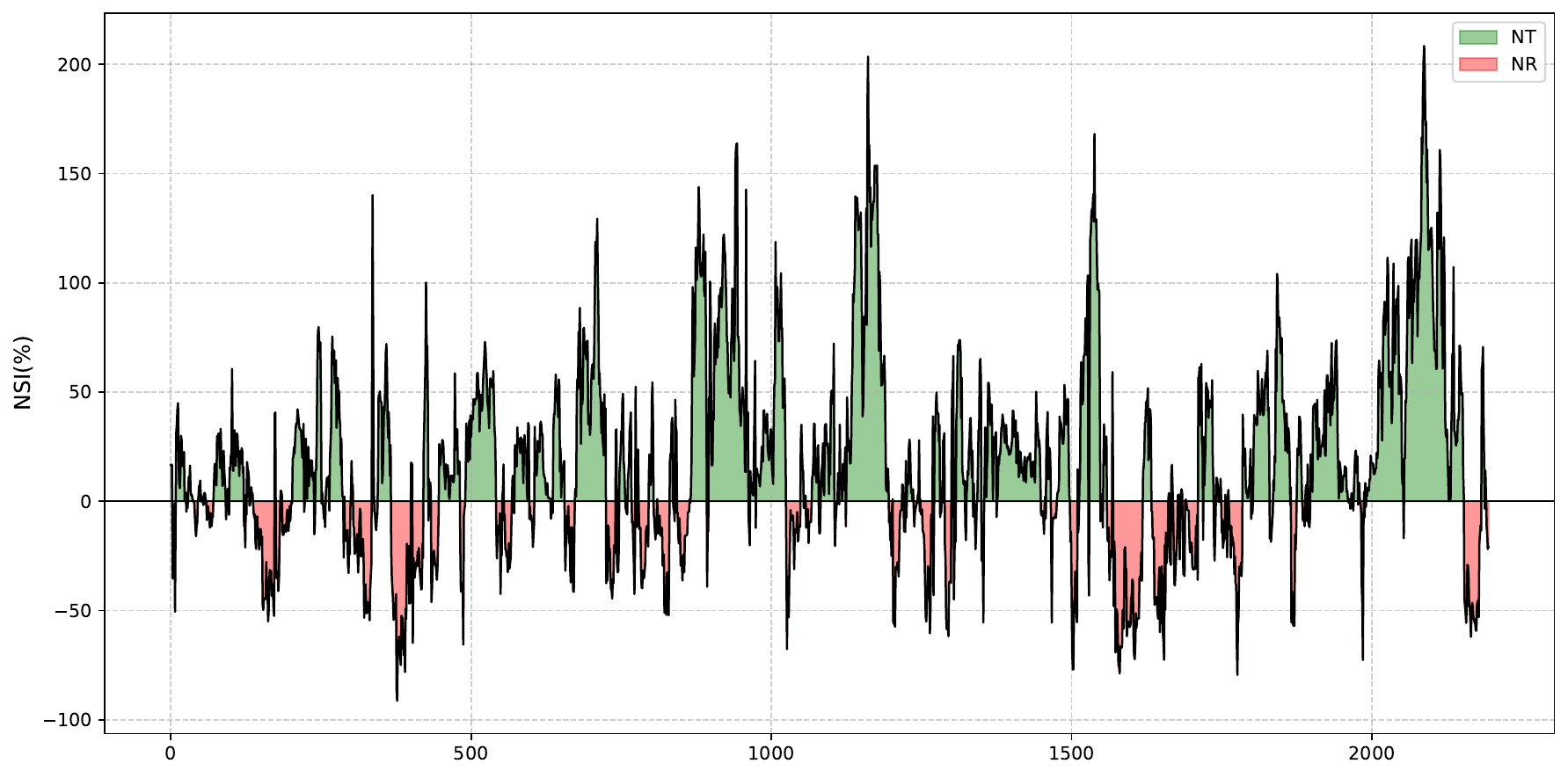}\label{fig:rsp_eth_mid}}\hfill
    \subfigure[ETH, $\tau=0.95$]{\includegraphics[width=0.32\linewidth]{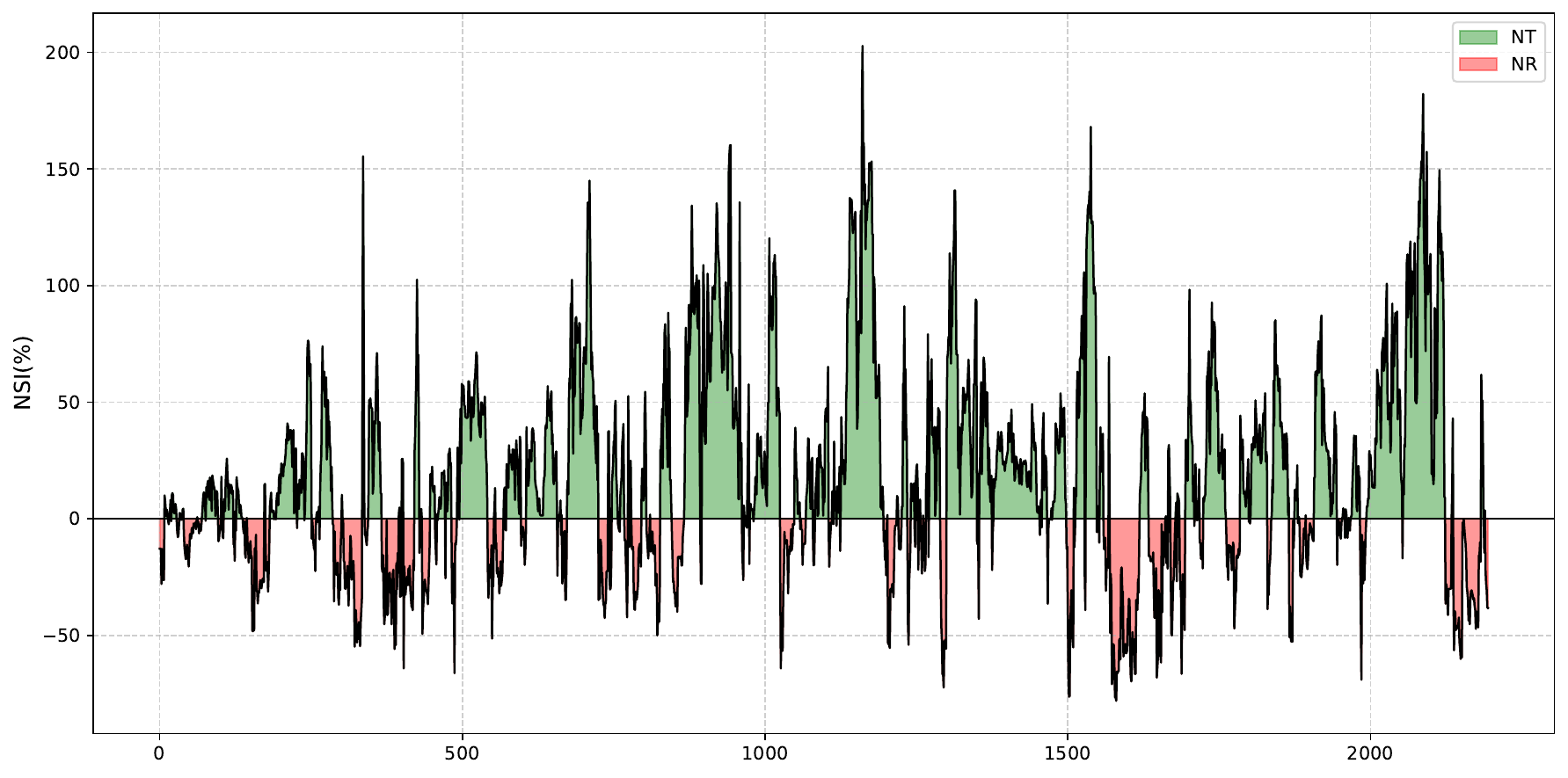}\label{fig:rsp_eth_high}}
    \vspace{0.3cm}
    \subfigure[LTC, $\tau=0.05$]{\includegraphics[width=0.32\linewidth]{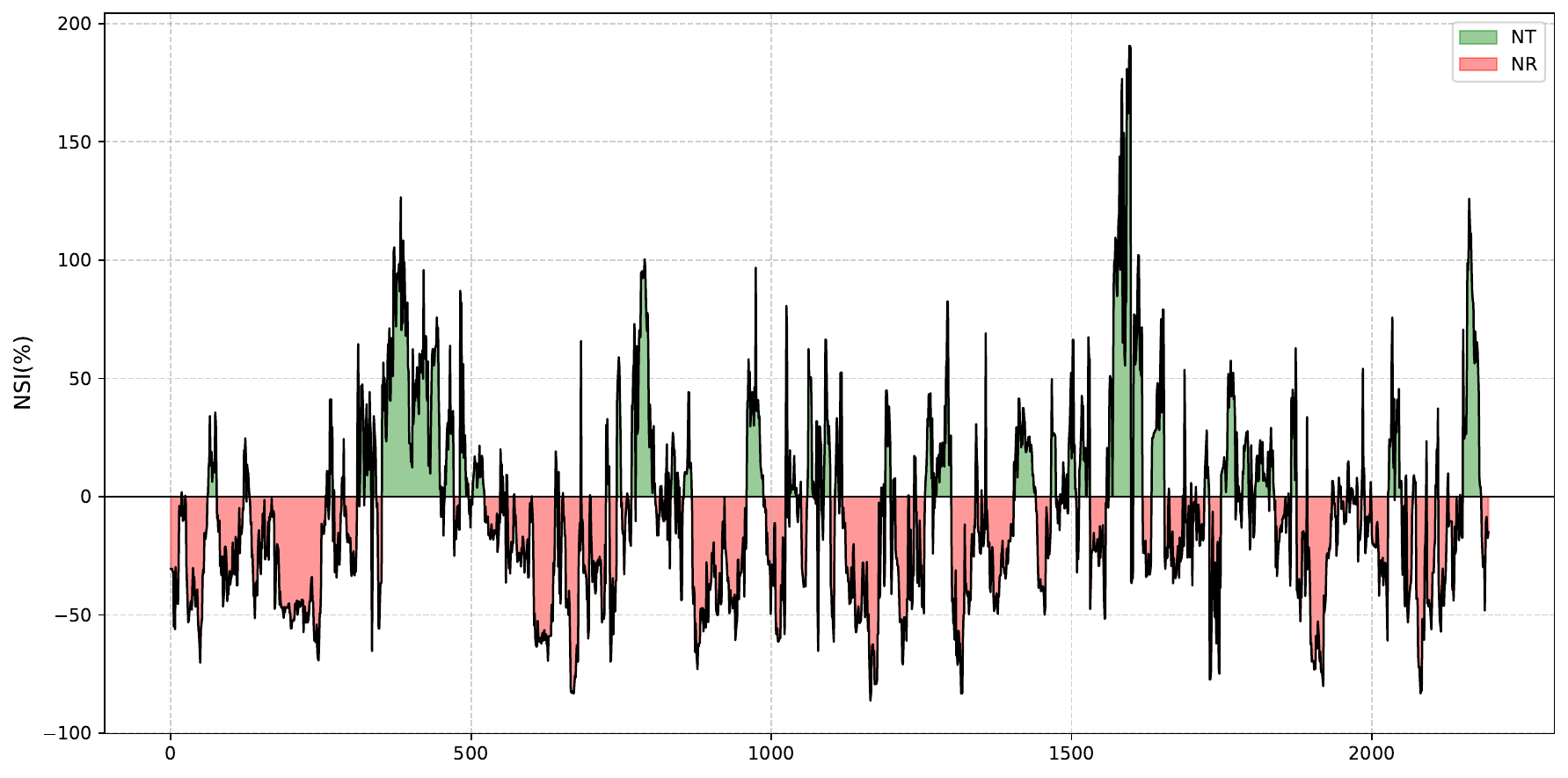}\label{fig:rsp_ltc_low}}\hfill
    \subfigure[LTC, $\tau=0.50$]{\includegraphics[width=0.32\linewidth]{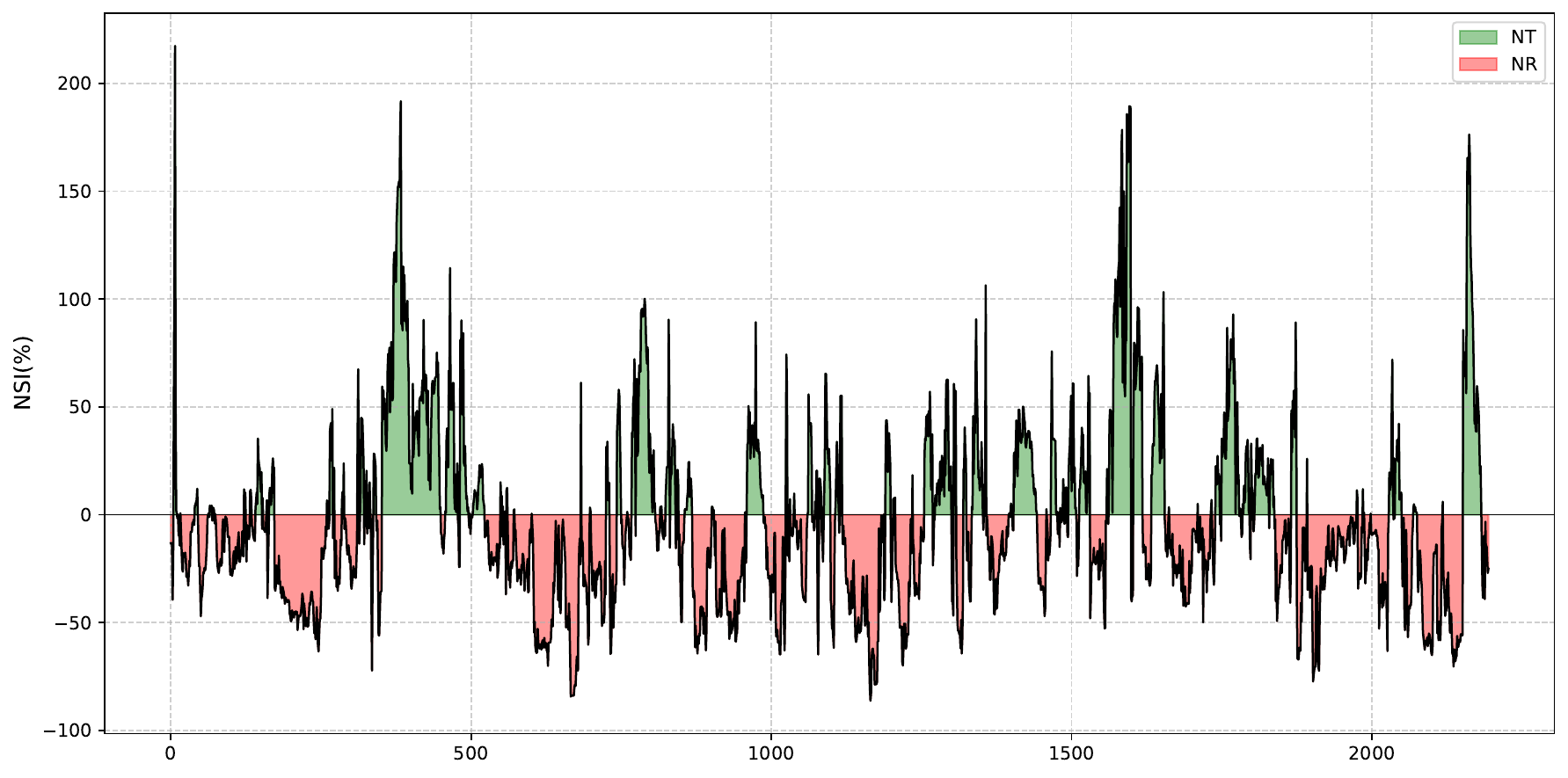}\label{fig:rsp_ltc_mid}}\hfill
    \subfigure[LTC, $\tau=0.95$]{\includegraphics[width=0.32\linewidth]{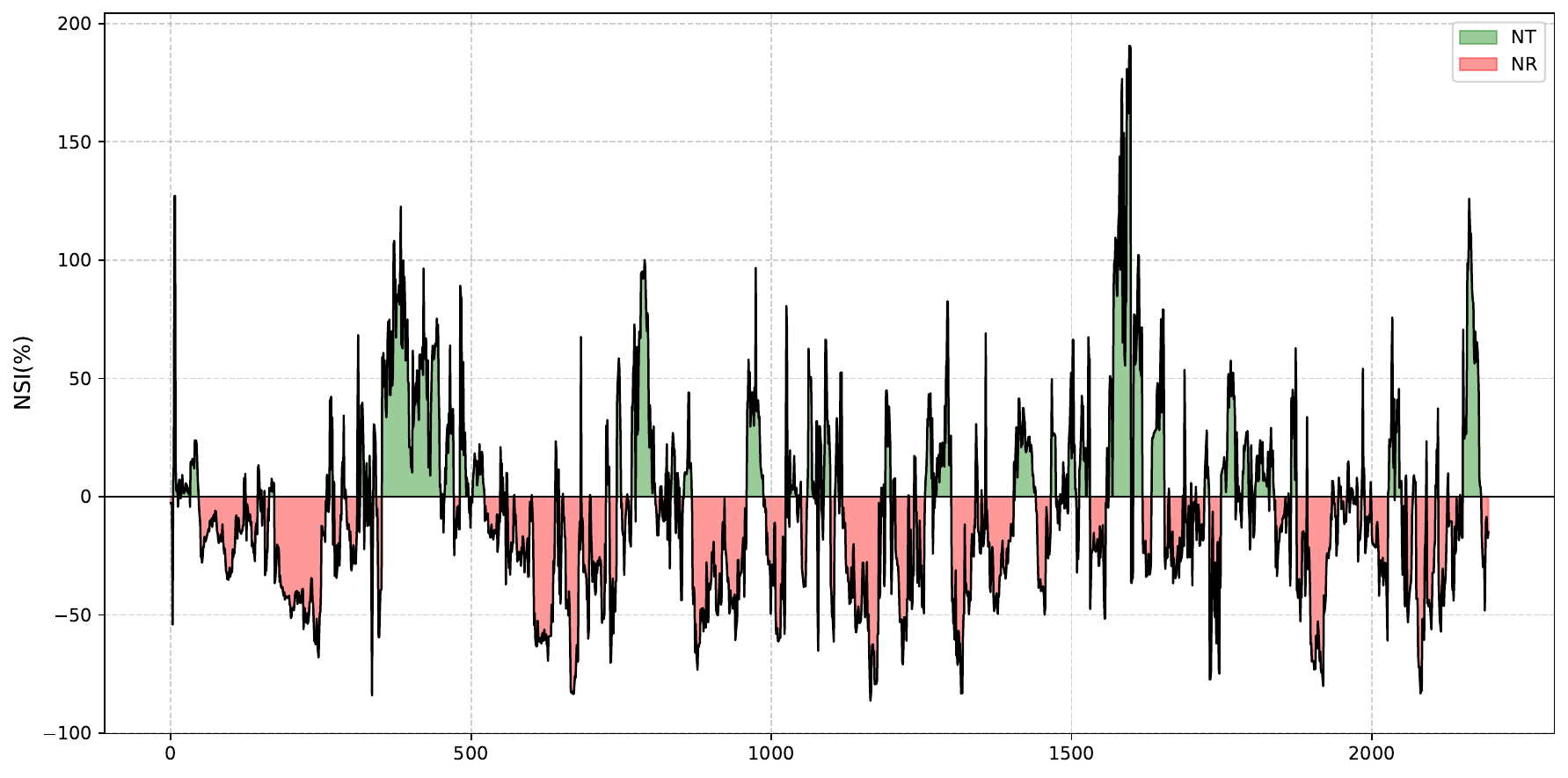}\label{fig:rsp_ltc_high}}
    \vspace{0.3cm}
    \subfigure[XLM, $\tau=0.05$]{\includegraphics[width=0.32\linewidth]{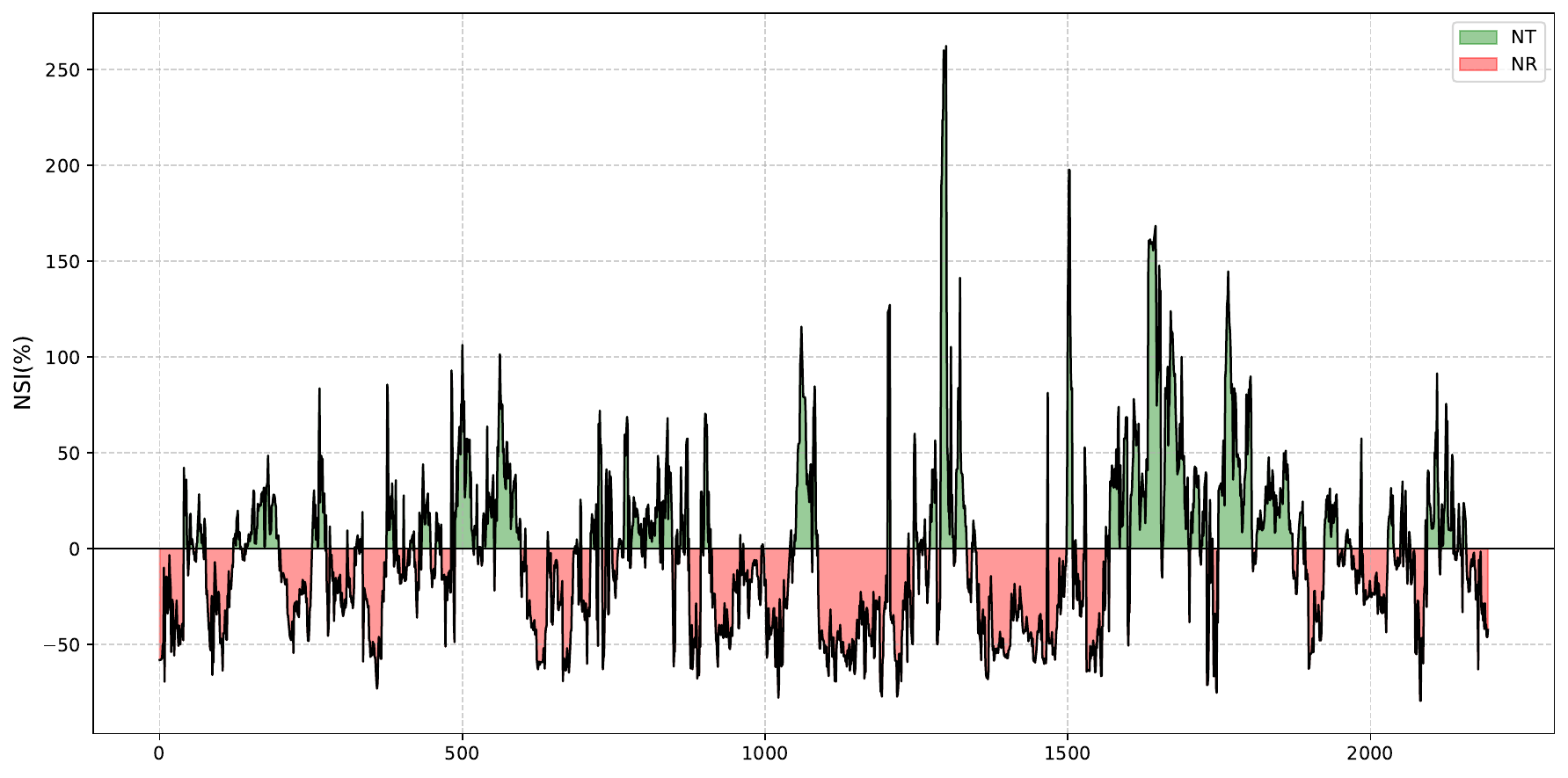}\label{fig:rsp_xlm_low}}\hfill
    \subfigure[XLM, $\tau=0.50$]{\includegraphics[width=0.32\linewidth]{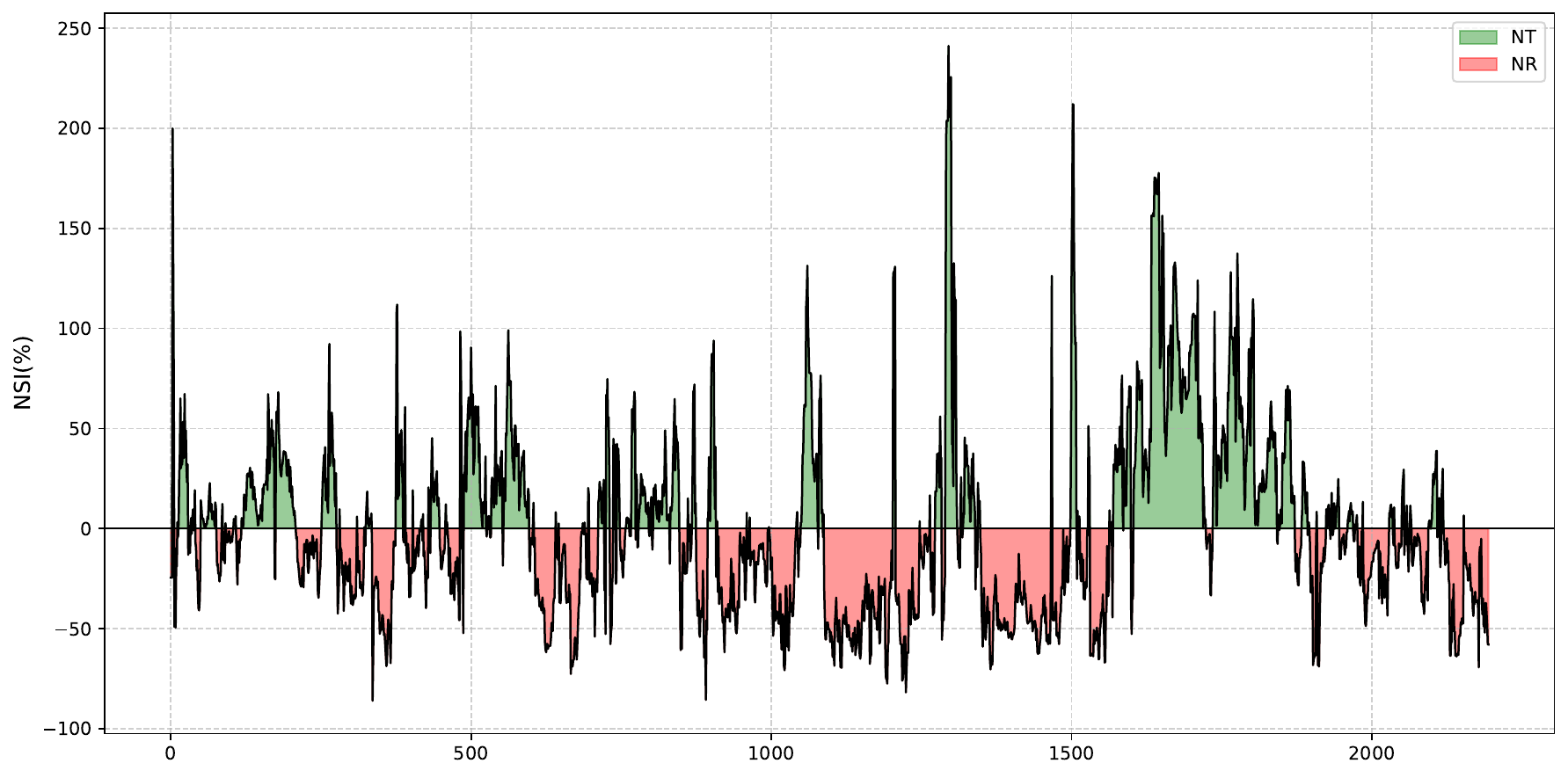}\label{fig:rsp_xlm_mid}}\hfill
    \subfigure[XLM, $\tau=0.95$]{\includegraphics[width=0.32\linewidth]{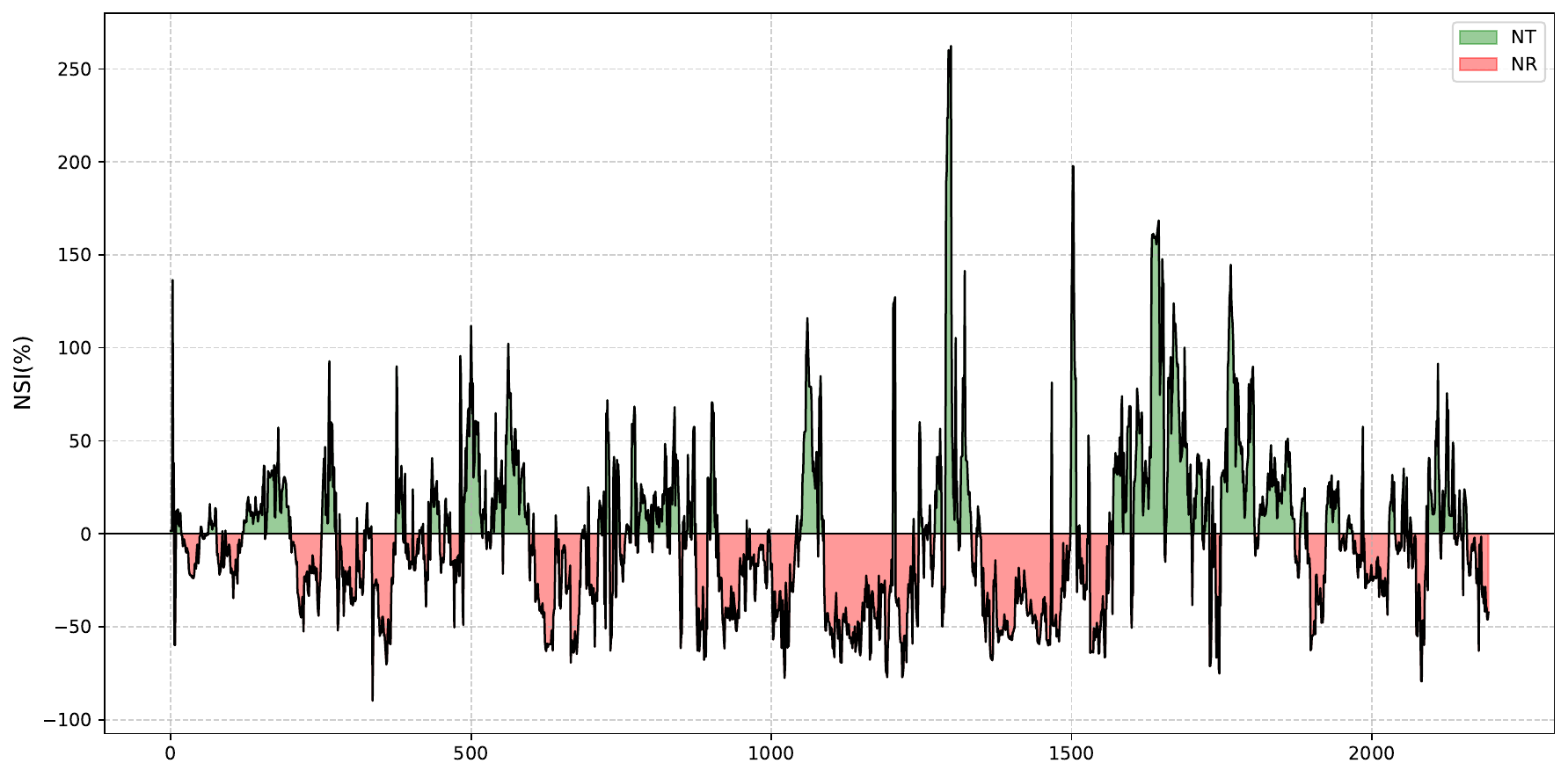}\label{fig:rsp_xlm_high}}
    \vspace{0.3cm}
    \subfigure[XRP, $\tau=0.05$]{\includegraphics[width=0.32\linewidth]{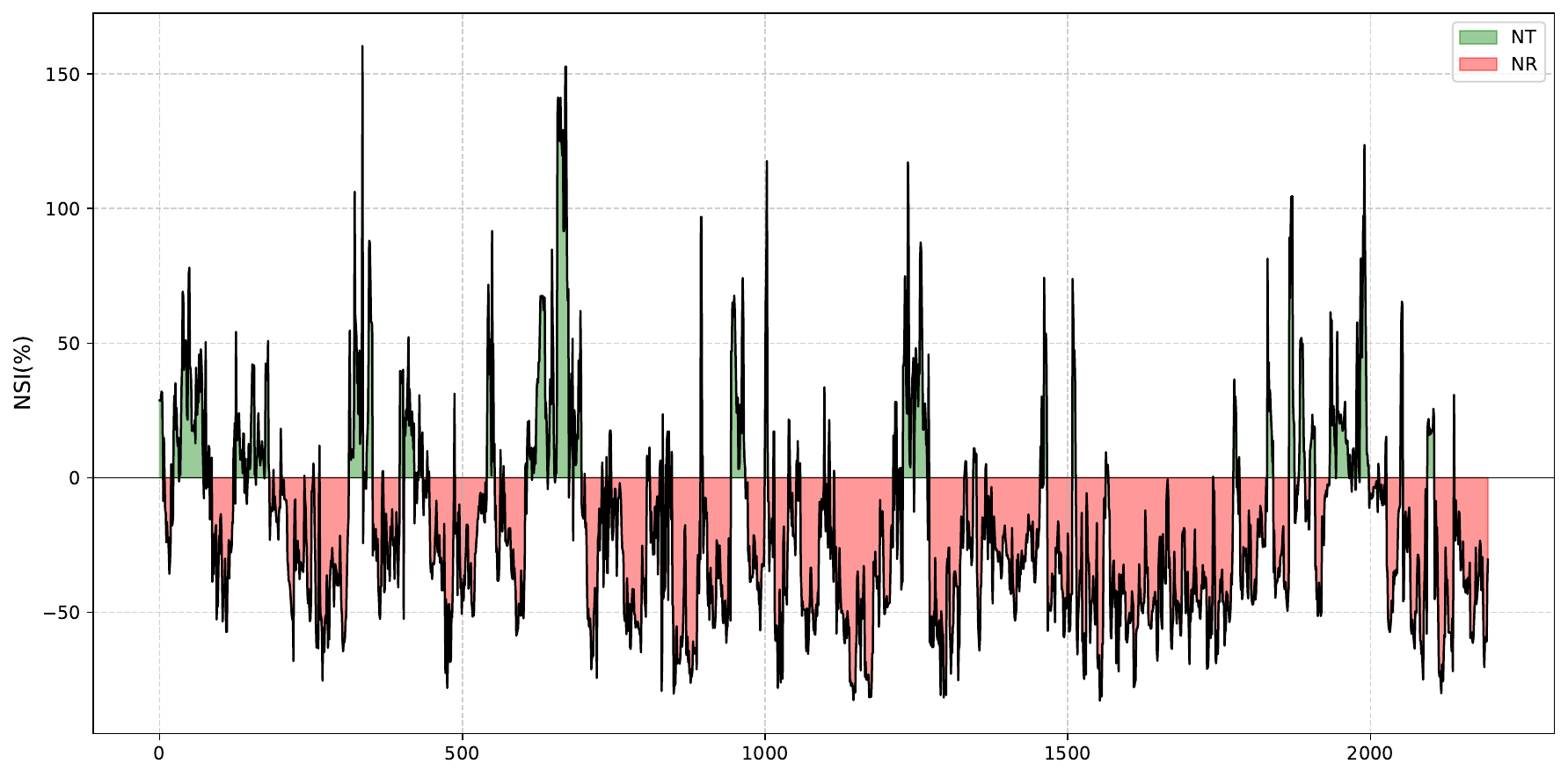}\label{fig:rsp_xrp_low}}\hfill
    \subfigure[XRP, $\tau=0.50$]{\includegraphics[width=0.32\linewidth]{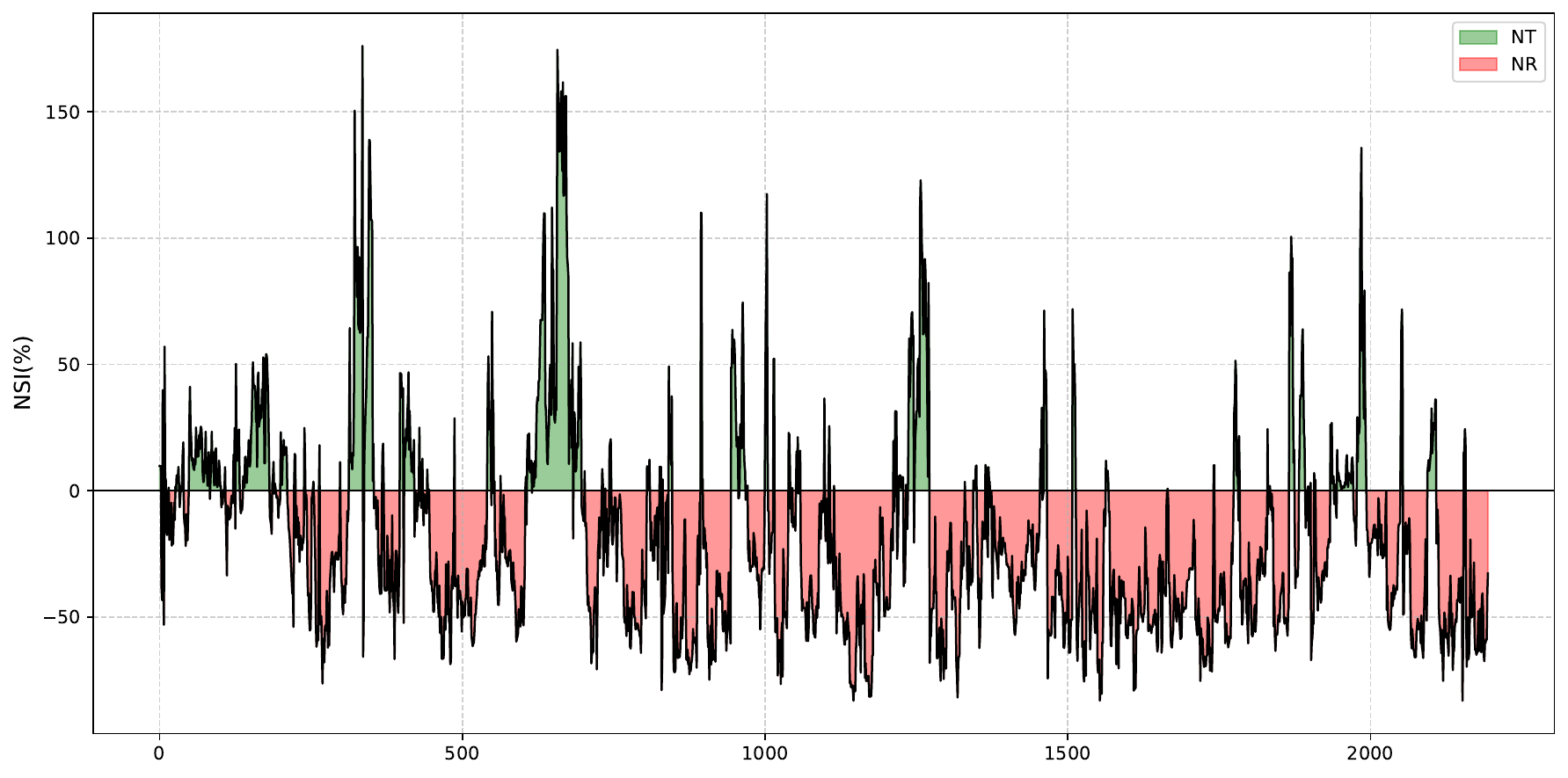}\label{fig:rsp_xrp_mid}}\hfill
    \subfigure[XRP, $\tau=0.95$]{\includegraphics[width=0.32\linewidth]{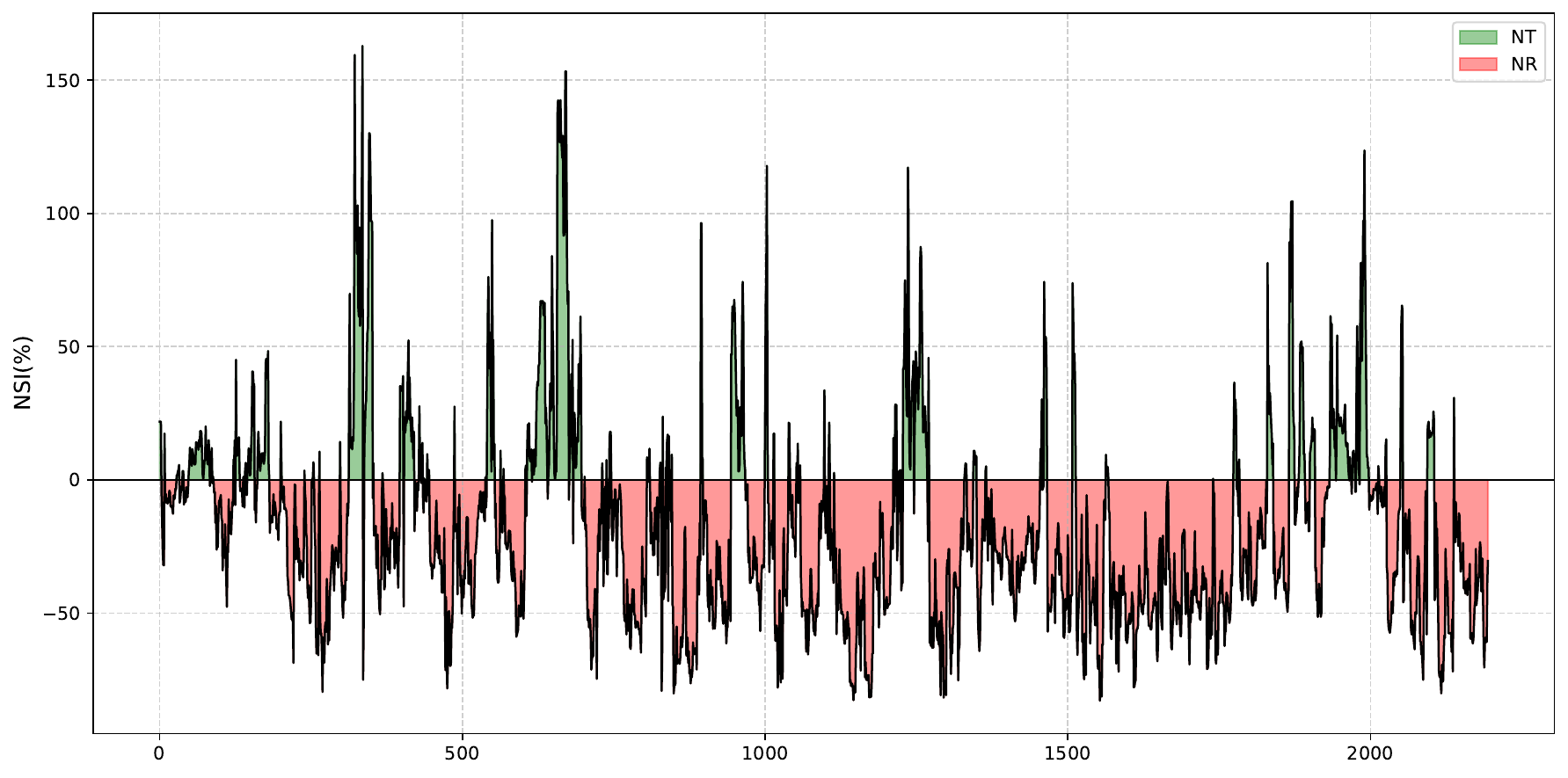}\label{fig:rsp_xrp_high}}
\end{figure}

\begin{figure}[p]
    \centering
    \caption{Quantile net spillovers for major cryptocurrencies using $RS^-$ as the feature variable.}
    \label{fig:rsm_net_spillover_by_coin}

    \subfigure[BTC, $\tau=0.05$]{\includegraphics[width=0.32\linewidth]{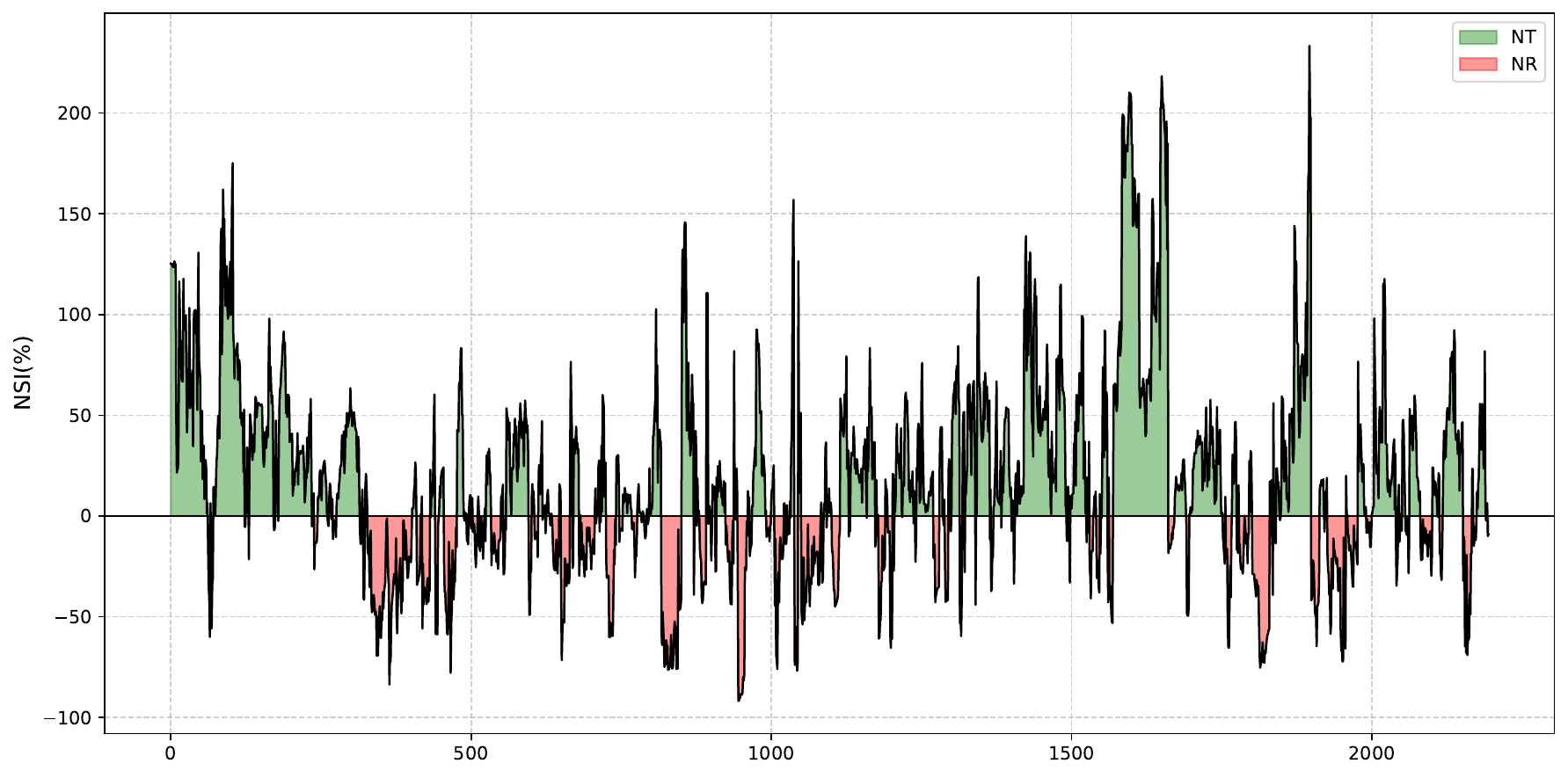}\label{fig:rsm_btc_low}}\hfill
    \subfigure[BTC, $\tau=0.50$]{\includegraphics[width=0.32\linewidth]{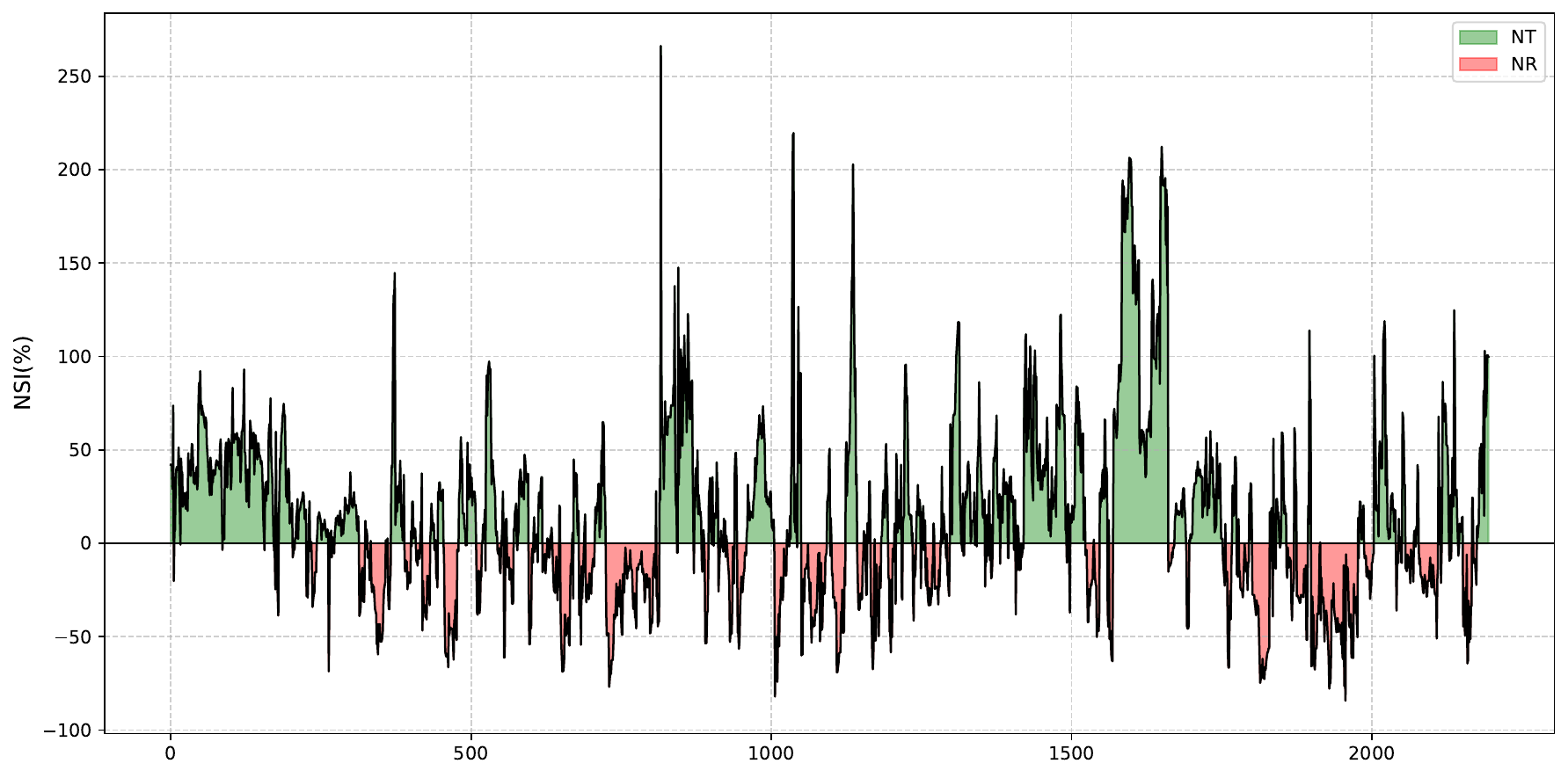}\label{fig:rsm_btc_mid}}\hfill
    \subfigure[BTC, $\tau=0.95$]{\includegraphics[width=0.32\linewidth]{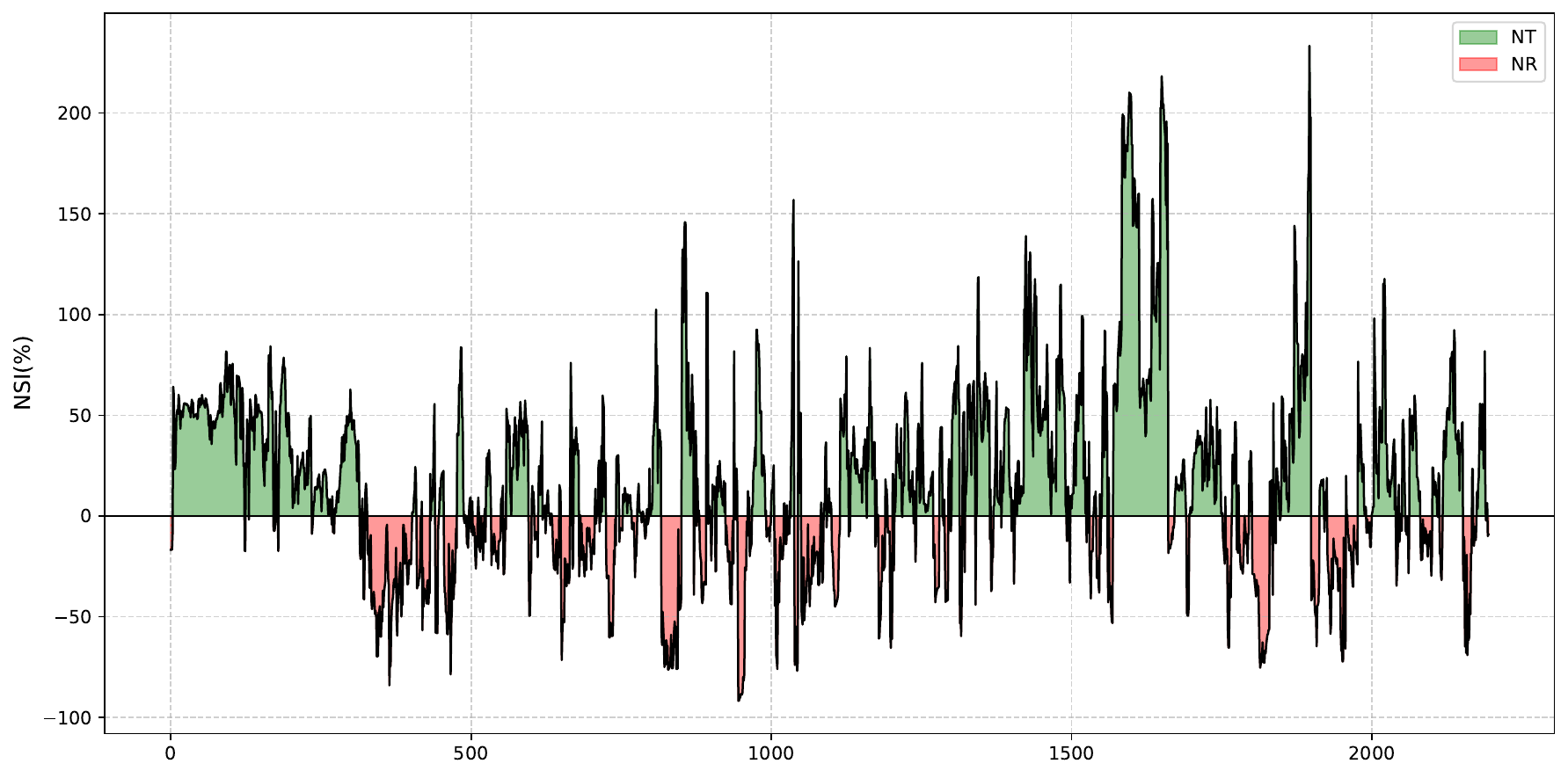}\label{fig:rsm_btc_high}}
    \vspace{0.3cm}
    \subfigure[DASH, $\tau=0.05$]{\includegraphics[width=0.32\linewidth]{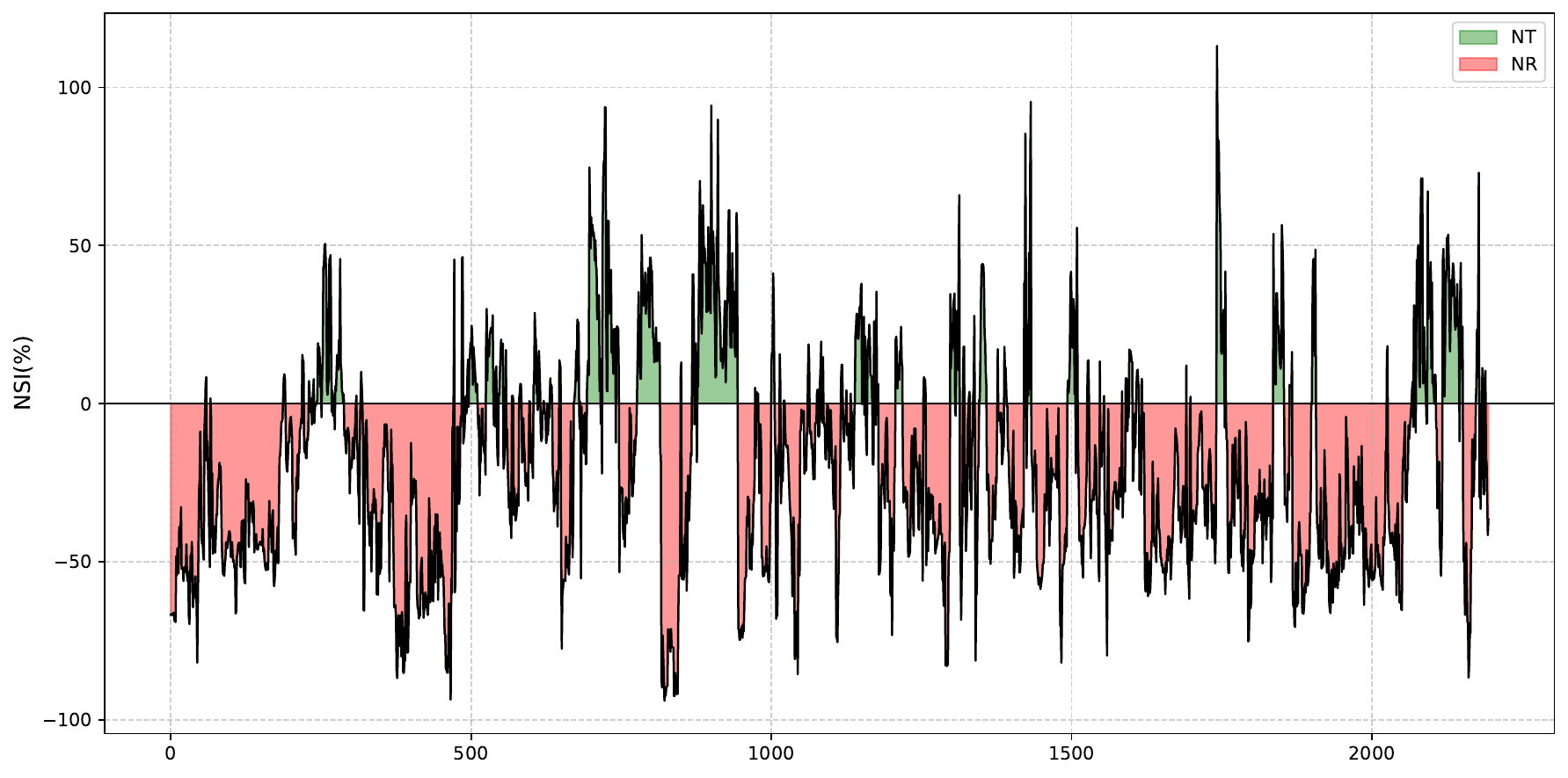}\label{fig:rsm_dash_low}}\hfill
    \subfigure[DASH, $\tau=0.50$]{\includegraphics[width=0.32\linewidth]{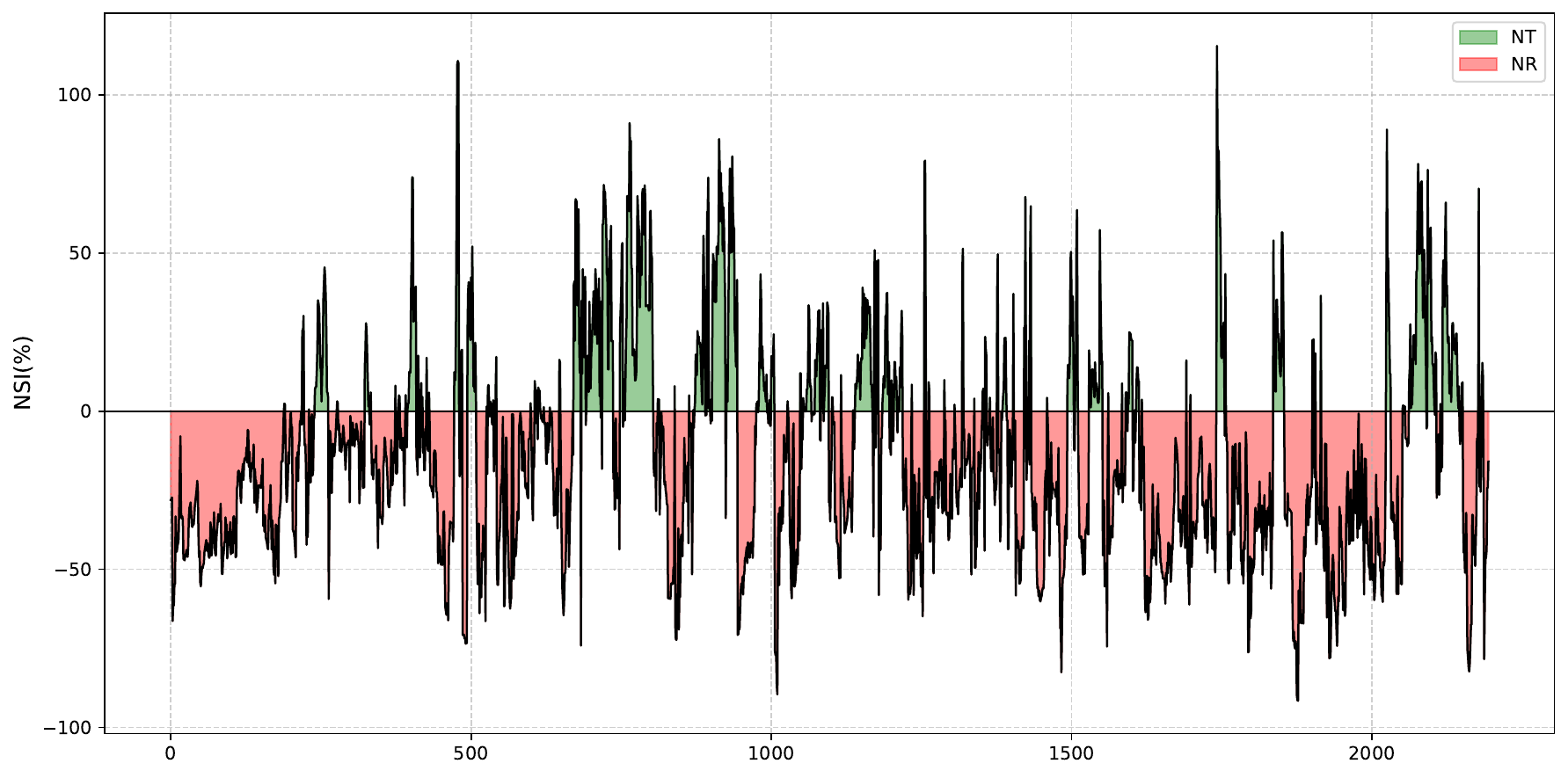}\label{fig:rsm_dash_mid}}\hfill
    \subfigure[DASH, $\tau=0.95$]{\includegraphics[width=0.32\linewidth]{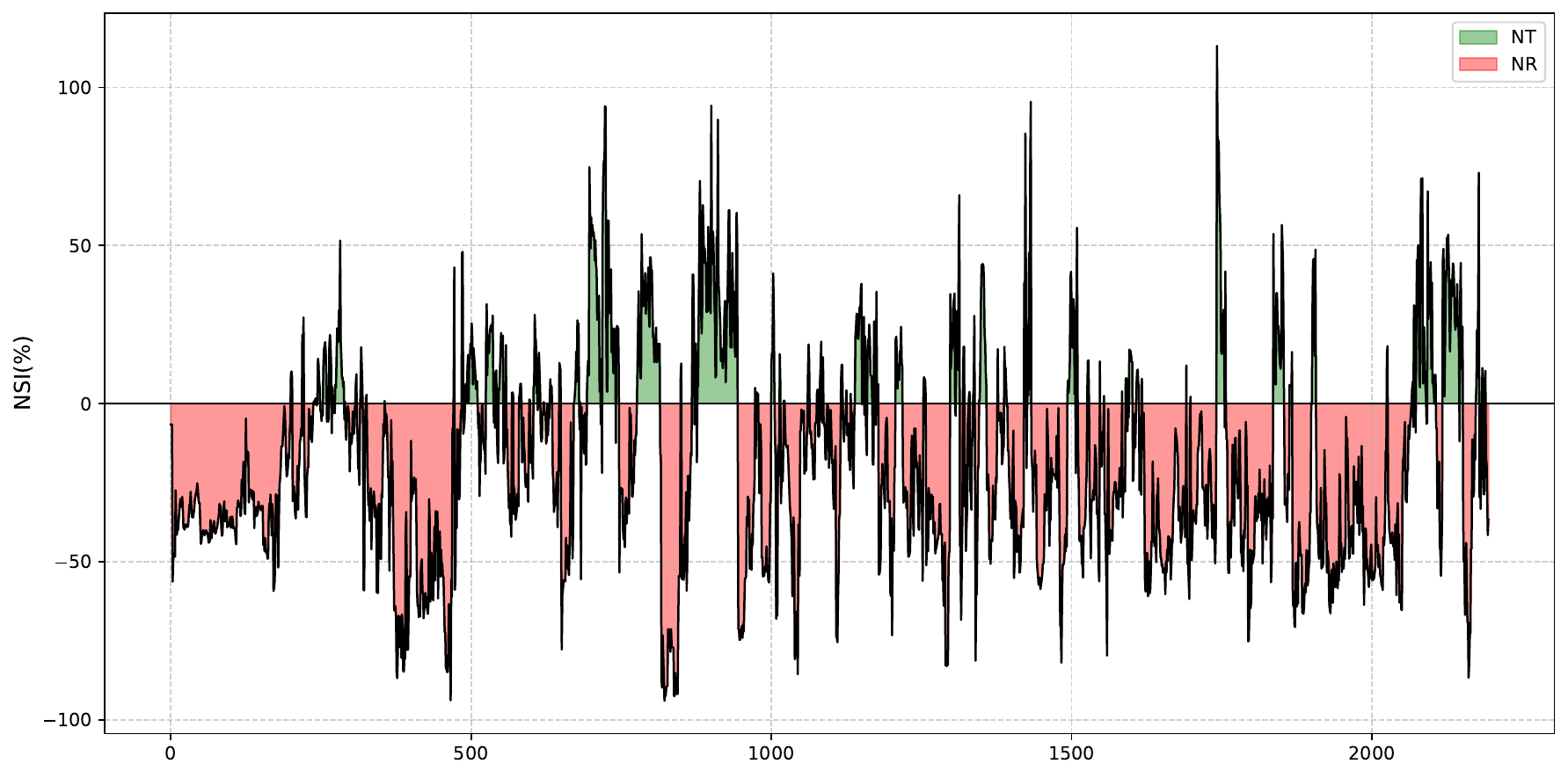}\label{fig:rsm_dash_high}}
    \vspace{0.3cm}
    \subfigure[ETH, $\tau=0.05$]{\includegraphics[width=0.32\linewidth]{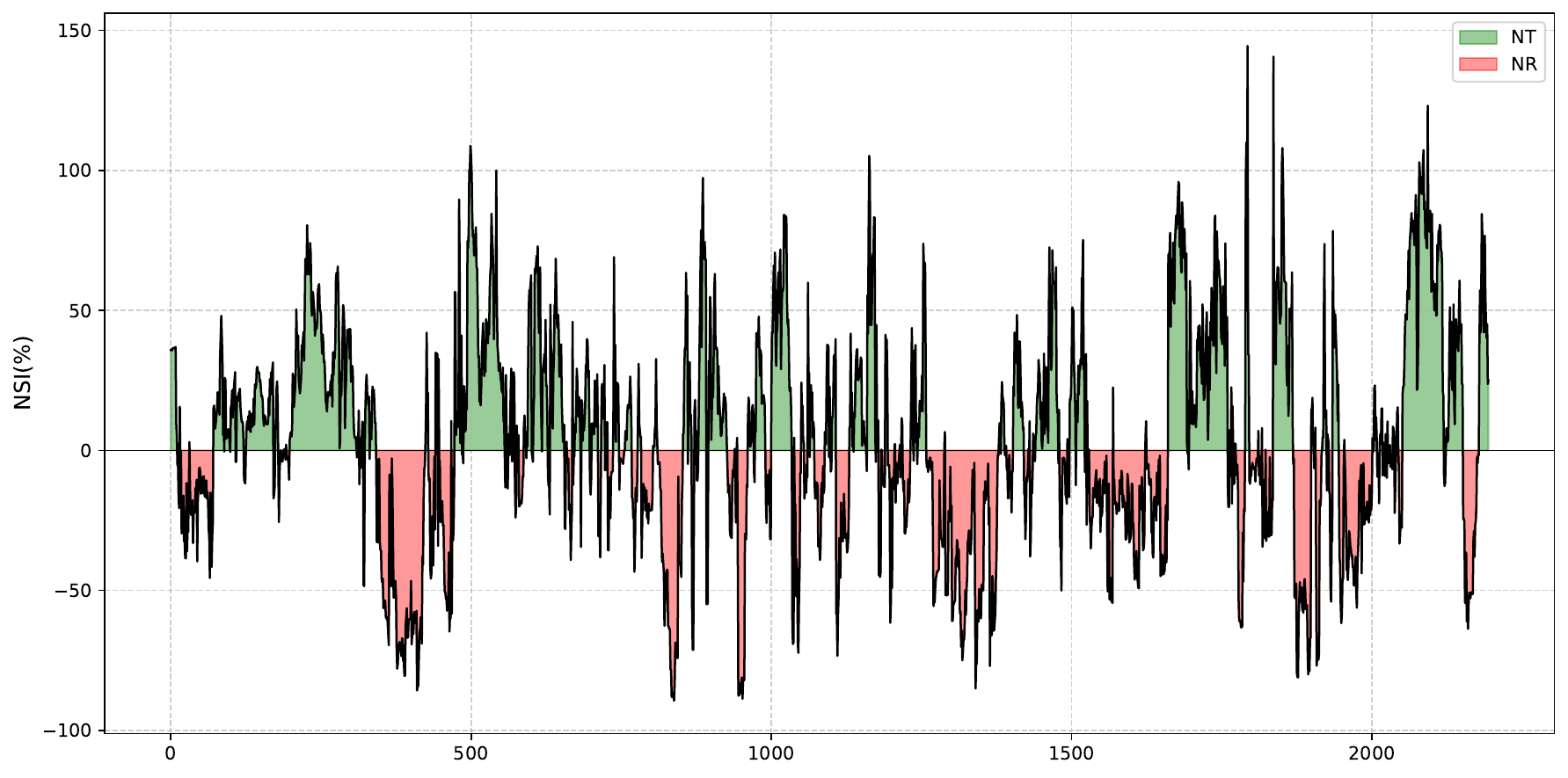}\label{fig:rsm_eth_low}}\hfill
    \subfigure[ETH, $\tau=0.50$]{\includegraphics[width=0.32\linewidth]{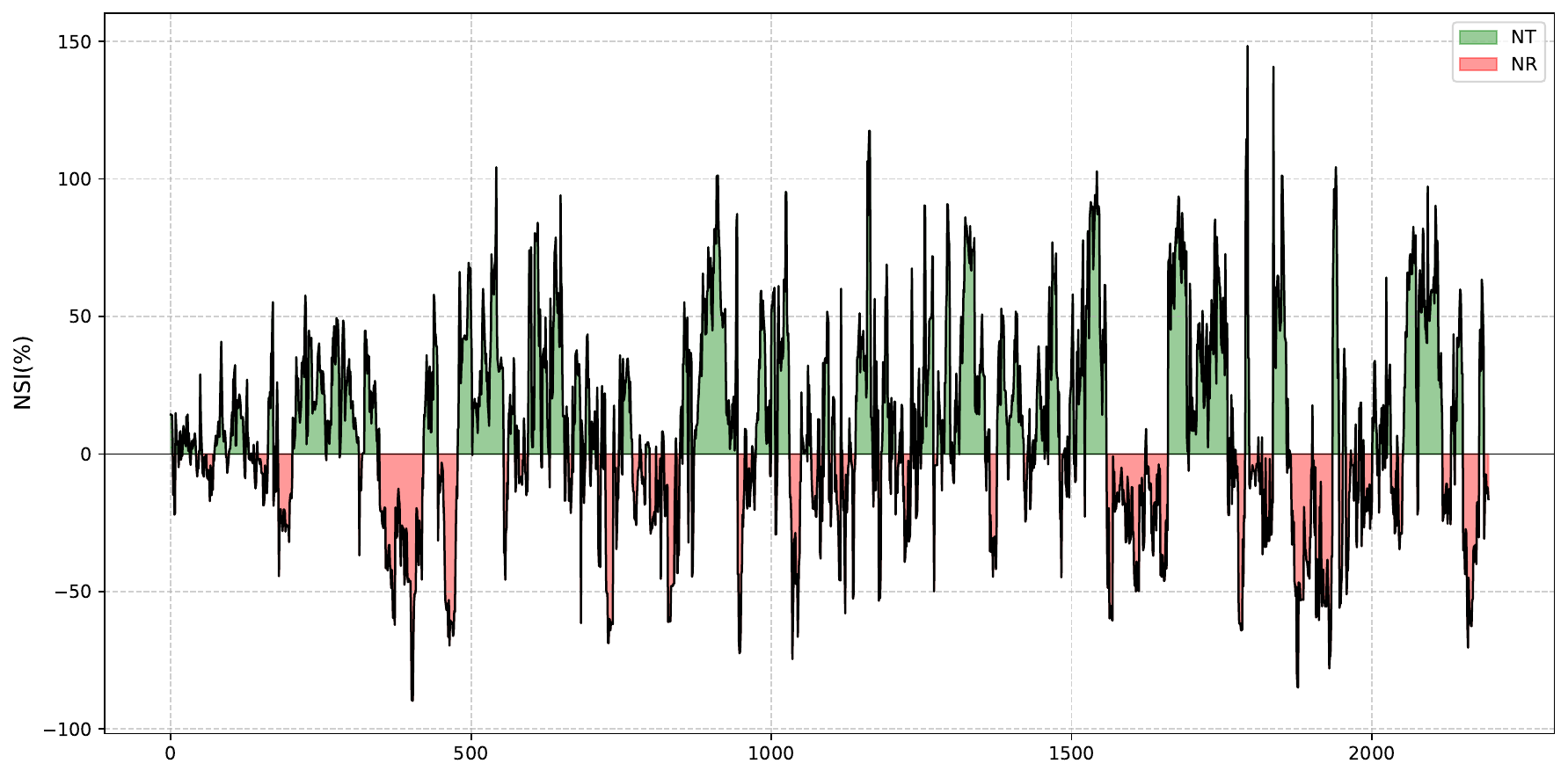}\label{fig:rsm_eth_mid}}\hfill
    \subfigure[ETH, $\tau=0.95$]{\includegraphics[width=0.32\linewidth]{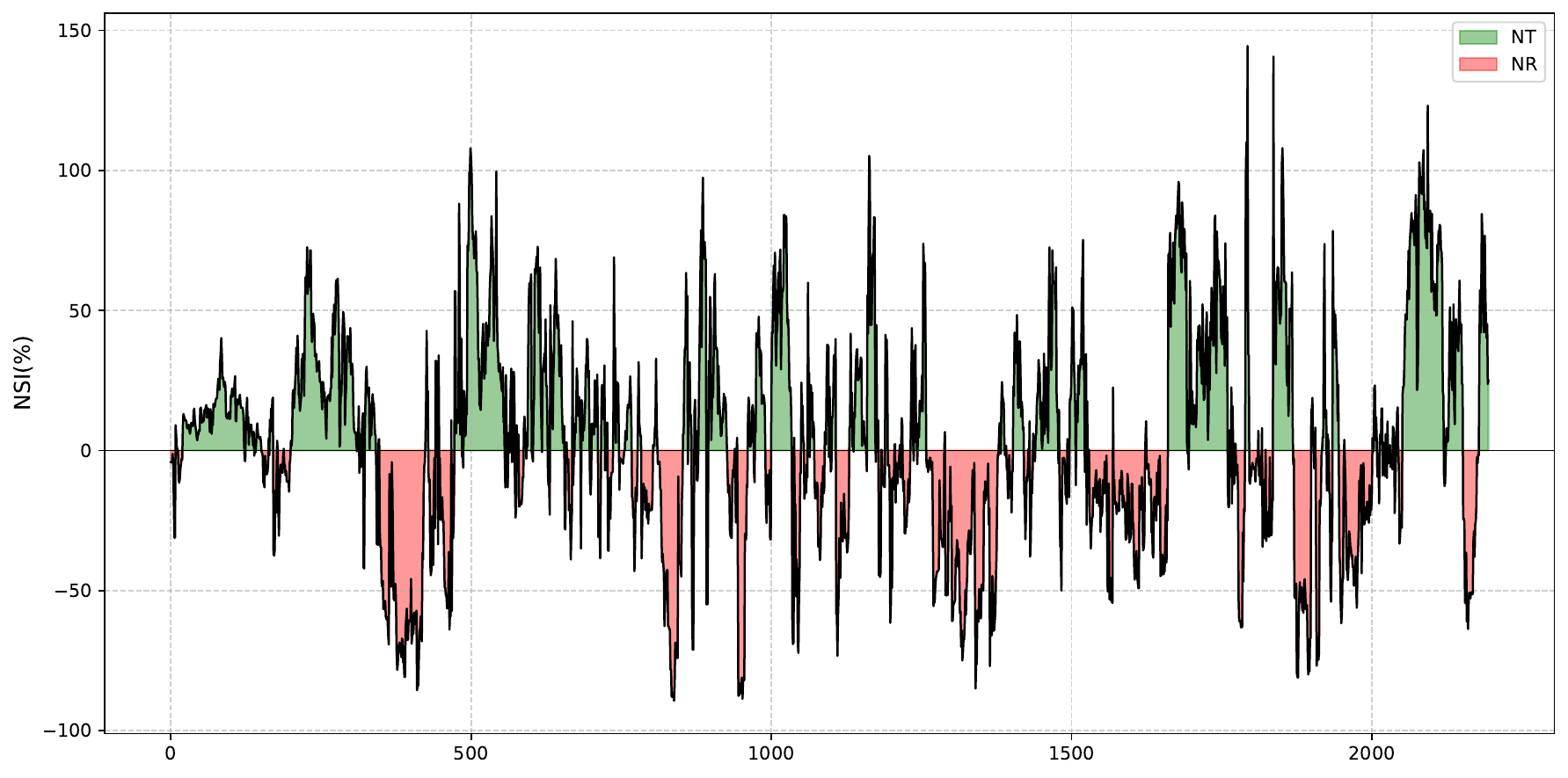}\label{fig:rsm_eth_high}}
    \vspace{0.3cm}
    \subfigure[LTC, $\tau=0.05$]{\includegraphics[width=0.32\linewidth]{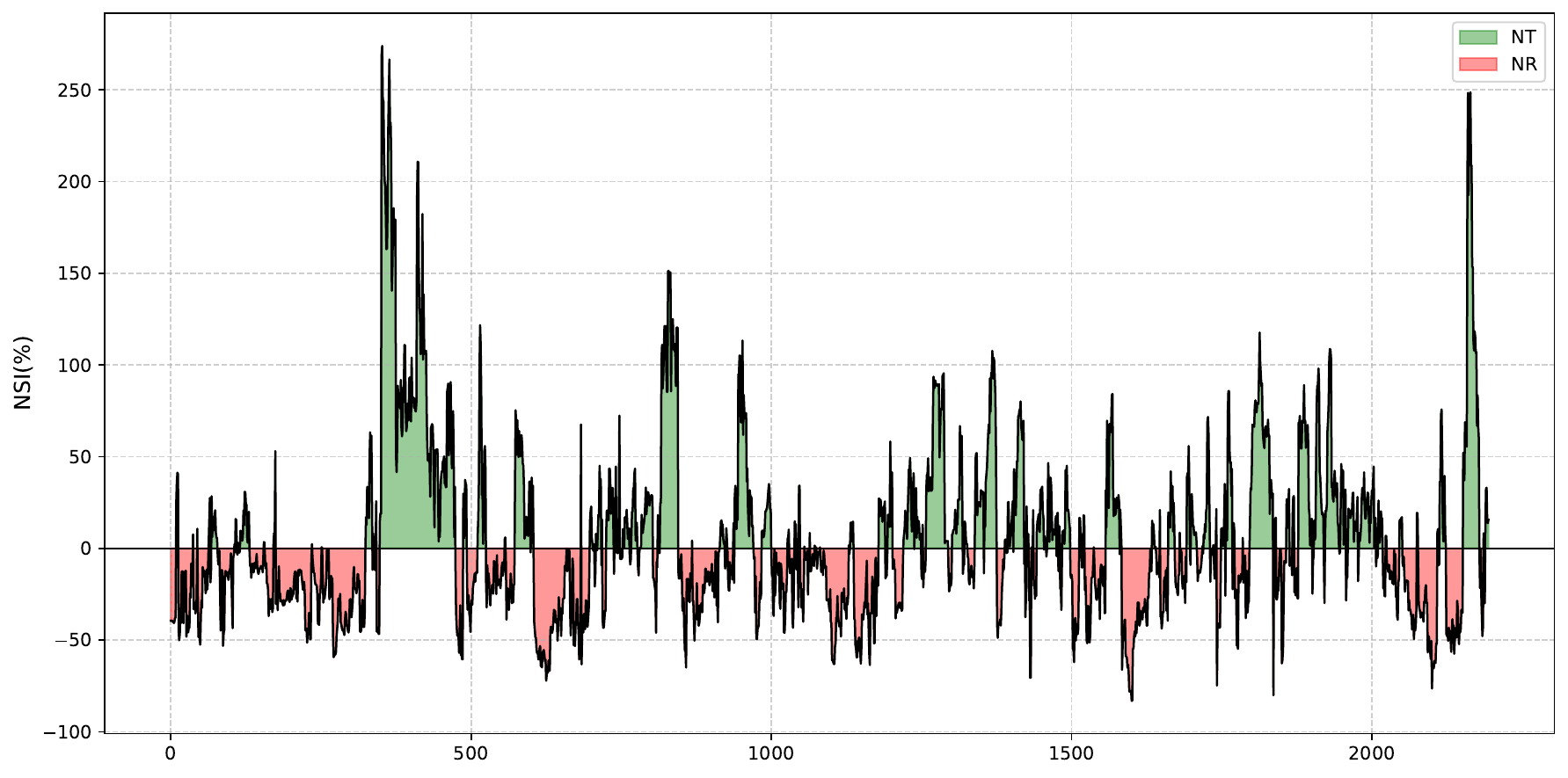}\label{fig:rsm_ltc_low}}\hfill
    \subfigure[LTC, $\tau=0.50$]{\includegraphics[width=0.32\linewidth]{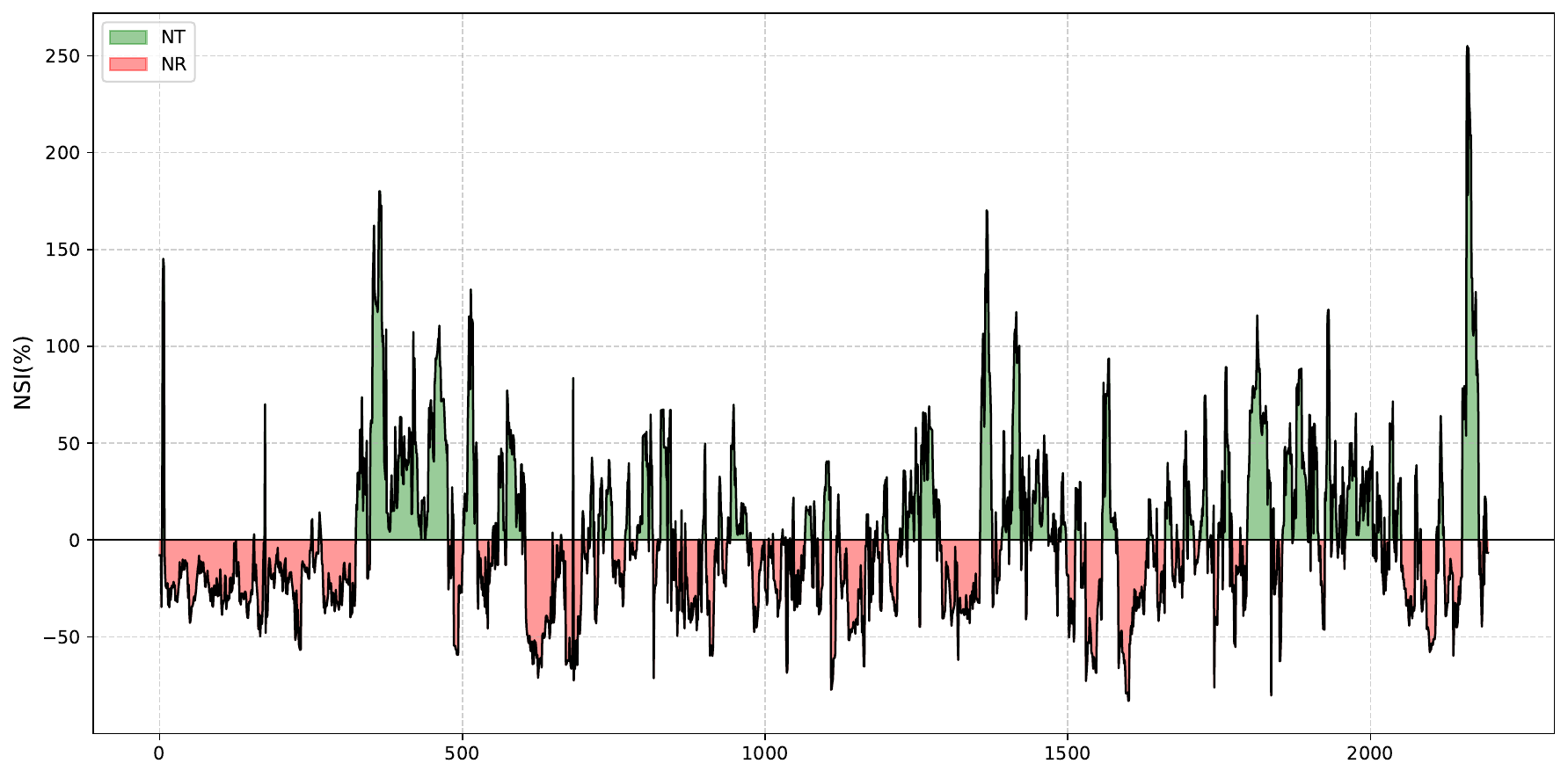}\label{fig:rsm_ltc_mid}}\hfill
    \subfigure[LTC, $\tau=0.95$]{\includegraphics[width=0.32\linewidth]{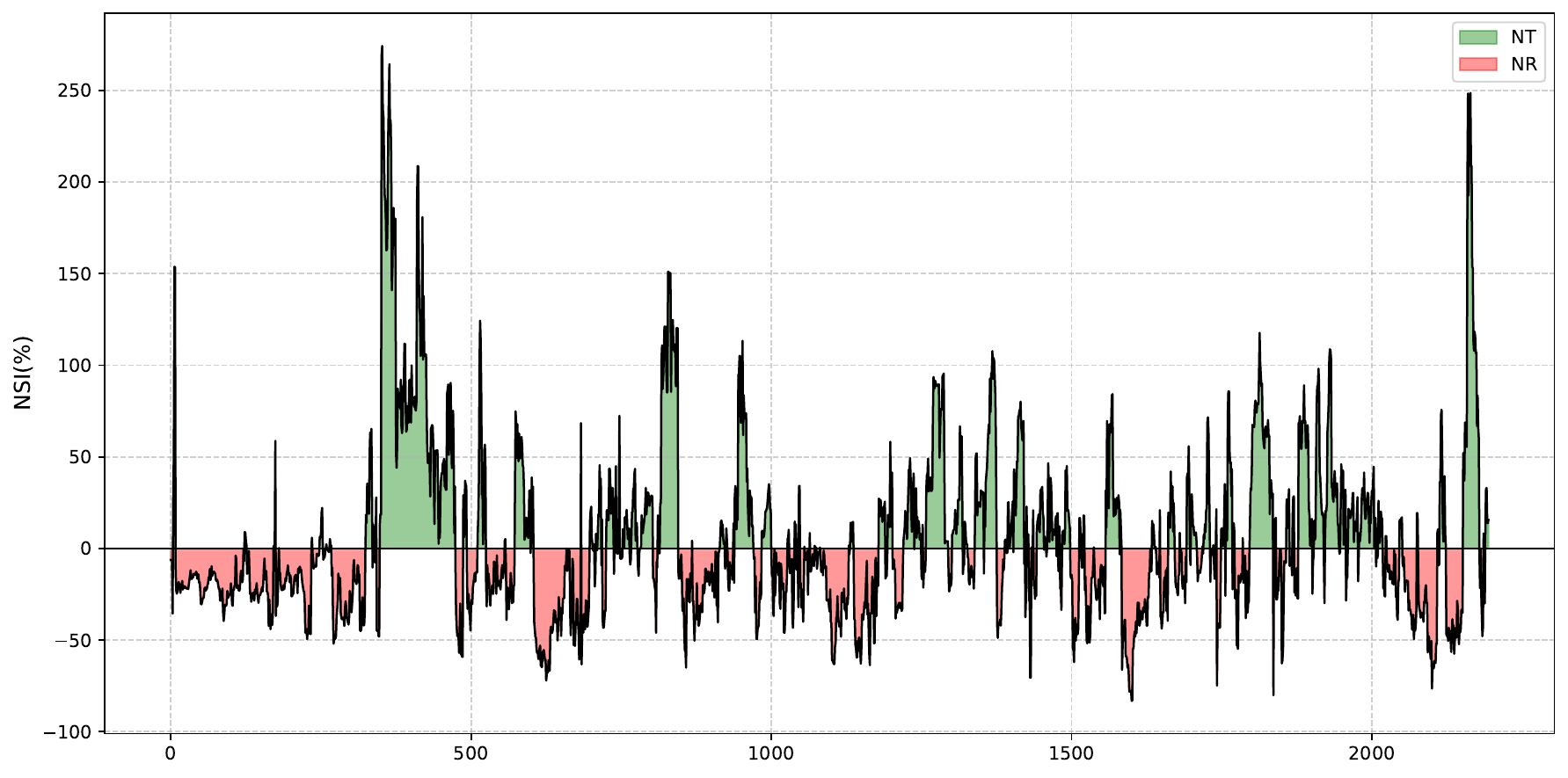}\label{fig:rsm_ltc_high}}
    \vspace{0.3cm}
    \subfigure[XLM, $\tau=0.05$]{\includegraphics[width=0.32\linewidth]{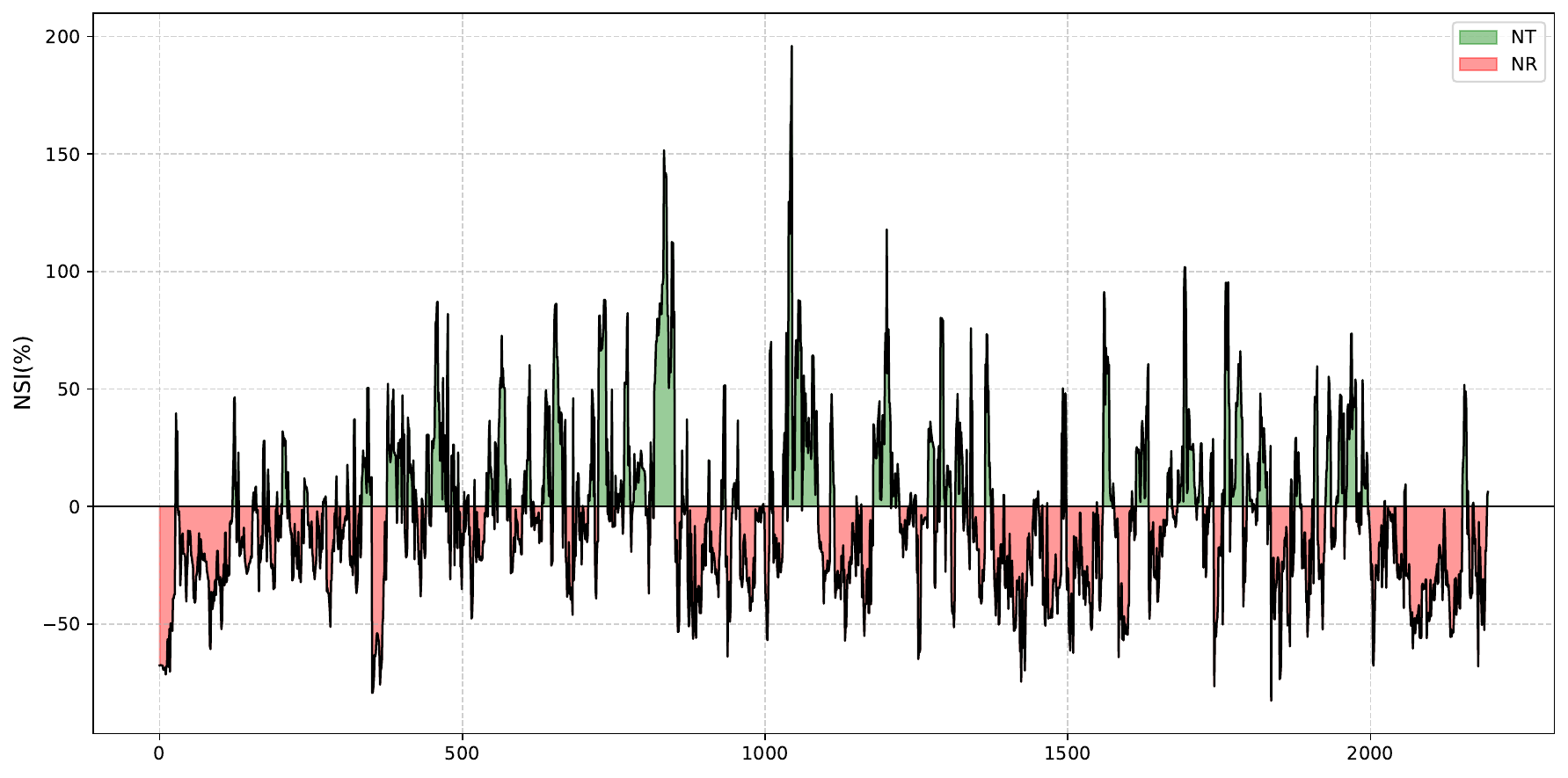}\label{fig:rsm_xlm_low}}\hfill
    \subfigure[XLM, $\tau=0.50$]{\includegraphics[width=0.32\linewidth]{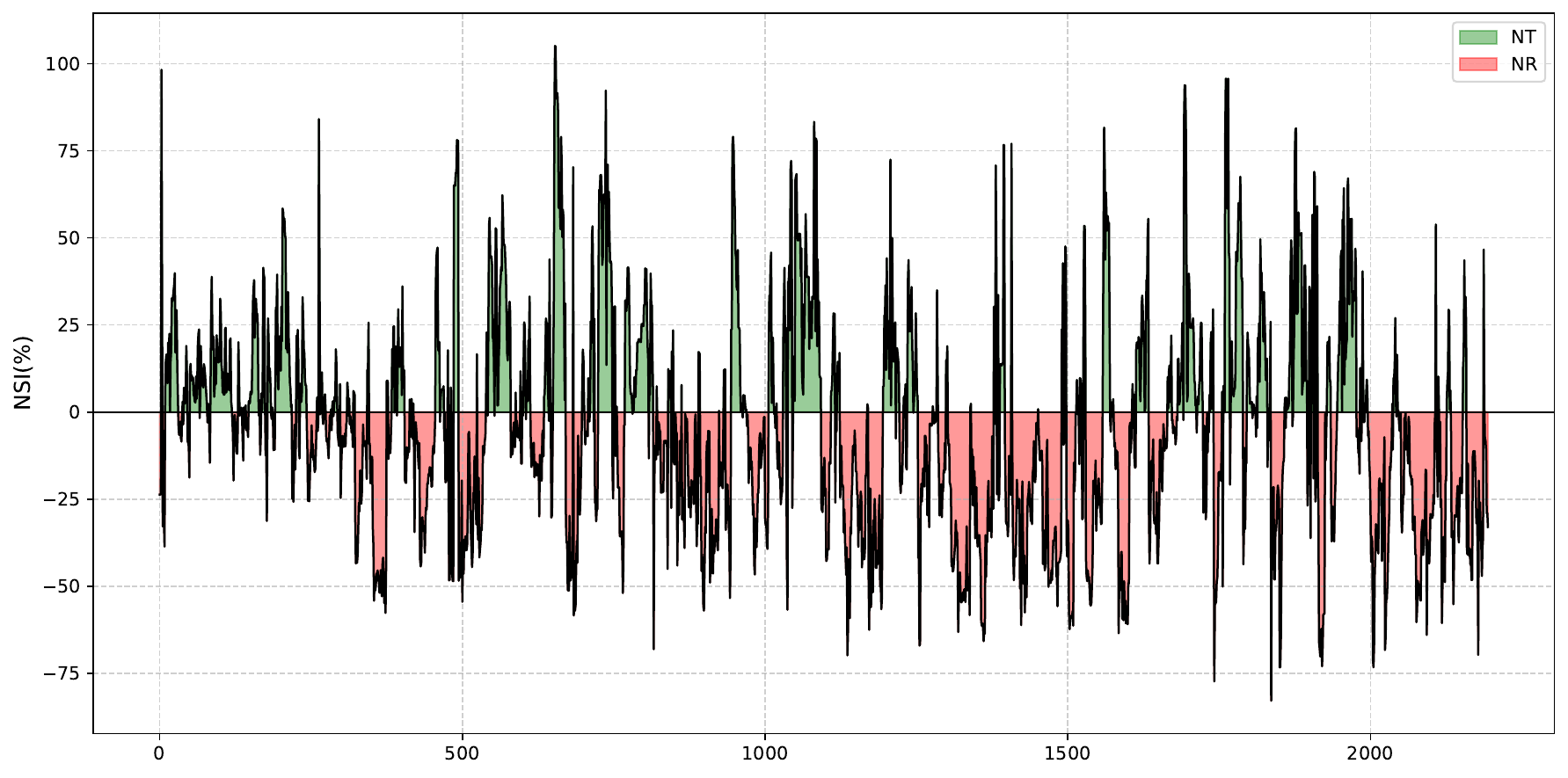}\label{fig:rsm_xlm_mid}}\hfill
    \subfigure[XLM, $\tau=0.95$]{\includegraphics[width=0.32\linewidth]{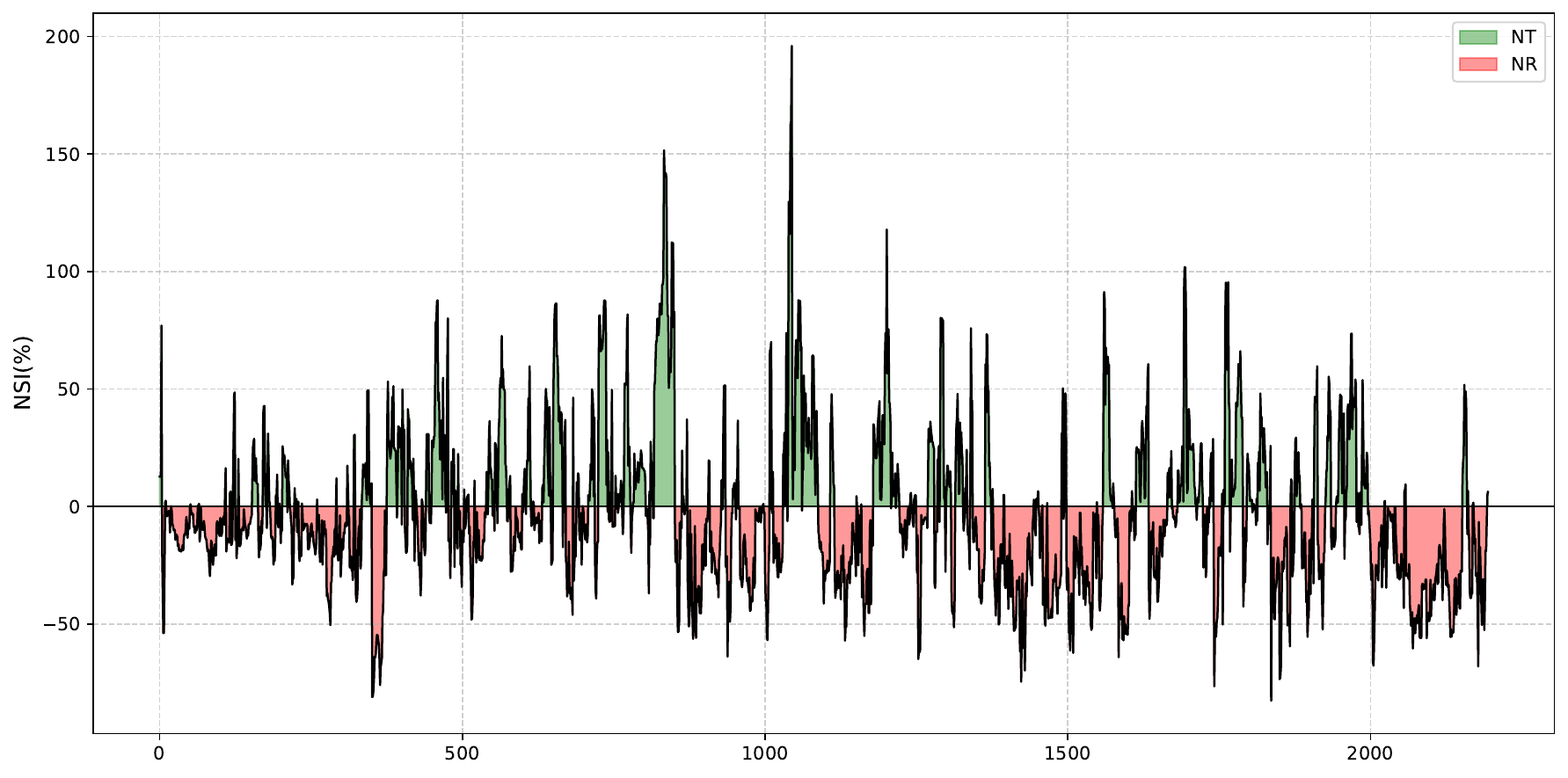}\label{fig:rsm_xlm_high}}
    \vspace{0.3cm}
    \subfigure[XRP, $\tau=0.05$]{\includegraphics[width=0.32\linewidth]{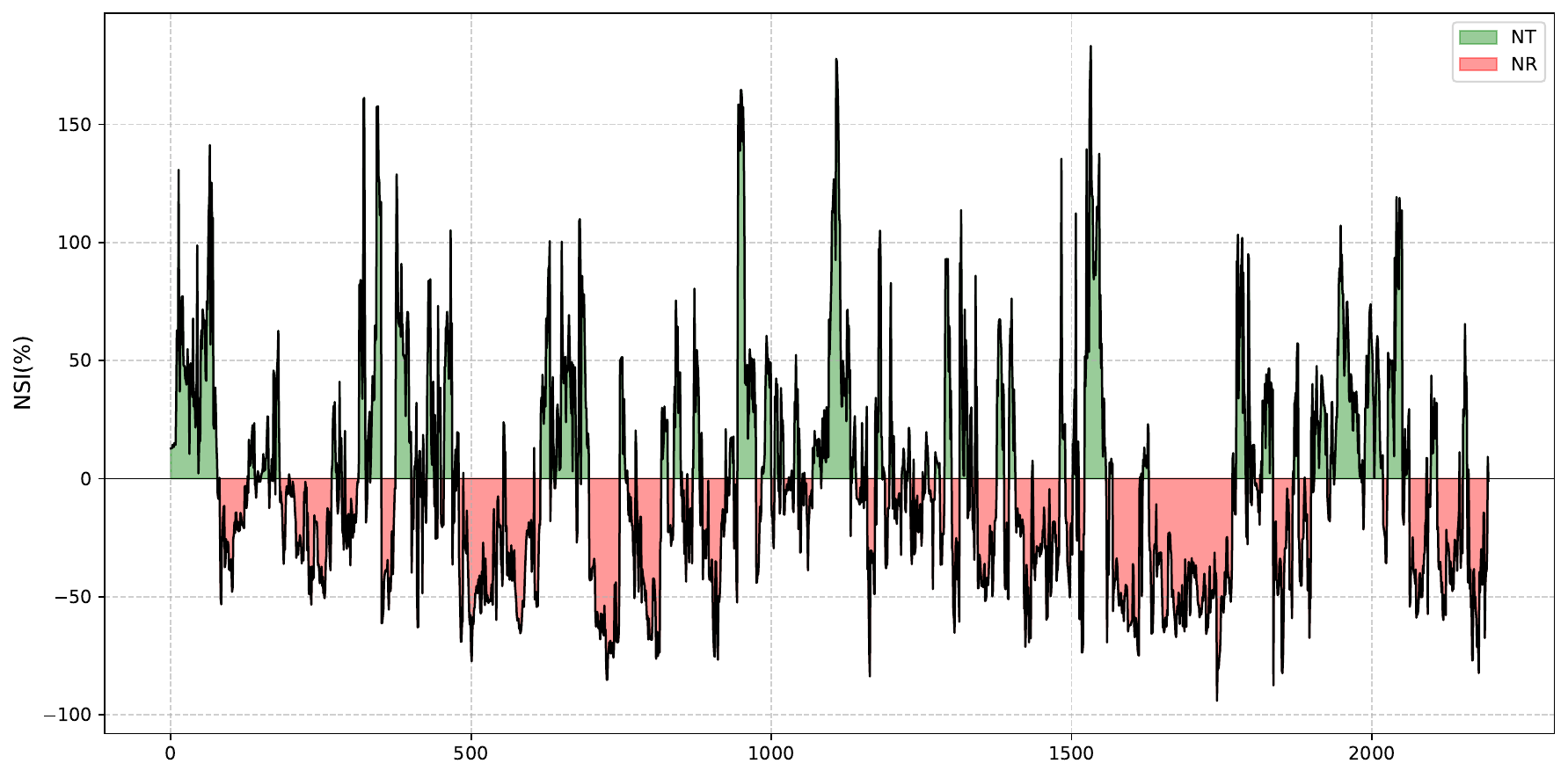}\label{fig:rsm_xrp_low}}\hfill
    \subfigure[XRP, $\tau=0.50$]{\includegraphics[width=0.32\linewidth]{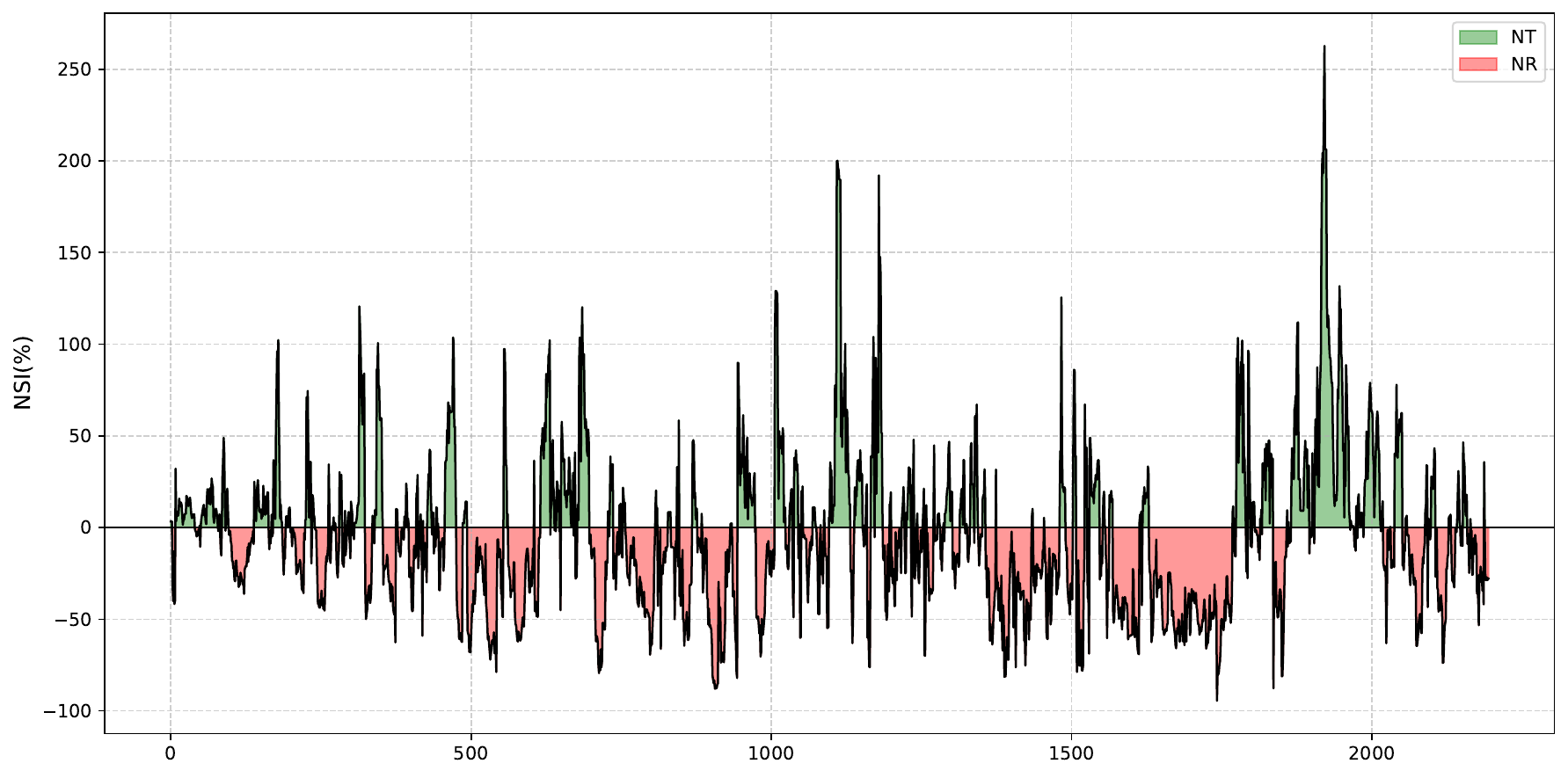}\label{fig:rsm_xrp_mid}}\hfill
    \subfigure[XRP, $\tau=0.95$]{\includegraphics[width=0.32\linewidth]{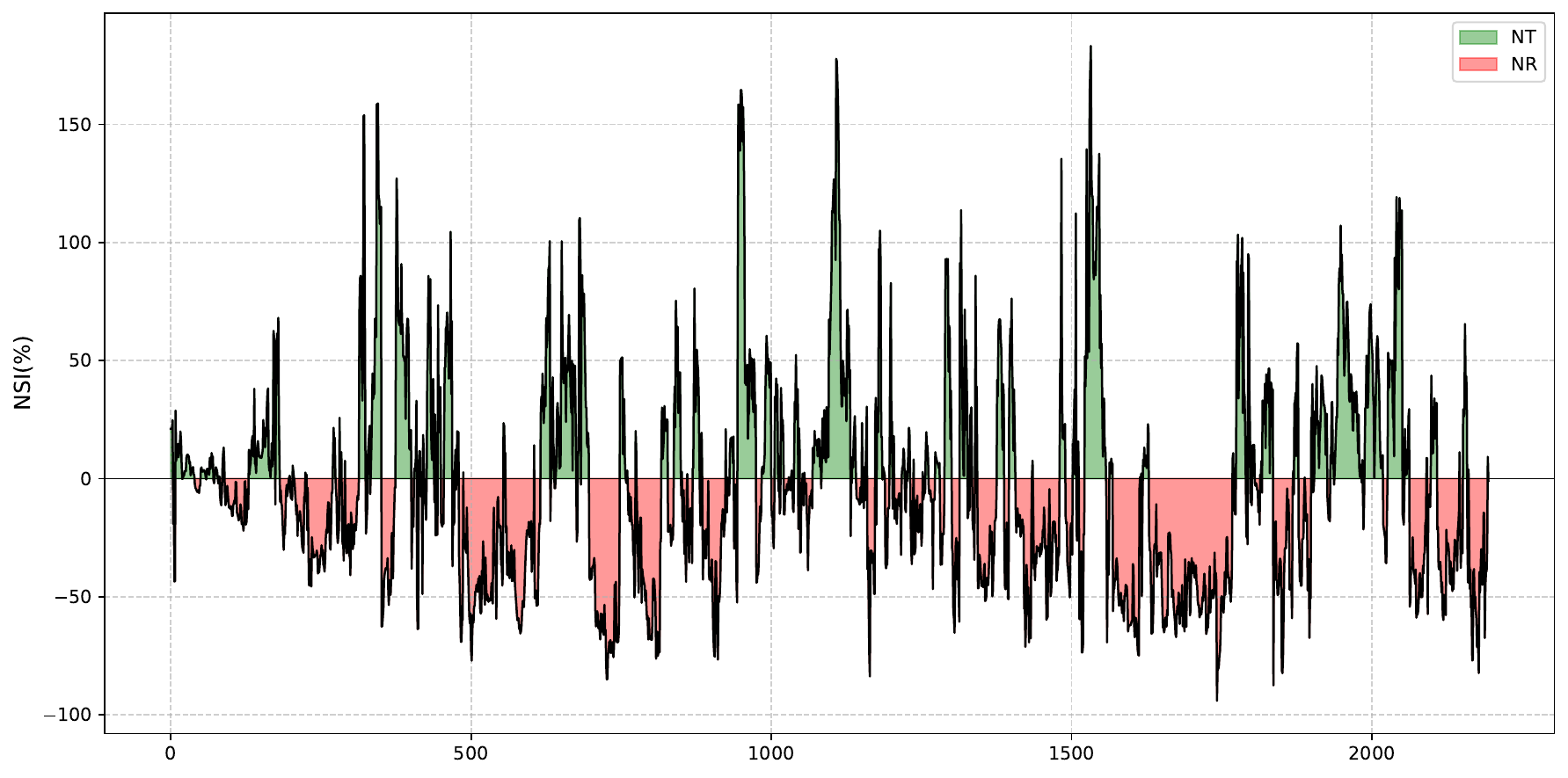}\label{fig:rsm_xrp_high}}
\end{figure}

\begin{figure}[p]
    \centering
    \caption{Quantile net spillovers for major cryptocurrencies using $REX^+$ as the feature variable.}
    \label{fig:rexp_net_spillover_by_coin}

    \subfigure[BTC, $\tau=0.05$]{\includegraphics[width=0.32\linewidth]{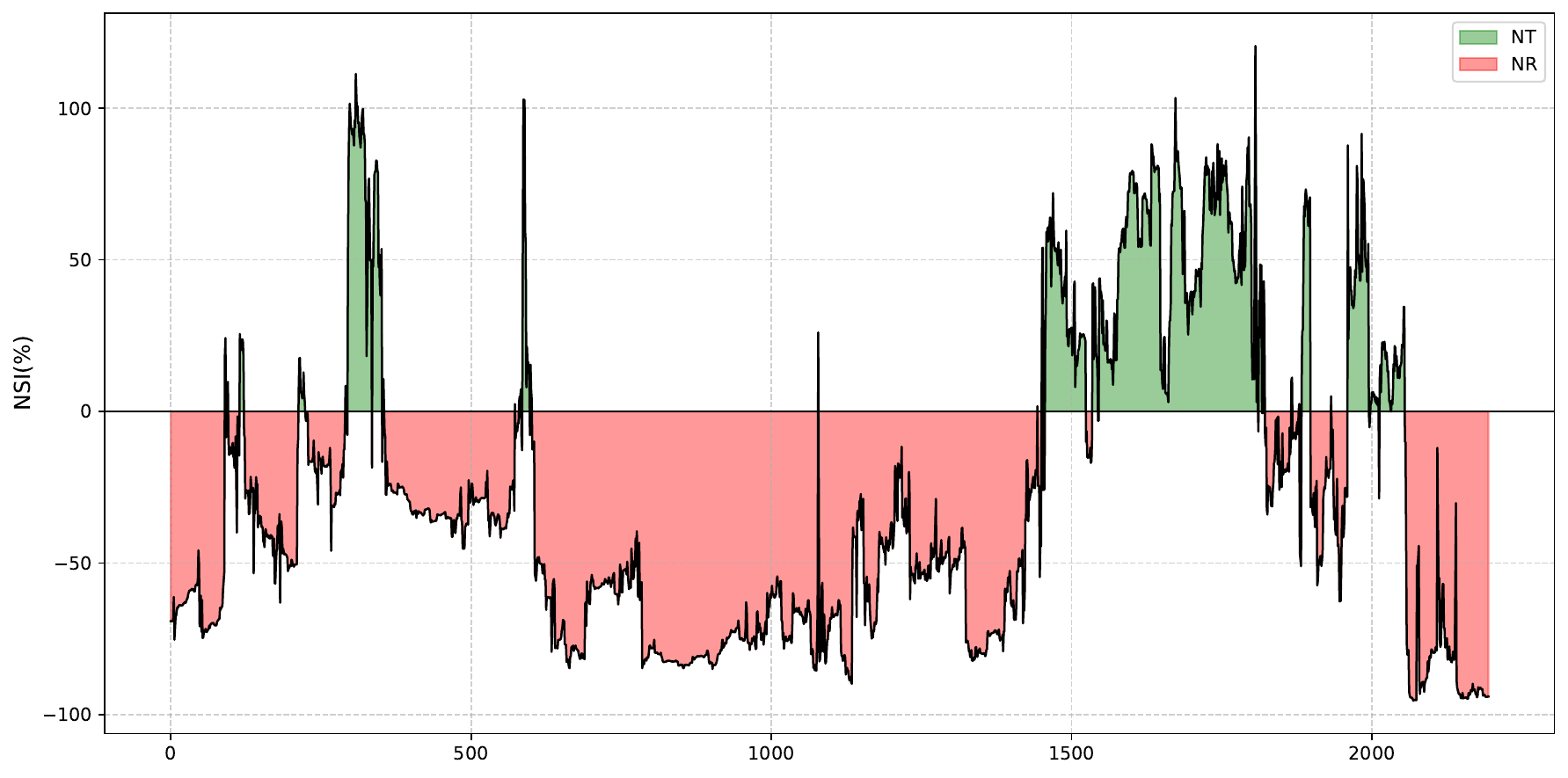}\label{fig:rexp_btc_low}}\hfill
    \subfigure[BTC, $\tau=0.50$]{\includegraphics[width=0.32\linewidth]{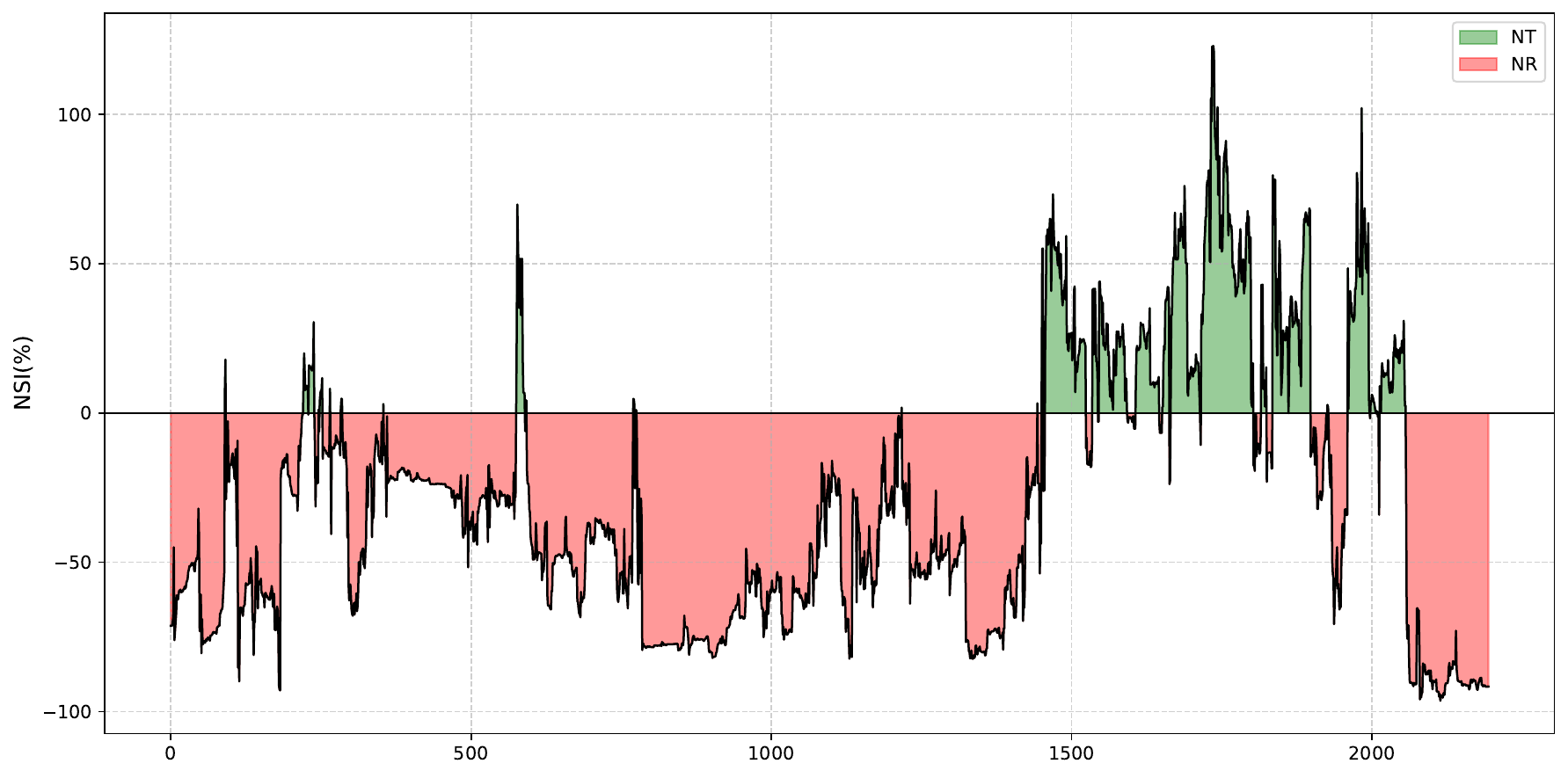}\label{fig:rexp_btc_mid}}\hfill
    \subfigure[BTC, $\tau=0.95$]{\includegraphics[width=0.32\linewidth]{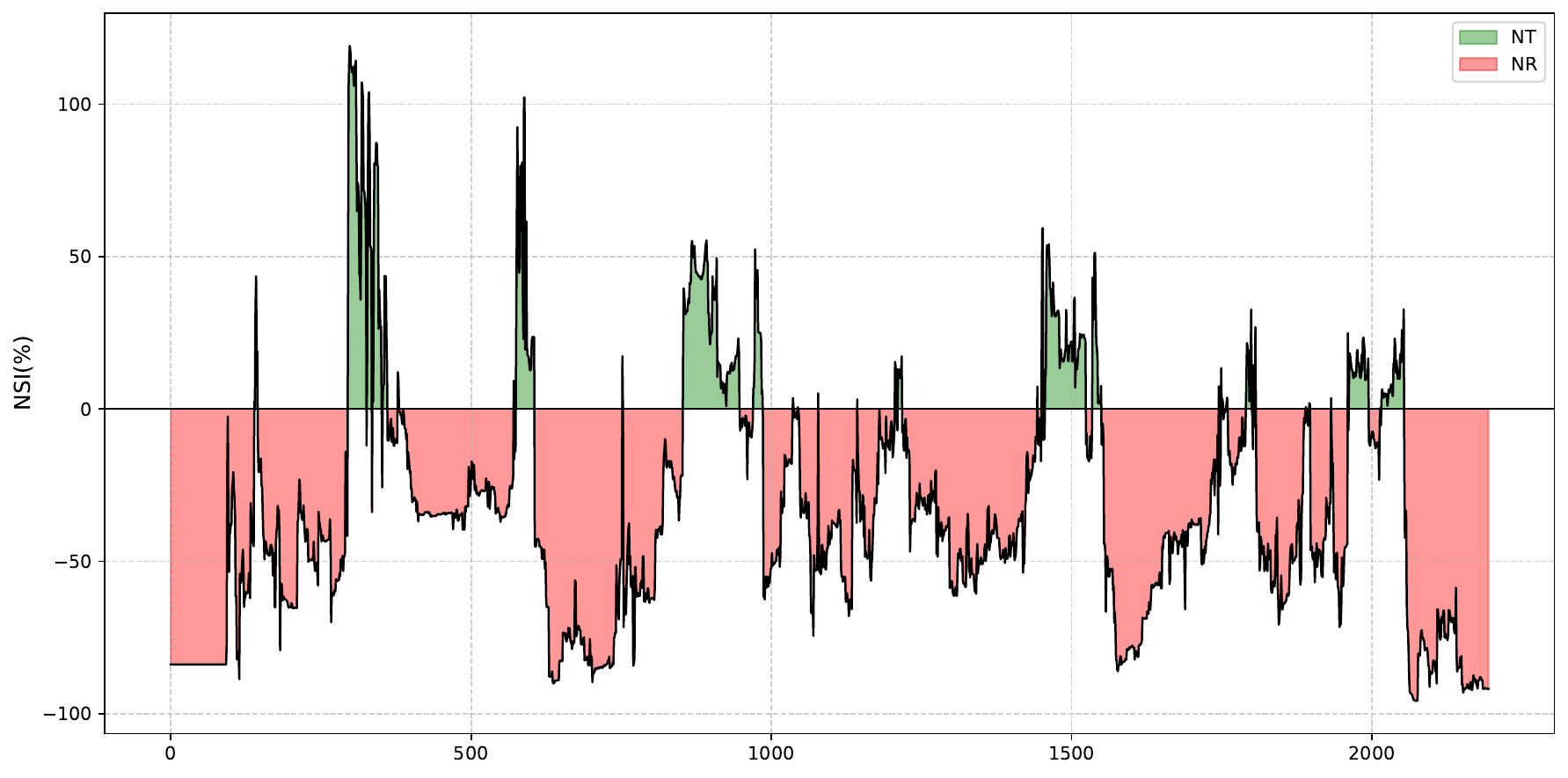}\label{fig:rexp_btc_high}}
    \vspace{0.3cm}
    \subfigure[DASH, $\tau=0.05$]{\includegraphics[width=0.32\linewidth]{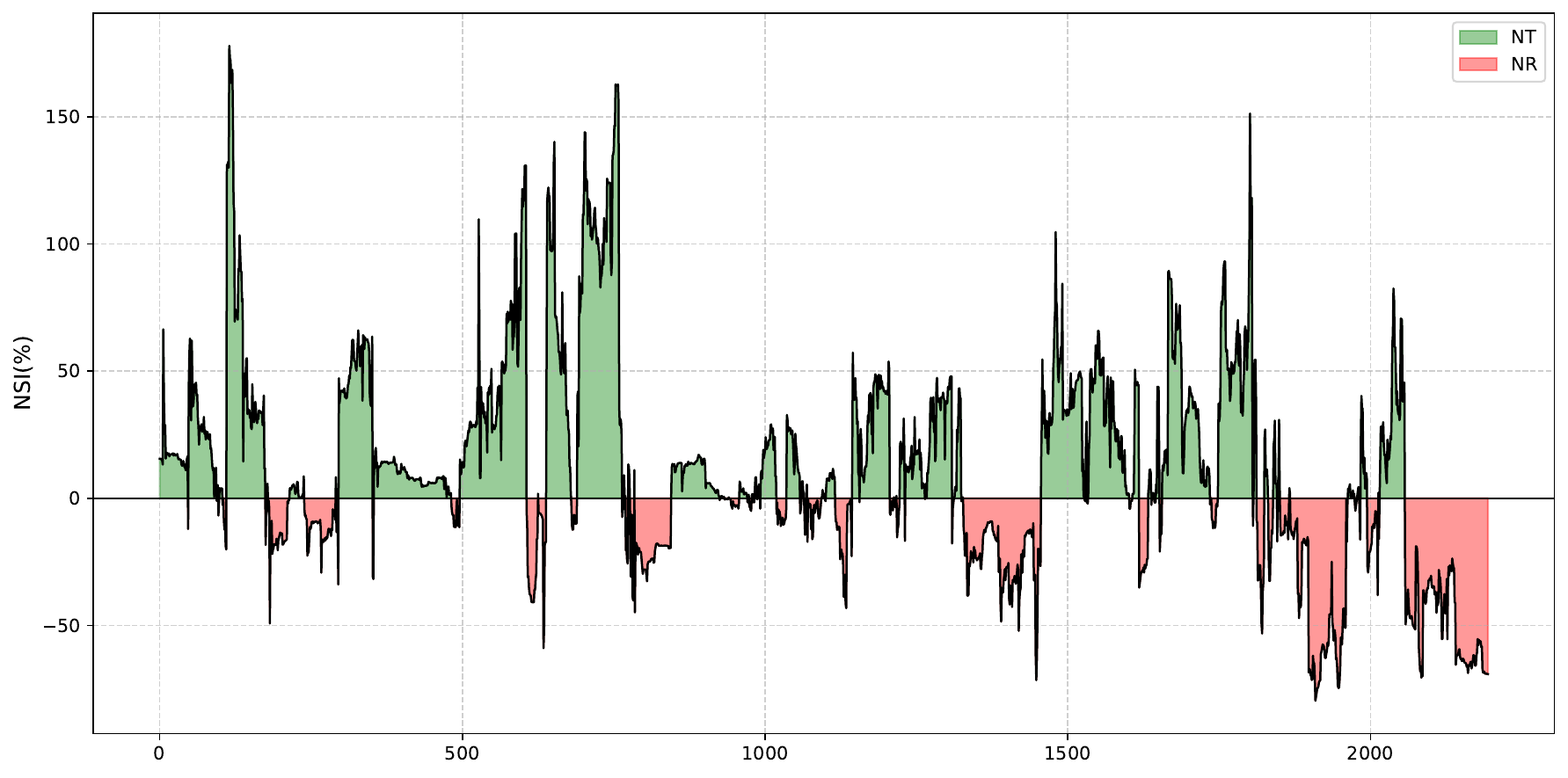}\label{fig:rexp_dash_low}}\hfill
    \subfigure[DASH, $\tau=0.50$]{\includegraphics[width=0.32\linewidth]{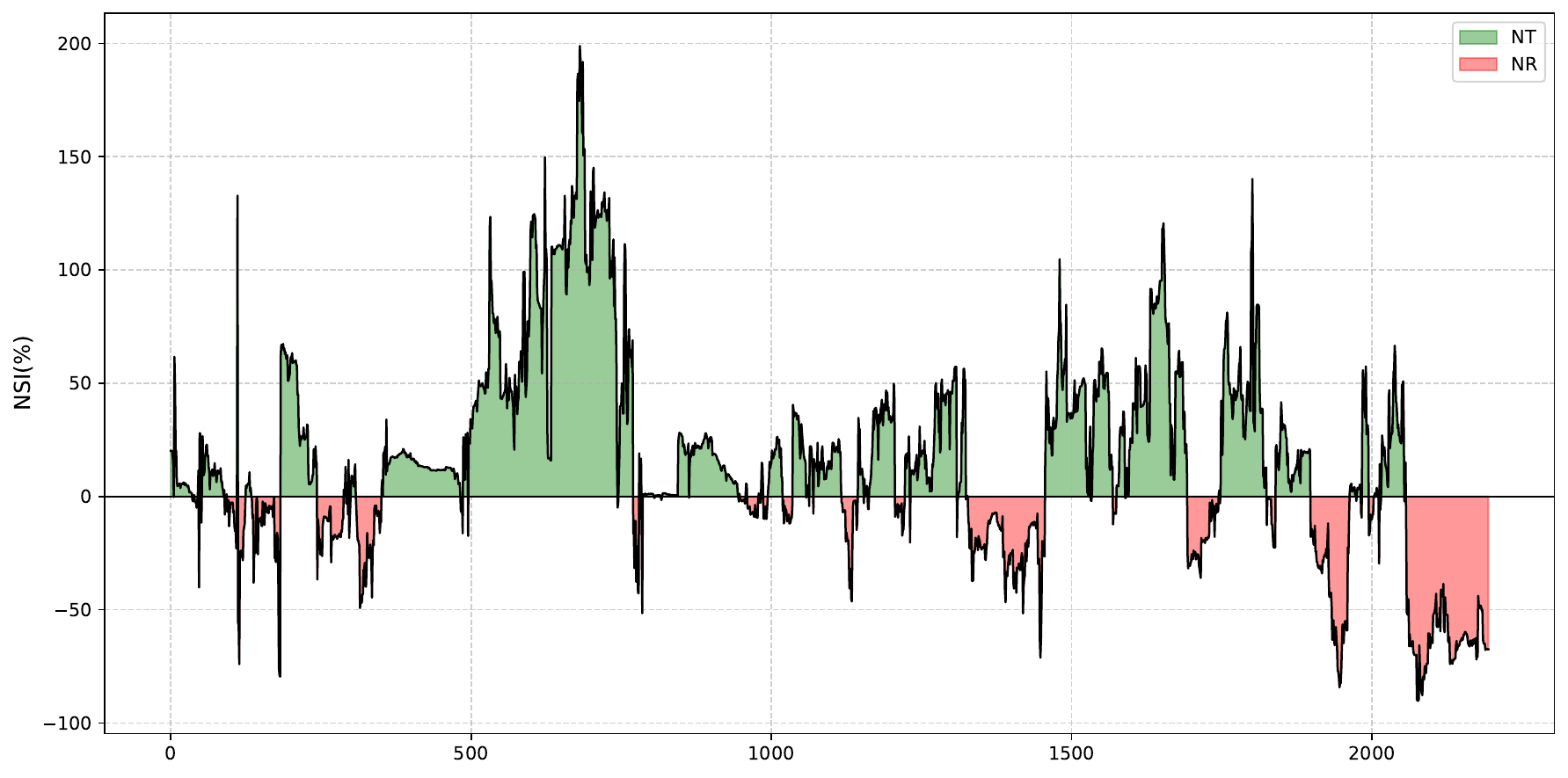}\label{fig:rexp_dash_mid}}\hfill
    \subfigure[DASH, $\tau=0.95$]{\includegraphics[width=0.32\linewidth]{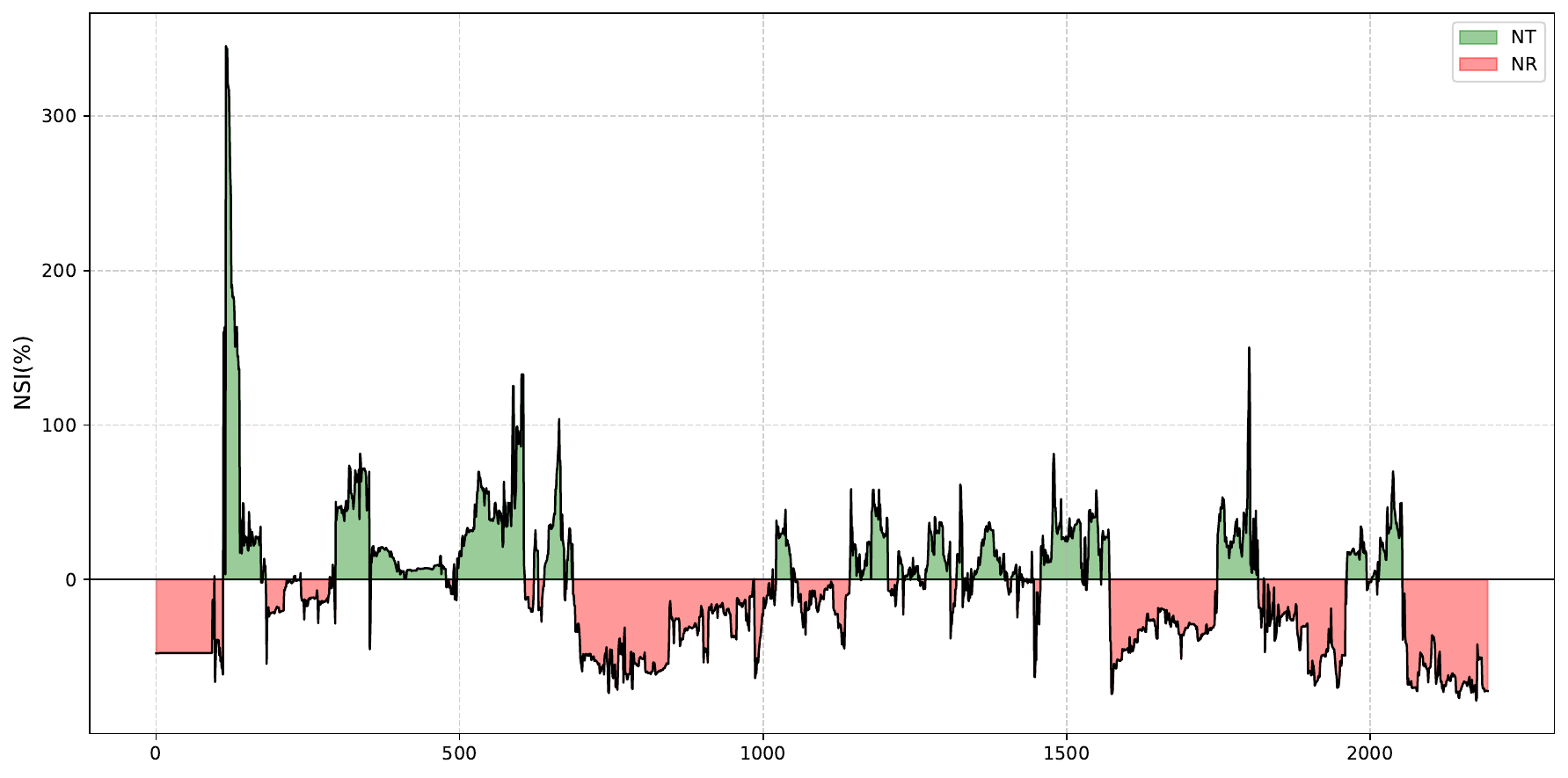}\label{fig:rexp_dash_high}}
    \vspace{0.3cm}
    \subfigure[ETH, $\tau=0.05$]{\includegraphics[width=0.32\linewidth]{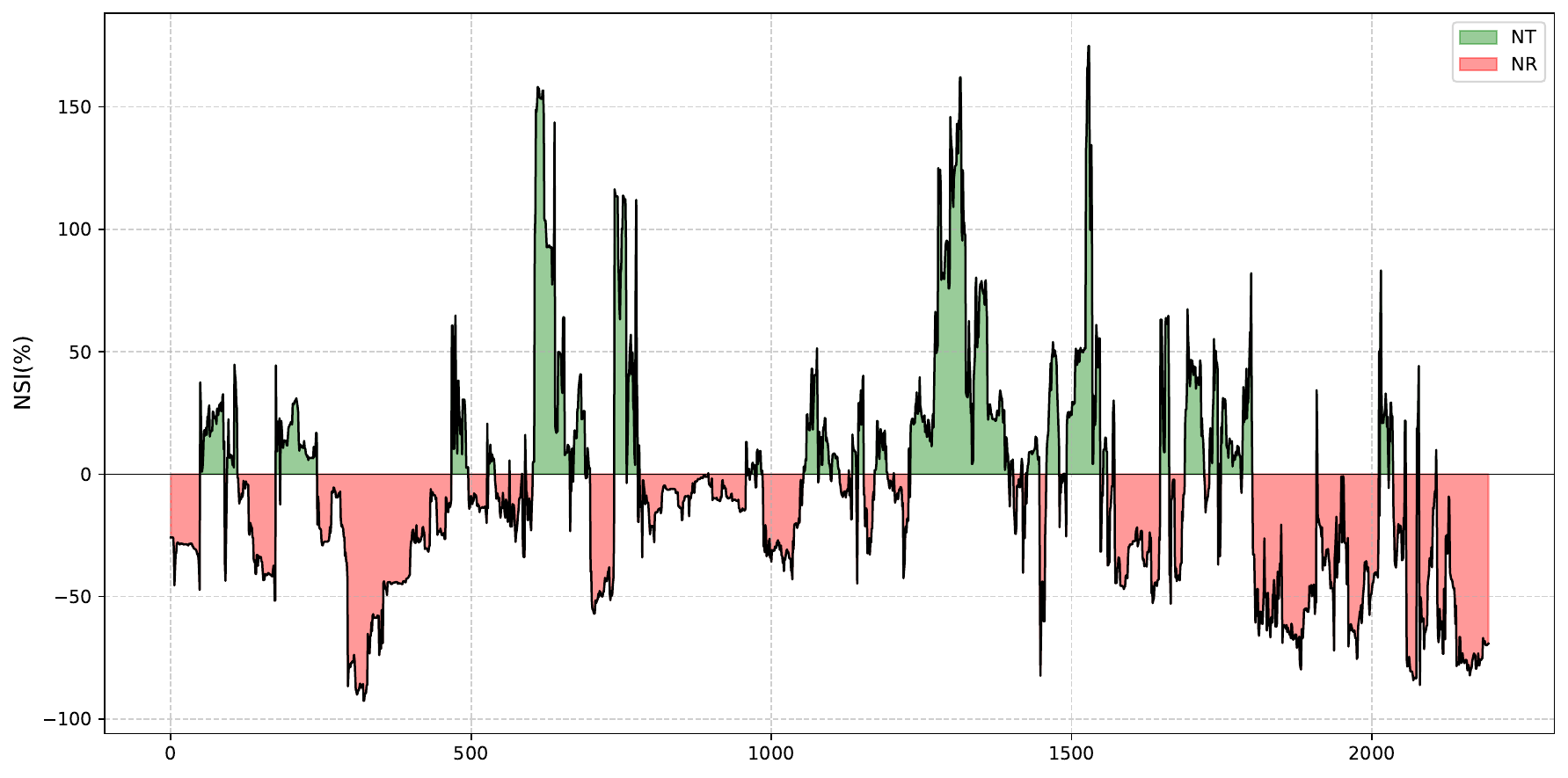}\label{fig:rexp_eth_low}}\hfill
    \subfigure[ETH, $\tau=0.50$]{\includegraphics[width=0.32\linewidth]{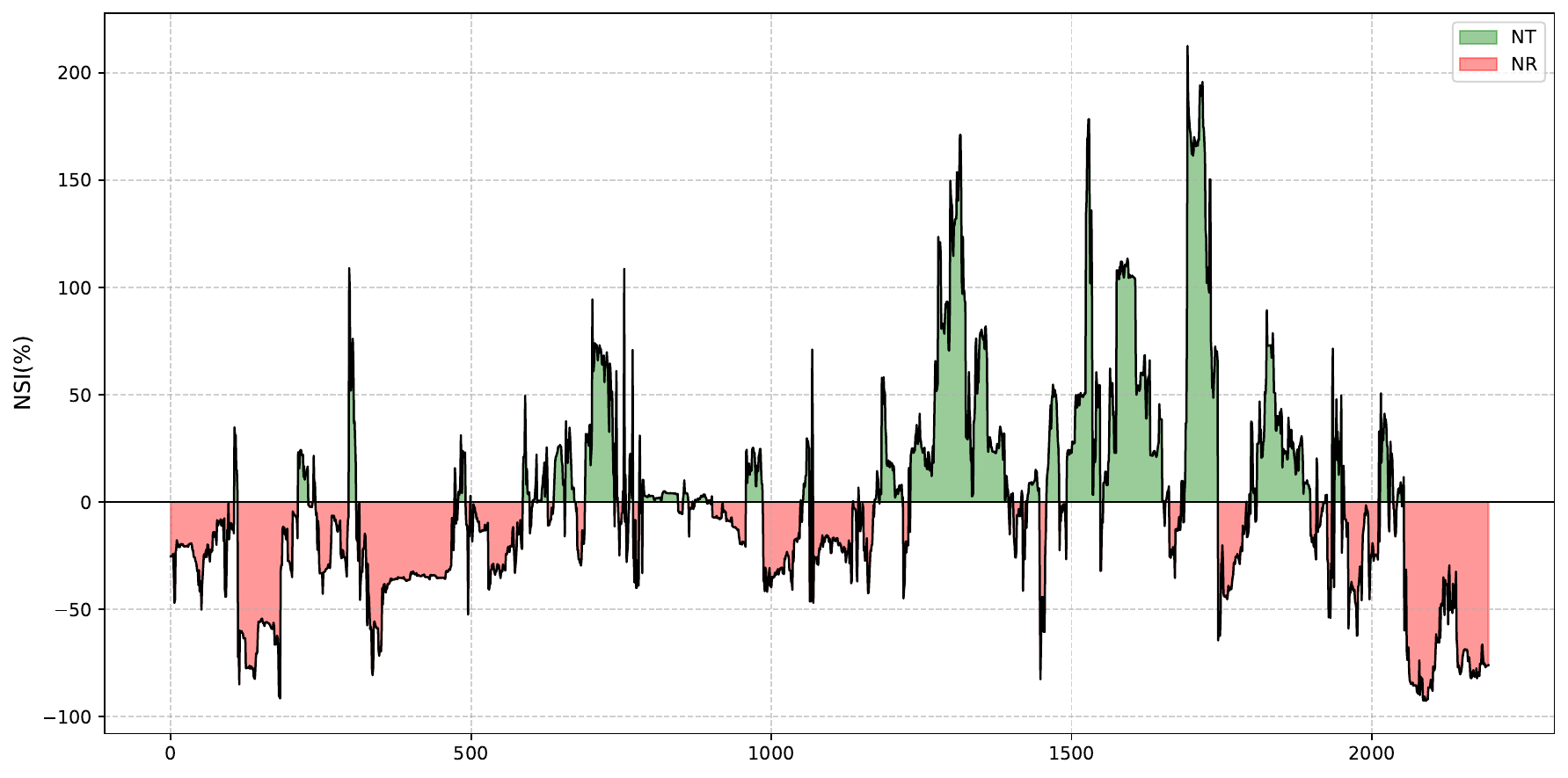}\label{fig:rexp_eth_mid}}\hfill
    \subfigure[ETH, $\tau=0.95$]{\includegraphics[width=0.32\linewidth]{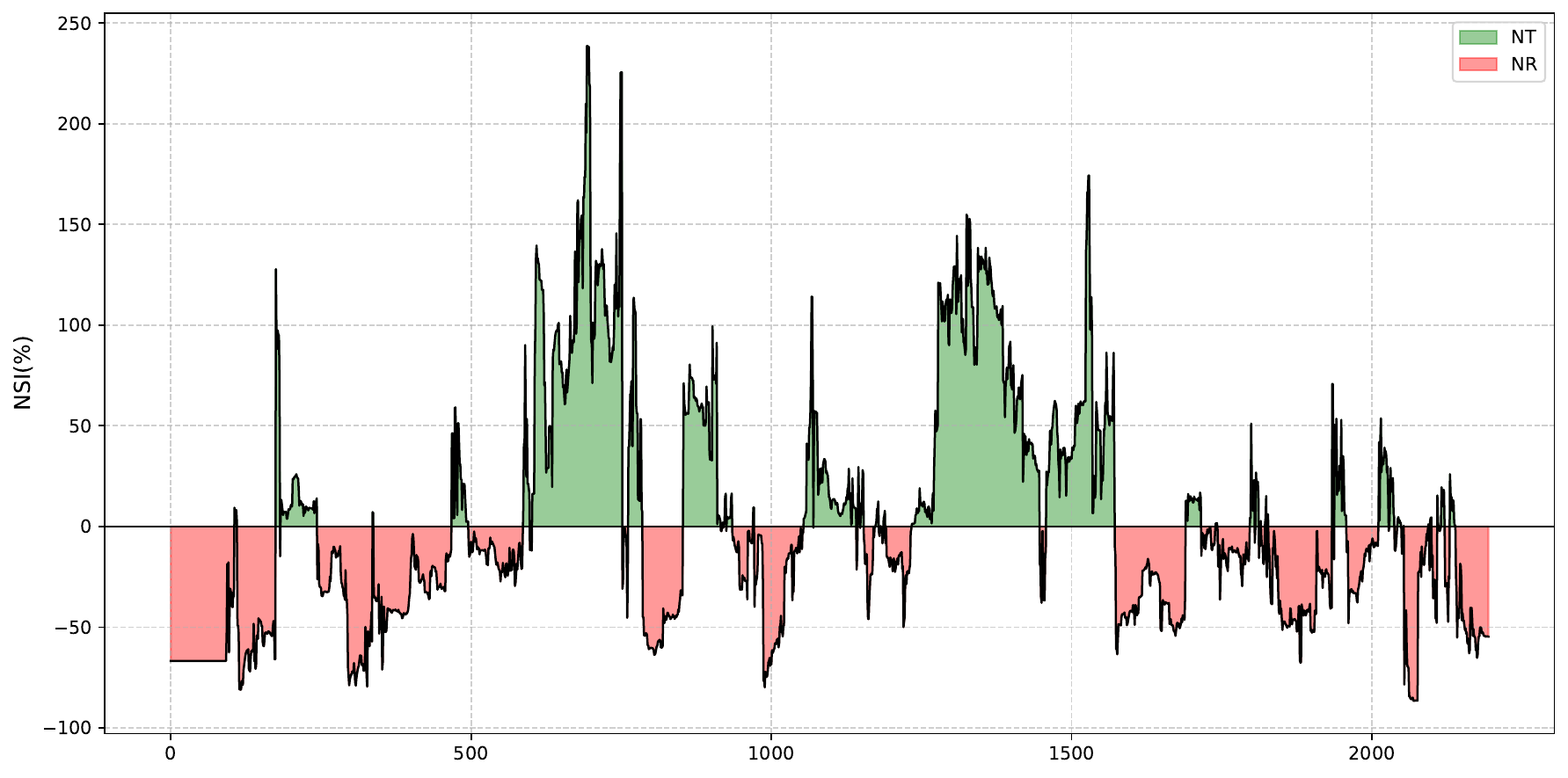}\label{fig:rexp_eth_high}}
    \vspace{0.3cm}
    \subfigure[LTC, $\tau=0.05$]{\includegraphics[width=0.32\linewidth]{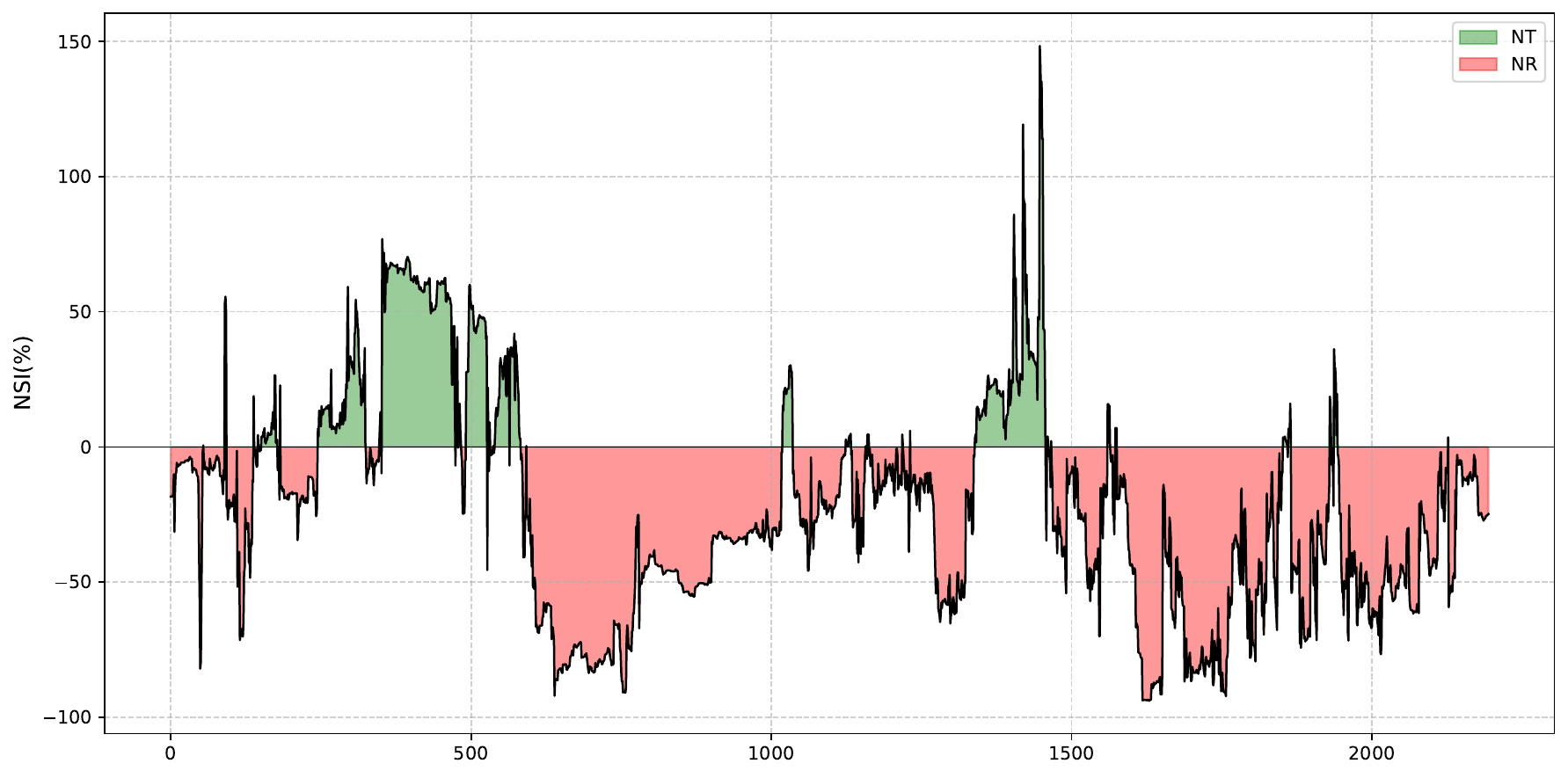}\label{fig:rexp_ltc_low}}\hfill
    \subfigure[LTC, $\tau=0.50$]{\includegraphics[width=0.32\linewidth]{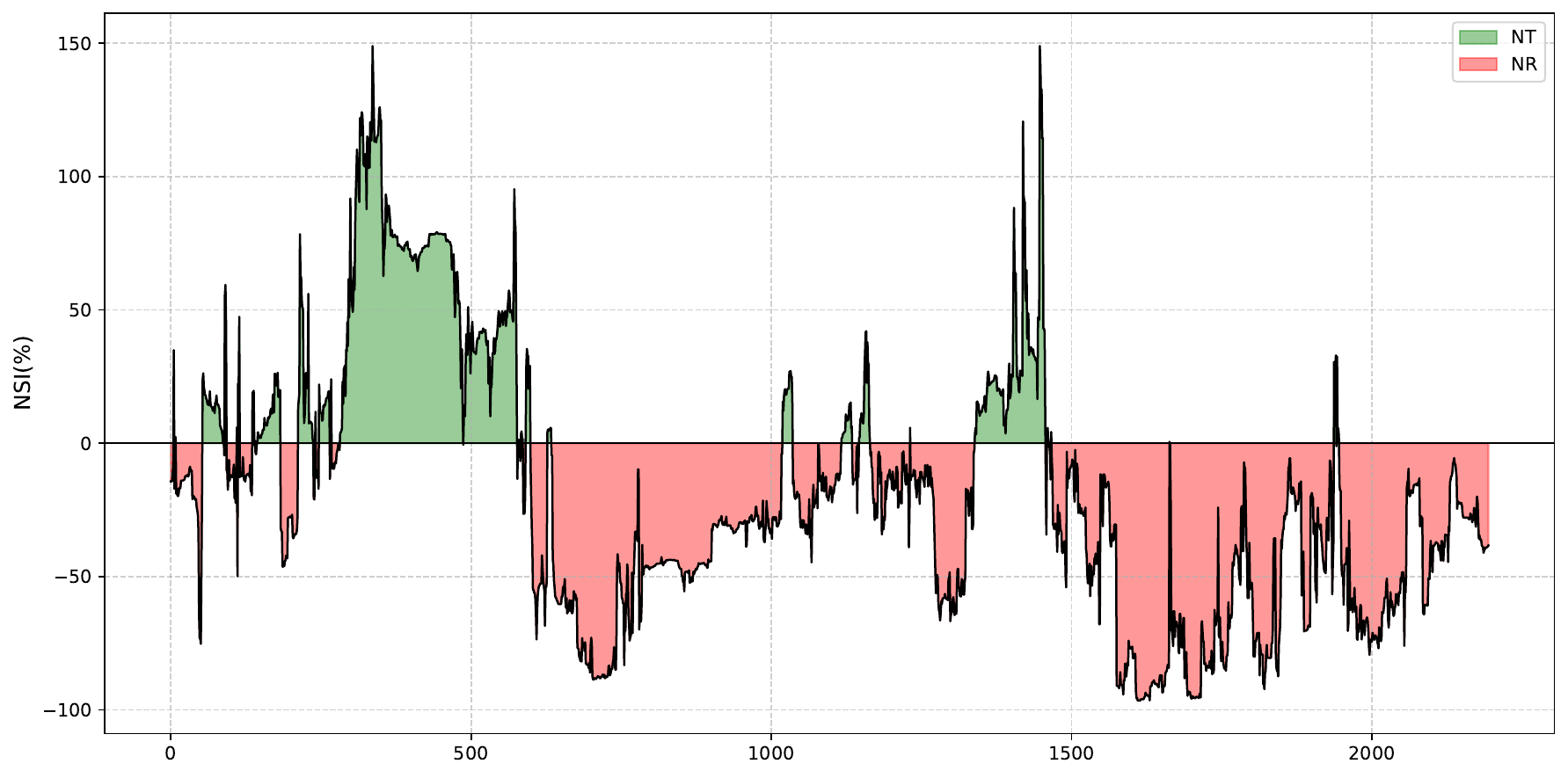}\label{fig:rexp_ltc_mid}}\hfill
    \subfigure[LTC, $\tau=0.95$]{\includegraphics[width=0.32\linewidth]{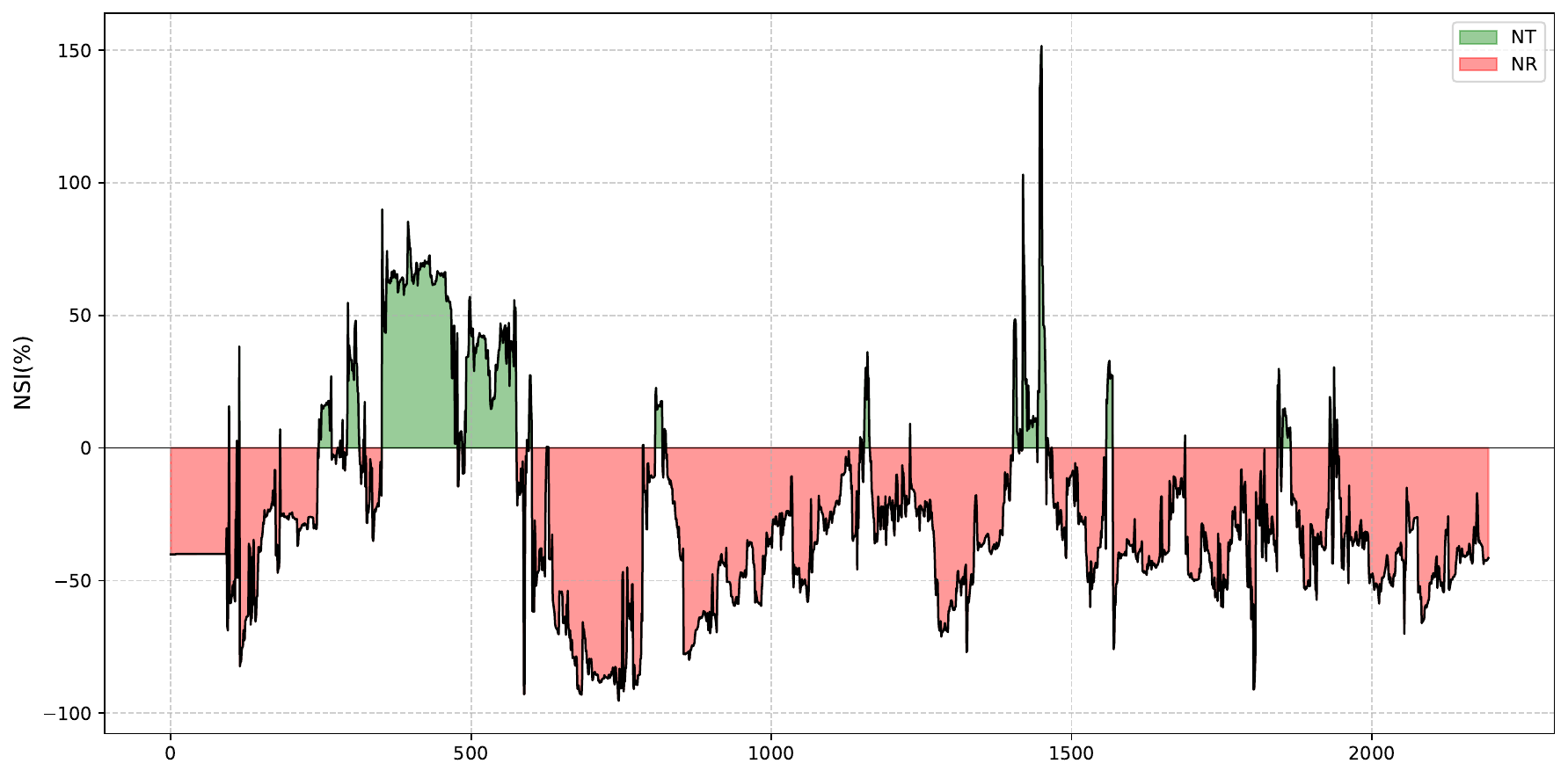}\label{fig:rexp_ltc_high}}
    \vspace{0.3cm}
    \subfigure[XLM, $\tau=0.05$]{\includegraphics[width=0.32\linewidth]{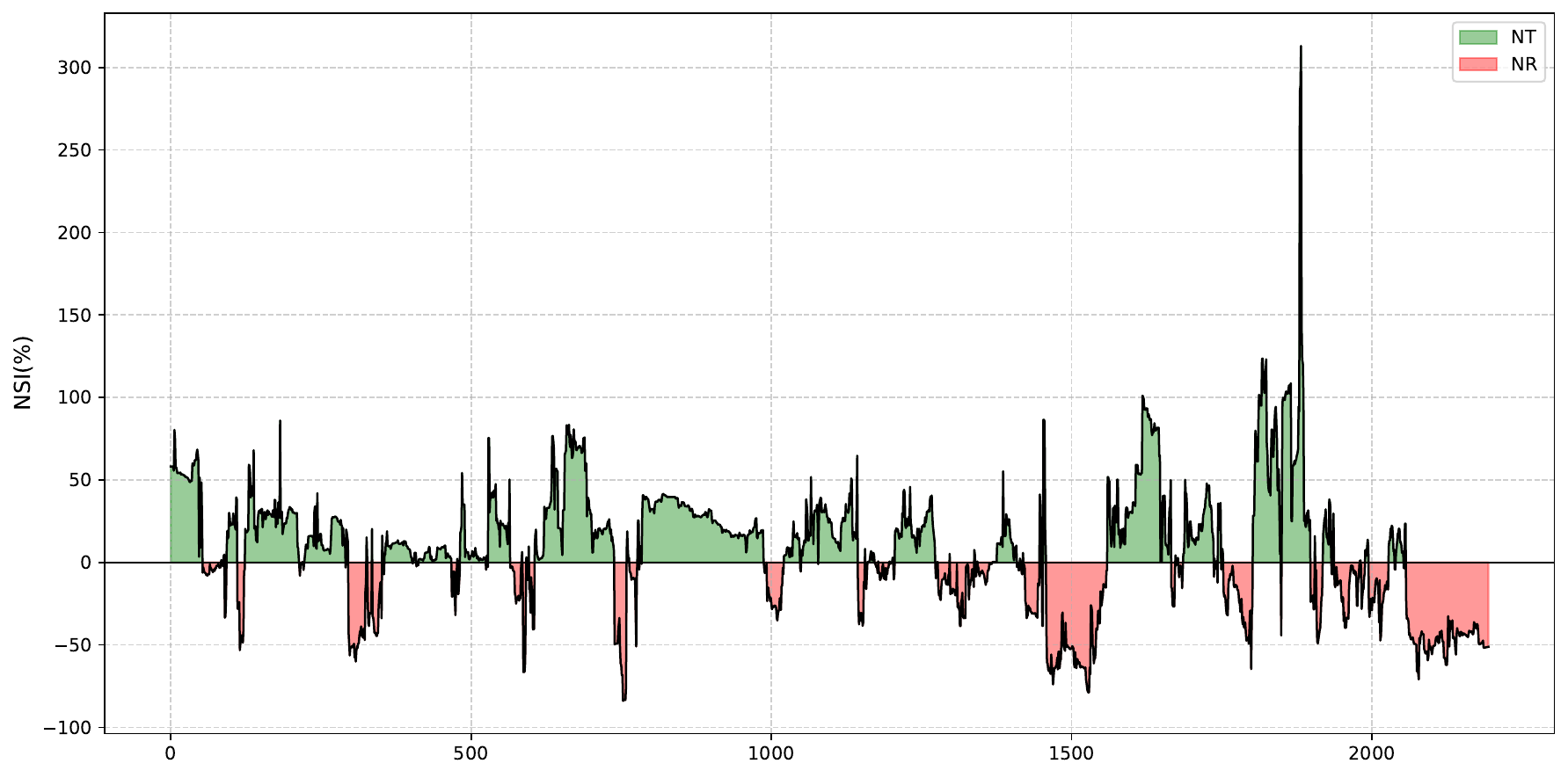}\label{fig:rexp_xlm_low}}\hfill
    \subfigure[XLM, $\tau=0.50$]{\includegraphics[width=0.32\linewidth]{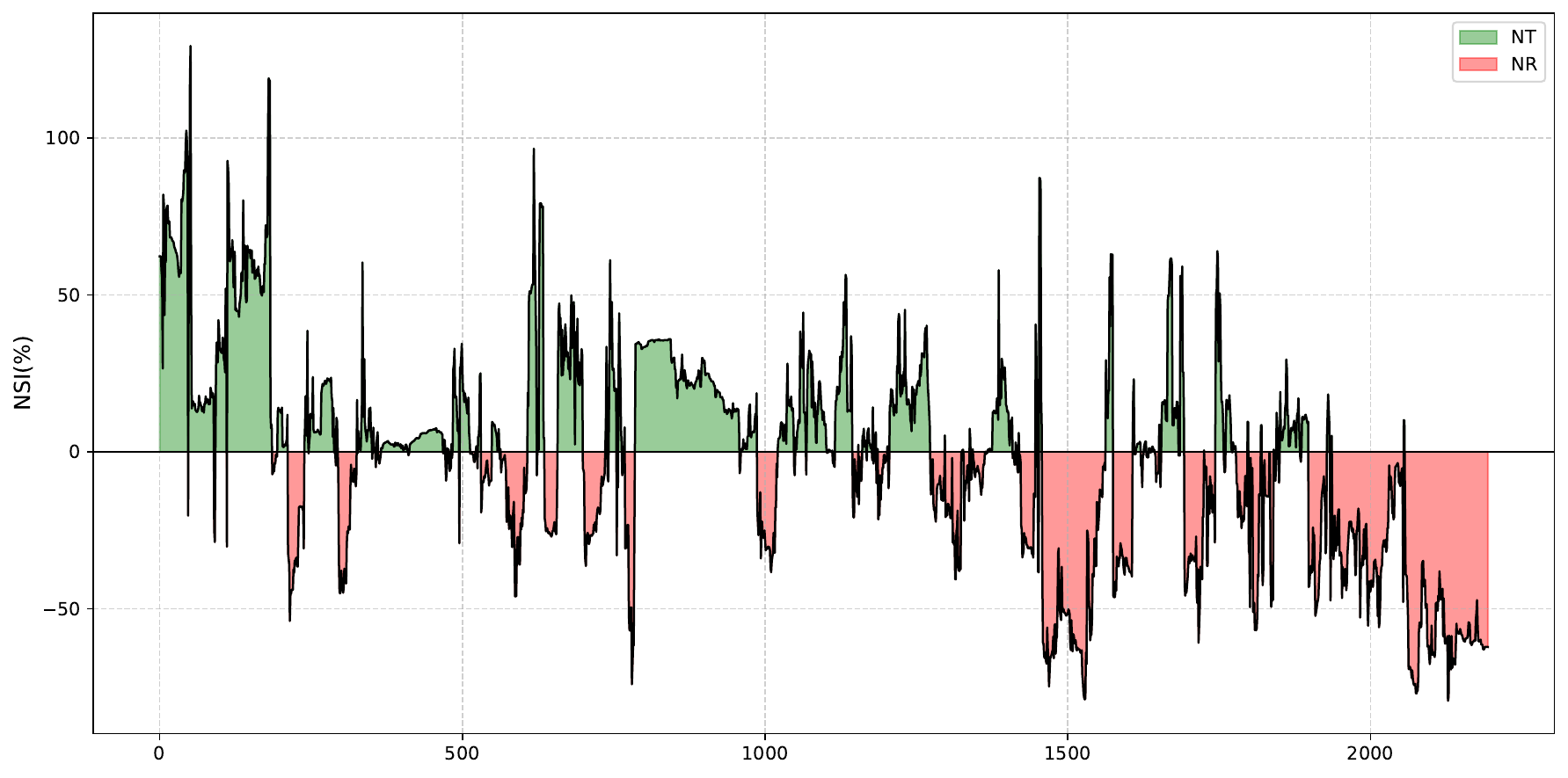}\label{fig:rexp_xlm_mid}}\hfill
    \subfigure[XLM, $\tau=0.95$]{\includegraphics[width=0.32\linewidth]{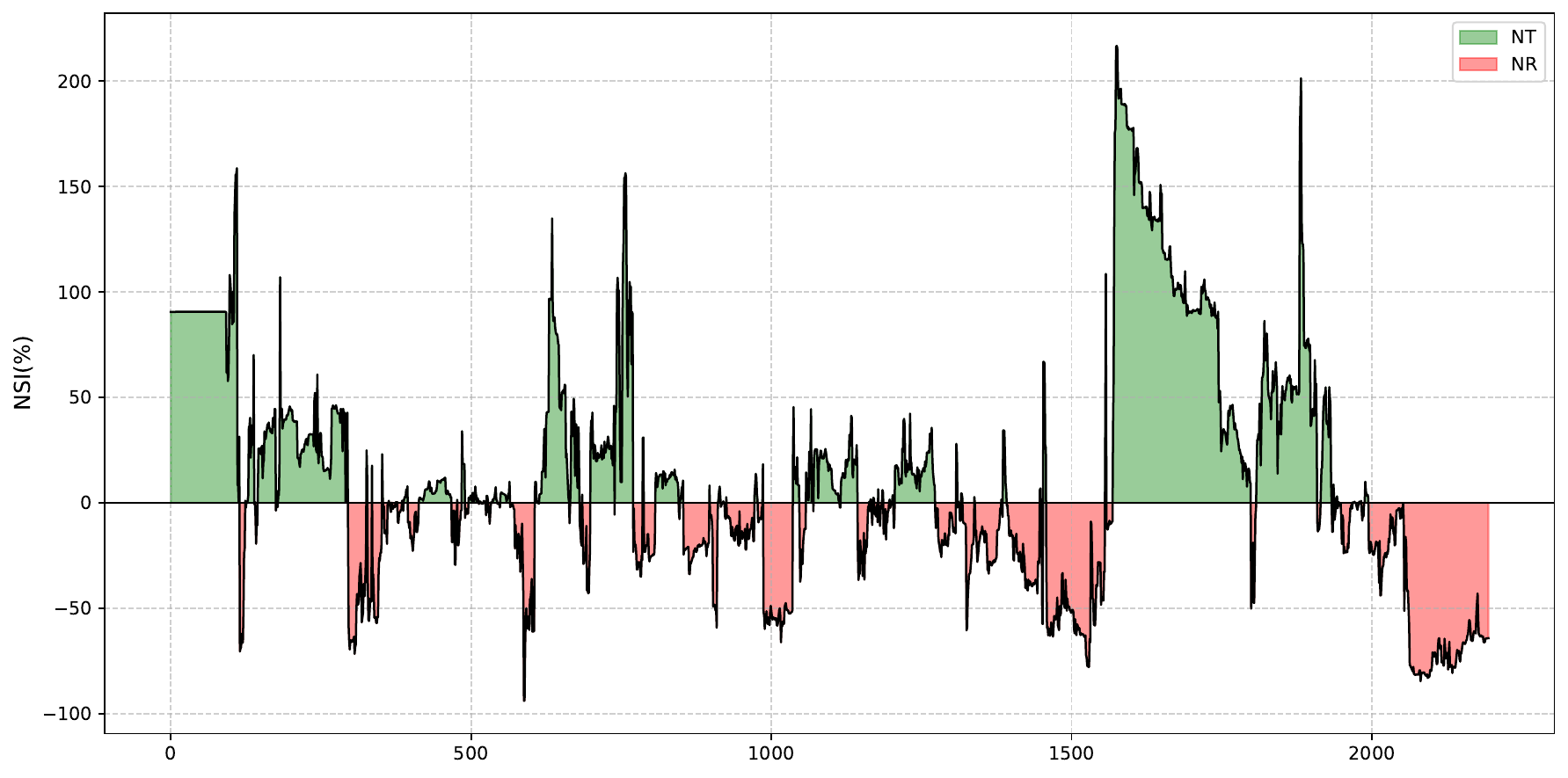}\label{fig:rexp_xlm_high}}
    \vspace{0.3cm}
    \subfigure[XRP, $\tau=0.05$]{\includegraphics[width=0.32\linewidth]{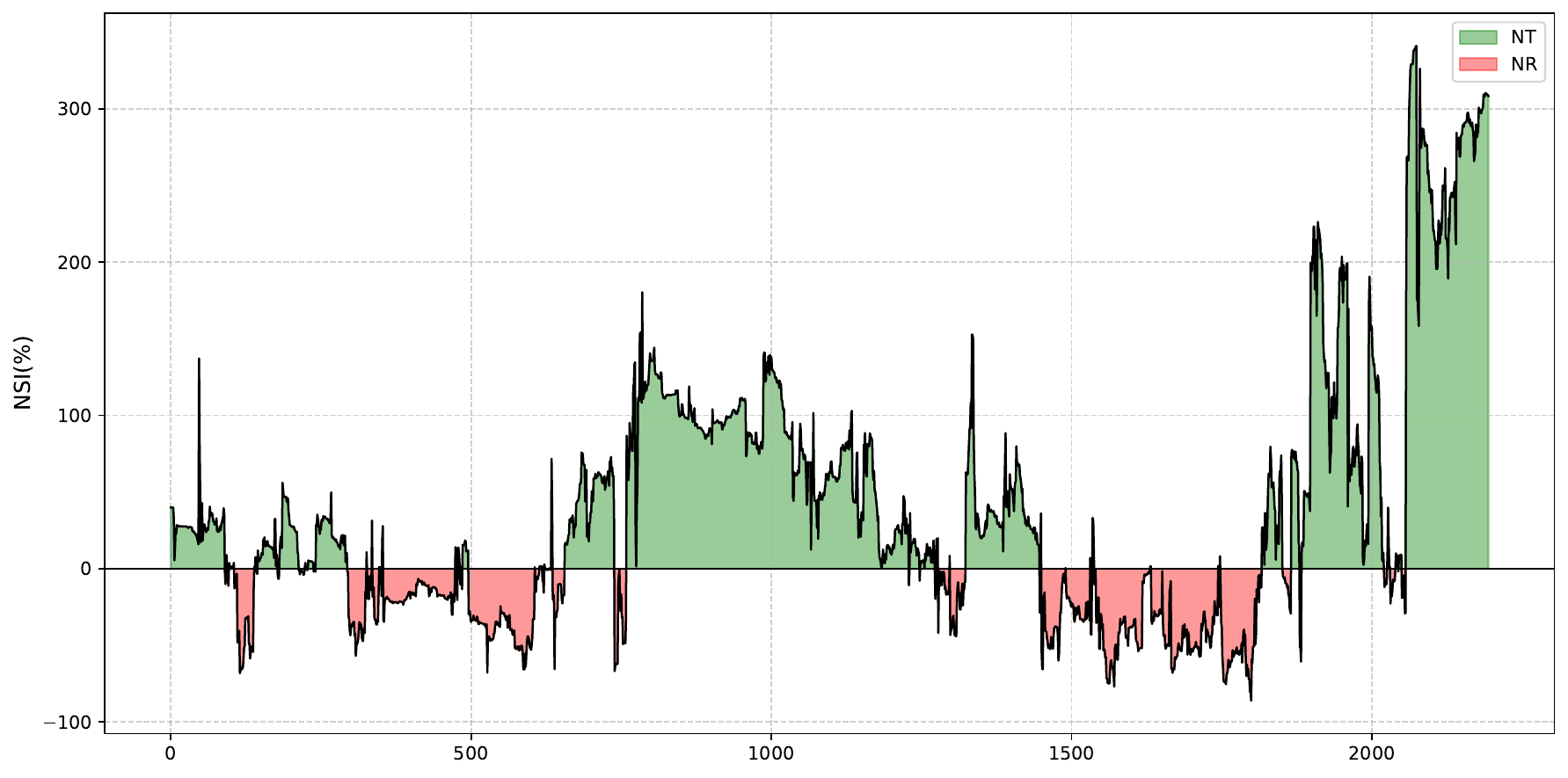}\label{fig:rexp_xrp_low}}\hfill
    \subfigure[XRP, $\tau=0.50$]{\includegraphics[width=0.32\linewidth]{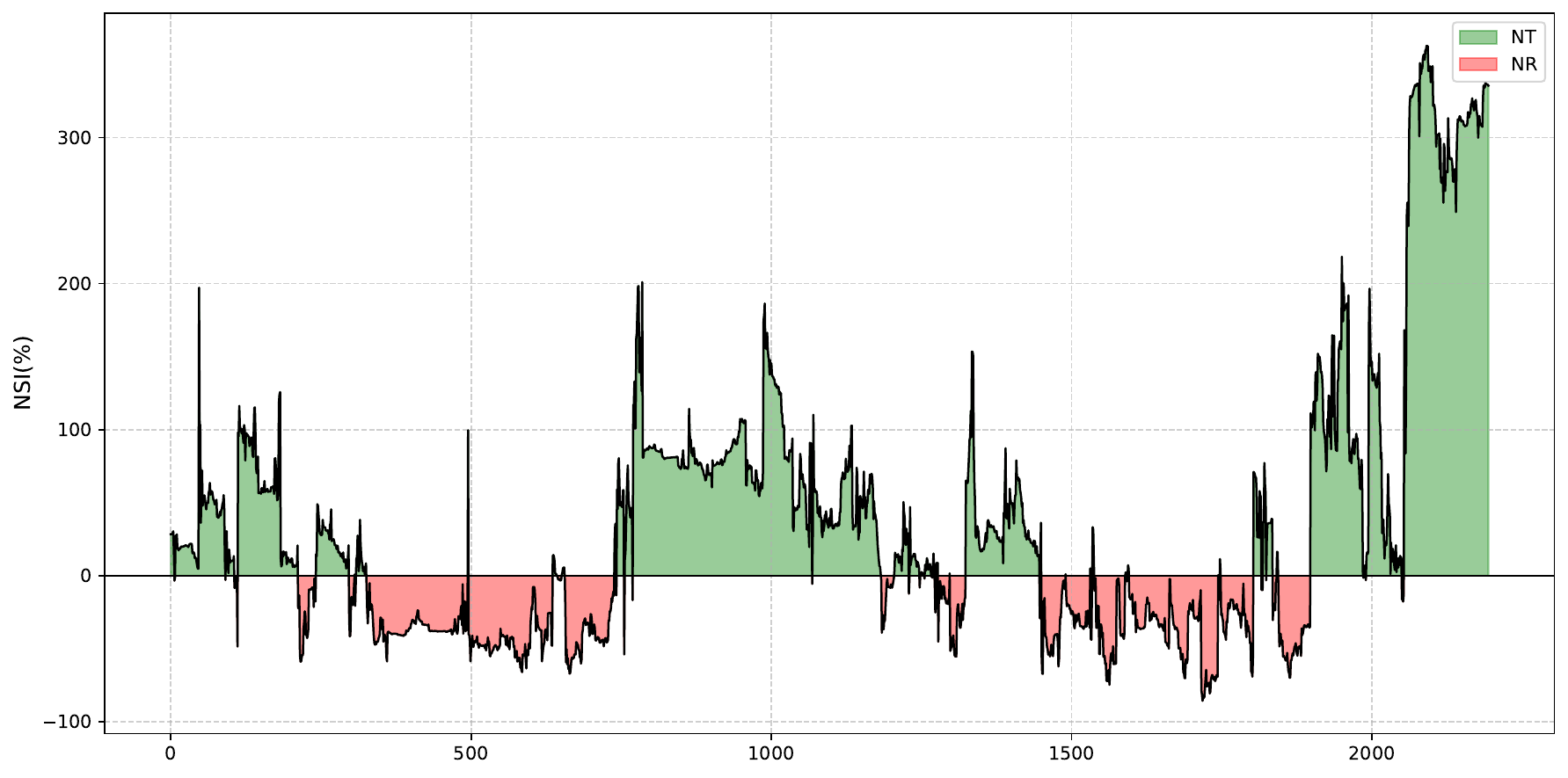}\label{fig:rexp_xrp_mid}}\hfill
    \subfigure[XRP, $\tau=0.95$]{\includegraphics[width=0.32\linewidth]{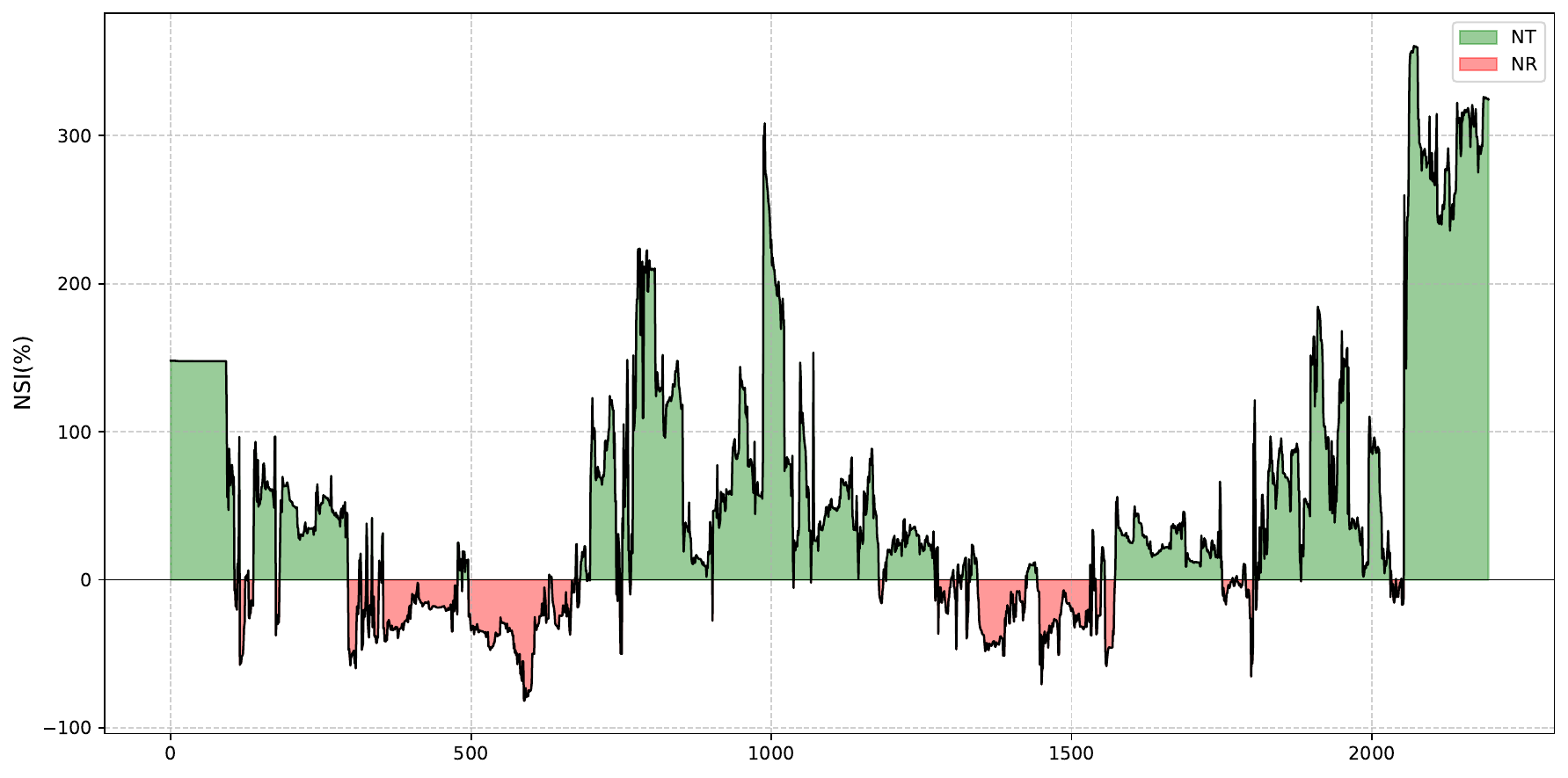}\label{fig:rexp_xrp_high}}
\end{figure}

\begin{figure}[p]
    \centering
    \caption{Quantile net spillovers for major cryptocurrencies using $REX^-$ as the feature variable.}
    \label{fig:rexm_neg_net_spillover_by_coin} 

    \subfigure[BTC, $\tau=0.05$]{\includegraphics[width=0.32\linewidth]{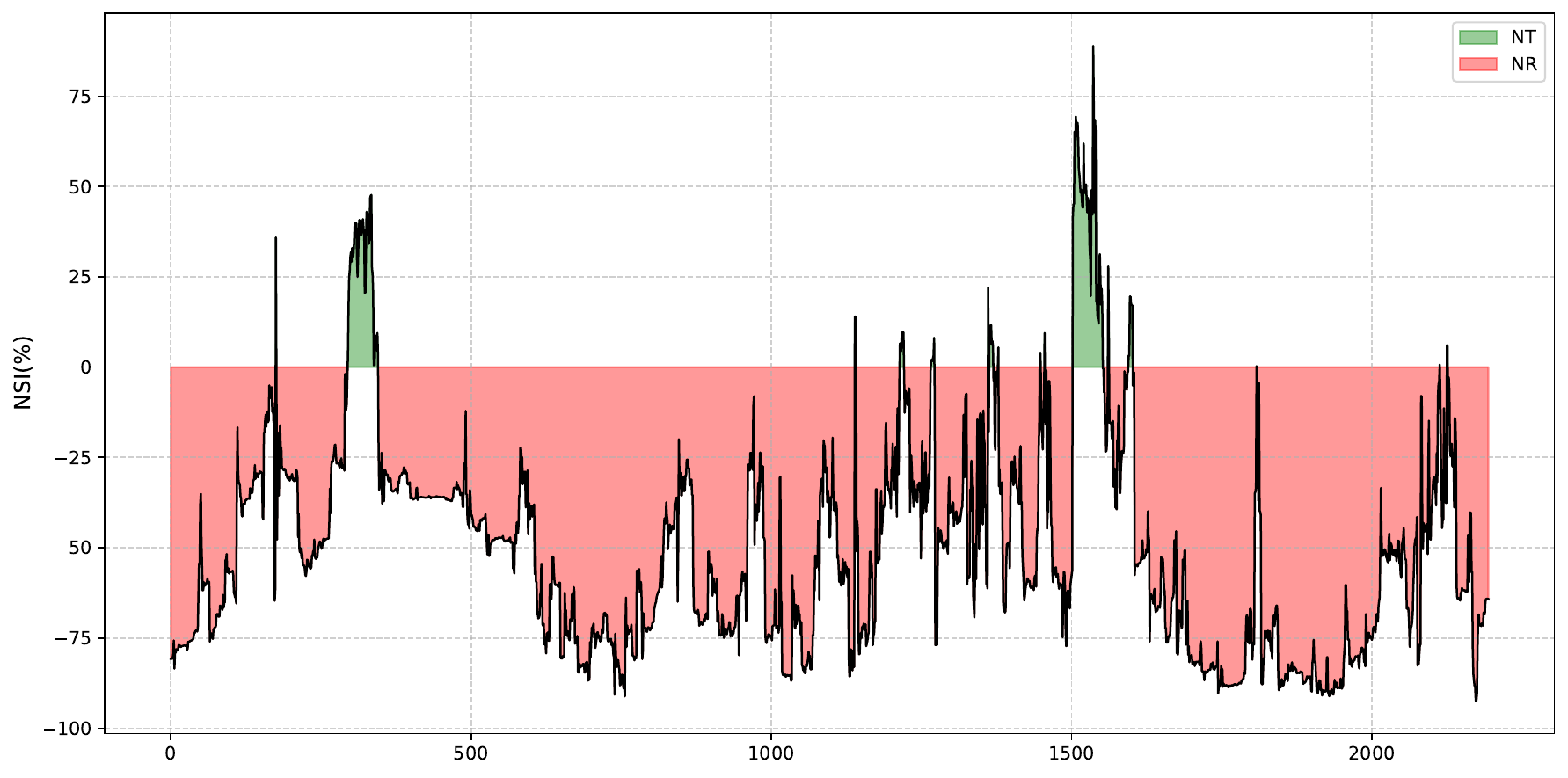}\label{fig:rexm_neg_btc_low}}\hfill
    \subfigure[BTC, $\tau=0.50$]{\includegraphics[width=0.32\linewidth]{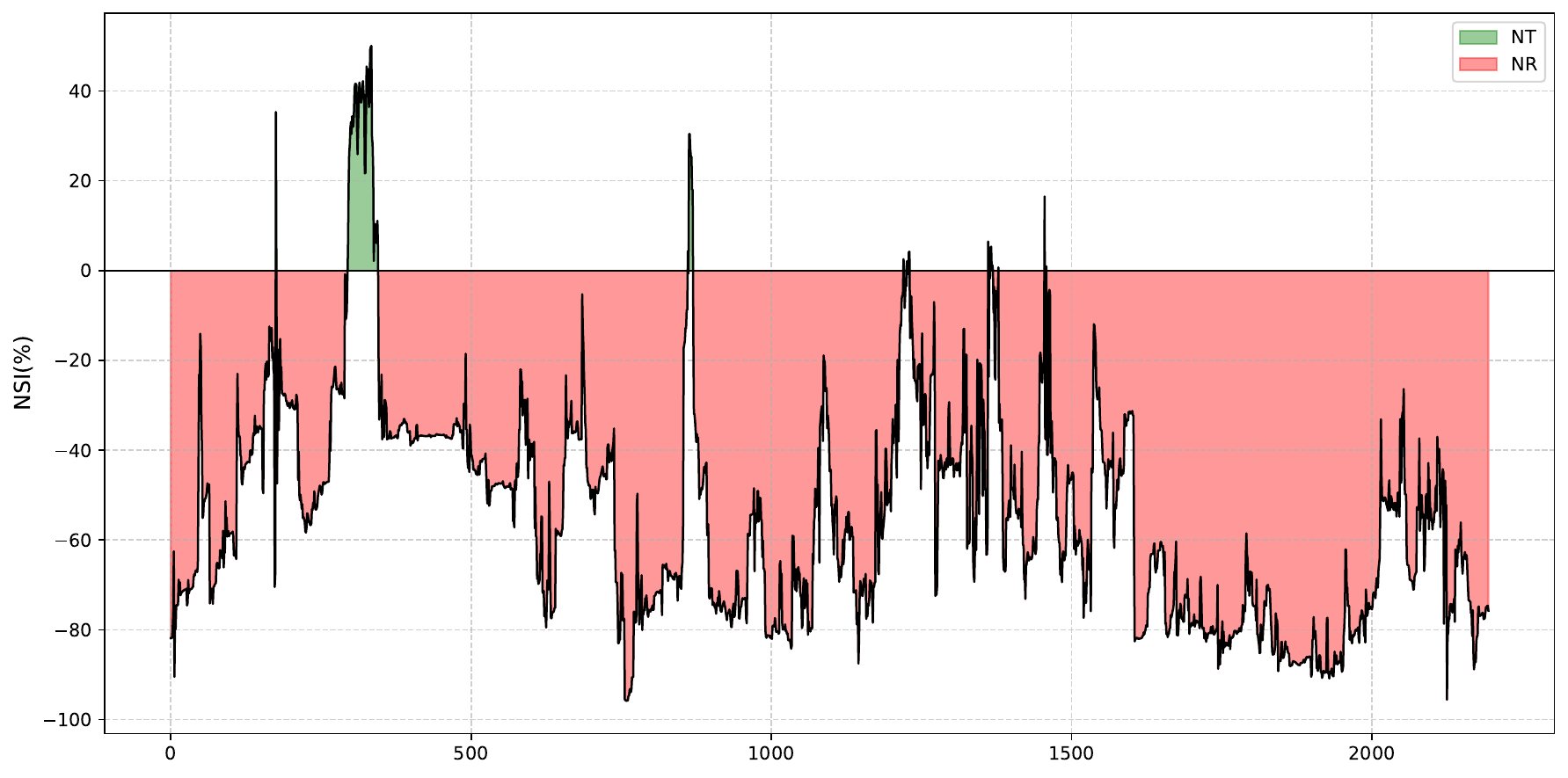}\label{fig:rexm_neg_btc_mid}}\hfill
    \subfigure[BTC, $\tau=0.95$]{\includegraphics[width=0.32\linewidth]{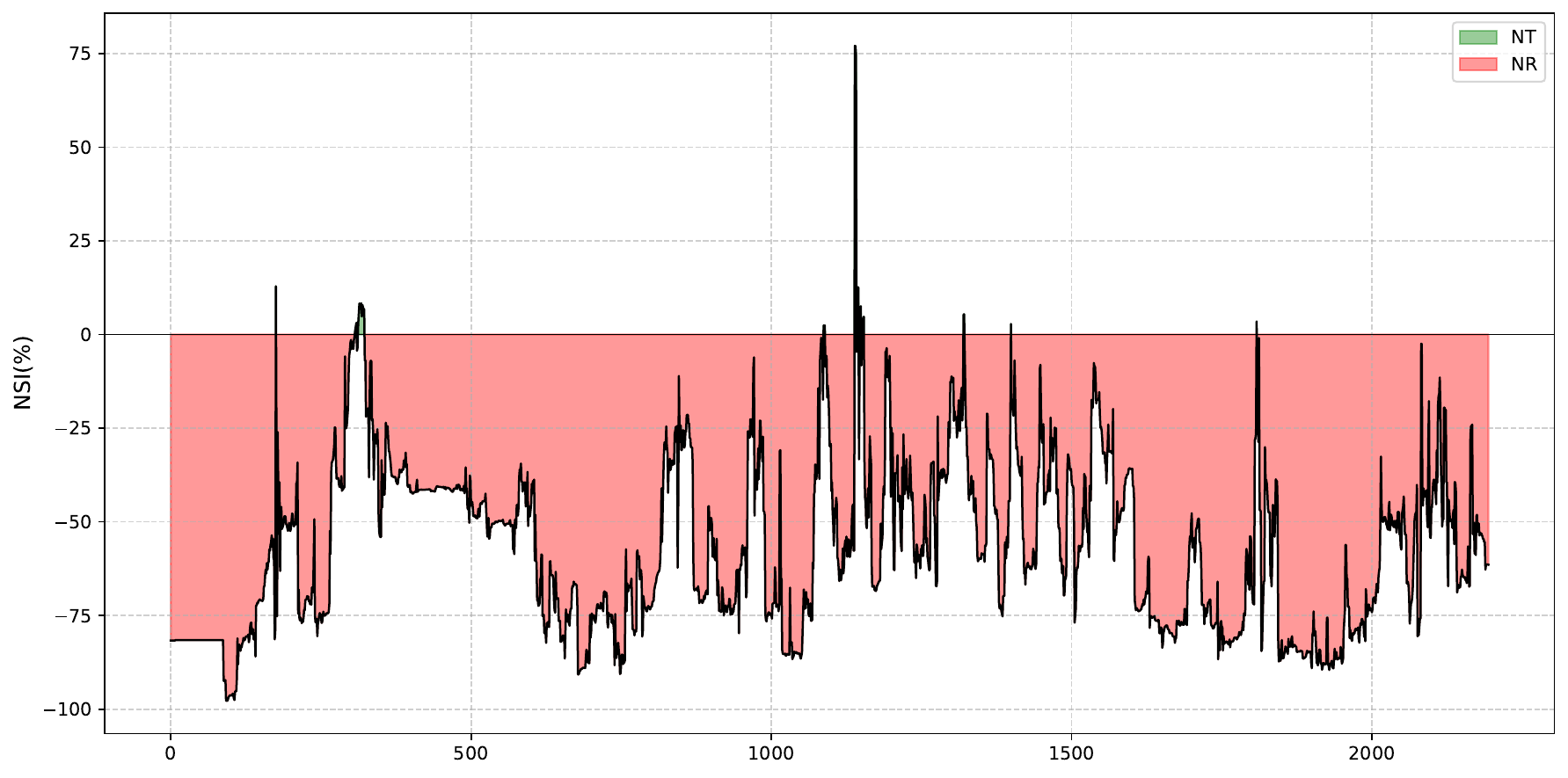}\label{fig:rexm_neg_btc_high}}
    \vspace{0.3cm}
    \subfigure[DASH, $\tau=0.05$]{\includegraphics[width=0.32\linewidth]{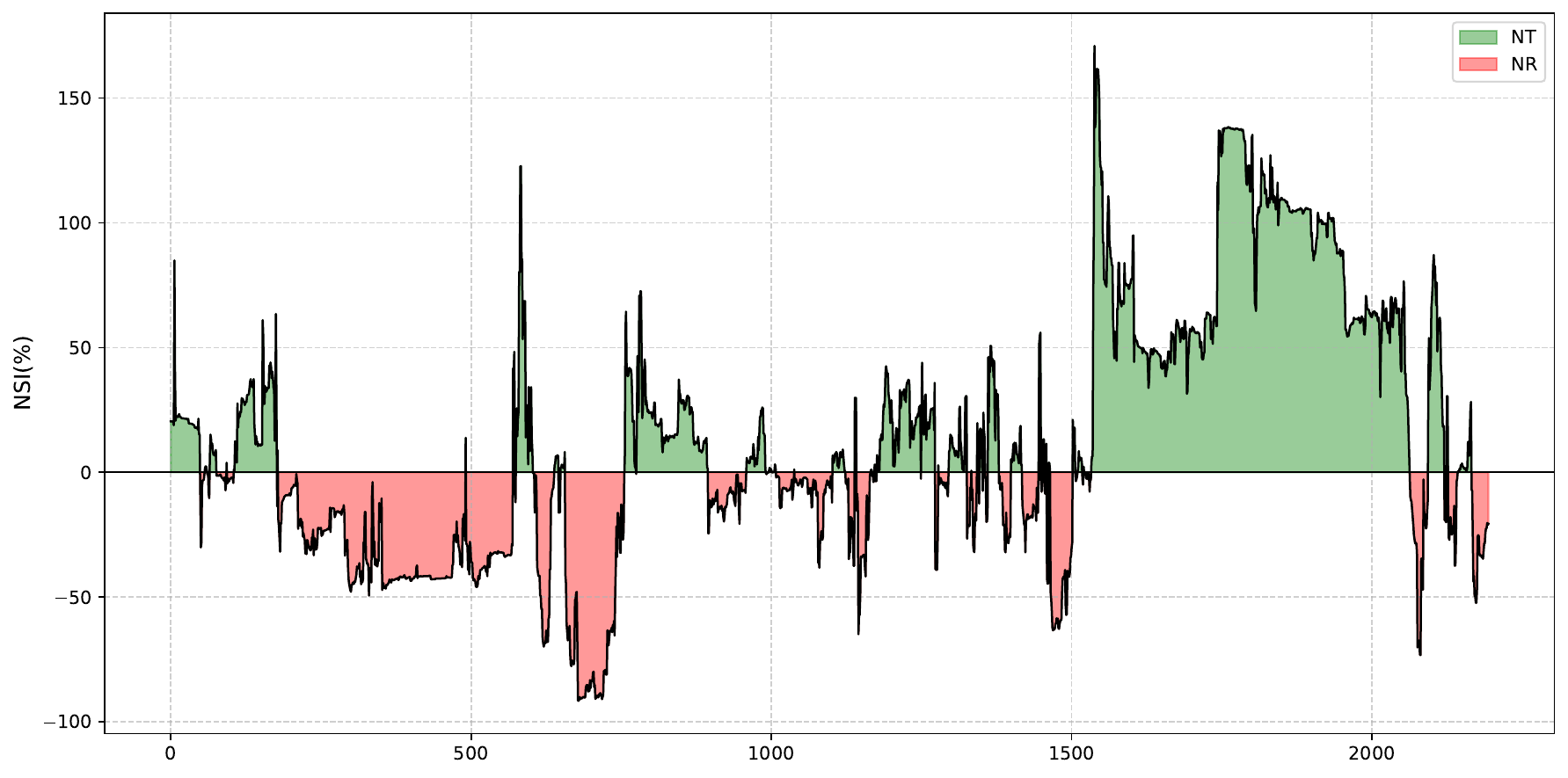}\label{fig:rexm_neg_dash_low}}\hfill
    \subfigure[DASH, $\tau=0.50$]{\includegraphics[width=0.32\linewidth]{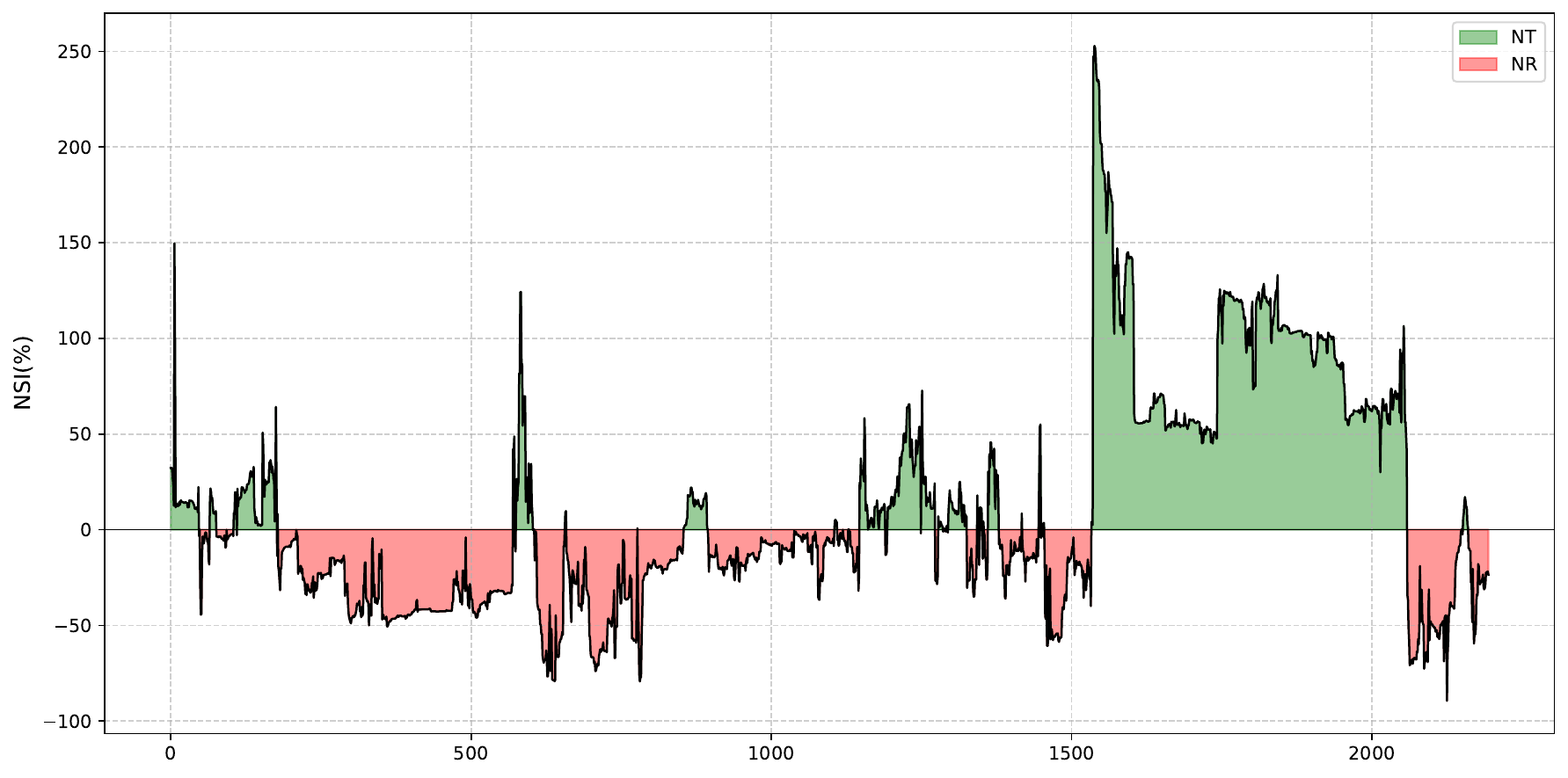}\label{fig:rexm_neg_dash_mid}}\hfill
    \subfigure[DASH, $\tau=0.95$]{\includegraphics[width=0.32\linewidth]{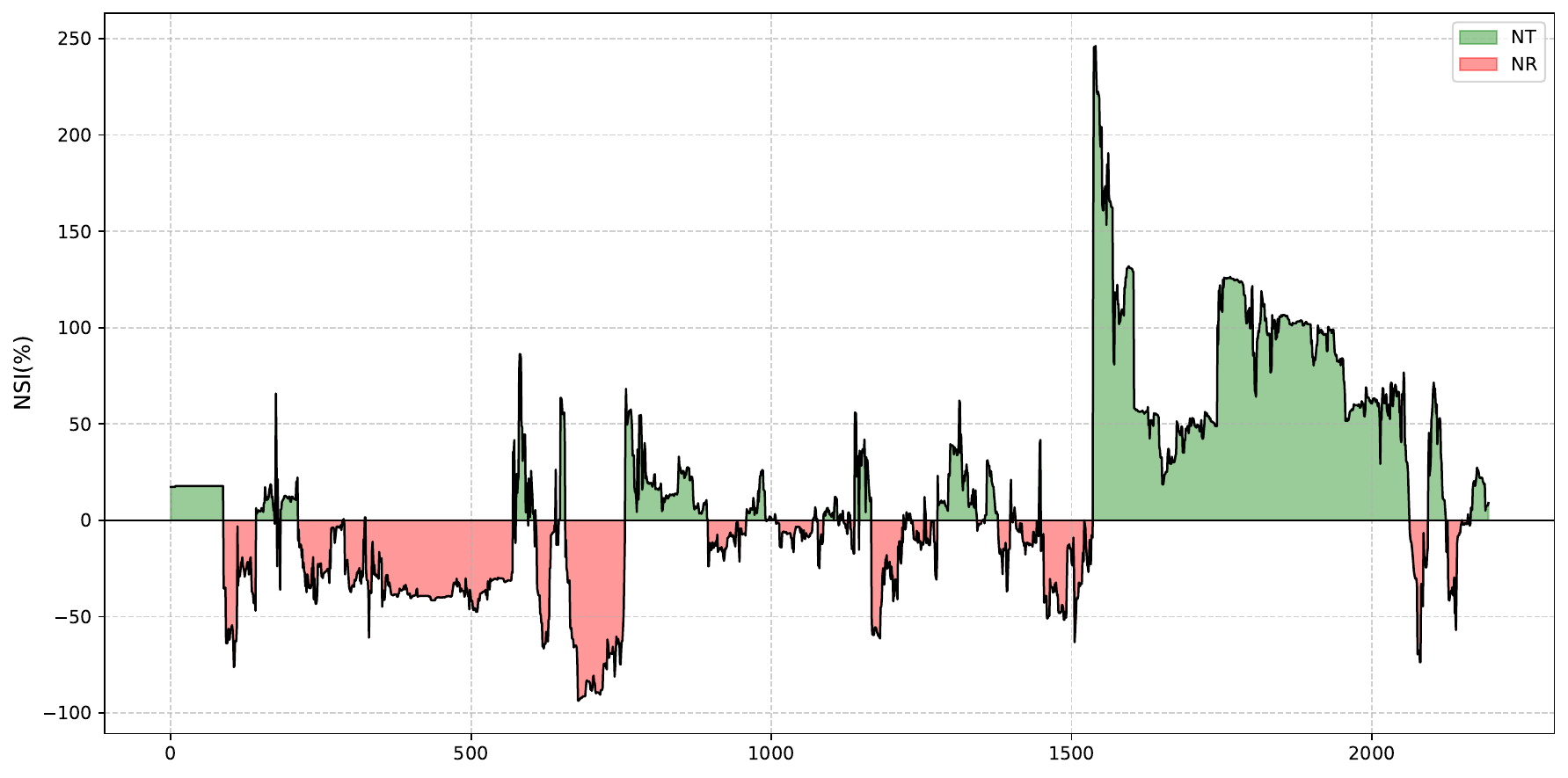}\label{fig:rexm_neg_dash_high}}
    \vspace{0.3cm}
    \subfigure[ETH, $\tau=0.05$]{\includegraphics[width=0.32\linewidth]{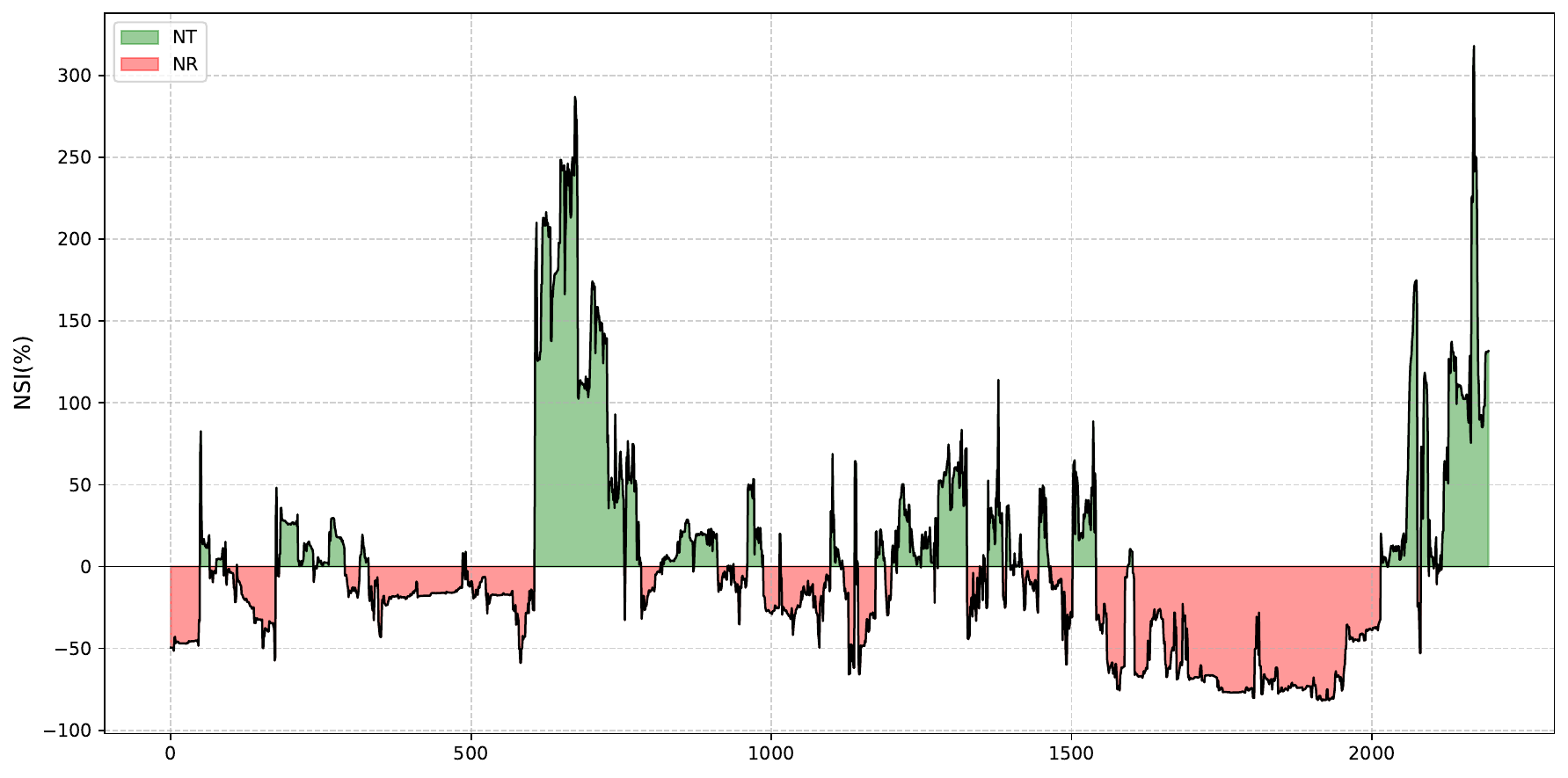}\label{fig:rexm_neg_eth_low}}\hfill
    \subfigure[ETH, $\tau=0.50$]{\includegraphics[width=0.32\linewidth]{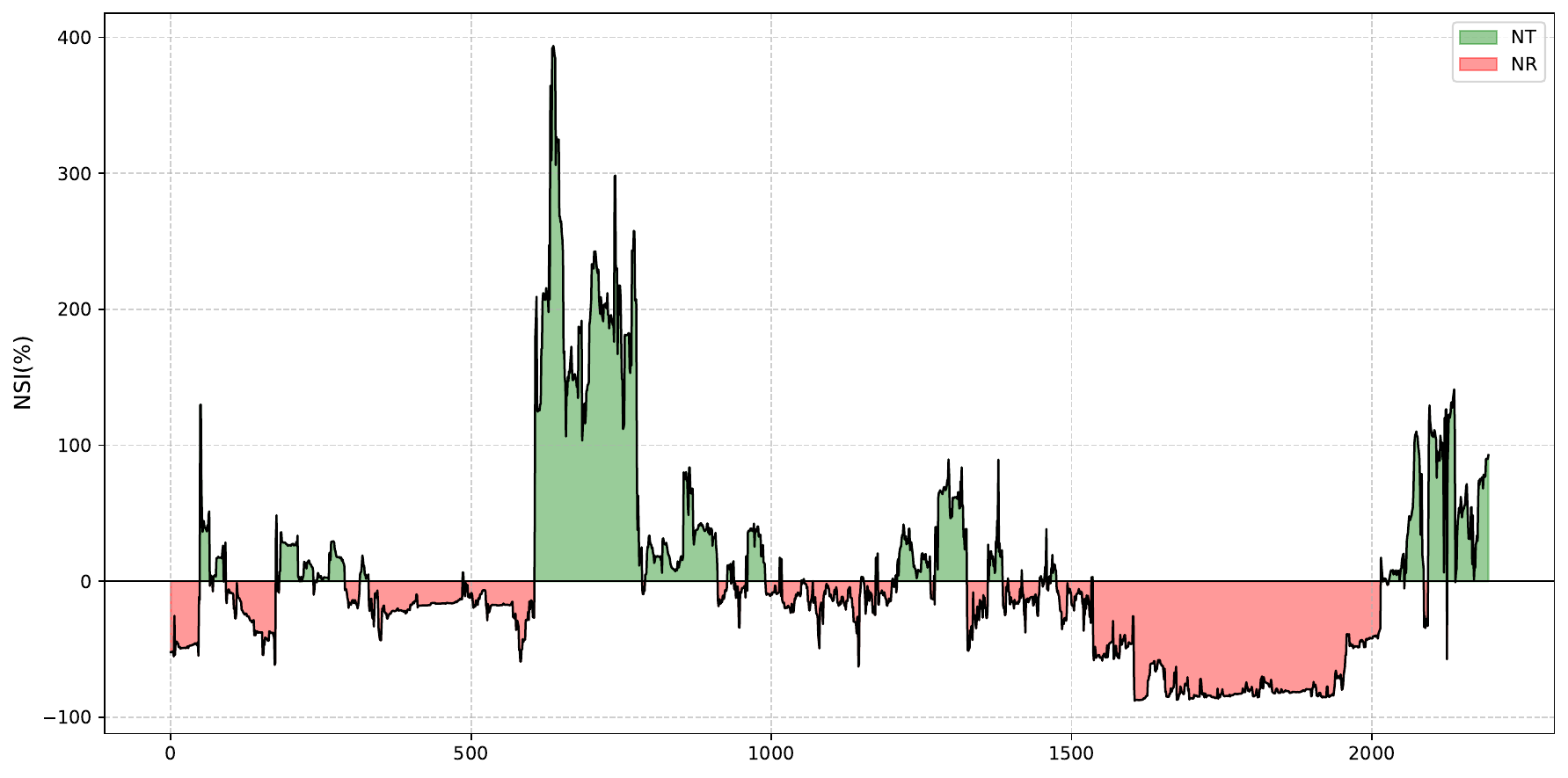}\label{fig:rexm_neg_eth_mid}}\hfill
    \subfigure[ETH, $\tau=0.95$]{\includegraphics[width=0.32\linewidth]{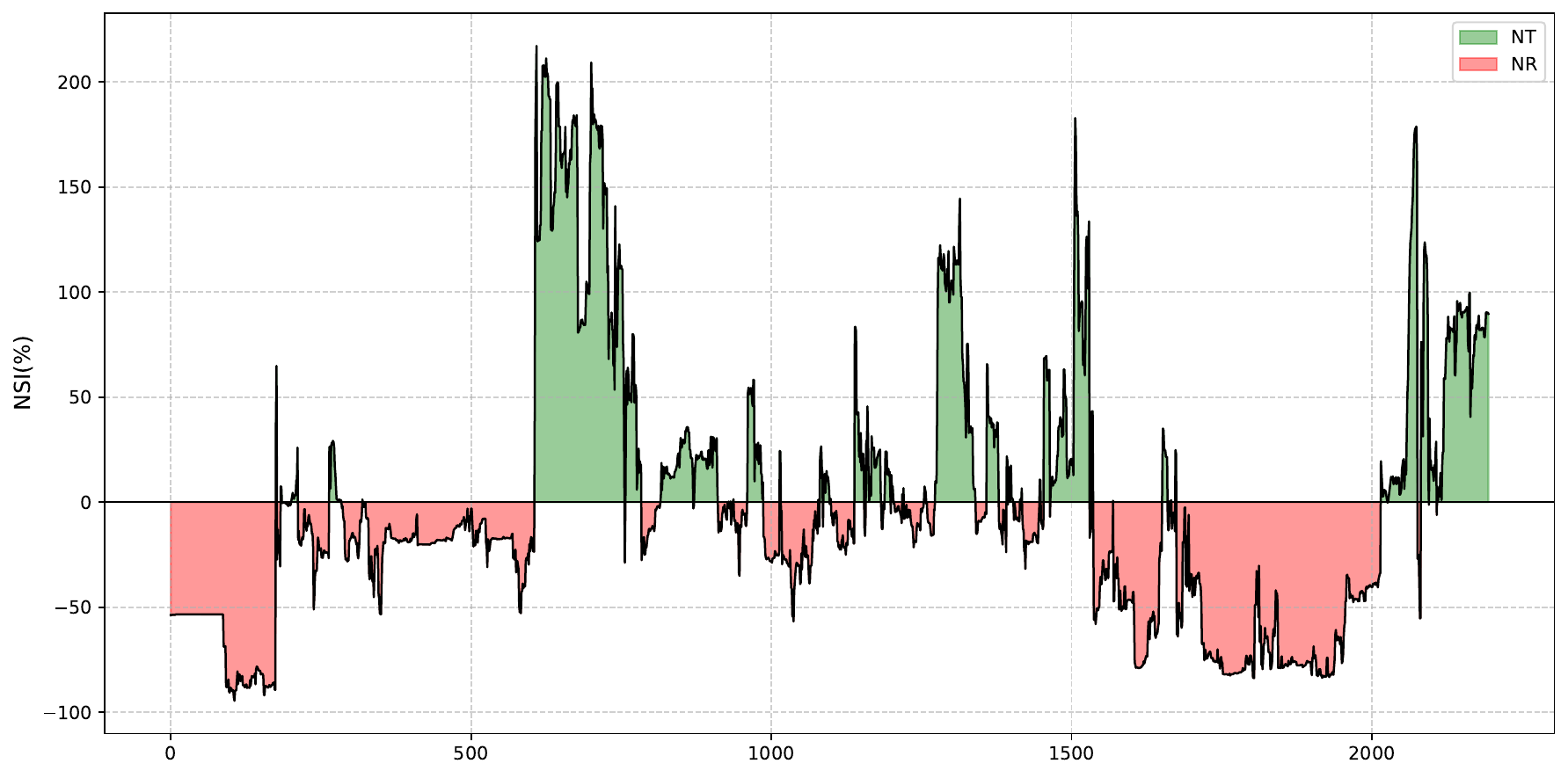}\label{fig:rexm_neg_eth_high}}
    \vspace{0.3cm}
    \subfigure[LTC, $\tau=0.05$]{\includegraphics[width=0.32\linewidth]{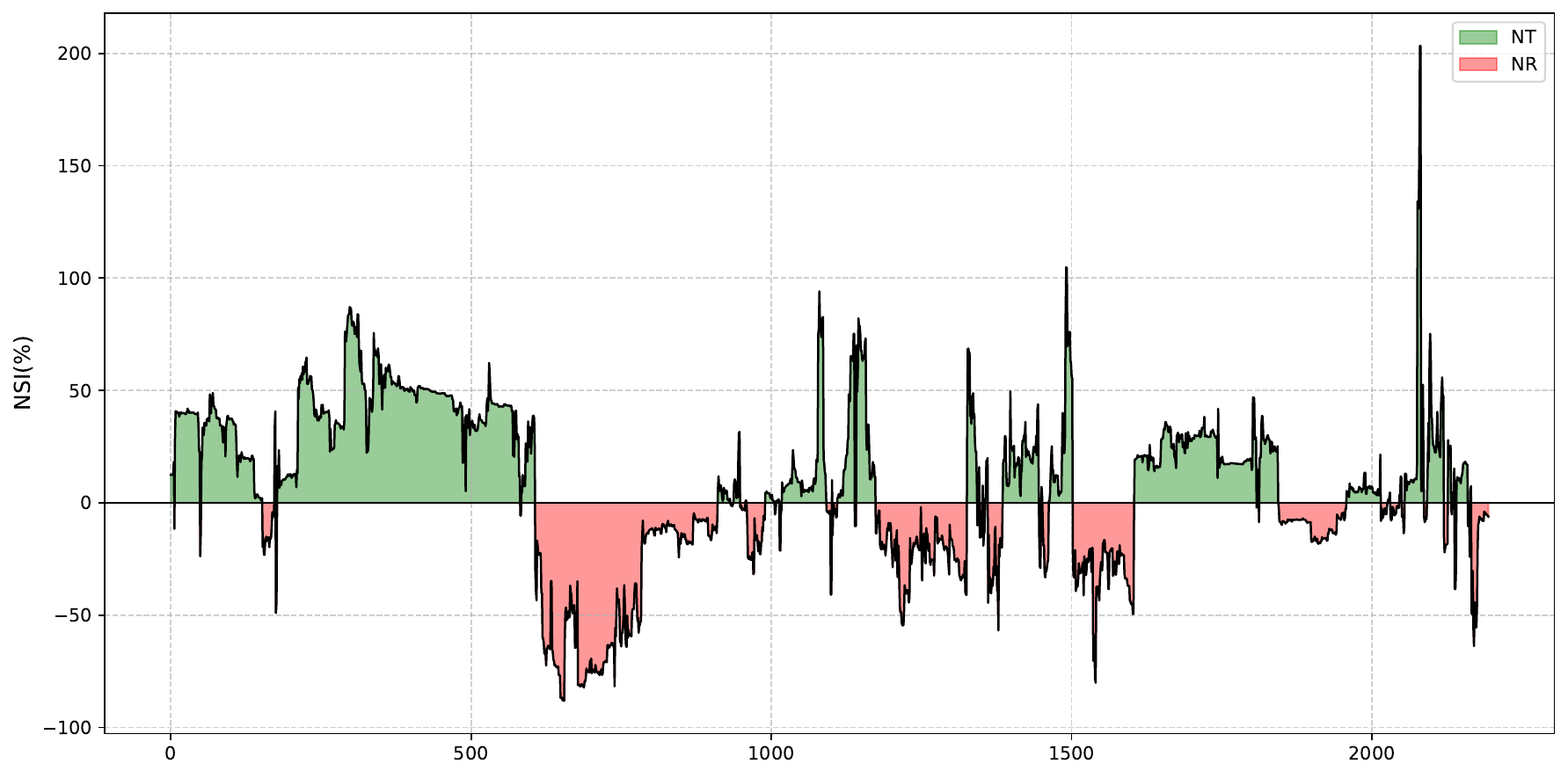}\label{fig:rexm_neg_ltc_low}}\hfill
    \subfigure[LTC, $\tau=0.50$]{\includegraphics[width=0.32\linewidth]{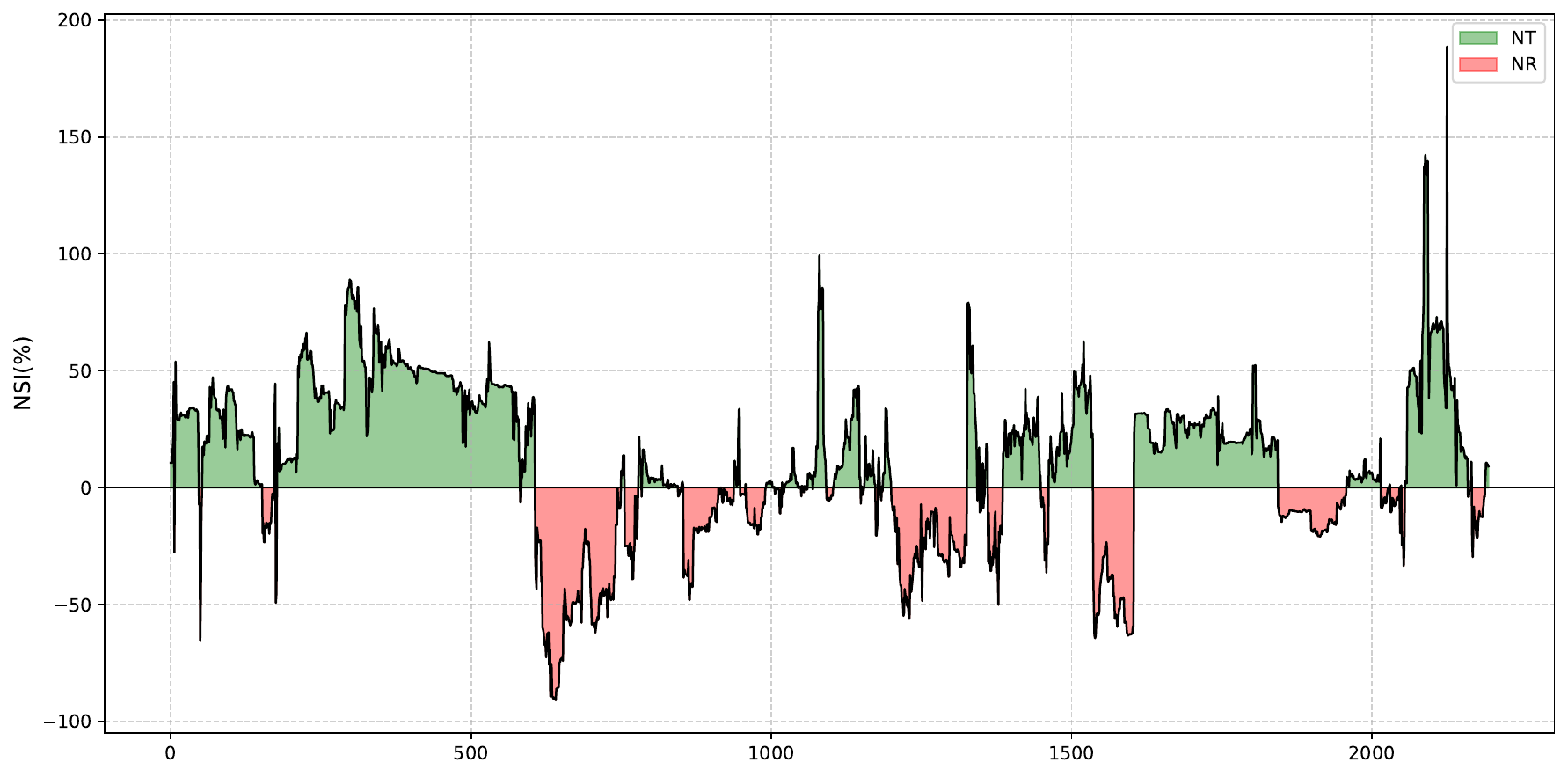}\label{fig:rexm_neg_ltc_mid}}\hfill
    \subfigure[LTC, $\tau=0.95$]{\includegraphics[width=0.32\linewidth]{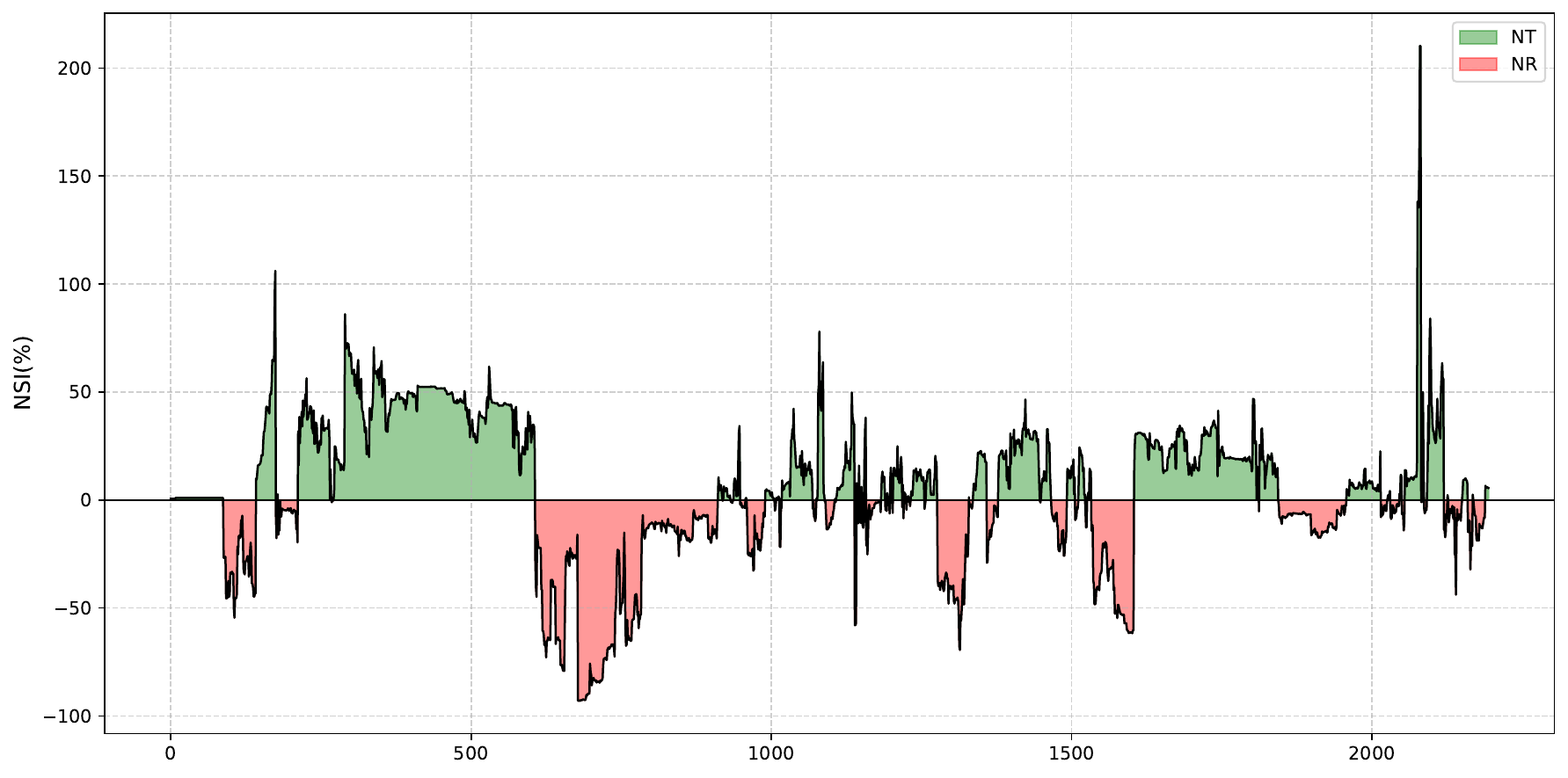}\label{fig:rexm_neg_ltc_high}}
    \vspace{0.3cm}
    \subfigure[XLM, $\tau=0.05$]{\includegraphics[width=0.32\linewidth]{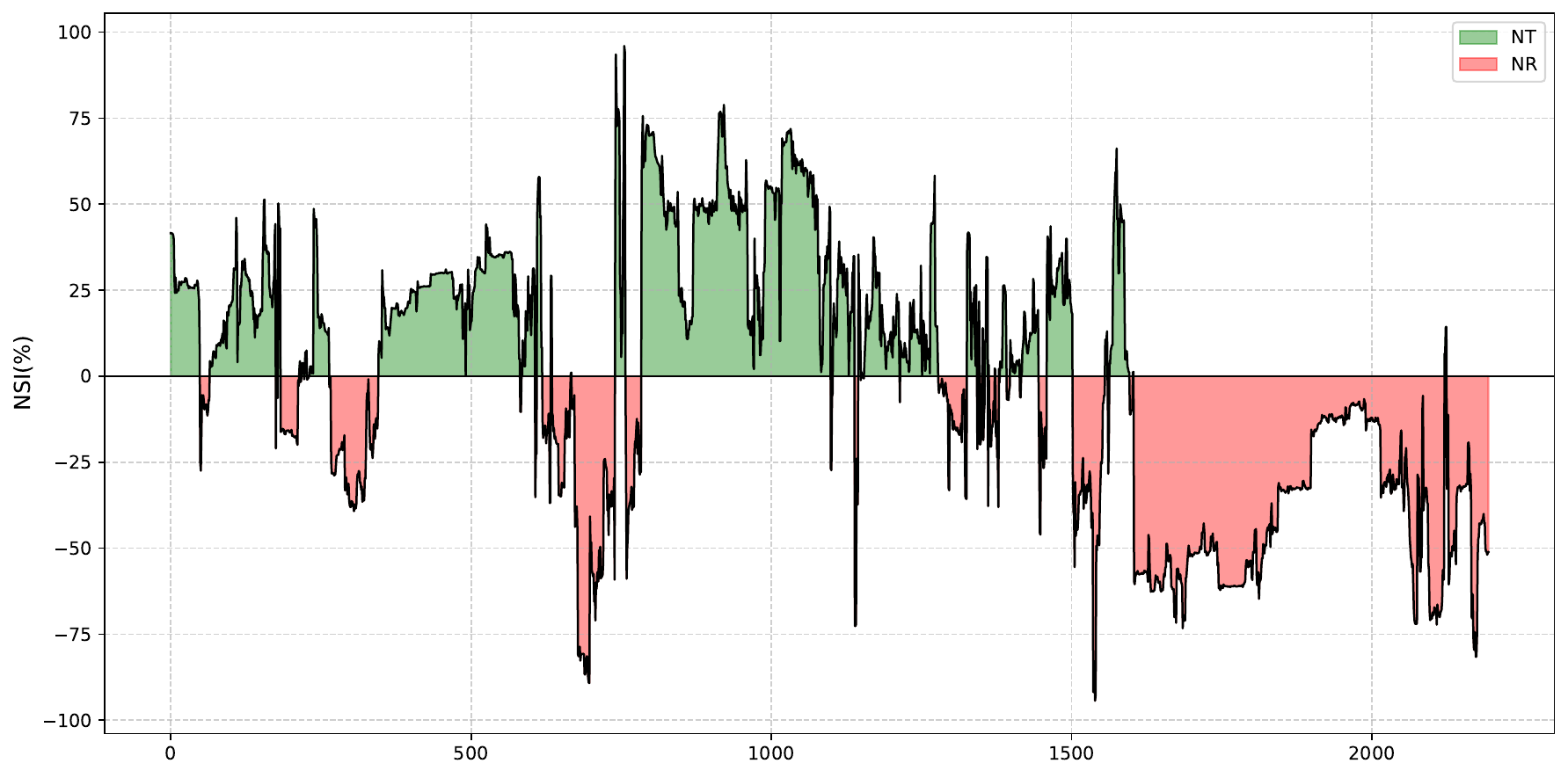}\label{fig:rexm_neg_xlm_low}}\hfill
    \subfigure[XLM, $\tau=0.50$]{\includegraphics[width=0.32\linewidth]{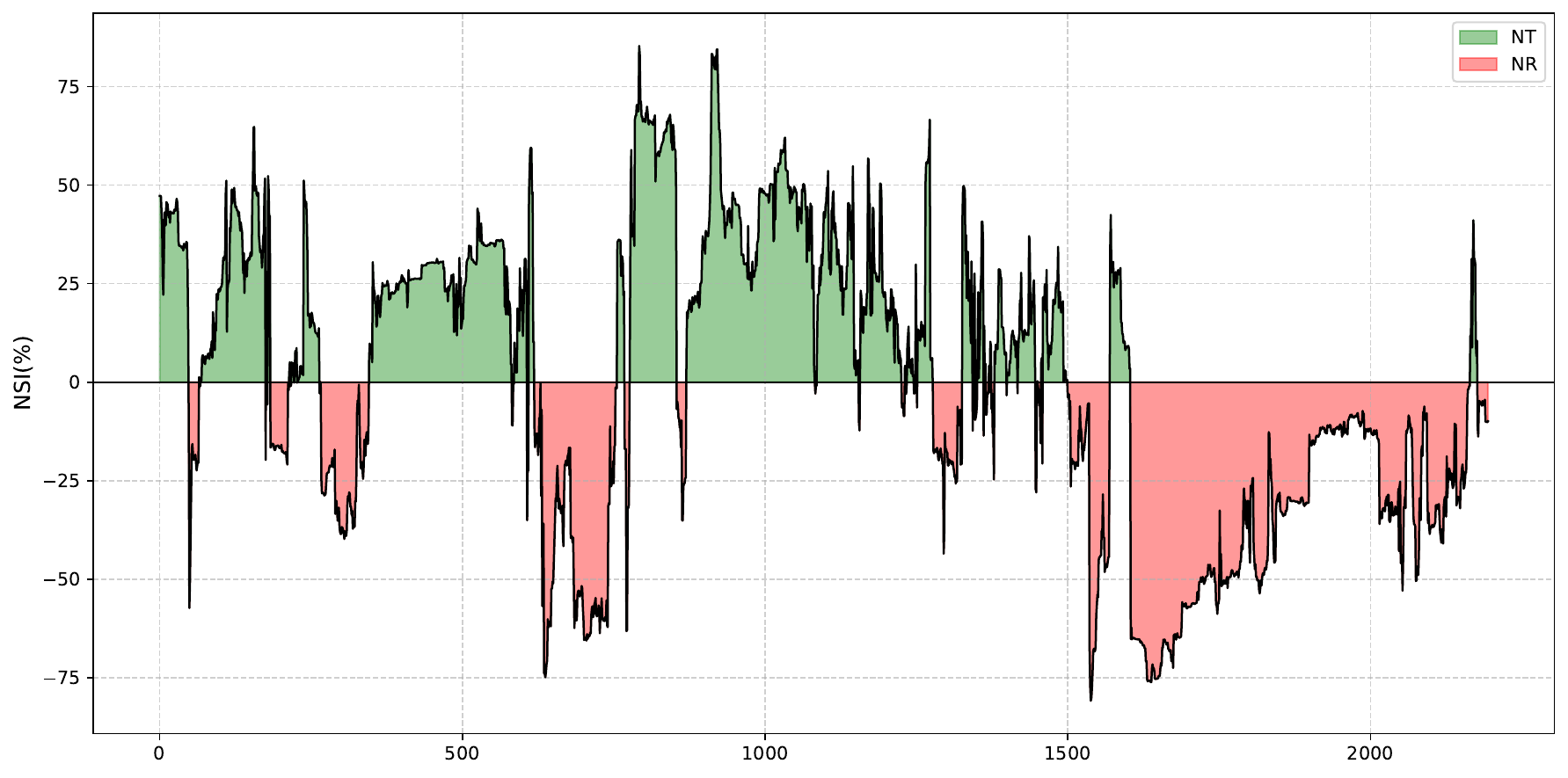}\label{fig:rexm_neg_xlm_mid}}\hfill
    \subfigure[XLM, $\tau=0.95$]{\includegraphics[width=0.32\linewidth]{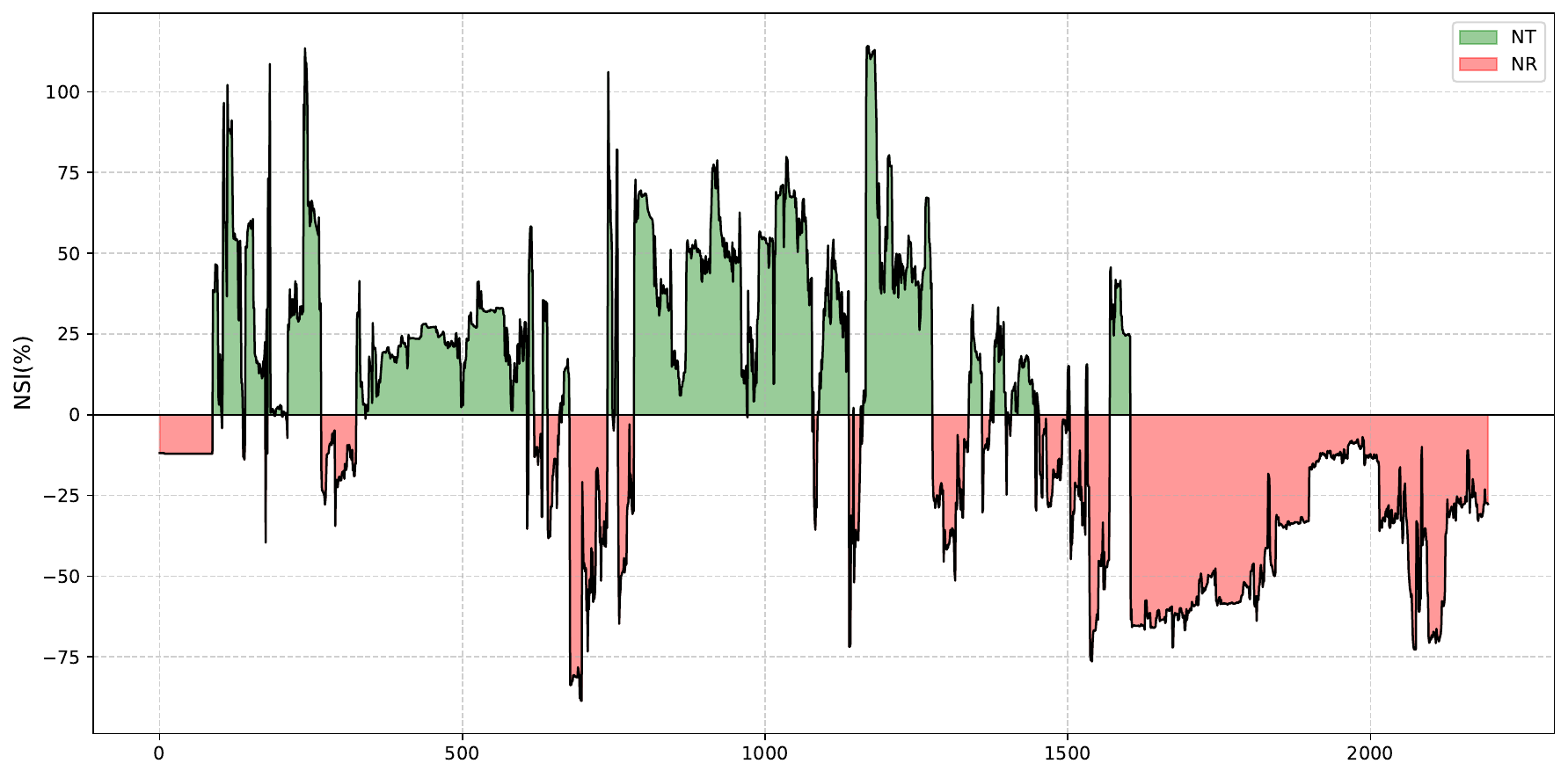}\label{fig:rexm_neg_xlm_high}}
    \vspace{0.3cm}
    \subfigure[XRP, $\tau=0.05$]{\includegraphics[width=0.32\linewidth]{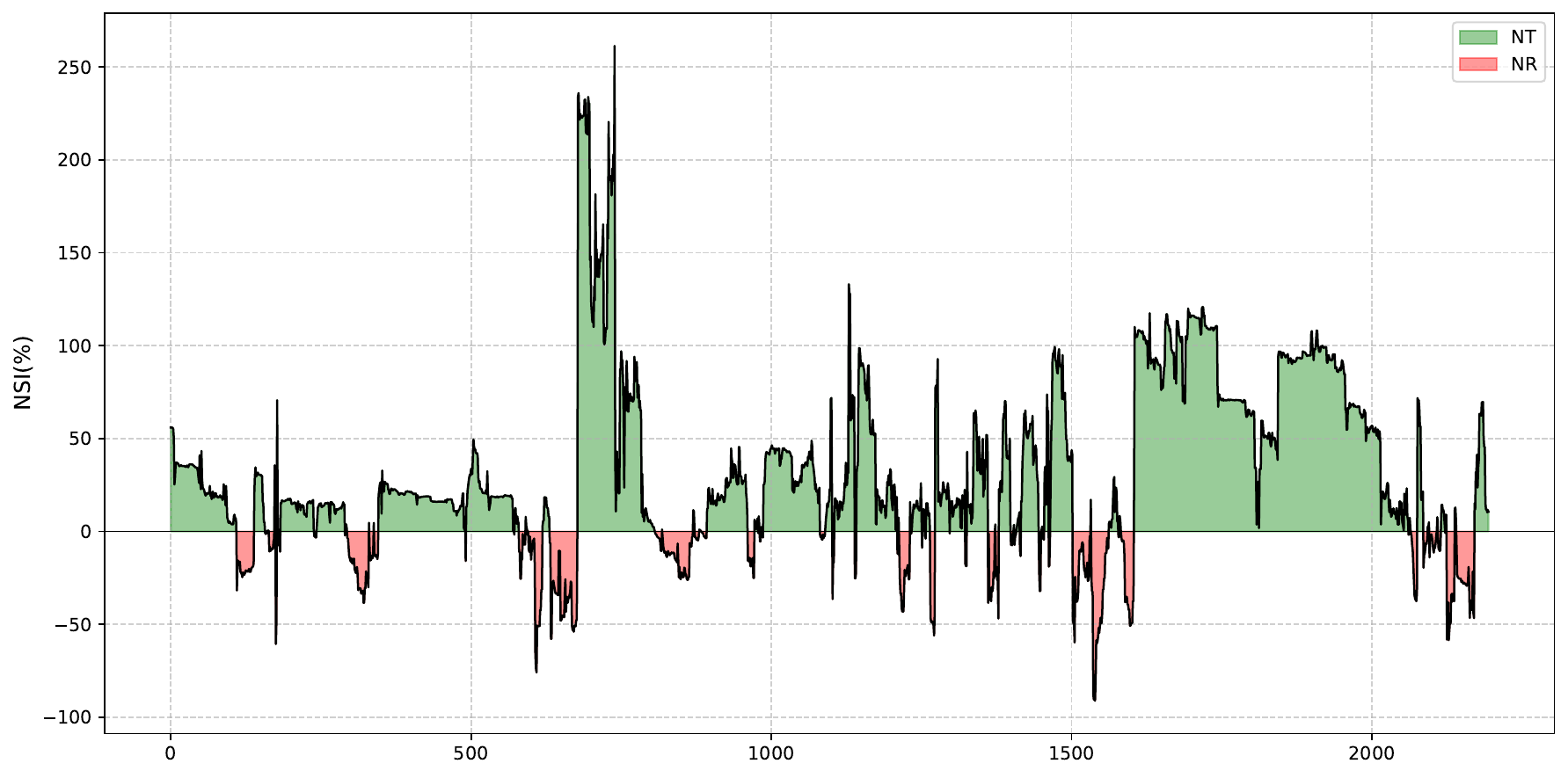}\label{fig:rexm_neg_xrp_low}}\hfill
    \subfigure[XRP, $\tau=0.50$]{\includegraphics[width=0.32\linewidth]{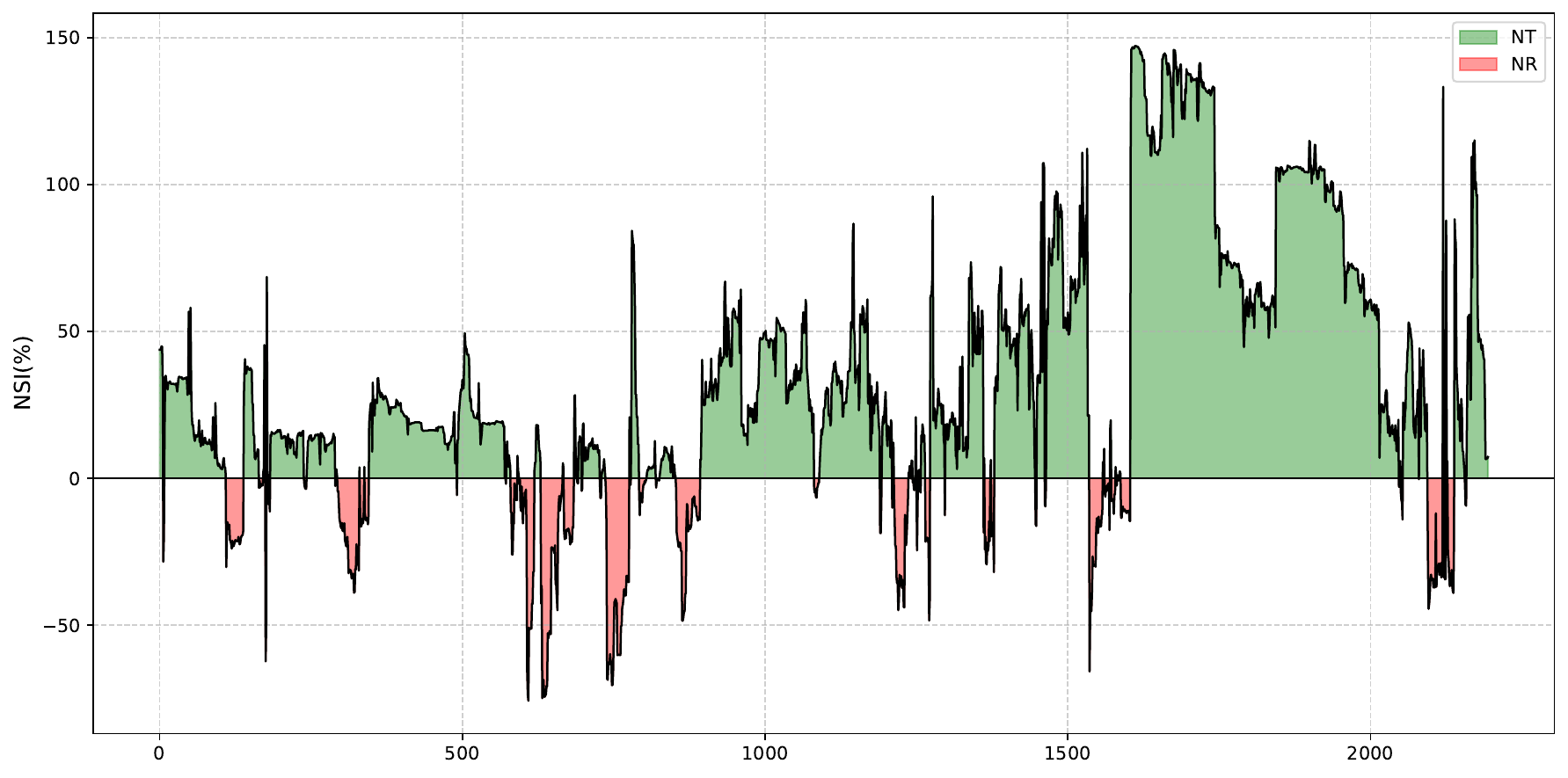}\label{fig:rexm_neg_xrp_mid}}\hfill
    \subfigure[XRP, $\tau=0.95$]{\includegraphics[width=0.32\linewidth]{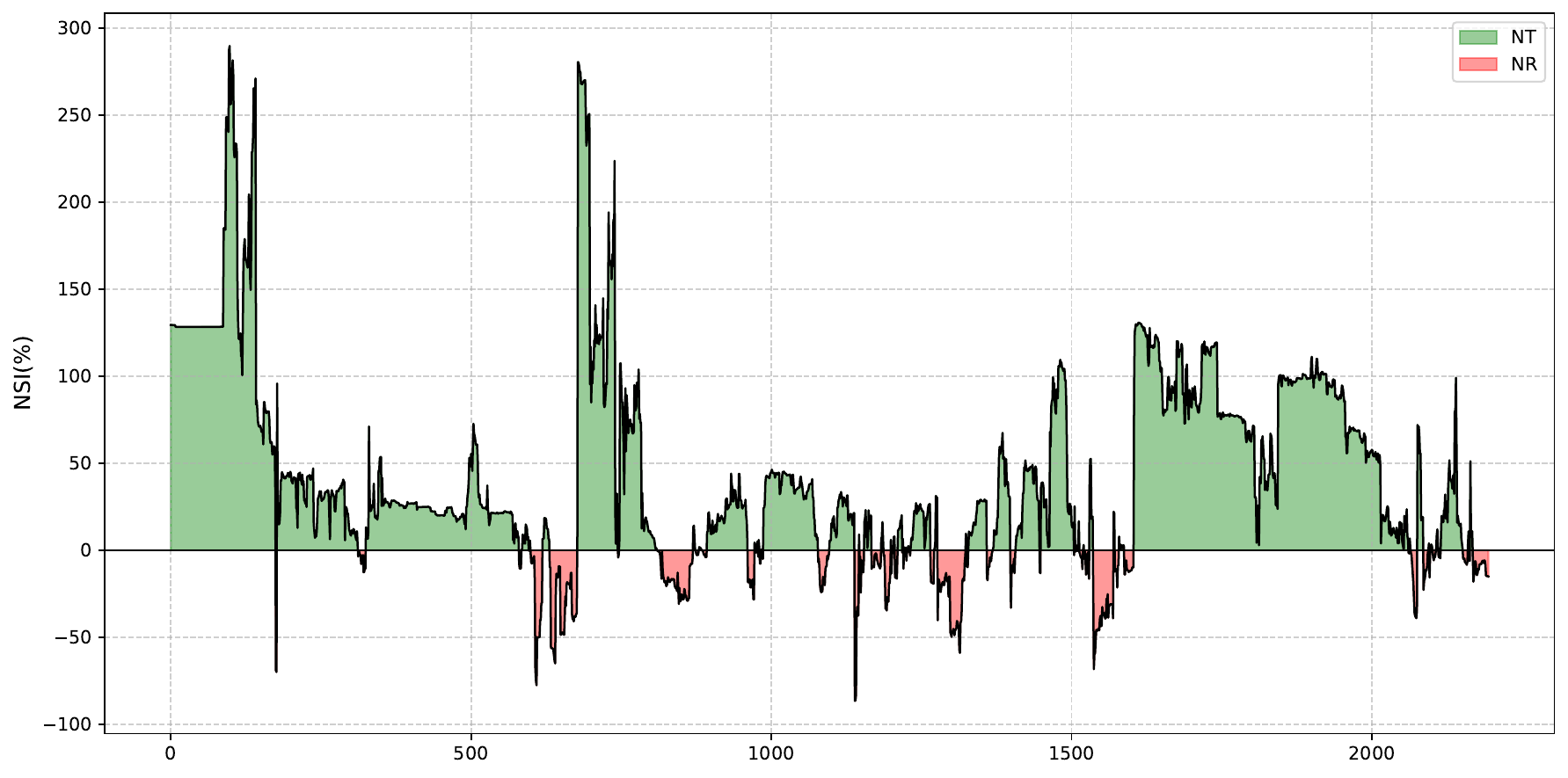}\label{fig:rexm_neg_xrp_high}}
\end{figure}

\begin{figure}[p]
    \centering
    \caption{Quantile net spillovers for major cryptocurrencies using $REX^m$ as the feature variable.}
    \label{fig:rexmod_net_spillover_by_coin}

    \subfigure[BTC, $\tau=0.05$]{\includegraphics[width=0.32\linewidth]{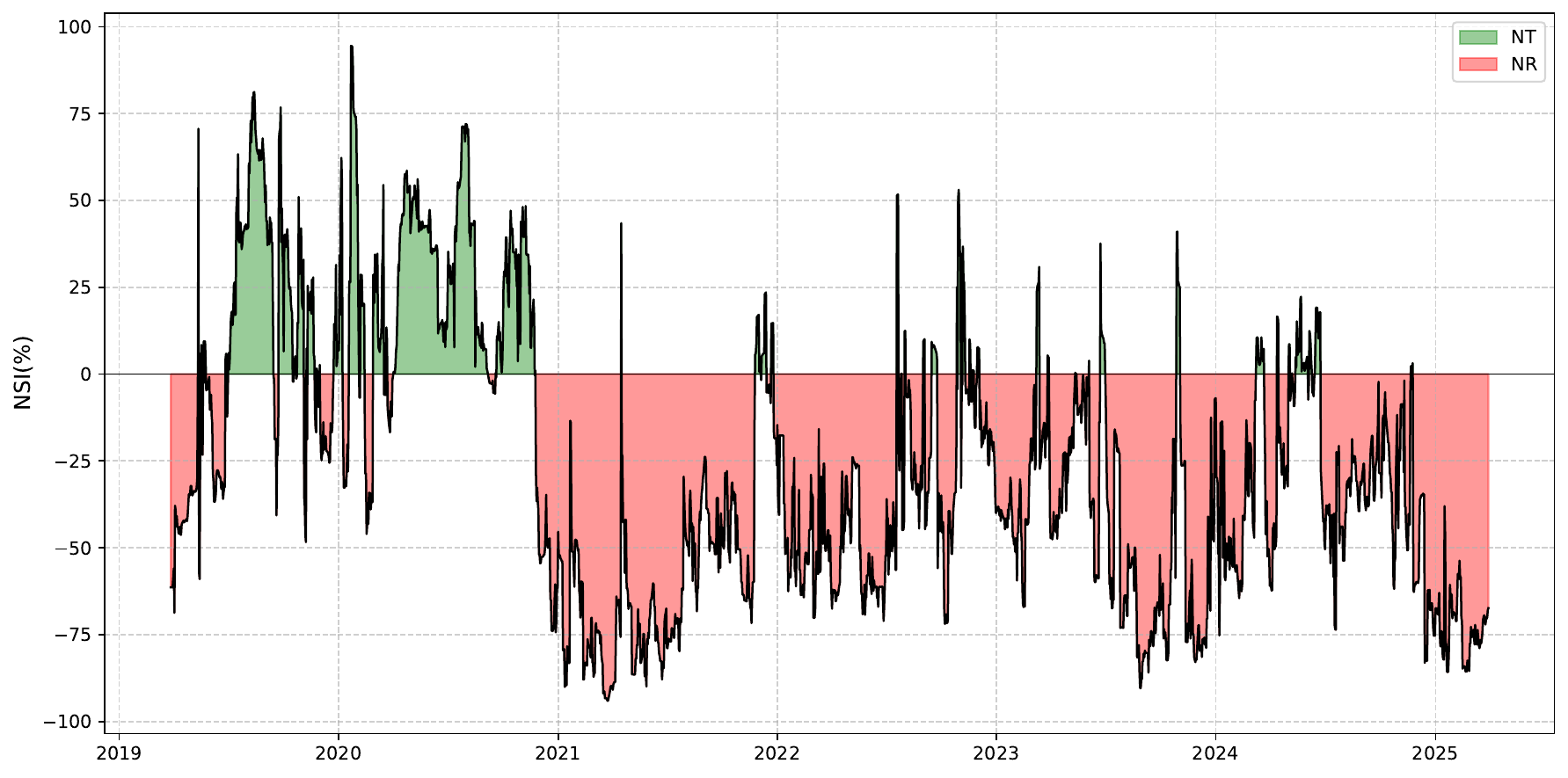}\label{fig:rexmod_btc_low}}\hfill
    \subfigure[BTC, $\tau=0.50$]{\includegraphics[width=0.32\linewidth]{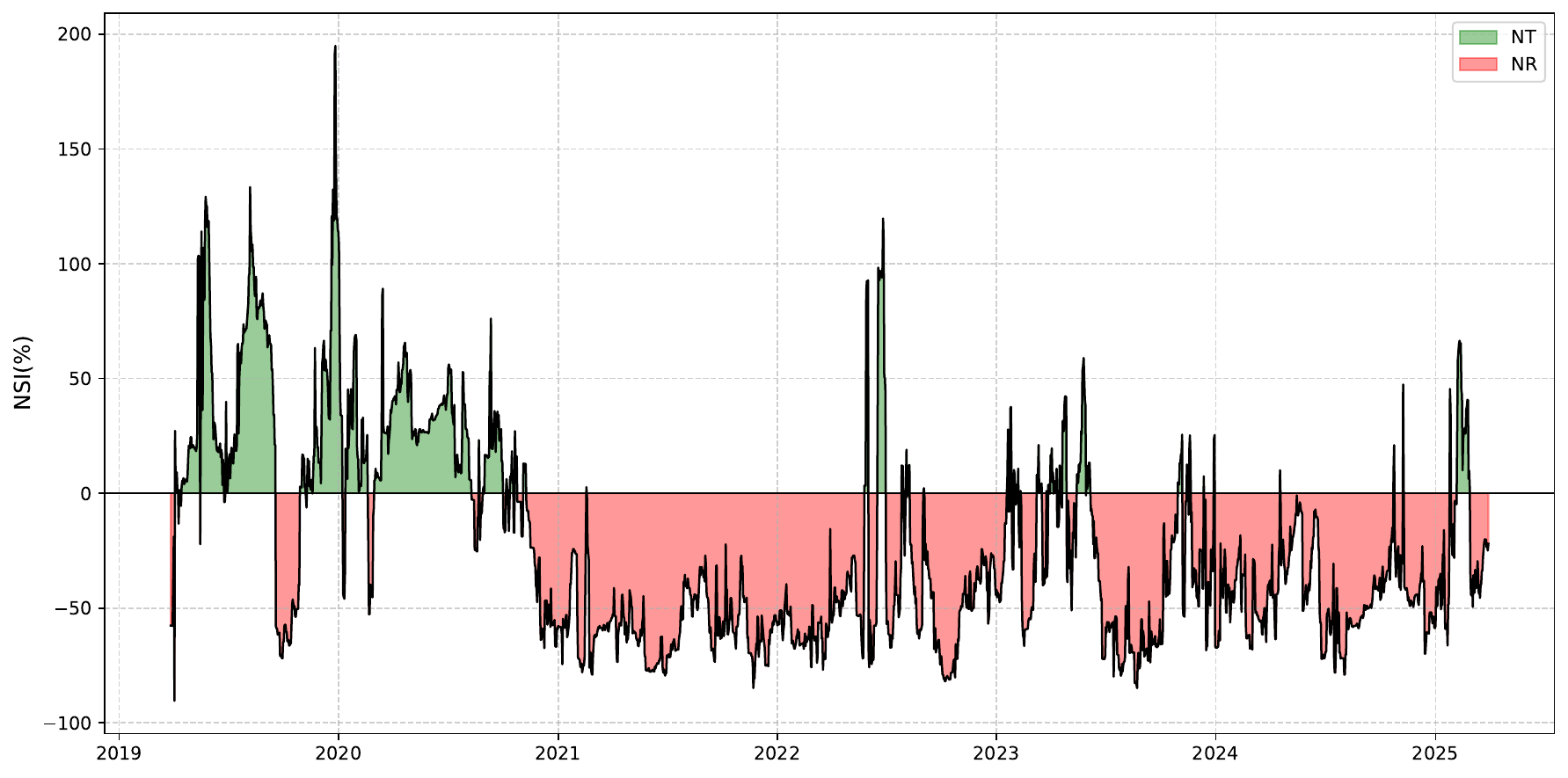}\label{fig:rexmod_btc_mid}}\hfill
    \subfigure[BTC, $\tau=0.95$]{\includegraphics[width=0.32\linewidth]{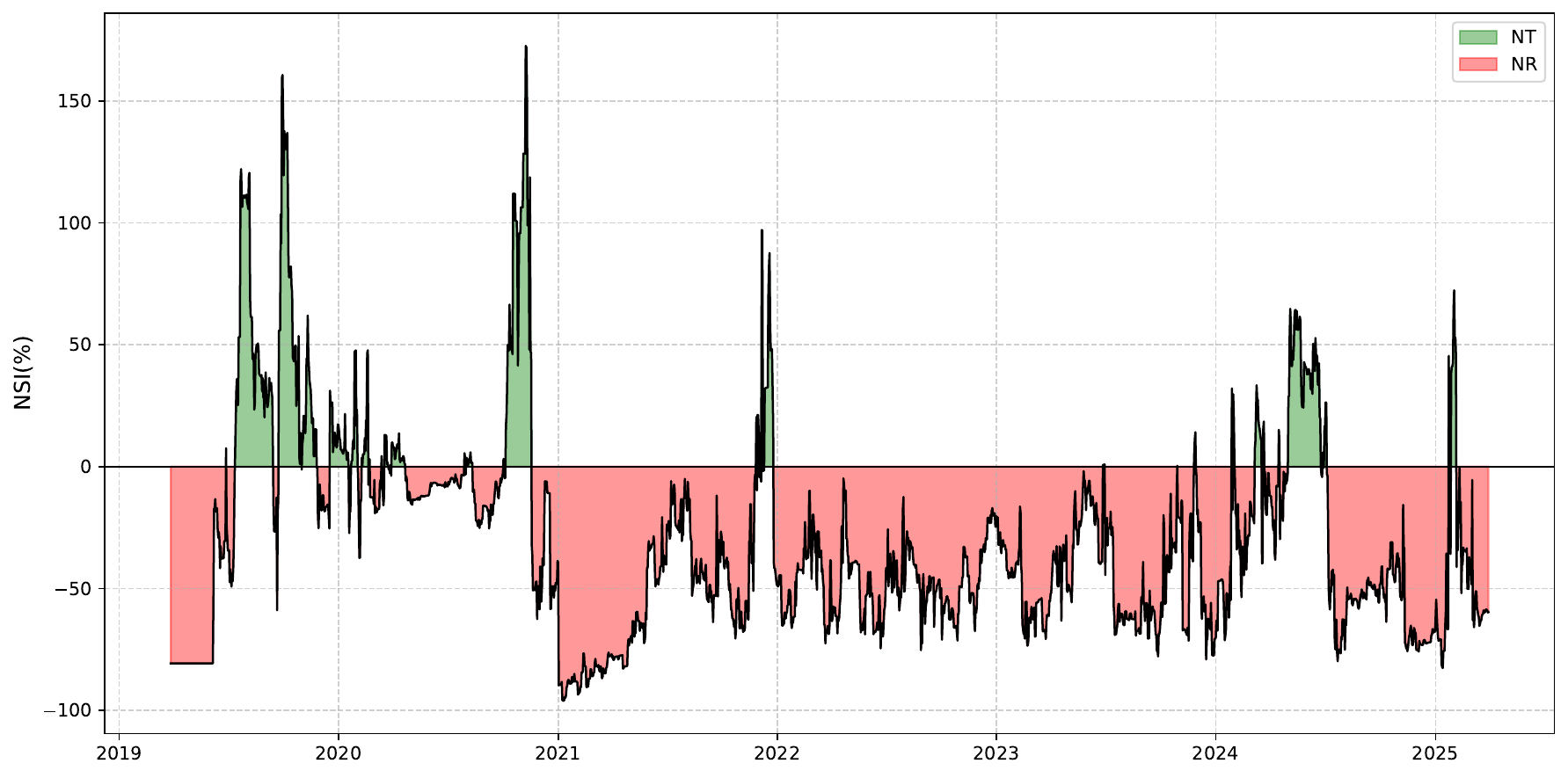}\label{fig:rexmod_btc_high}}
    \vspace{0.3cm}
    \subfigure[DASH, $\tau=0.05$]{\includegraphics[width=0.32\linewidth]{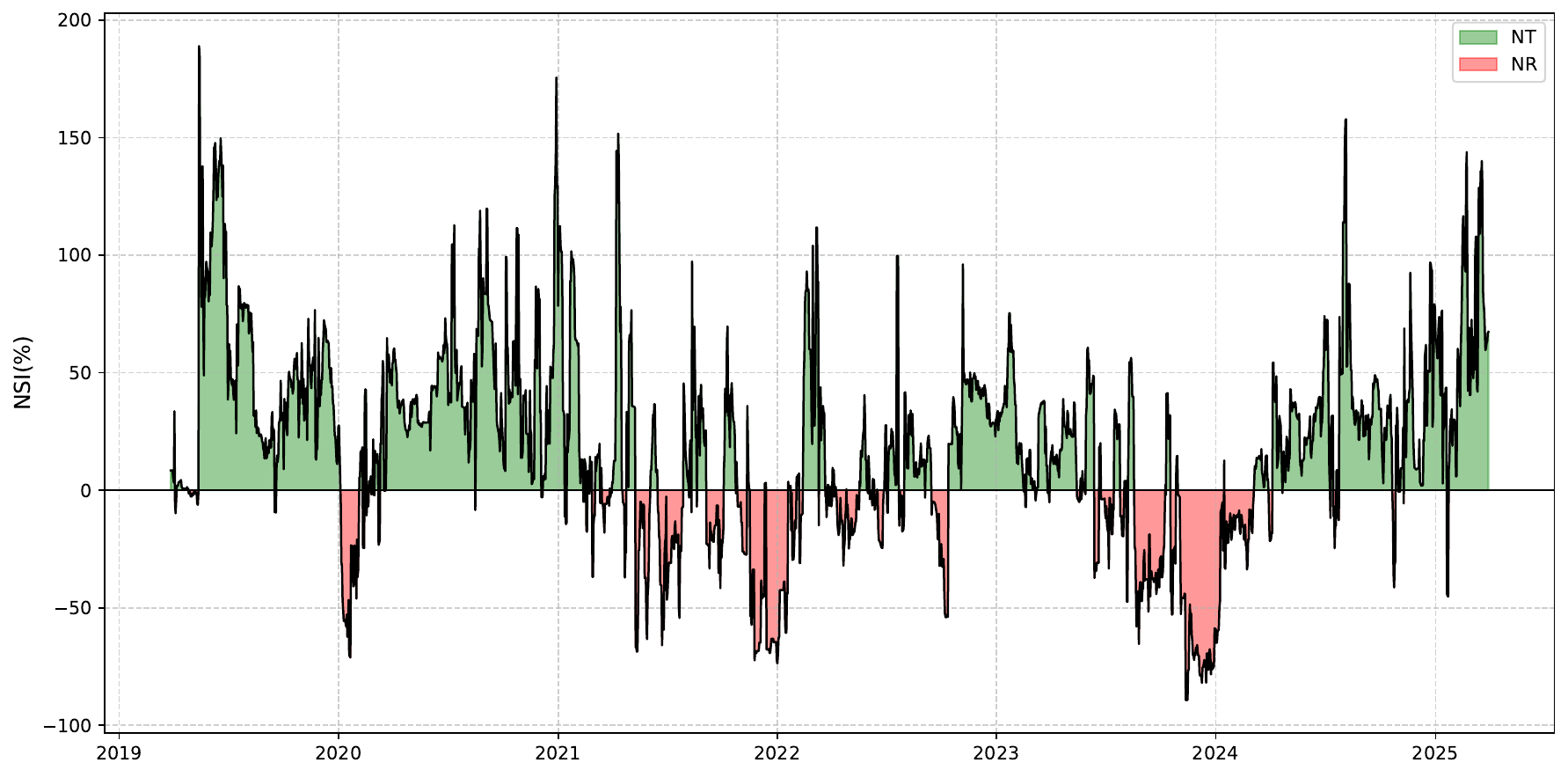}\label{fig:rexmod_dash_low}}\hfill
    \subfigure[DASH, $\tau=0.50$]{\includegraphics[width=0.32\linewidth]{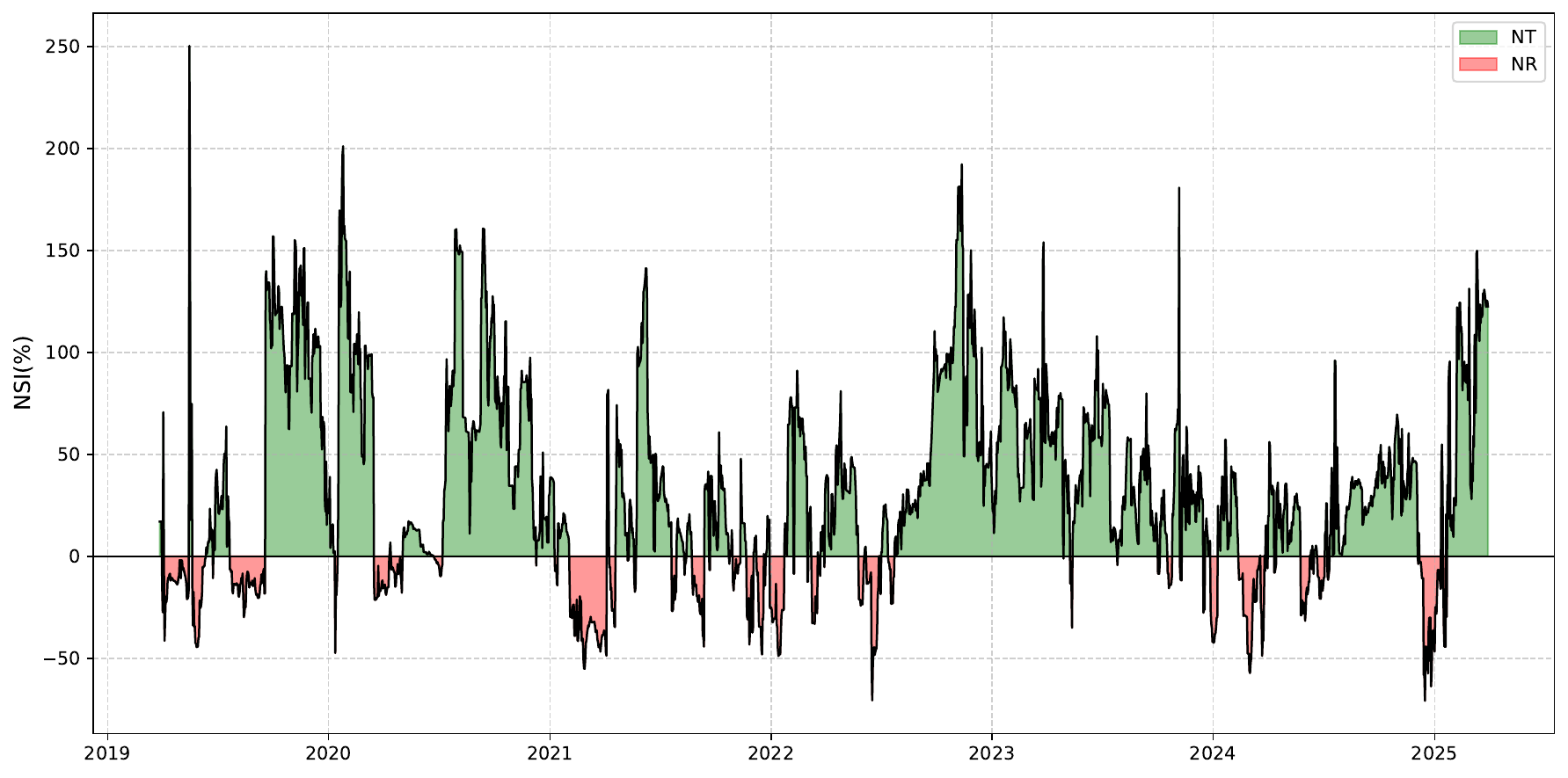}\label{fig:rexmod_dash_mid}}\hfill
    \subfigure[DASH, $\tau=0.95$]{\includegraphics[width=0.32\linewidth]{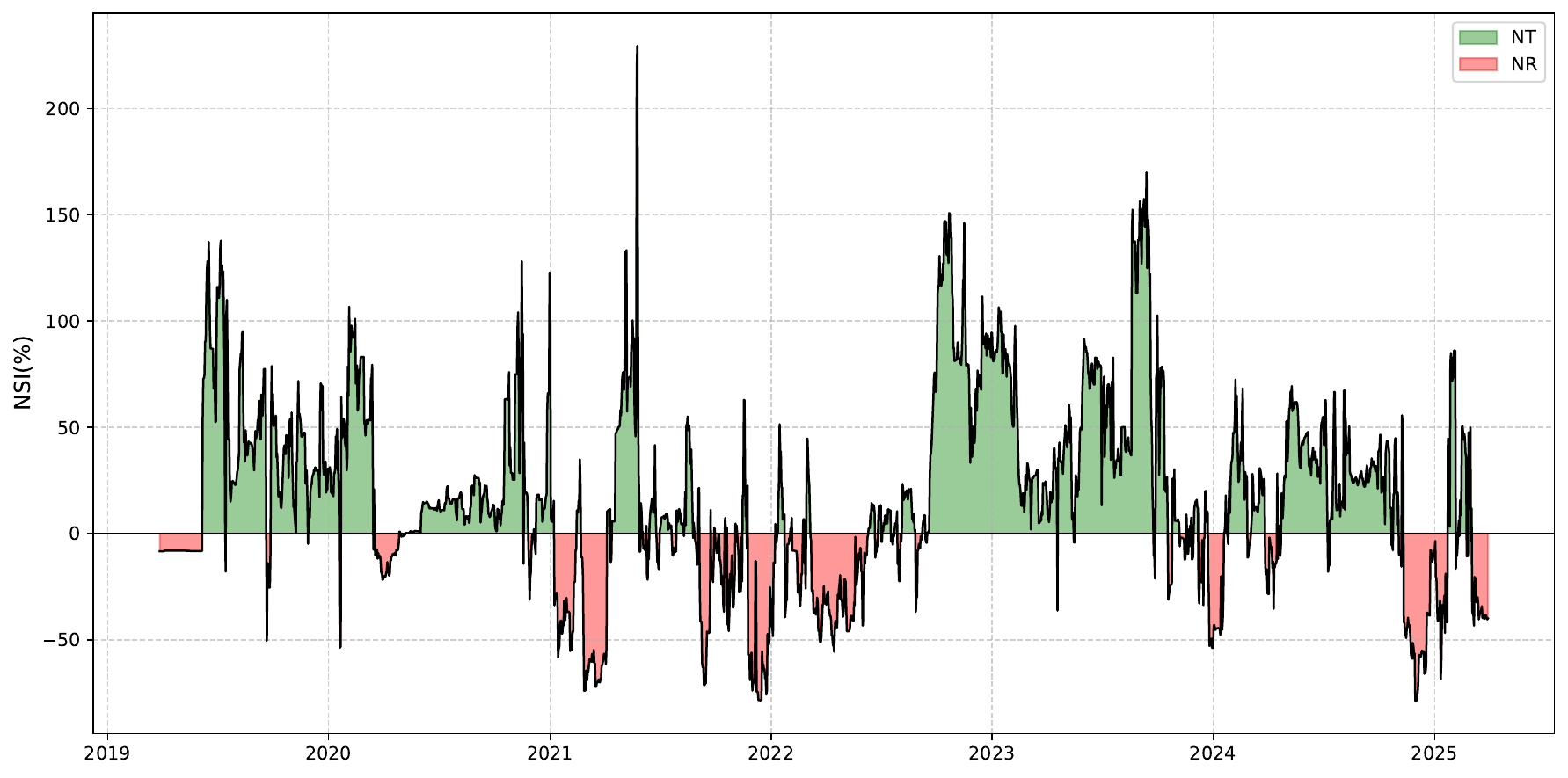}\label{fig:rexmod_dash_high}}
    \vspace{0.3cm}
    \subfigure[ETH, $\tau=0.05$]{\includegraphics[width=0.32\linewidth]{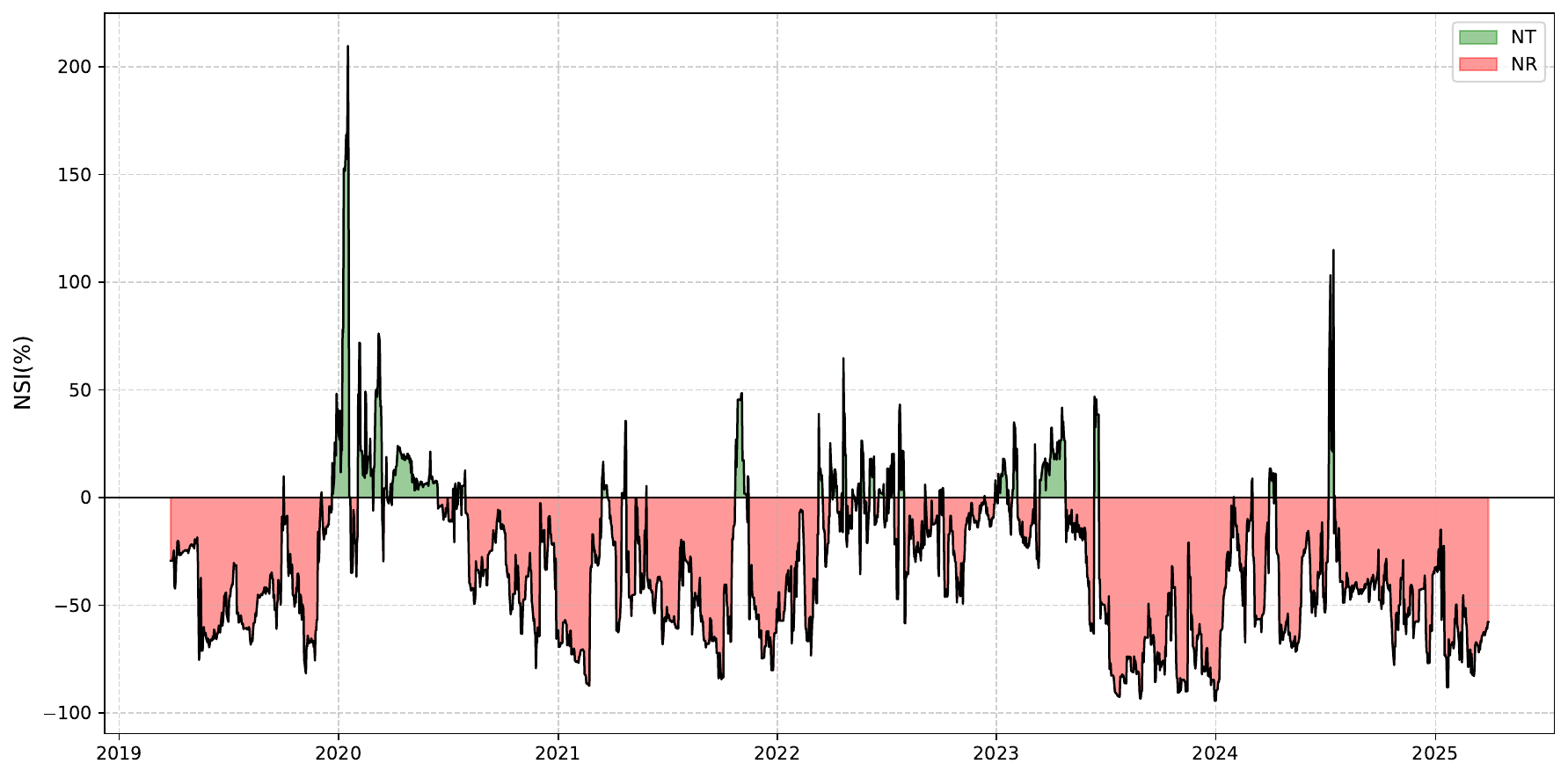}\label{fig:rexmod_eth_low}}\hfill
    \subfigure[ETH, $\tau=0.50$]{\includegraphics[width=0.32\linewidth]{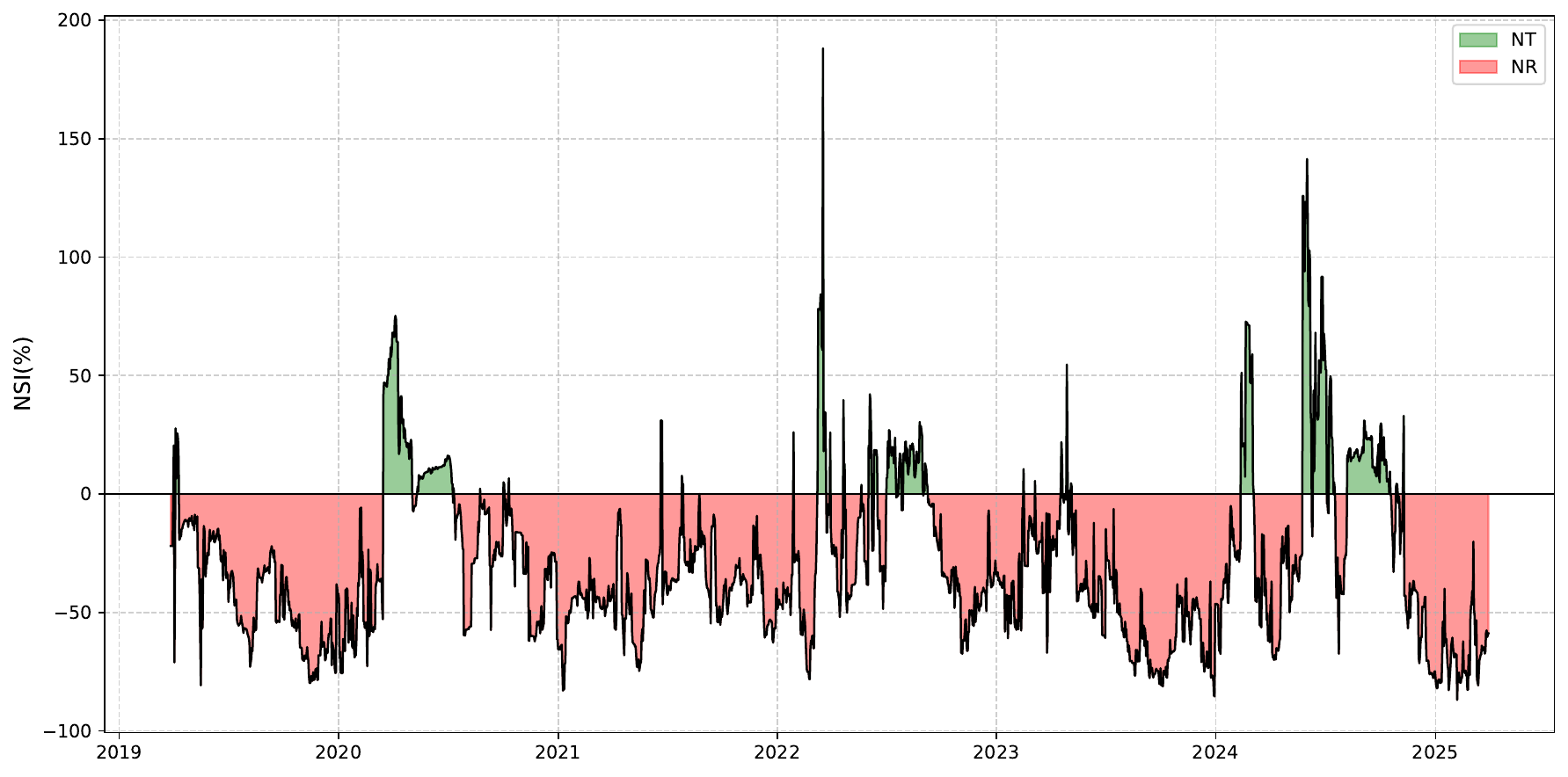}\label{fig:rexmod_eth_mid}}\hfill
    \subfigure[ETH, $\tau=0.95$]{\includegraphics[width=0.32\linewidth]{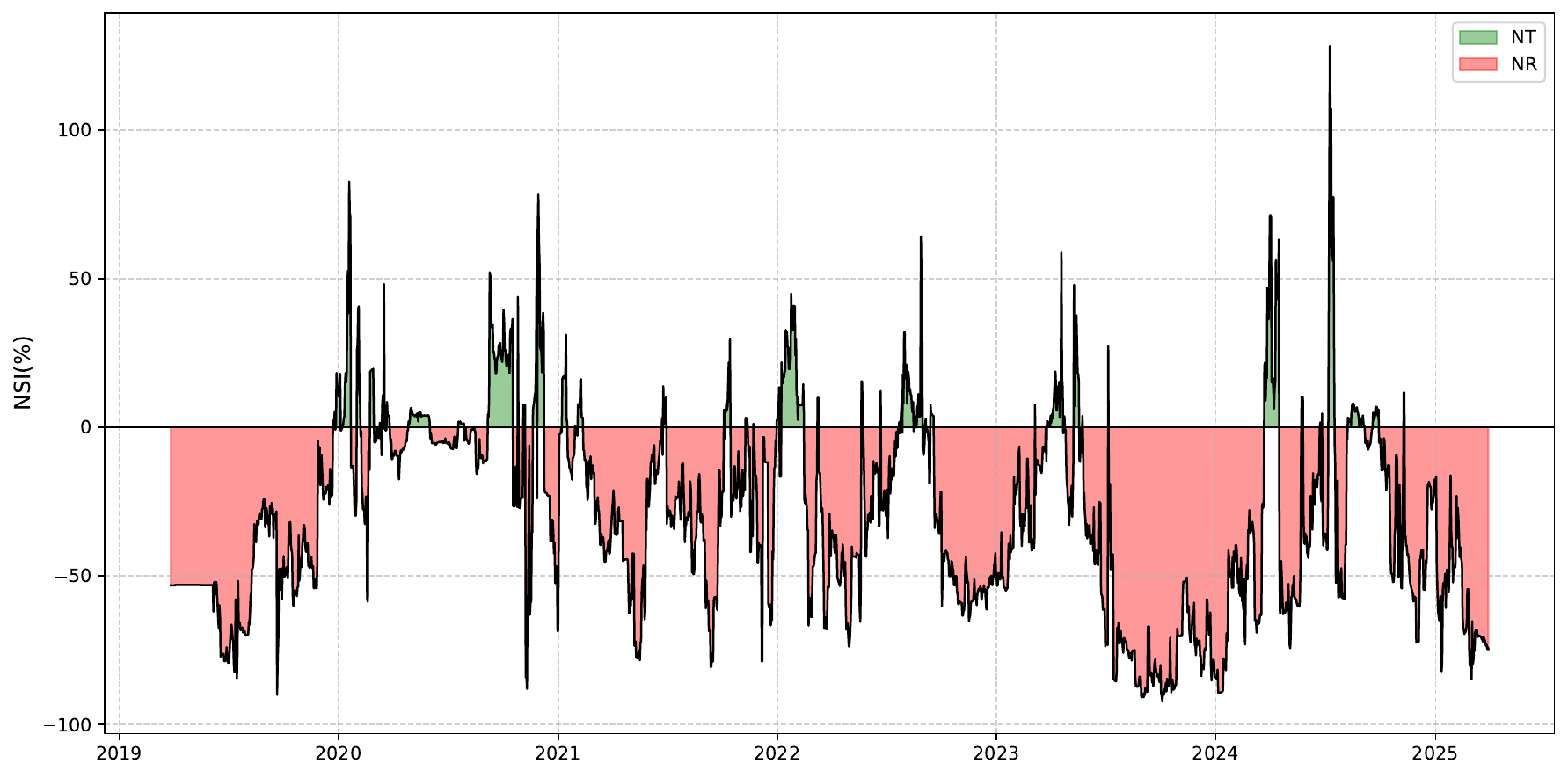}\label{fig:rexmod_eth_high}}
    \vspace{0.3cm}
    \subfigure[LTC, $\tau=0.05$]{\includegraphics[width=0.32\linewidth]{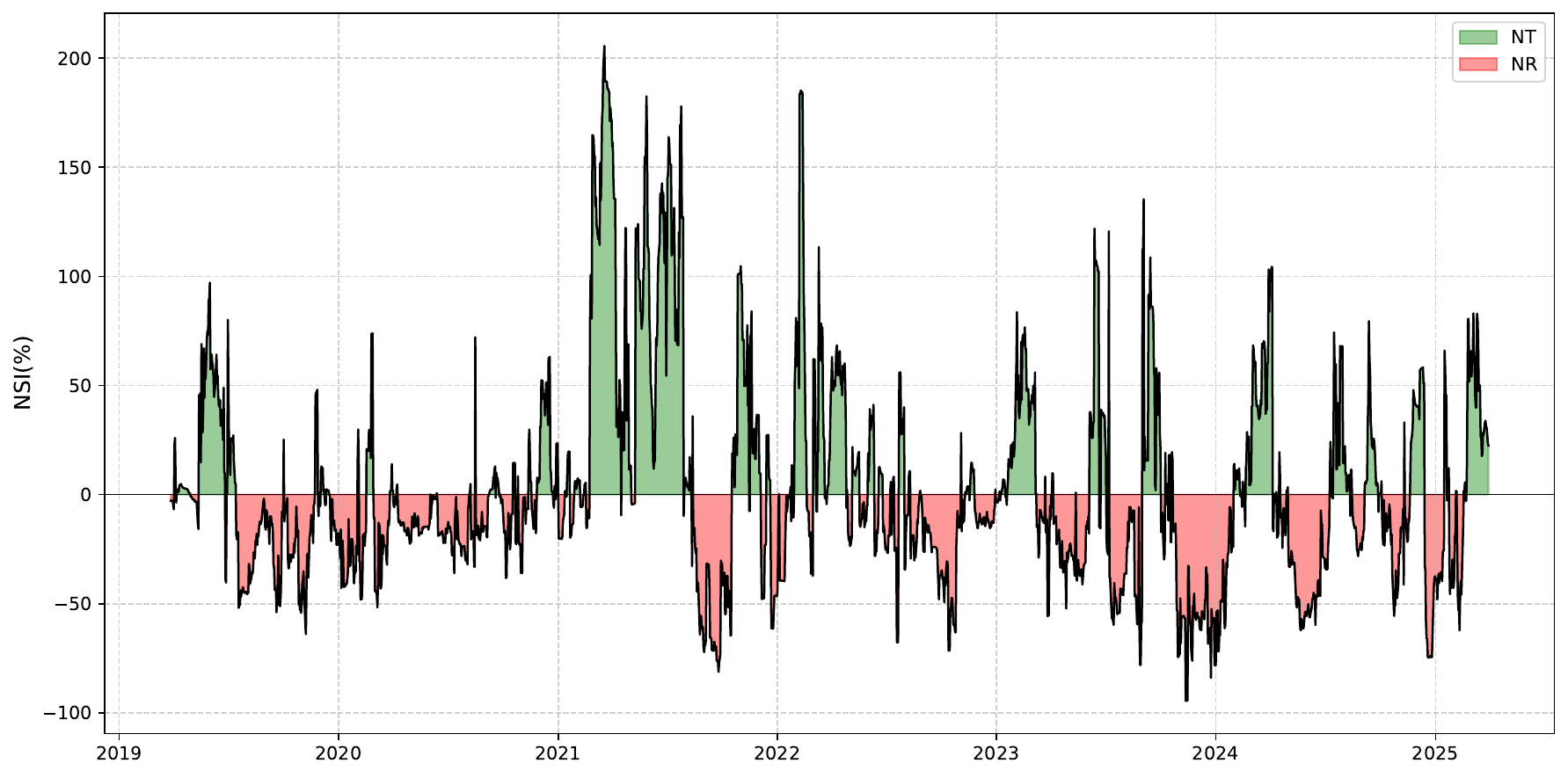}\label{fig:rexmod_ltc_low}}\hfill
    \subfigure[LTC, $\tau=0.50$]{\includegraphics[width=0.32\linewidth]{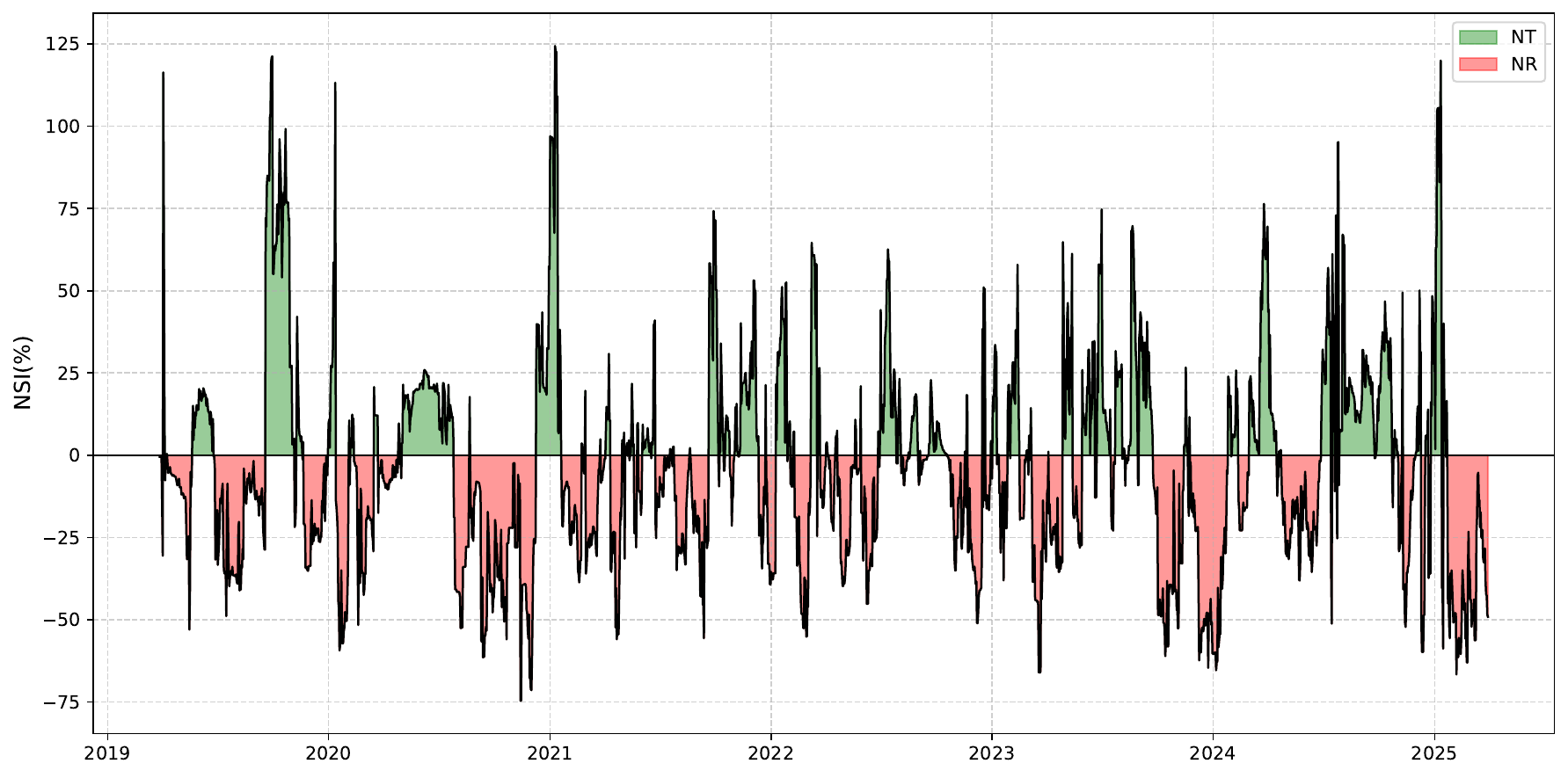}\label{fig:rexmod_ltc_mid}}\hfill
    \subfigure[LTC, $\tau=0.95$]{\includegraphics[width=0.32\linewidth]{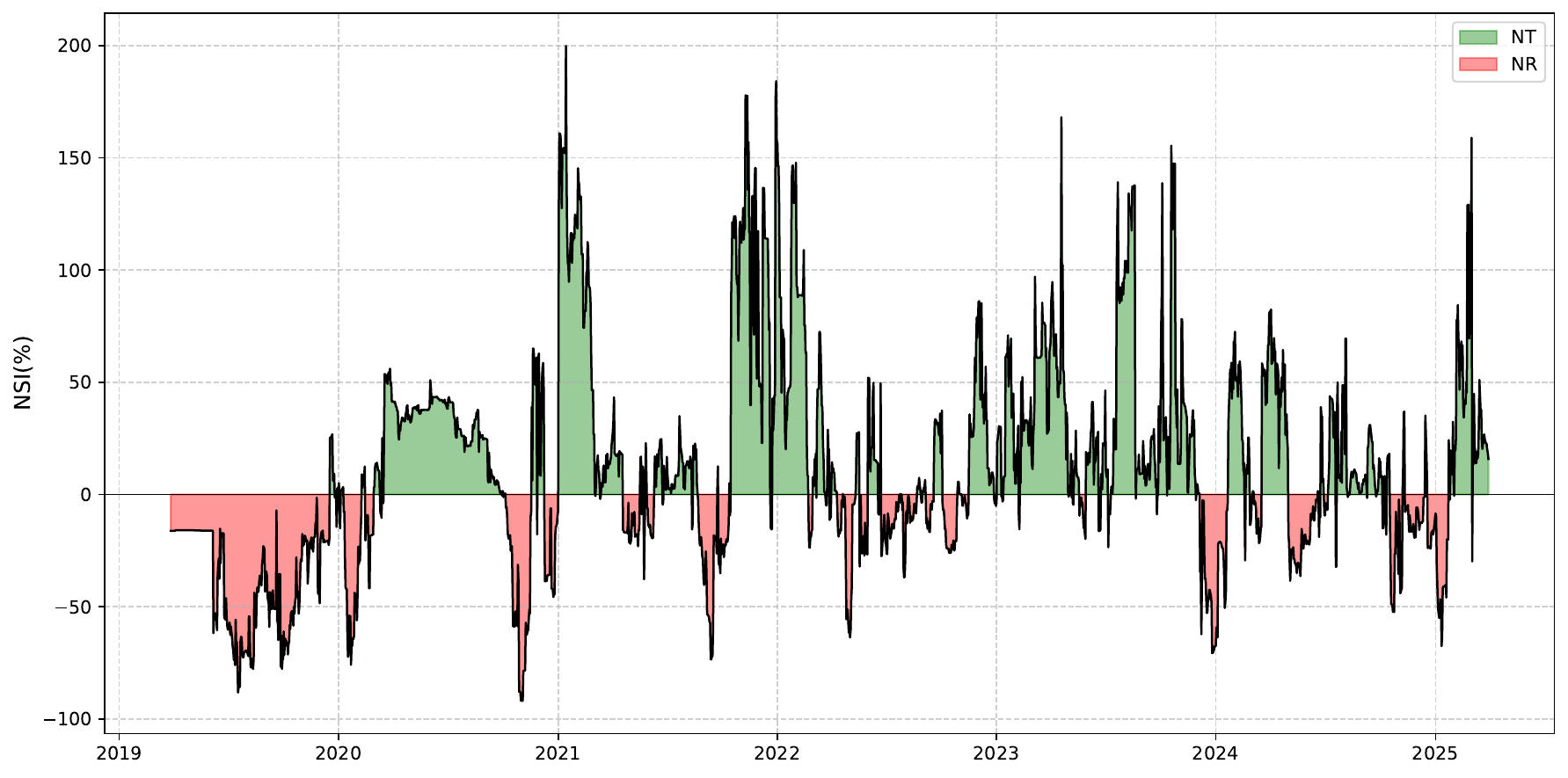}\label{fig:rexmod_ltc_high}}
    \vspace{0.3cm}
    \subfigure[XLM, $\tau=0.05$]{\includegraphics[width=0.32\linewidth]{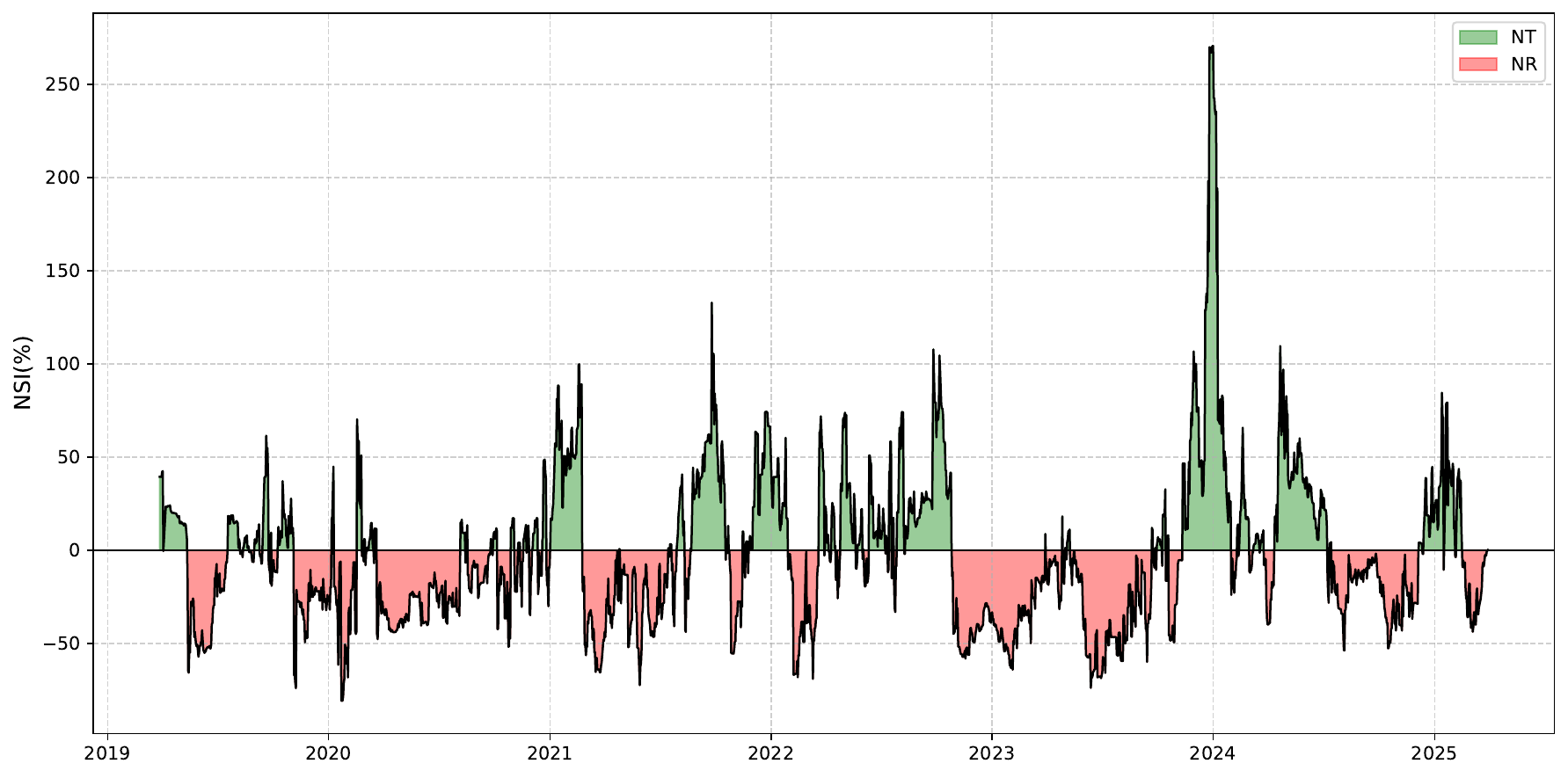}\label{fig:rexmod_xlm_low}}\hfill
    \subfigure[XLM, $\tau=0.50$]{\includegraphics[width=0.32\linewidth]{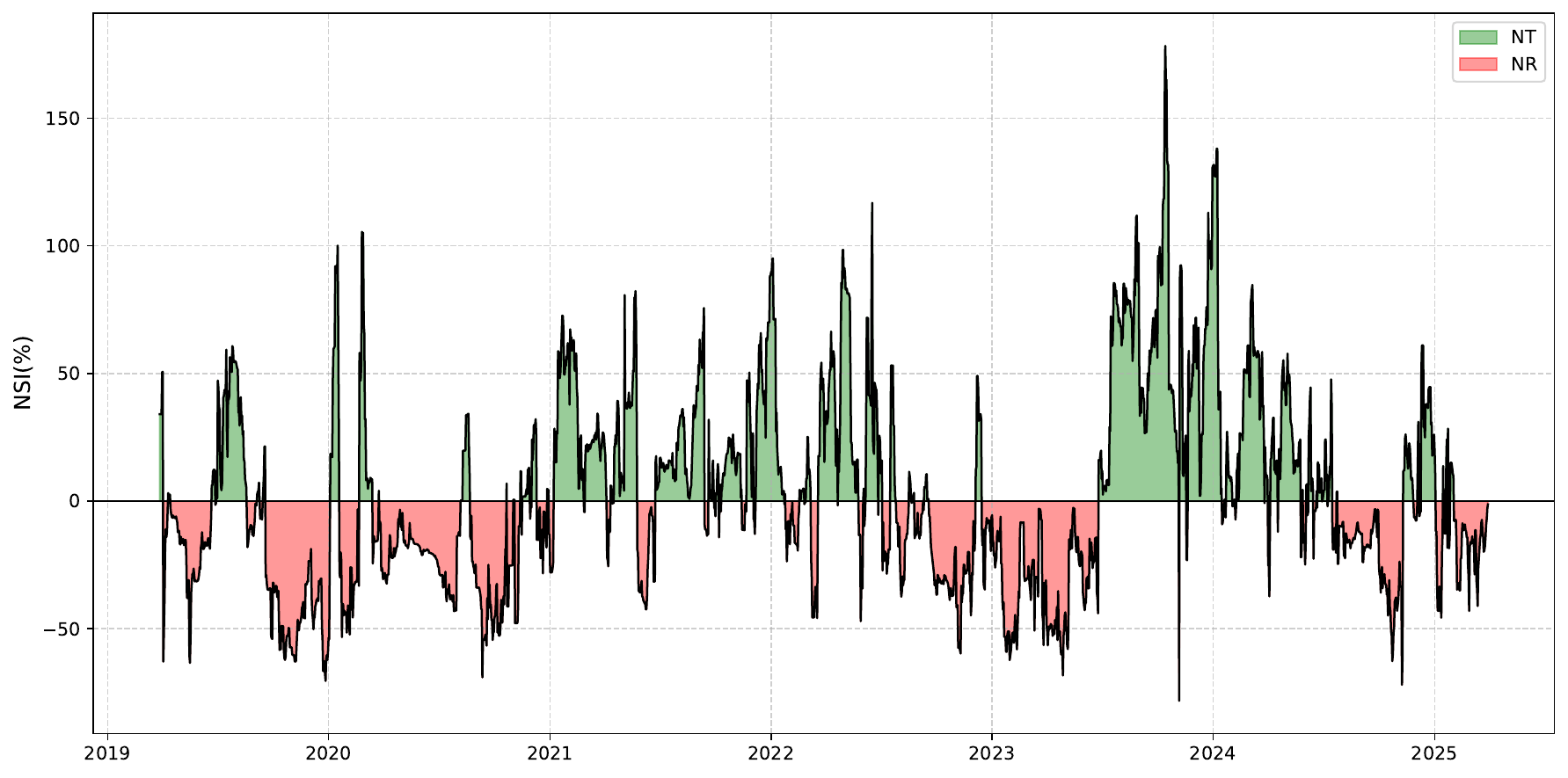}\label{fig:rexmod_xlm_mid}}\hfill
    \subfigure[XLM, $\tau=0.95$]{\includegraphics[width=0.32\linewidth]{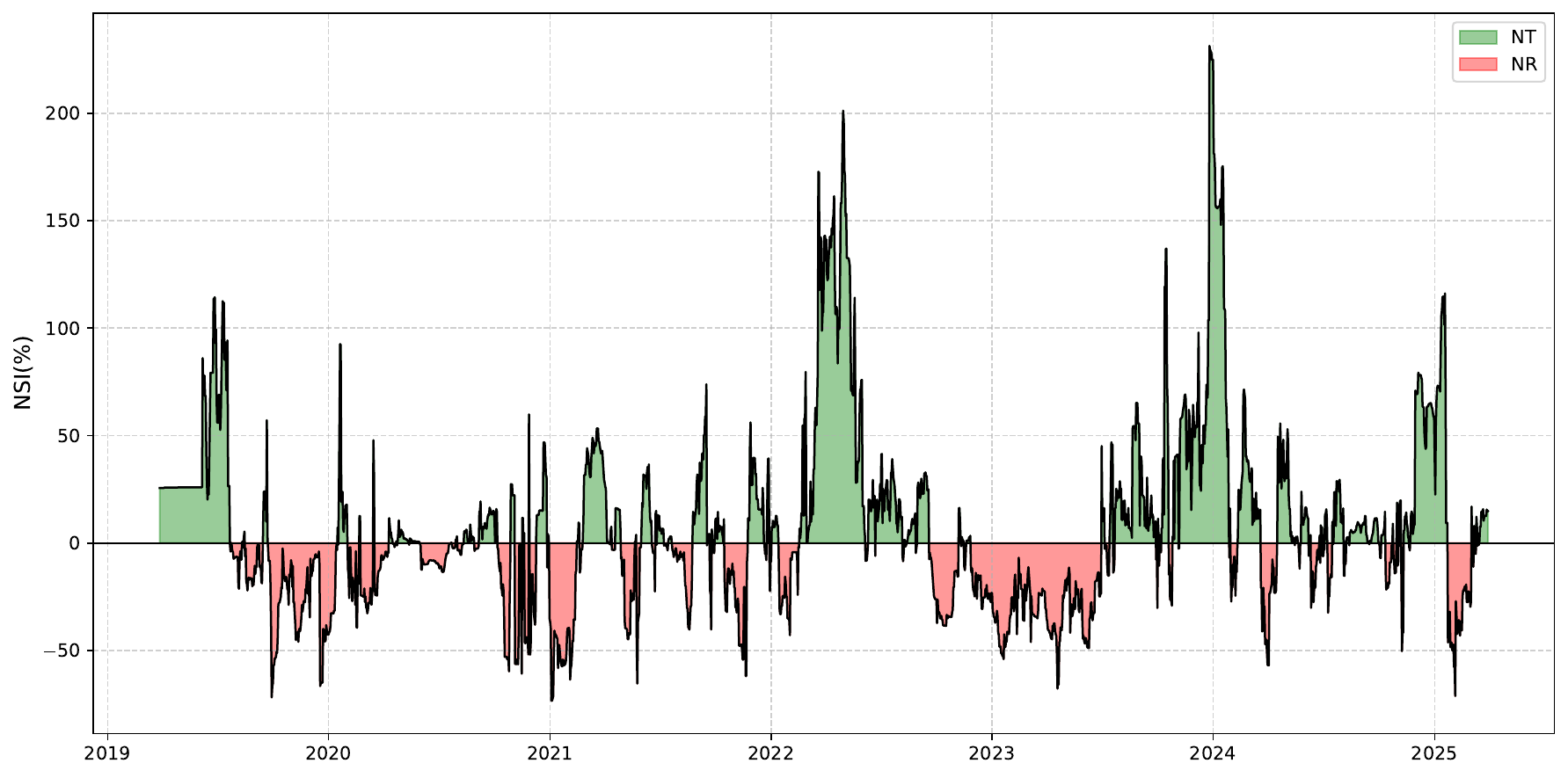}\label{fig:rexmod_xlm_high}}
    \vspace{0.3cm}
    \subfigure[XRP, $\tau=0.05$]{\includegraphics[width=0.32\linewidth]{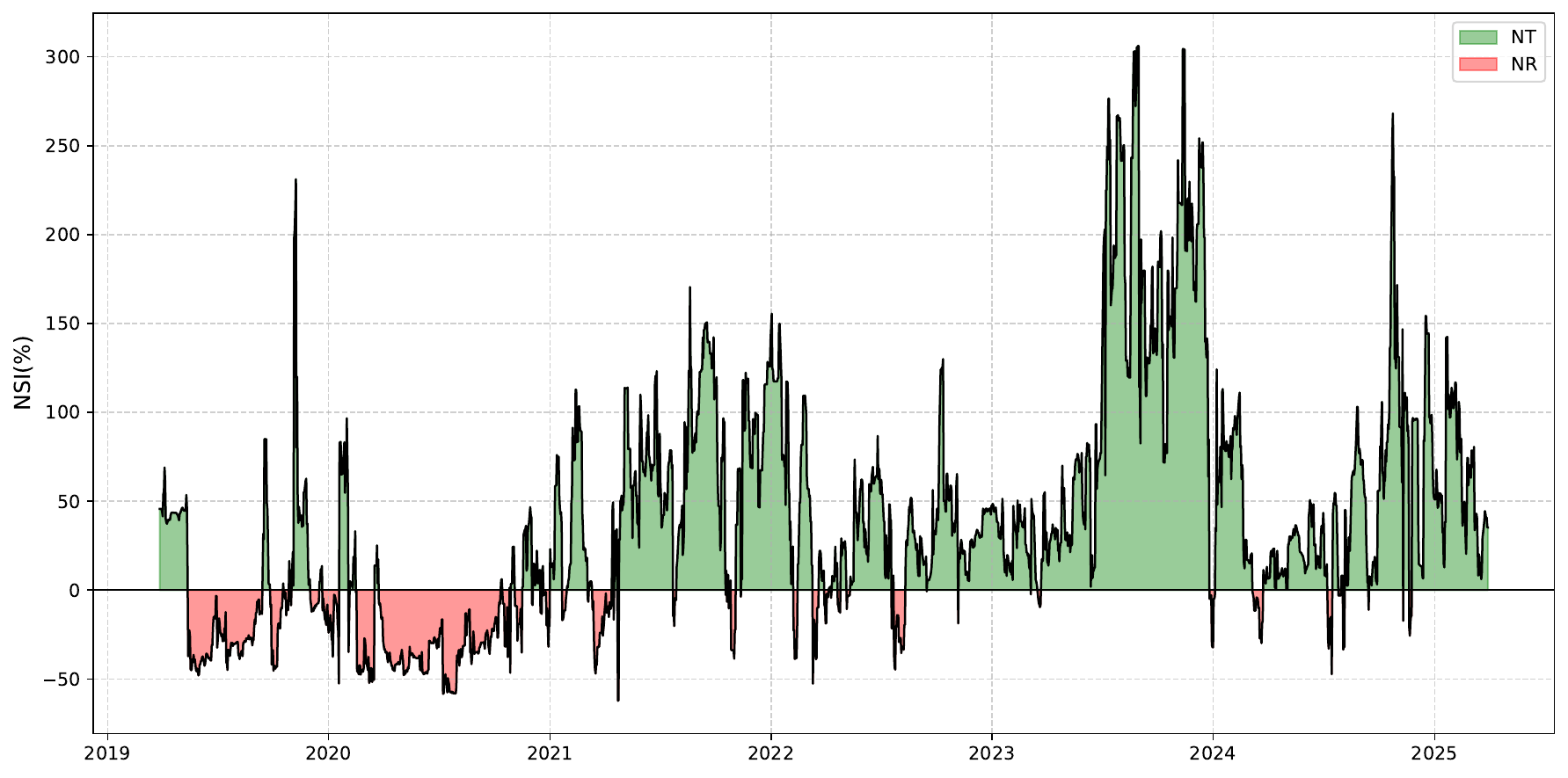}\label{fig:rexmod_xrp_low}}\hfill
    \subfigure[XRP, $\tau=0.50$]{\includegraphics[width=0.32\linewidth]{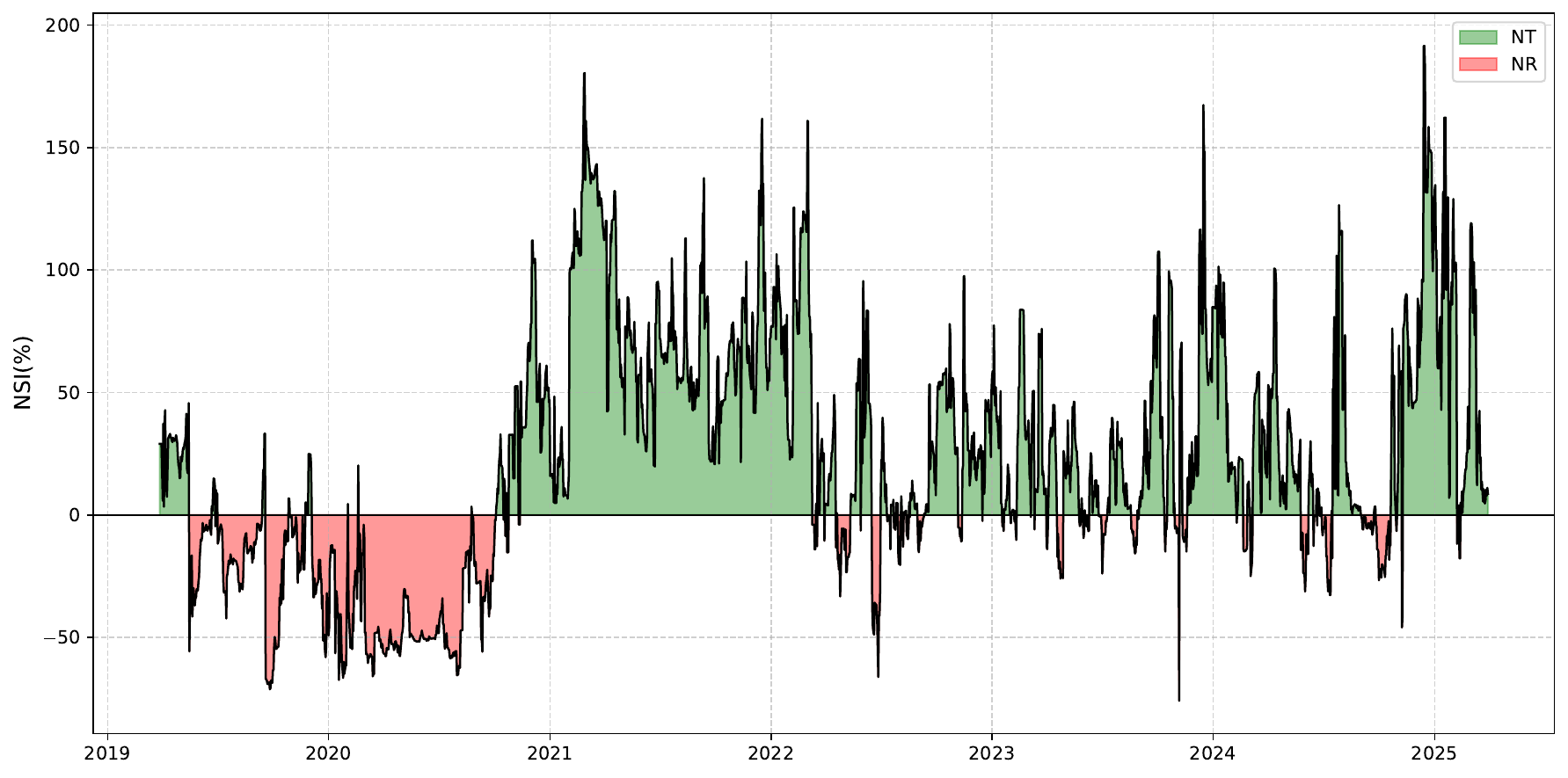}\label{fig:rexmod_xrp_mid}}\hfill
    \subfigure[XRP, $\tau=0.95$]{\includegraphics[width=0.32\linewidth]{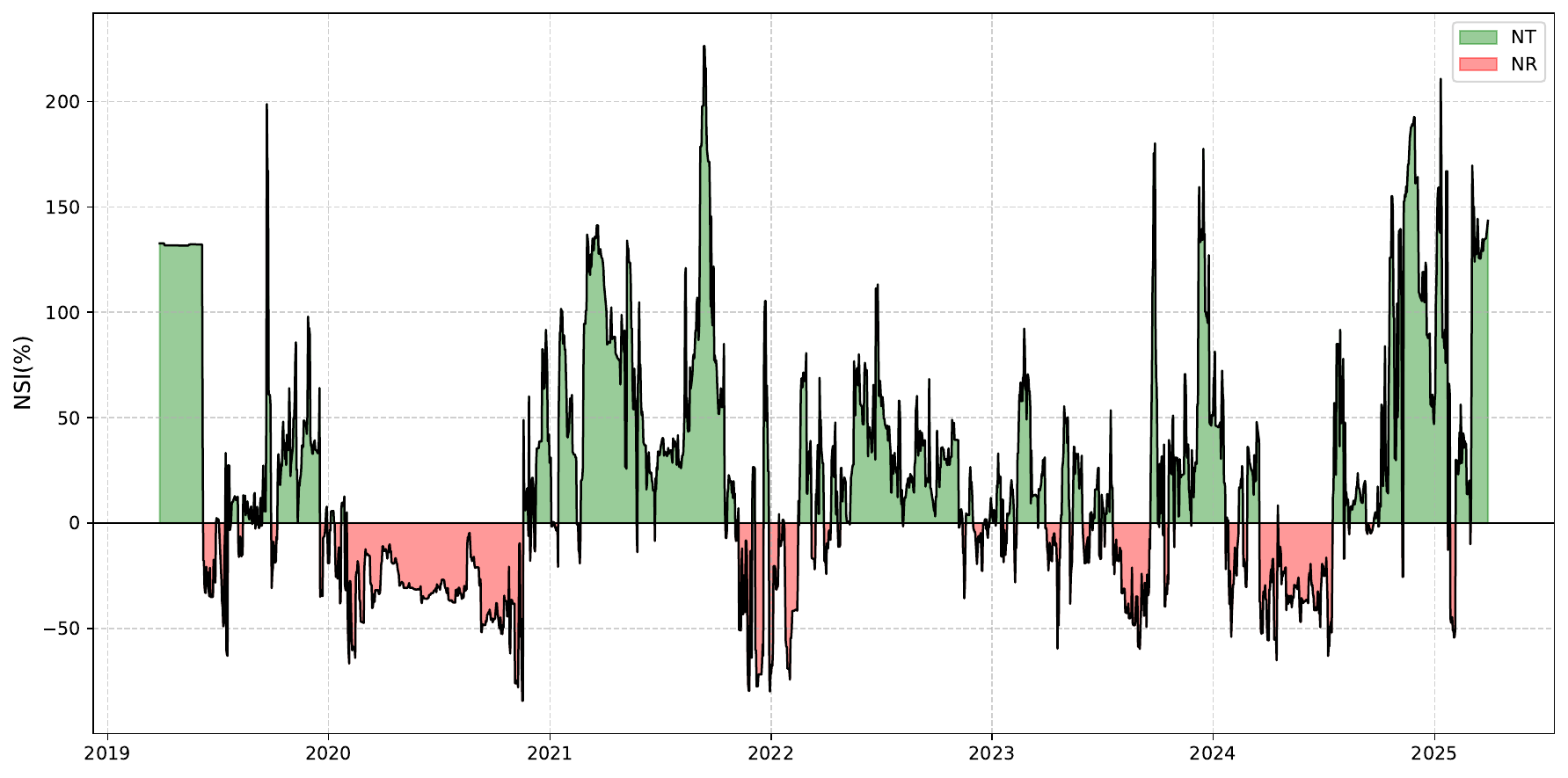}\label{fig:rexmod_xrp_high}}
\end{figure}

\section{Net spillover contribution rates of volatility measures}
\begin{figure}[htbp]
\centering
\caption{Net Spillover Contributions of Various Volatility Measures.}
\label{fig:contribute}

\subfigure[Realized kernel ($RK$)]{%
    \includegraphics[width=0.32\textwidth]{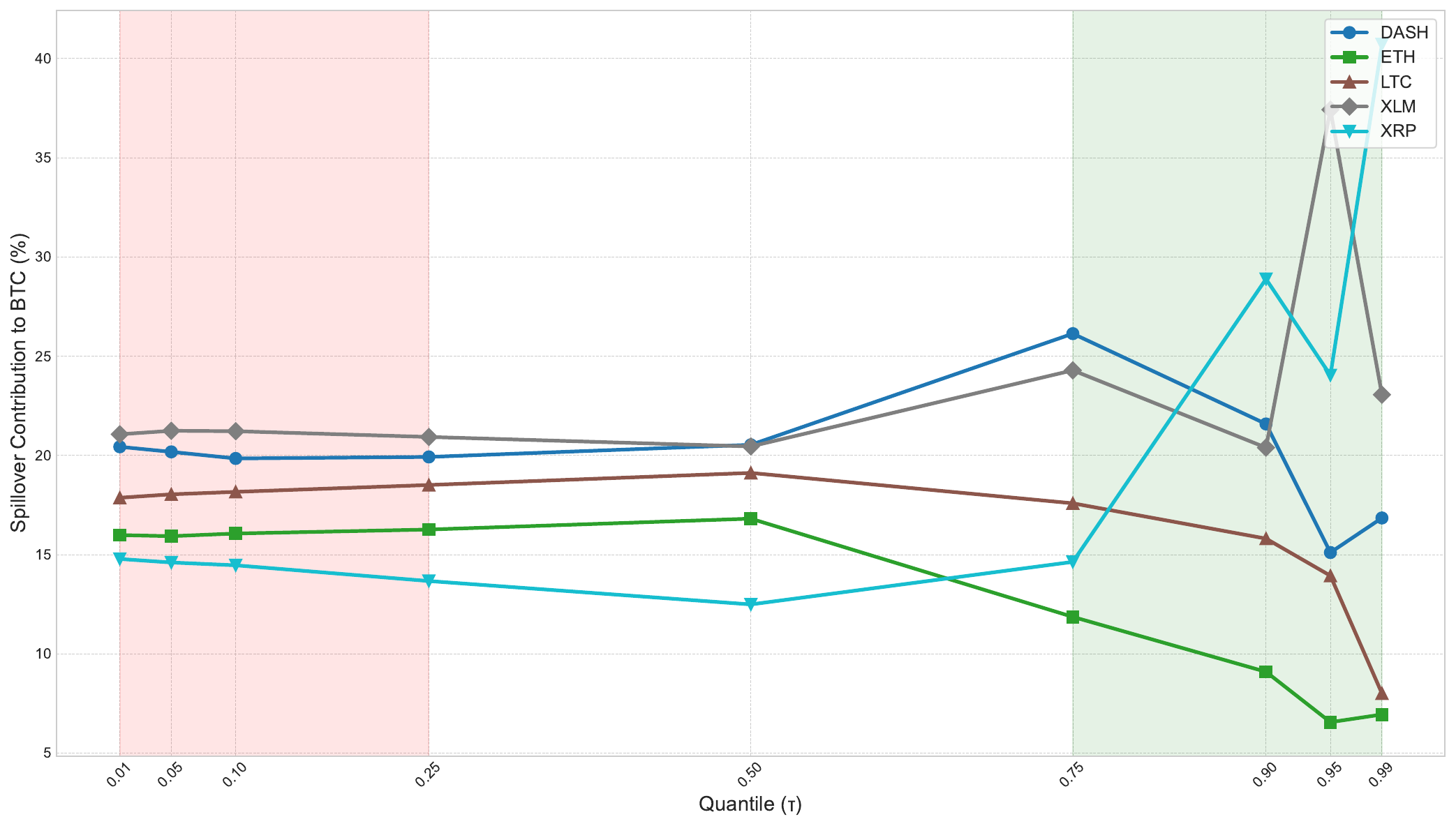}%
    \label{fig:contribute_rk}%
}\hfill
\subfigure[Realized volatility ($RV$)]{%
    \includegraphics[width=0.32\textwidth]{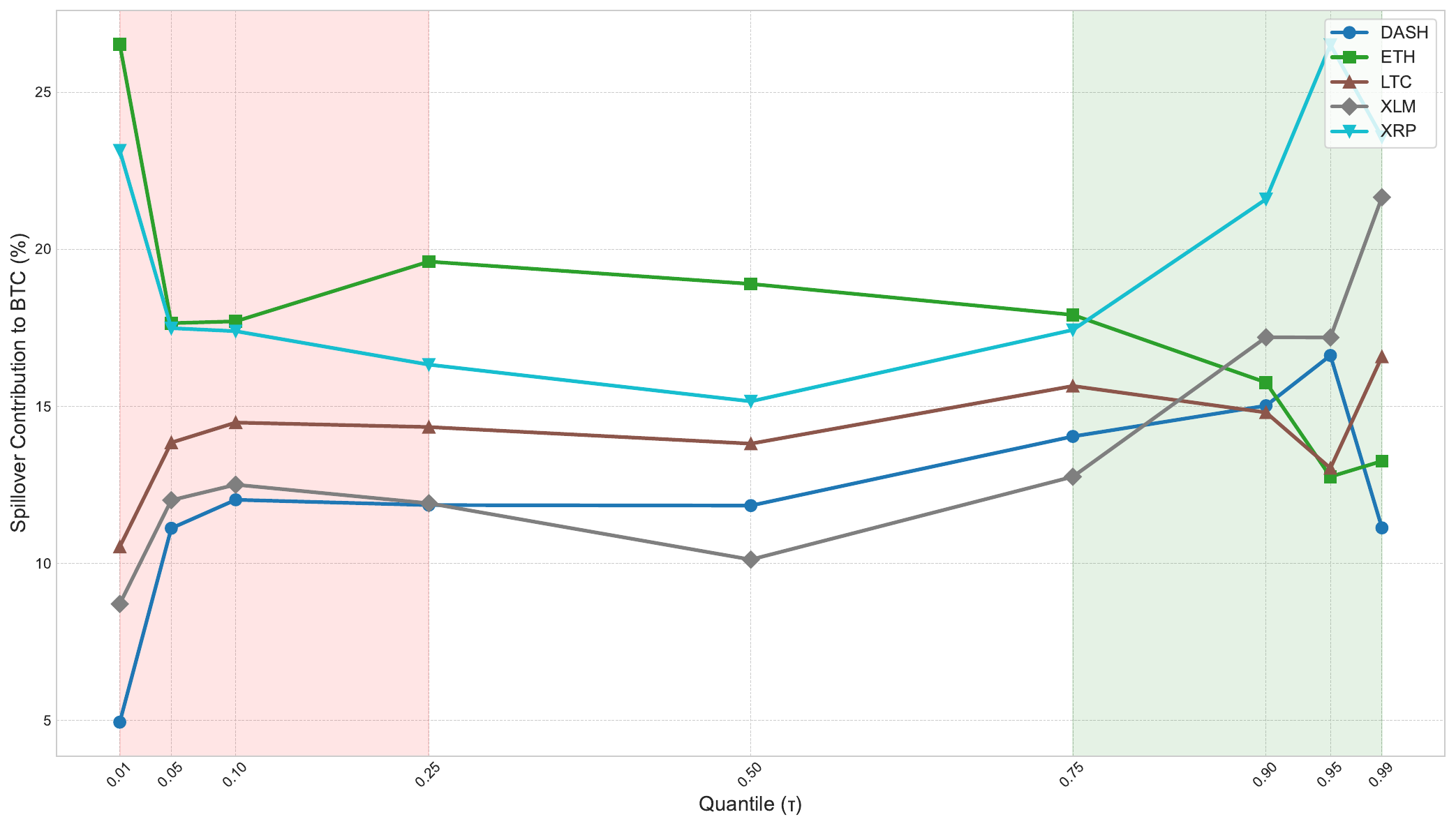}%
    \label{fig:contribute_rv}%
}\hfill
\subfigure[Continuous component ($CV$)]{%
    \includegraphics[width=0.32\textwidth]{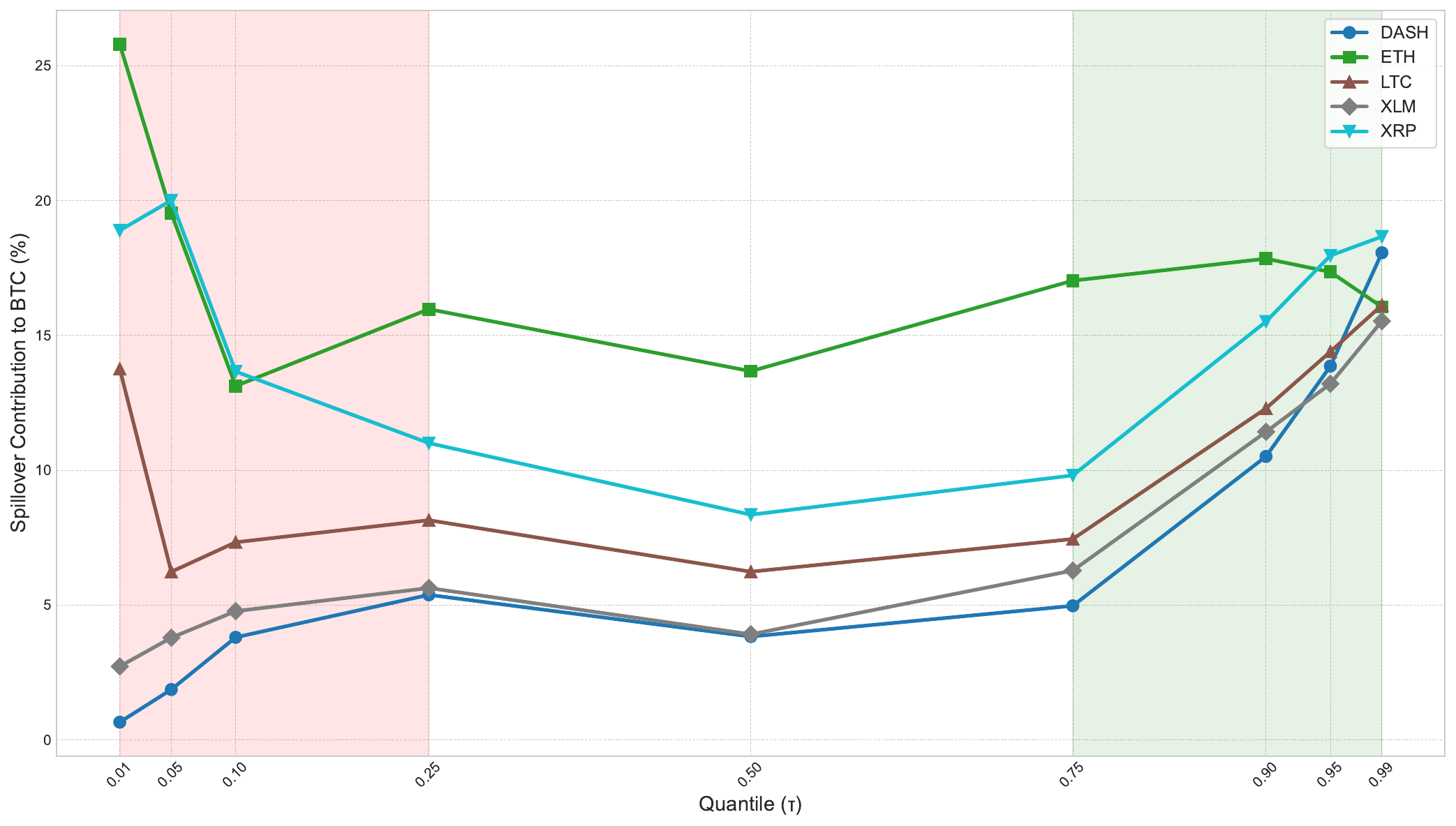}%
    \label{fig:contribute_ct}%
}

\vspace{0.5cm} 

\subfigure[Jump component ($CJ$)]{%
    \includegraphics[width=0.32\textwidth]{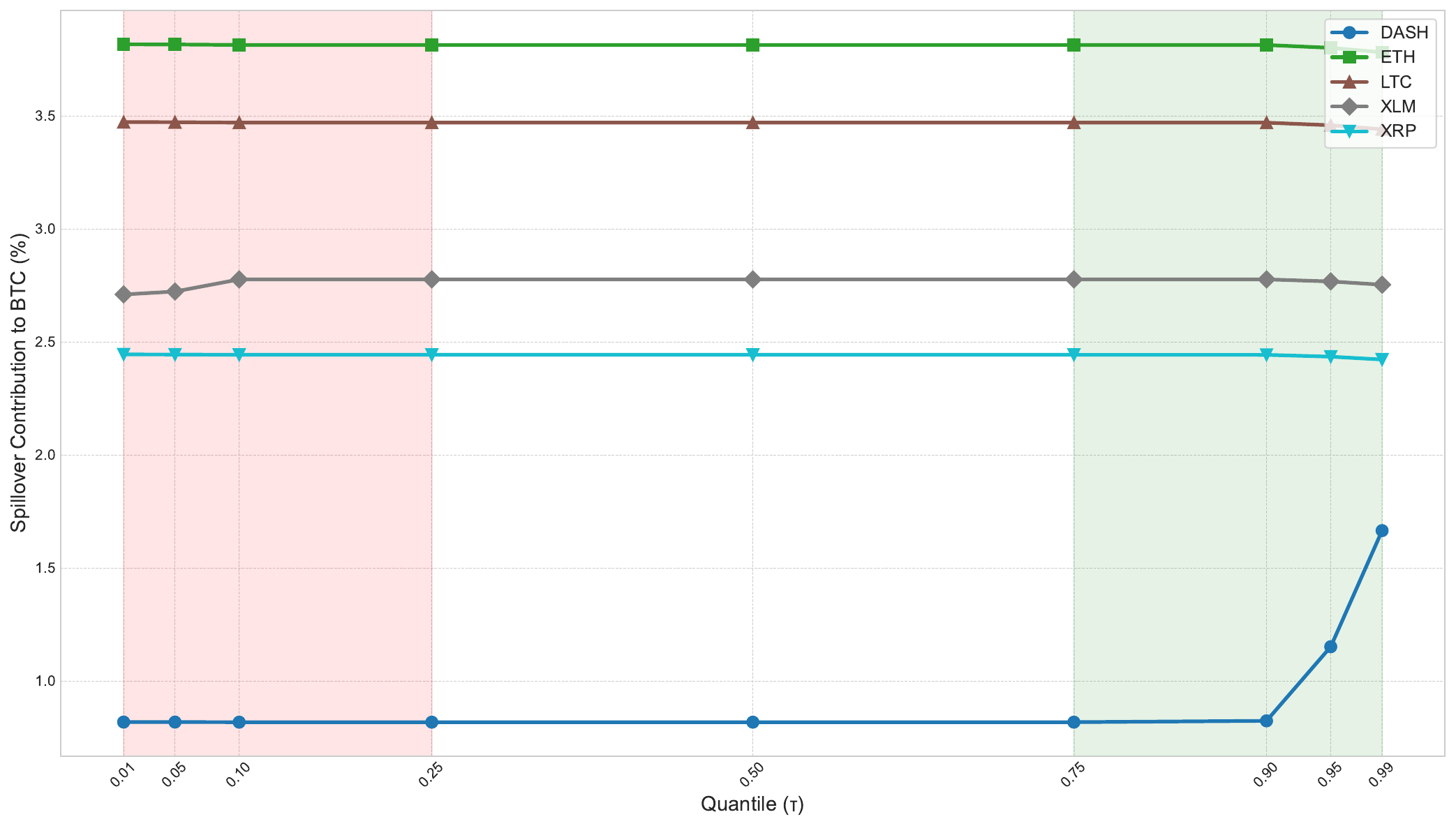}%
    \label{fig:contribute_jv}%
}\hfill
\subfigure[Positive semi-variance ($RS^{+}$)]{%
    \includegraphics[width=0.32\textwidth]{"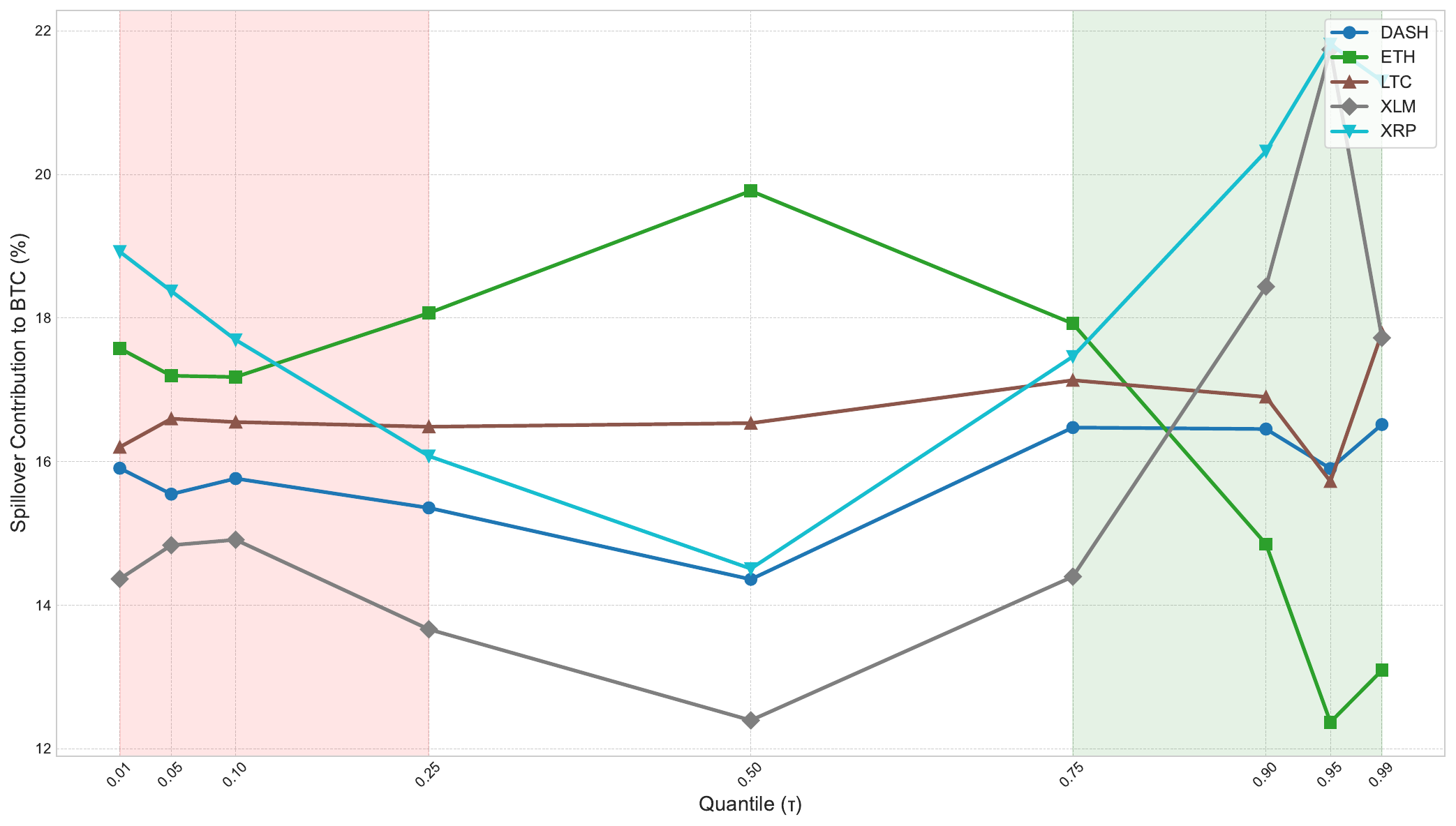"}%
    \label{fig:contribute_rs_plus}%
}\hfill
\subfigure[Negative semi-variance ($RS^{-}$)]{%
    \includegraphics[width=0.32\textwidth]{"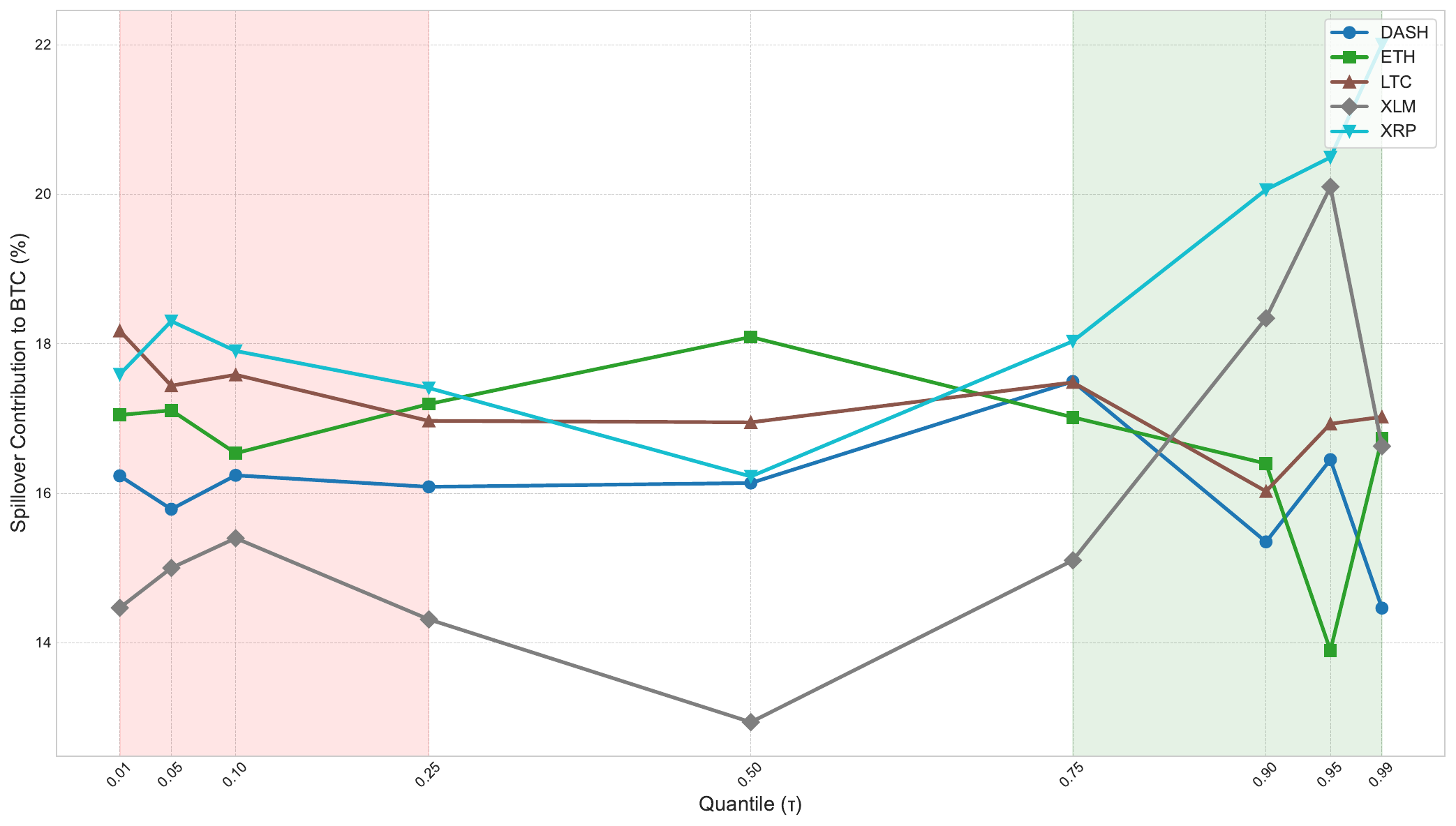"}%
    \label{fig:contribute_rs_minus}%
}

\vspace{0.5cm} 

\subfigure[Positive extreme volatility ($REX^{+}$)]{%
    \includegraphics[width=0.32\textwidth]{"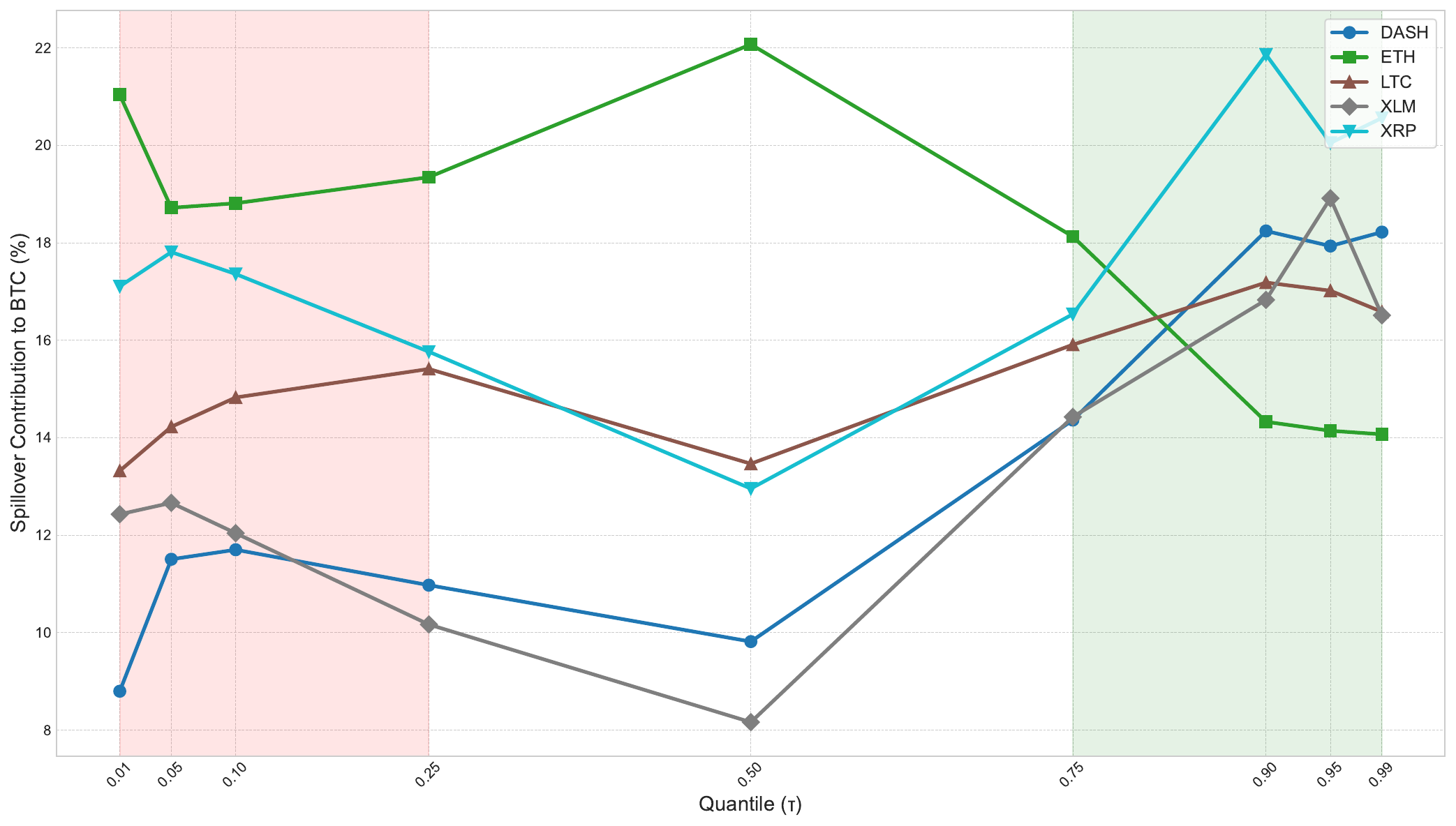"}%
    \label{fig:contribute_rex_p}%
}\hfill
\subfigure[Negative extreme volatility ($REX^{-}$)]{%
    \includegraphics[width=0.32\textwidth]{"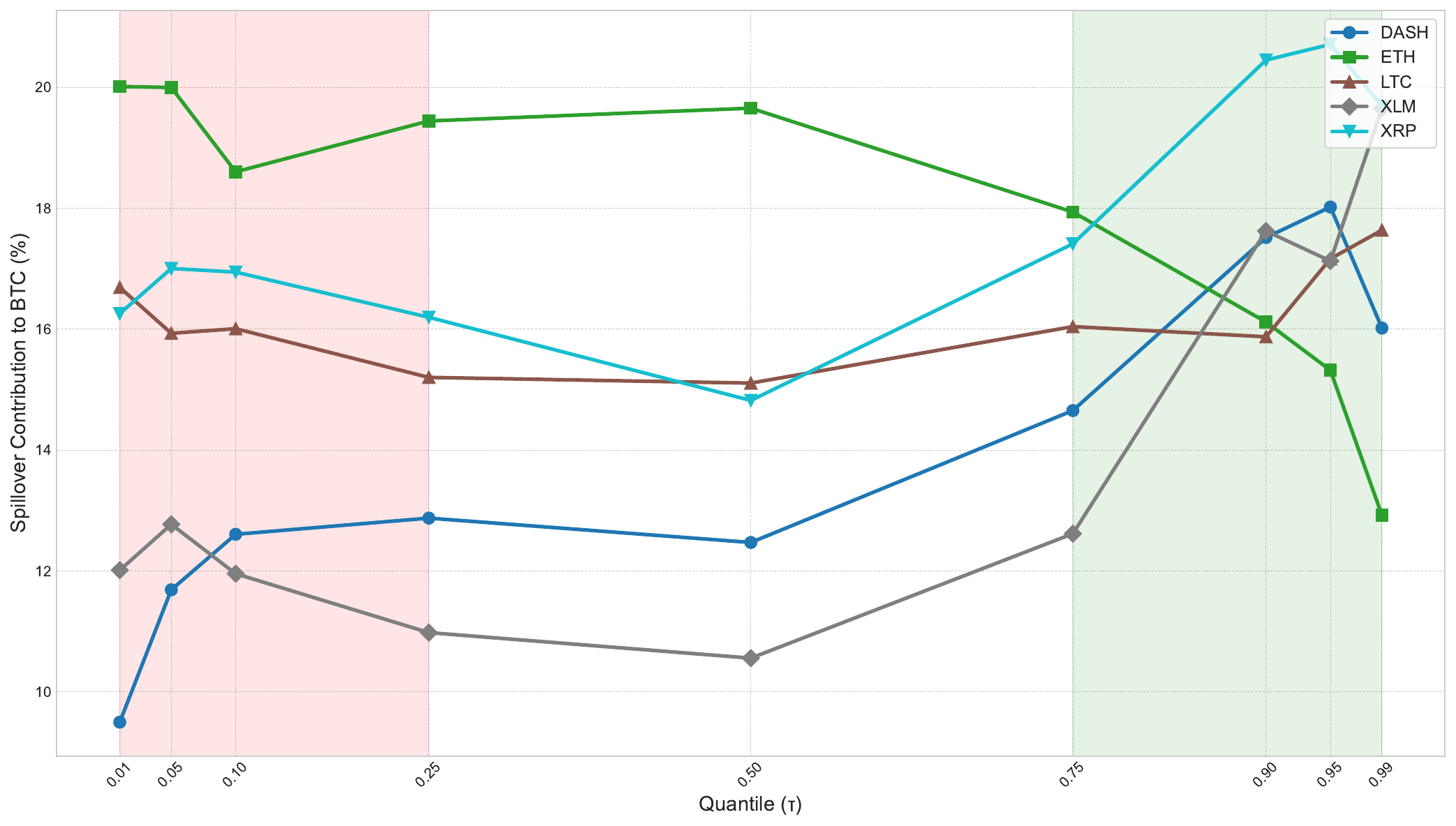"}%
    \label{fig:contribute_rex_d}%
}\hfill
\subfigure[Moderate extreme volatility ($REX^{m}$)]{%
    \includegraphics[width=0.32\textwidth]{"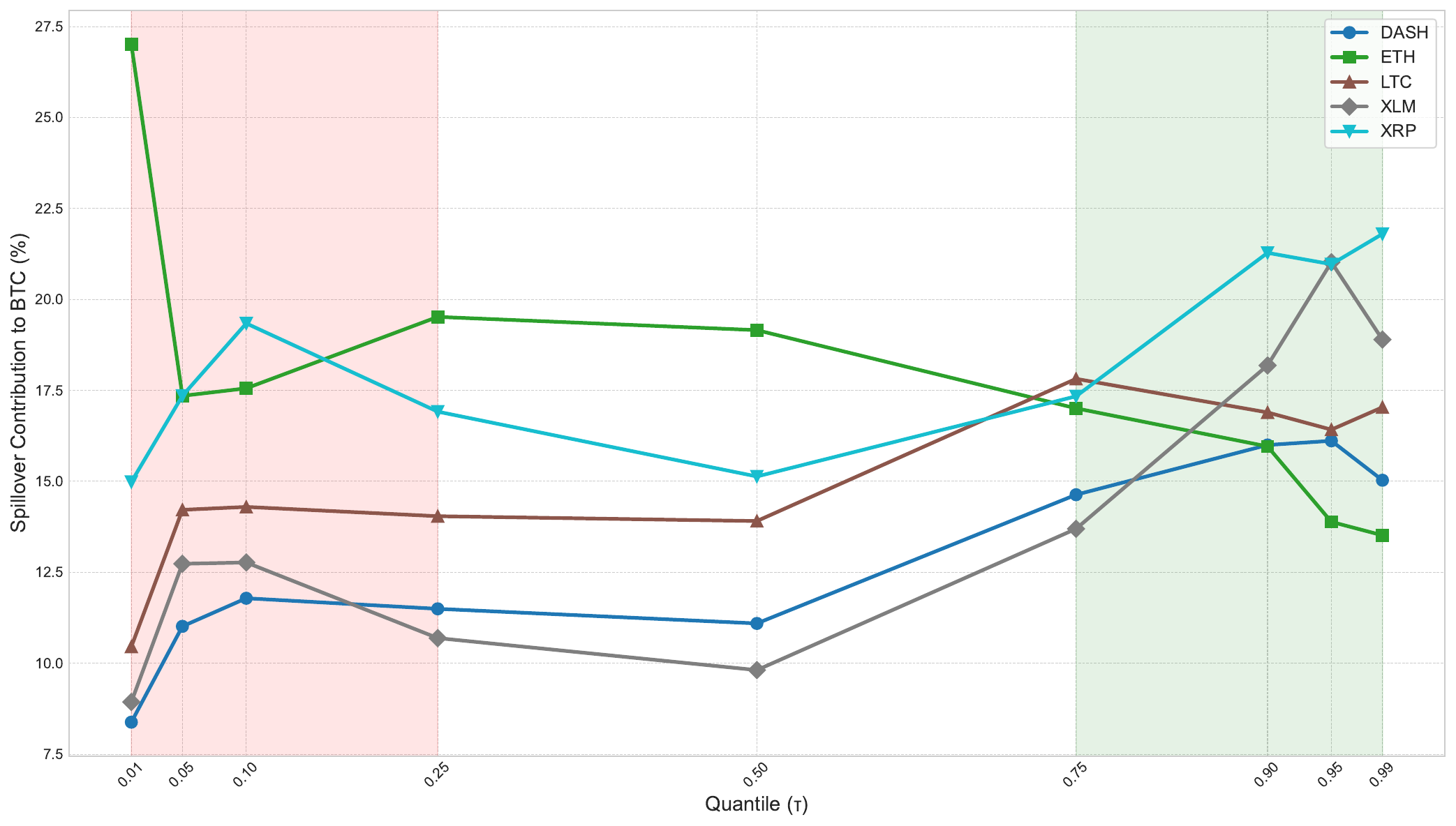"}%
    \label{fig:contribute_rex_m}%
}
\end{figure}

\section{References}\label{references}


\begin{thebibliography}{}
\bibitem[Andersen, Bollerslev, and Diebold(2007)]{And07}
Andersen, T. G., T. Bollerslev, and F. X. Diebold. 2007. ``Roughing it up: Including jump components in the measurement, modeling, and forecasting of return volatility.'' \emph{The Review of Economics and Statistics} 89 (4): 701--720.

\bibitem[Antonakakis, Chatziantoniou, and Gabauer(2020)]{Ant20}
Antonakakis, N., I. Chatziantoniou, and D. Gabauer. 2020. ``Refined measures of dynamic connectedness based on time-varying parameter vector autoregressions.'' \emph{Journal of Risk and Financial Management} 13 (4): 84.

\bibitem[Barndorff-Nielsen et~al.(2008)]{Bar08a}
Barndorff-Nielsen, O. E., P. R. Hansen, A. Lunde, et al. 2008. ``Designing realized kernels to measure the ex post variation of equity prices in the presence of noise.'' \emph{Econometrica} 76 (6): 1481--1536.

\bibitem[Barndorff-Nielsen and Shephard(2008)]{Bar08b}
Barndorff-Nielsen, O. E., and N. Shephard. 2008. ``Measuring downside risk-realised semivariance.'' CREATES Research Paper 2008-42.

\bibitem[Bollerslev(1986)]{Bol86}
Bollerslev, T. 1986. ``Generalized autoregressive conditional heteroskedasticity.'' \emph{Journal of Econometrics} 31 (3): 307--327.

\bibitem[Borri(2019)]{Bor19}
Borri, N. 2019. ``Conditional tail-risk in cryptocurrency markets.'' \emph{Journal of Empirical Finance} 50: 1--19.

\bibitem[Campbell and Thompson(2008)]{Cam08}
Campbell, J. Y., and S. B. Thompson. 2008. ``Predicting excess stock returns out of sample: Can anything beat the historical average?'' \emph{The Review of Financial Studies} 21 (4): 1509--1531.

\bibitem[Clements and Rodrigo(2019)]{Cle19}
Clements, A., and H. Rodrigo. 2019. ``Moderate and extreme volatility: Do the magnitude of returns matter for forecasting?'' SSRN working paper 3443259.

\bibitem[Corsi(2009)]{Cor09}
Corsi, F. 2009. ``A simple approximate long-memory model of realized volatility.'' \emph{Journal of Financial Econometrics} 7 (2): 174--196.

\bibitem[Diebold and Yilmaz(2009)]{Die09}
Diebold, F. X., and K. Yilmaz. 2009. ``Measuring financial asset return and volatility spillovers, with application to global equity markets.'' \emph{The Economic Journal} 119 (534): 158--171.

\bibitem[Engle(1982)]{Eng82}
Engle, R. F. 1982. ``Autoregressive conditional heteroscedasticity with estimates of the variance of United Kingdom inflation.'' \emph{Econometrica} 50 (4): 987--1007.

\bibitem[Gupta and Pierdzioch(2023)]{Gup23}
Gupta, R., and C. Pierdzioch. 2023. ``Do US economic conditions at the state level predict the realized volatility of oil-price returns? A quantile machine-learning approach.'' \emph{Financial Innovation} 9 (1): 24.

\bibitem[Hansen, Lunde, and Nason(2003)]{Han03}
Hansen, P. R., A. Lunde, and J. M. Nason. 2003. ``Choosing the best volatility models: the model confidence set approach.'' \emph{Oxford Bulletin of Economics and Statistics} 65: 839--861.

\bibitem[Lahmiri and Bekiros(2018)]{Lah18}
Lahmiri, S., and S. Bekiros. 2018. ``Chaos, randomness and multi-fractality in Bitcoin market.'' \emph{Chaos, Solitons \& Fractals} 106: 28--34.

\bibitem[Yang et~al.(2023)]{Yan23}
Yang, M. Y., C. Wang, Z. G. Wu, et al. 2023. ``Influential risk spreaders and their contribution to the systemic risk in the cryptocurrency network.'' \emph{Finance Research Letters} 57: 104225.
\end{thebibliography}
\end{document}